\documentstyle[11pt]{article}
\begin{document}

\begin{flushright}
{hep-ph/0008029}\\

\end{flushright}
\vspace{1cm}
\begin{center}
{\Large \bf  COMPLETE ANALYSIS ON THE NLO SUSY-QCD CORRECTIONS}
\vspace{0.5cm}
{\Large\bf TO
$B^0-\overline{B}^0$ MIXING}\\
\vspace{.2cm}
\end{center}
\vspace{1cm}
\begin{center}
{Tai-Fu Feng$^{a,b,d}$ \hspace{0.5cm} Xue-Qian Li$^{a,b,d}$ \hspace{0.5cm}
Wen-Gan Ma$^{a,c,d}$ \hspace{0.5cm}and \hspace{0.5cm} Feng
Zhang$^{b}$}\\
\vspace{.5cm}

{$^a$CCAST (World Laboratory), P.O.Box 8730,
Beijing 100080, China}\\
{$^b$Department of Physics, Nankai
University, Tianjin 300070, China}\footnote{The post address}\\
{$^c$Department of Modern Physics,
University of Science and Technology of China, Hefei 230026, China}\\
{$^d$Institute of Theoretical Physics, Academia Sinica, P.O. Box
2735, Beijing 100080, China}\\

\vspace{.5cm}

\end{center}
\hspace{3in}

\begin{center}
\begin{minipage}{11cm}
{\large\bf Abstract}

{\small  We present a complete next-to-leading-order calculation of
the QCD corrections
to $B^0-\overline{B}^0$ ($K^0-\overline{K}^0$) mixing in the framework of
the minimal flavor violating (MFV) supersymmetry. We take
into account the contributions from the gluino and find that the
gluino-mediated corrections modify the LO result obviously
even when the mass of gluino $m_{\tilde{g}} \gg m_{\rm w}$.
In general, one cannot neglect gluino contributions.
}
\end{minipage}
\end{center}

\vspace{4mm}
\begin{center}
\end{center}


\baselineskip 22pt

\section{Introduction \label{introduction}}

As the most popular candidate for new physics beyond the standard model
(SM), the supersymmetry\cite{haber} has been studied extensively
during the last two decades. Even so, there is no
experimental evidence for any of the new particles predicted by
various supersymmetry (SUSY) models at present. Before  new colliders
are available searching for new physics, we should focus on indirect
probes of the phenomena induced by SUSY at low energies. At this point, the
most promising processes that we can depend on are the Flavor
Changing Neutral Current (FCNC) processes, especially $b\rightarrow s
\gamma $ and oscillations of neutral mesons.
For weak decays at presence of  the strong
interaction, evaluating such processes requires special technique.
The main tool to calculate concerned quantities of those
processes is the effective Hamiltonian theory. It is a two
step program, starting with an operator product expansion
(OPE)\cite{wilson,witten} and performing a renormalization
group equation (RGE)\cite{rge-buras}
analysis afterwards. The necessary machinery has been developed over
years\cite{Ali-Greub,Buras-etal,Cella,Misiak,Grinstein}.
The new physics effects on the rare B processes
are discussed in literature. The calculation of the rate of inclusive
decay $B\rightarrow X_s\gamma$ is presented by authors of \cite{ciuchini1,
ciafaloni, borzumati} in the two-Higgs doublet model (THDM).
The supersymmetric effect on $B\rightarrow X_s\gamma$ is discussed
in \cite{bertolini, barbieri, borzumati1} and the NLO
QCD corrections are given in \cite{ciuchini2}. The transition
$b\rightarrow s\gamma\gamma$ in the supersymmetric extension
of standard model is computed in \cite{bertolini1}. The hadronic
B decays\cite{cottingham} and CP-violation in those processes\cite{barenboim}
have been discussed also. The authors of \cite{hewett} have discussed
possibility of observing supersymmetric effects in
the rare decays $B\rightarrow X_s\gamma$ and $B\rightarrow X_se^+e^-$
in the B-factory. Studies on decays $B\rightarrow
(K, K^*)l^+l^-$ in SM and supersymmetric model have been carried out
in \cite{ali}. The SUSY effects on these processes are very
interesting and studies on them may shed some light on the
general characteristics of the SUSY model.
A relevant review can be found in \cite{masiero}.
For oscillations of $B_{0}-\overline{B}_{0}$ ($K_{0}-\overline{K}_{0}$),
calculations have been done in the
Standard Model (SM) and THDM. As for the supersymmetric
extension of SM, the calculation
involving the gluino contributions
should be re-studied carefully for gluino has non-zero mass.
In this paper, we will present a complete
analysis of SUSY-QCD corrections to the oscillations of
$B_{0}-\overline{B}_{0}$ ($K_{0}-\overline{K}_{0}$) in the supersymmetric
extension of SM with the minimal flavor
violation, i.e. the flavor violation occurs only via the charged
current at the tree level.

Our main results can be summarized as follows:
\begin{itemize}
\item We give a complete computation of
supersymmetric QCD-corrections to $B_{0}-\overline{B}_{0}$ oscillations
up to NLO, our technique can be applied
to compute SUSY-QCD corrections to other rare B-decay processes
(such as $b\rightarrow s\gamma$ etc.) up to NLO.
\item Additionally, we find that the
gluino contribution to
the NLO-QCD corrections grows as $\ln x_{\tilde{g}{\rm w}}$
when gluino is heavier than the lightest up-type scalar quark,
where $x_{\tilde{g}{\rm w}}=\frac{m_{\tilde{g}}^{2}}{m_{\rm w}^{2}}$.
\end{itemize}

At the next-to-leading order approximation, the QCD corrections to the
$B^0-\overline B^0$ mixing in the SUSY model have been discussed
recently. The authors of \cite{ciuchini3, contino} applied the
mass-insertion method to estimate QCD correction effects on
the $B^0-\overline B^0$ mixing. However, in their work, the
contribution from gluino ($\tilde g$) was ignored. It was
thought that at high energies, $\alpha_s/4\pi\ll 1$, so that the
correction induced by $\tilde g$ might be discarded.
The authors of \cite{Soff} noticed the significance of the
gluino contributions, however they only gave a general discussion
without carrying out any concrete calculation on the gluino
contributions. The calculation including the gluino-mediated QCD
corrections needs new technique for $m_{\tilde g}>m_t$.
In this work, our method is
analogous to that employed in \cite{Soff}, but
we develop the technique and handle all problems carefully, then
draw our conclusion about the size of the
gluino contributions through a reliable calculation.

The paper is organized as follows. In section 2, we display the
necessary parts of the MSSM-Feynman rules and give the effective
Hamiltonian without QCD-corrections. In section 3, we  discuss the
features of the NLO calculation with special focus on the explicit QCD-
corrections and matching-procedure. Furthermore, we show that
reasonably removing the contributions from SUSY-particles and the
physical charged Higgs $H^{\pm}$, our
result turns back to the SM result\cite{Buras1}.
In section 4 we give the numerical results of the NLO
and scan the extent of the parameter space
in the MSSM with minimal flavor violation. We close this
paper with conclusions and discussions. Some technical
details are collected in the long appendices.

\section{Notation and the box-diagram results\label{mssm}}

\subsection{Notation and the Feynman-Rules}

Throughout this paper we adopt the notation of \cite{rosiek},
the expressions of the concerned propagators and vertices can be found
in the Appendix
of \cite{rosiek}. For convenience, we give the superpotential
and relevant mixing matrices. The most general form of the
superpotential which does not violate gauge invariance and the
conservation laws in SM is
\begin{eqnarray}
&&{\cal W}=\mu \epsilon_{ij}\hat{H}_{i}^{1}\hat{H}_{j}^{2}+
\epsilon_{ij}h_{l}^{IJ}\hat{H}_{i}^{1}\hat{L}^{I}_{j}\hat{R}^{J}+
\epsilon_{ij}h_{d}^{IJ}\hat{H}_{i}^{1}\hat{Q}^{I}_{j}\hat{D}^{J}+
\epsilon_{ij}h_{u}^{IJ}\hat{H}_{i}^{2}\hat{Q}^{I}_{j}\hat{U}^{J}.
\label{superpotential}
\end{eqnarray}
Here $\hat{H}^{1}$, $\hat{H}^{2}$ are Higgs superfields;
$\hat{Q}^{I}$ and $\hat{L}^{I}$ are quark and lepton superfields
in doublets of the weak SU(2) group,
where I=1, 2, 3 are the indices of generations;
the rest superfields:
$\hat{U}^{I}$ and $\hat{D}^{I}$ being quark superfields
of u- and d-types, and $\hat{R}^{I}$ charged leptons are
in singlets of the weak SU(2). The indices i, j are contracted
in a general way for the SU(2) group, and $h_{l}$, $h_{u,d}$
are the Yukawa couplings. Taking into account of the soft breaking terms,
we can study the phenomenology within the minimal supersymmetric extension of
the standard model (MSSM). One difference between the MSSM and
SM is the Higgs sector. There are four charged scalars,
two of them are physical massive Higgs bosons and other are
massless Goldstones. The mixing matrix can be written as:
\begin{equation}
{\cal Z}_{H}=\left(
\begin{array}{cc}
\sin\beta & -\cos\beta \\
\cos\beta & \sin\beta \end{array}
\right) \label{zh} \end{equation}
with $\tan\beta=\frac{\upsilon_{2}}{\upsilon_{1}}$ and $v_1,v_2$ are
the vacuum-expectation values of the two Higgs scalars.
Another matrix that we will use is the chargino mixing
matrix. The SUSY partners of charged Higgs and $W^{\pm}$
combine to give four Dirac fermions: $\kappa^{\pm}_{1}$,
$\kappa^{\pm}_{2}$. The two mixing matrices ${\cal Z}^{\pm}$
appearing in the Lagrangian are defined as
\begin{eqnarray}
&&({\cal Z}^{-})^{T}{\cal M}_{c}{\cal Z}^{+} = diag(m_{\kappa_{1}},
m_{\kappa_{2}}),
\label{zpm}
\end{eqnarray}
where ${\cal M}_{c}$ is the mass matrix of charginos.
In a similar way, $Z_{U,D}$ diagonalize the mass
matrices of the up- and down-type squarks respectively:
\begin{eqnarray}
&& {\cal Z}_{{\tiny U,D}}^{\dag}{\cal M}^2_{\tilde{q}}
{\cal Z}_{{\tiny U,D}}=
diag(m^2_{\tilde{{\tiny U}},\tilde{{\tiny D}}}).
\label{zud}
\end{eqnarray}

\begin{figure}
\setlength{\unitlength}{1mm}
\begin{picture}(140,200)(30,30)
\put(-10,0){\includegraphics{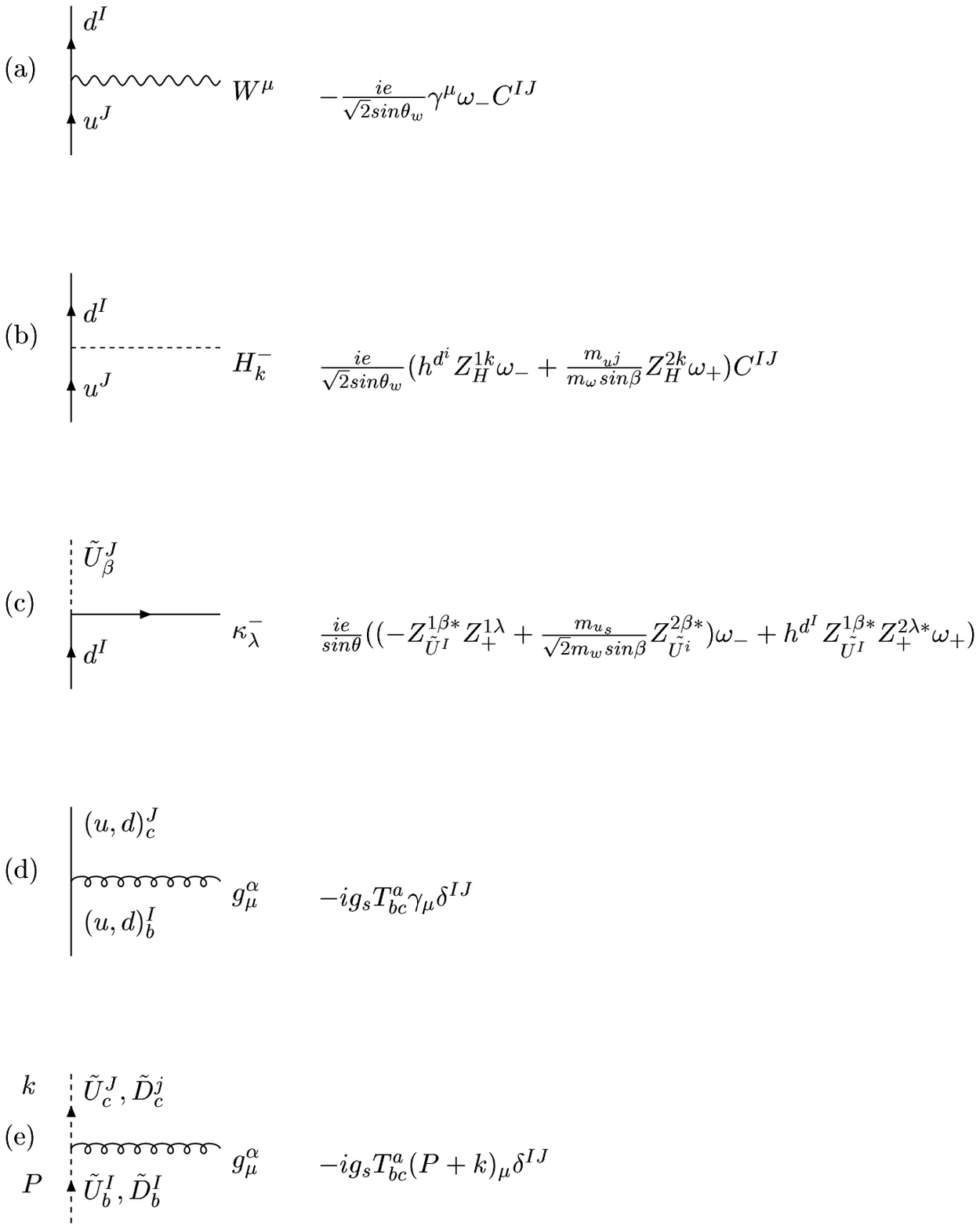}}
\end{picture}
\caption[]{The Feynman-rules which are adopted in the calculations (Part I).}
\label{fig1}
\end{figure}

\begin{figure}
\setlength{\unitlength}{1mm}
\begin{picture}(140,200)(30,30)
\put(-10,0){\includegraphics{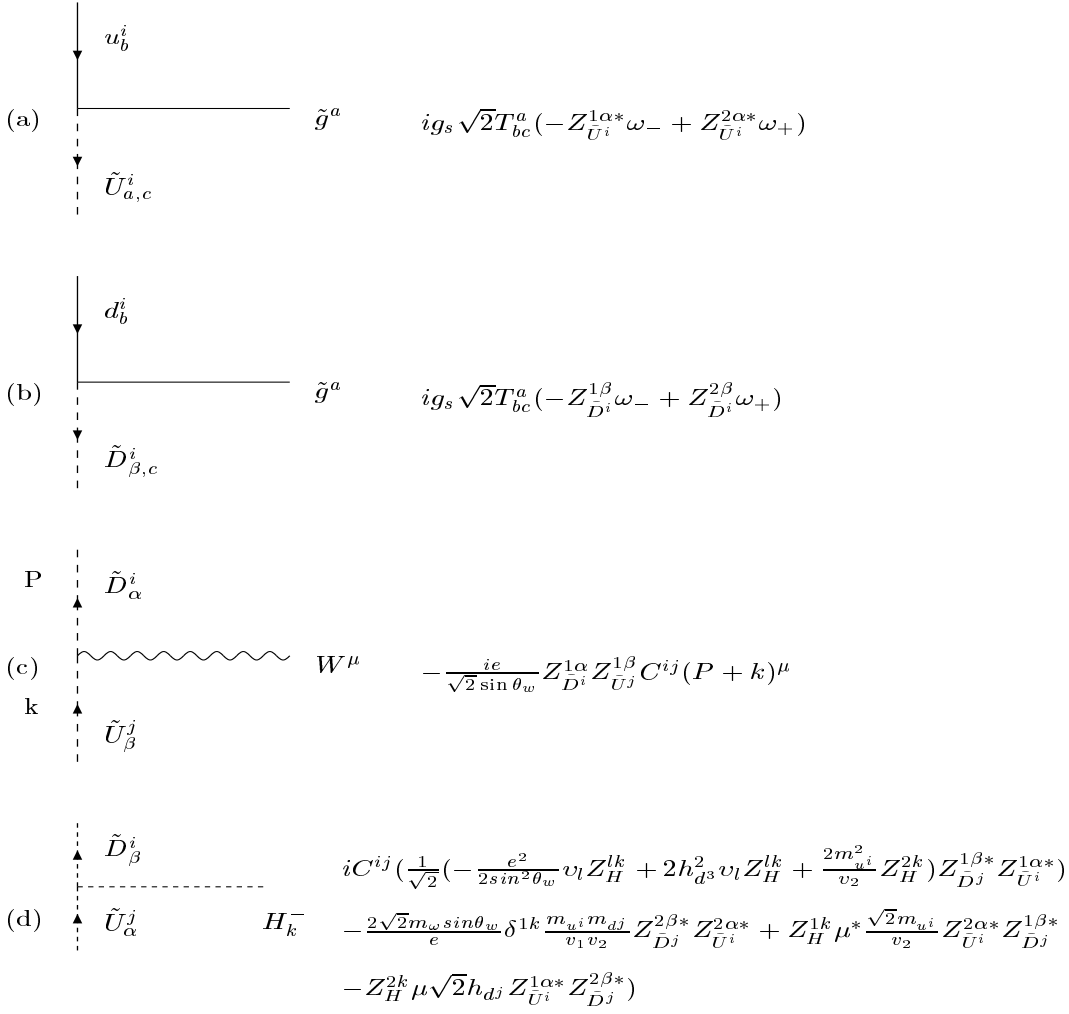}}
\end{picture}
\caption[]{The Feynman-rules which are adopted in the calculations (Part II).}
\label{fig2}
\end{figure}

We present the relevant vertices
in Fig.\ref{fig1} and Fig.\ref{fig2}. $a$, $b$, $c$
are the indices of SU(3)
group in appropriate representations. We have explicitly
written down the Yukawa-type couplings for the up-type
quarks. For the down-type quarks, we use the symbol $h_{d^{I}}
=\frac{m_{d^{I}}}{\cos\beta m_{\rm w}}$ to represent the Yukawa
couplings and
the short-hand notation $\omega_{\pm} = \frac{1\pm \gamma_{5}}
{2}$ for the left- and right-handed projectors.

\subsection{Box-diagram results}

\begin{figure}
\setlength{\unitlength}{1mm}
\begin{picture}(140,200)(30,30)
\put(-10,0){\includegraphics{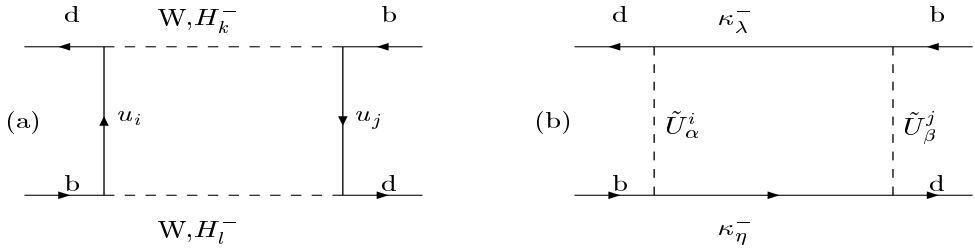}}
\end{picture}
\caption[]{The box-diagrams contributing to the $B^0-\overline{B}^0$
mixing in the supersymmetric model with minimial flavor violation.
In the calculations, the crossed diagrams should be included.}
\label{fig3}
\end{figure}

At absence of QCD corrections, the effective Hamiltonian for
the $B^0-\overline{B}^0$ mixing is obtained by evaluating the
box diagrams (Fig.\ref{fig3}).
Neglecting external momenta and masses, the effective Hamiltonian
for $\Delta B=2$ transitions at the weak-scale is\cite{Inami-Lim}
\begin{eqnarray}
&&H_{eff}^{0}=\frac{G_{F}^{2}}{4\pi^{2}}m_{\rm w}^{2}\sum\limits_{ij}
\sum\limits_{\alpha}\lambda_{i}\lambda_{j}S_{\alpha}{\cal O}_{\alpha}
\label{eff0}
\end{eqnarray}
where $\lambda_{i}=V_{ib}V_{id}^*$ ($V_{ij}$ are the elements of the CKM
matrix with $i,\; j=1,\; 2,\; 3$) and the operators ${\cal O}_{\alpha}$
are defined as
\begin{eqnarray}
&&{\cal O}_{1}=\overline{d}\gamma_{\mu}\omega_{-}b\overline{d}
\gamma^{\mu}\omega_{-}b,\nonumber \\
&&{\cal O}_{2}=\overline{d}\gamma_{\mu}\omega_{-}b\overline{d}
\gamma^{\mu}\omega_{+}b,\nonumber \\
&&{\cal O}_{3}=\overline{d}\omega_{-}b\overline{d}\omega_{+}b,\nonumber \\
&&{\cal O}_{4}=\overline{d}\omega_{-}b\overline{d}\omega_{-}b,\nonumber \\
&&{\cal O}_{5}=\overline{d}\sigma_{\mu\nu}\omega_{-}b\overline{d}
\sigma^{\mu\nu}\omega_{-}b,\nonumber \\
&&{\cal O}_{6}=\overline{d}\gamma_{\mu}\omega_{+}b\overline{d}
\gamma^{\mu}\omega_{+}b,\nonumber \\
&&{\cal O}_{7}=\overline{d}\omega_{+}b\overline{d}\omega_{+}b,\nonumber \\
&&{\cal O}_{8}=\overline{d}\sigma_{\mu\nu}\omega_{+}b\overline{d}
\sigma^{\mu\nu}\omega_{+}b,
\label{ope}
\end{eqnarray}
with the parameter $x_{i{\rm w}}=\frac{m_{i}^{2}}{m_{\rm w}^{2}}$,
the coefficients $S_{\alpha}$ are given as
\begin{eqnarray}
&&S_{1}=\bigg(\Big(f_{a}(x_{i{\rm w}},x_{j{\rm w}},1,1)-
2\frac{{\cal Z}_{H}^{2k}{\cal Z}_{H}^{2k*}}{\sin^{2}\beta}x_{i{\rm w}}
x_{j{\rm w}}f_{b}(x_{i{\rm w}},x_{j{\rm w}},1,x_{{\tiny H}_{k}^{-}{\rm w}})\nonumber \\
&&\hspace{1.2cm}+\frac{x_{i{\rm w}}x_{j{\rm w}}}{4\sin^{4}\beta}
{\cal Z}_{H}^{2k}{\cal Z}_{H}^{2k*}
{\cal Z}_{H}^{2l}{\cal Z}_{H}^{2l*}f_{a}(x_{i{\rm w}},x_{j{\rm w}},
x_{{\tiny H}_{k}^{-}{\rm w}},x_{{\tiny H}_{l}^{-}{\rm w}})\Big)\nonumber \\
&&\hspace{1.2cm}-\frac{1}{4}a_{+,i}^{km}b_{-,j}^{kn}
a_{+,j}^{ln}b_{-,i}^{lm}
f_{a}(x_{\tilde{\tiny U^{i}}_{m}{\rm w}},x_{\tilde{\tiny U^{j}}_{n}{\rm w}},
x_{\kappa_{k}^{-}{\rm w}},x_{\kappa_{l}^{-}{\rm w}})\bigg),\nonumber \\
&&S_{2}=\frac{1}{4}\bigg(\frac{h_{d}h_{d}}{2\sin^{2}\beta}\Big(
{\cal Z}_{H}^{1k}{\cal Z}_{H}^{2k*}{\cal Z}_{H}^{2l}
{\cal Z}_{H}^{1l*}x_{j{\rm w}}+{\cal Z}_{H}^{2k}{\cal Z}_{H}^{1k*}
{\cal Z}_{H}^{1l}{\cal Z}_{H}^{2l*}x_{i{\rm w}}\Big)f_{a}(x_{i{\rm w}},
x_{j{\rm w}},x_{{\tiny H}_{k}^{-}{\rm w}},x_{{\tiny H}_{l}^{-}{\rm w}})\nonumber \\
&&\hspace{1.2cm}+\Big(a_{+,i}^{km}b_{+,j}^{kn}
a_{-,j}^{ln}b_{-,i}^{lm}+
a_{-,i}^{km}b_{-,j}^{kn}a_{+,j}^{ln}b_{+,i}^{lm}\Big)\sqrt{
x_{\kappa_{k}{\rm w}}x_{\kappa_{l}{\rm w}}}f_{b}(x_{\tilde{\tiny U^{i}}_{m}{\rm w}},
x_{\tilde{\tiny U^{j}}_{n}{\rm w}},x_{\kappa_{k}^{-}{\rm w}},
x_{\kappa_{l}^{-}{\rm w}})\bigg)\nonumber \\
&&\hspace{1.2cm}-\frac{1}{4}\bigg(-2h_{d}h_{b}
{\cal Z}_{H}^{1k}{\cal Z}_{H}^{1k*}
f_{a}(x_{i{\rm w}},x_{j{\rm w}},1,x_{{\tiny H}_{k}^{-}{\rm w}})\nonumber \\
&&\hspace{1.2cm}+\frac{x_{i{\rm w}}x_{j{\rm w}}}{\sin^{2}\beta}h_{d}h_{b}\Big(
{\cal Z}_{H}^{2k}{\cal Z}_{H}^{2k*}{\cal Z}_{H}^{1l}
{\cal Z}_{H}^{1l*}+{\cal Z}_{H}^{1k}{\cal Z}_{H}^{1k*}
{\cal Z}_{H}^{2l}{\cal Z}_{H}^{2l*}\Big)f_{b}(x_{i{\rm w}},
x_{j{\rm w}},x_{{\tiny H}_{k}^{-}{\rm w}},x_{{\tiny H}_{l}^{-}{\rm w}})\nonumber \\
&&\hspace{1.2cm}+\frac{1}{2}\Big(a_{+,i}^{km}b_{-,j}^{kn}
a_{-,j}^{ln}b_{+,i}^{lm}+
a_{-,i}^{km}b_{+,j}^{kn}a_{+,j}^{ln}b_{-,i}^{lm}\Big)
f_{a}(x_{\tilde{\tiny U^{i}}_{m}{\rm w}},
x_{\tilde{\tiny U^{j}}_{n}{\rm w}},x_{\kappa_{k}^{-}{\rm w}},
x_{\kappa_{l}^{-}{\rm w}})\bigg), \nonumber \\
&&S_3=-2S_2, \nonumber \\
&&S_{4}=\frac{1}{4}\bigg(\frac{x_{i{\rm w}}x_{j{\rm w}}}{\sin^{2}\beta}h_{d}^{2}
{\cal Z}_{H}^{1k}{\cal Z}_{H}^{2k*}{\cal Z}_{H}^{2l*}
{\cal Z}_{H}^{1l}f_{b}(x_{i{\rm w}},x_{j{\rm w}},x_{{\tiny H}_{k}^{-}{\rm w}},
x_{{\tiny H}_{l}^{-}{\rm w}})\nonumber \\
&&\hspace{1.2cm}+\frac{1}{2}a_{-,i}^{km}b_{-,j}^{kn}
a_{-,j}^{ln}b_{-,i}^{lm}f_{b}(x_{\tilde{\tiny U^{i}}_{m}{\rm w}},
x_{\tilde{\tiny U^{j}}_{n}{\rm w}},x_{\kappa_{k}^{-}{\rm w}},
x_{\kappa_{l}^{-}{\rm w}})\bigg)\nonumber \\
&&\hspace{1.2cm}-\frac{3}{8}a_{-,i}^{km}b_{-,j}^{kn}
a_{-,j}^{ln}b_{-,i}^{lm}f_{b}(x_{\tilde{\tiny U^{i}}_{m}{\rm w}},
x_{\tilde{\tiny U^{j}}_{n}{\rm w}},x_{\kappa_{k}^{-}{\rm w}},
x_{\kappa_{l}^{-}{\rm w}}),\nonumber \\
&&S_5=\frac{1}{4}S_4, \nonumber \\
&&S_6=\bigg(\frac{1}{4}h_{d}^{2}h_{b}^{2}{\cal Z}_{H}^{1k}
{\cal Z}_{H}^{1k*}{\cal Z}_{H}^{1l}{\cal Z}_{H}^{1l*}
f_{a}(x_{i{\rm w}},x_{j{\rm w}},x_{{\tiny H}_{k}^{-}{\rm w}},
x_{{\tiny H}_{l}^{-}{\rm w}})\nonumber \\
&&\hspace{1.2cm}-\frac{1}{4}a_{-,i}^{km}b_{+,j}^{kn}
a_{-,j}^{ln}b_{+,i}^{lm}f_{a}(x_{\tilde{\tiny U^{i}}_{m}{\rm w}},
x_{\tilde{\tiny U^{j}}_{n}{\rm w}},x_{\kappa_{k}^{-}{\rm w}},
x_{\kappa_{l}^{-}{\rm w}})\bigg), \nonumber \\
&&S_{7}=\frac{1}{4}\bigg(\frac{x_{i{\rm w}}x_{j{\rm w}}}{\sin^{2}\beta}h_{b}^{2}
{\cal Z}_{H}^{2k}{\cal Z}_{H}^{1k*}{\cal Z}_{H}^{1l*}
{\cal Z}_{H}^{2l}f_{b}(x_{i{\rm w}},x_{j{\rm w}},x_{{\tiny H}_{k}^{-}{\rm w}},
x_{{\tiny H}_{l}^{-}{\rm w}})\nonumber \\
&&\hspace{1.2cm}+\frac{1}{2}a_{+,i}^{km}b_{+,j}^{kn}
a_{+,j}^{ln}b_{+,i}^{lm}f_{b}(x_{\tilde{\tiny U^{i}}_{m}{\rm w}},
x_{\tilde{\tiny U^{j}}_{n}{\rm w}},x_{\kappa_{k}^{-}{\rm w}},
x_{\kappa_{l}^{-}{\rm w}})\bigg)\nonumber \\
&&\hspace{1.2cm}-\frac{3}{8}a_{+,i}^{km}b_{+,j}^{kn}
a_{+,j}^{ln}b_{+,i}^{lm}f_{b}(x_{\tilde{\tiny U^{i}}_{m}{\rm w}},
x_{\tilde{\tiny U^{j}}_{n}{\rm w}},x_{\kappa_{k}^{-}{\rm w}},
x_{\kappa_{l}^{-}{\rm w}}),\nonumber \\
&&S_8=\frac{1}{4}S_7.
\label{1loop}
\end{eqnarray}
The functions $f_{a,b}(x_{1},x_{2},x_{3},x_{4})$ are given in
the appendix.\ref{oloop} and the new symbols $a_{\pm}$, $b_{\pm}$
are defined as
\begin{eqnarray}
&&a_{+,i}^{j,k}=-{\cal Z}_{\tilde{U}^{i}}^{1j}{\cal Z}_{1k}^{+*}
+\frac{x_{i{\rm w}}}{\sqrt{2}\sin\beta}{\cal Z}_{\tilde{U}^{i}}^{2j}
{\cal Z}_{2k}^{+*},\nonumber \\
&&a_{-,i}^{j,k}=h_{d}{\cal Z}_{\tilde{U}^{i}}^{1j}
{\cal Z}_{2k}^{-},\nonumber \\
&&b_{+,i}^{j,k}=h_{b}{\cal Z}_{\tilde{U}^{i}}^{1j}
{\cal Z}_{2k}^{-*},\nonumber \\
&&b_{-,i}^{j,k}=-{\cal Z}_{\tilde{U}^{i}}^{1j*}{\cal Z}_{1k}^{+}
+\frac{x_{i{\rm w}}}{\sqrt{2}\sin\beta}{\cal Z}_{\tilde{U}^{i}}^{2j*}
{\cal Z}_{2k}^{+}.
\label{abpm}
\end{eqnarray}
On purpose, we keep the Yukawa-couplings of the down-type
quarks explicitly in Eq.\ref{eff0}, so that we can discuss any
possible value of $\tan\beta$ in the Higgs sector. This is
different from some early works\cite{Urban, Soff}. Another point which
should be emphasized is that Eq.\ref{eff0} can recover the
one-loop result of \cite{Soff} as long as considering the unitarity of
the CKM matrix and discarding the Yukawa couplings of down-type quarks.

\section{Explicit QCD corrections to the box diagram \label{explicit}}

\subsection{The general method to compute the two-loop integral}
\begin{figure}
\setlength{\unitlength}{1mm}
\begin{picture}(140,200)(30,30)
\put(-10,0){\includegraphics{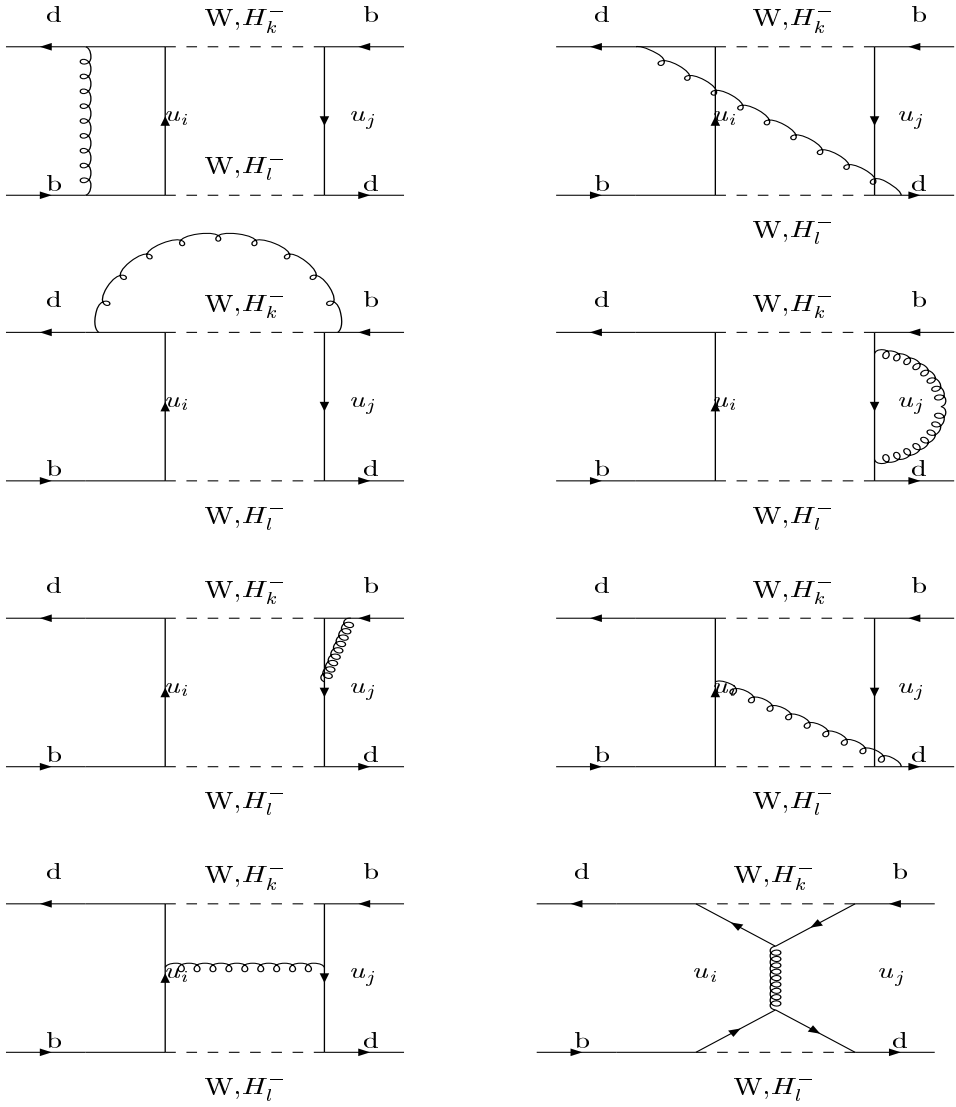}}
\end{picture}
\caption[]{The diagrams responsible for  QCD-corrections
in the framework of the SM and THDM. In the calculations, the crossed
diagrams should be included.}
\label{fig4}
\end{figure}

\begin{figure}
\setlength{\unitlength}{1mm}
\begin{picture}(140,200)(30,30)
\put(-10,0){\includegraphics{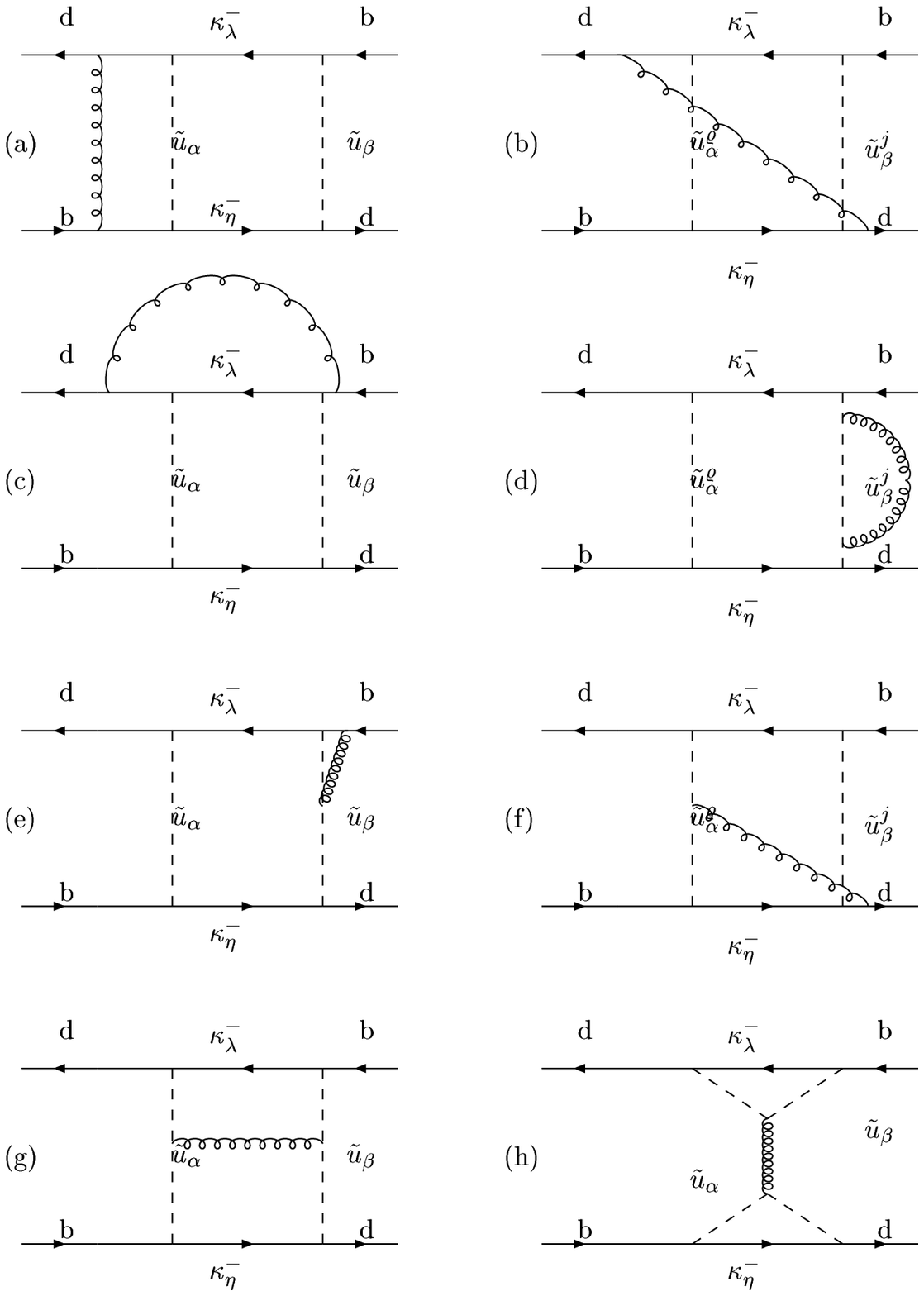}}
\end{picture}
\caption[]{The diagrams responsible for  QCD-corrections
caused by the gluon sector of the supersymmetric model with minimial
flavor violation. In the calculations, the crossed
diagrams should be included.}
\label{fig5}
\end{figure}

\begin{figure}
\setlength{\unitlength}{1mm}
\begin{picture}(140,200)(30,30)
\put(-10,0){\includegraphics{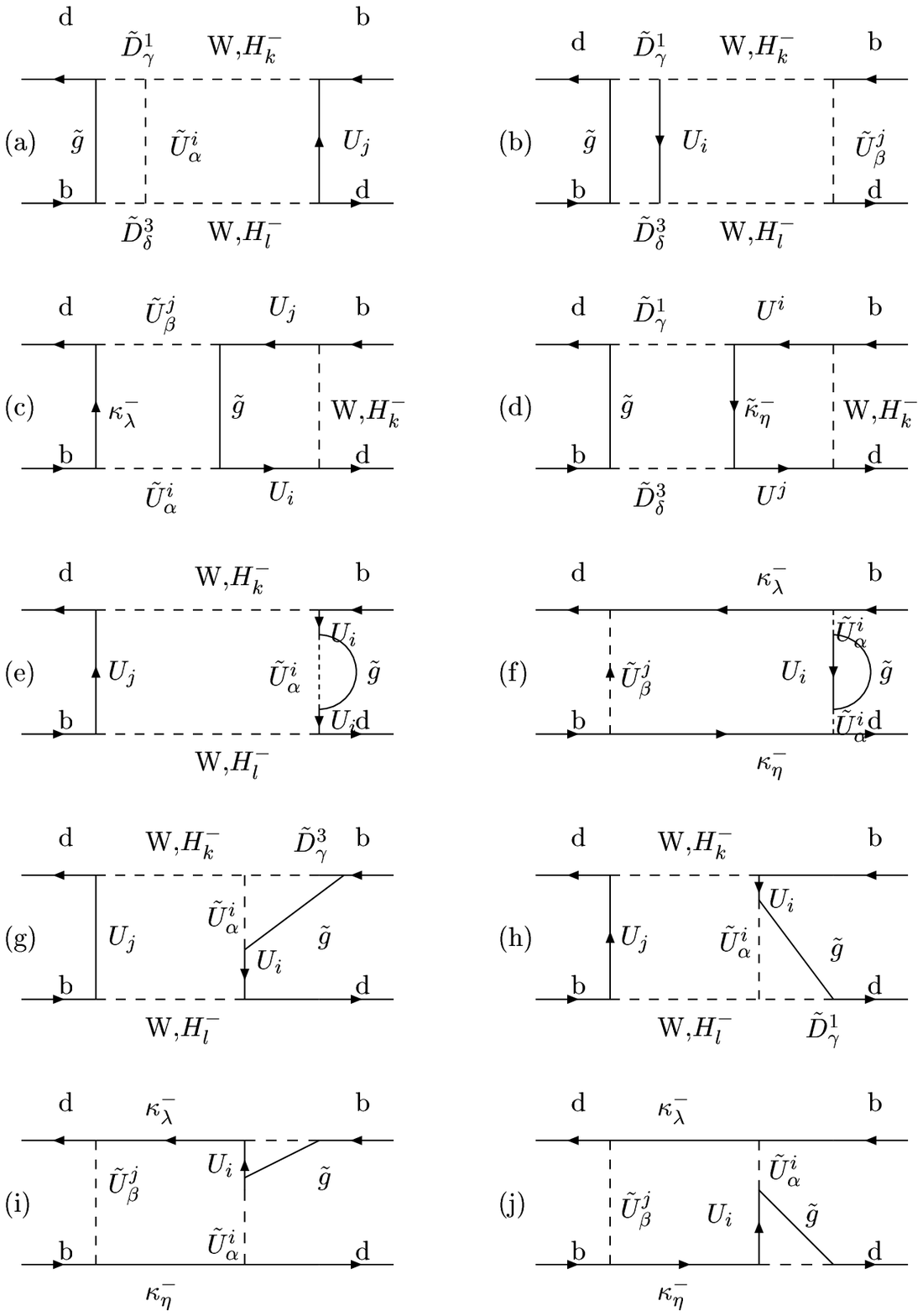}}
\end{picture}
\caption[]{The diagrams responsible for QCD-corrections
caused by the gluino sector of the supersymmetric theory with minimial
flavor violation. In the calculations, the crossed
diagrams should be included.}
\label{fig6}
\end{figure}

In this section, we will give the explicit perturbative QCD correction up
to ${\cal O}(\alpha_{s})$. The Feynman diagrams are drawn in
Fig.\ref{fig4}, Fig.\ref{fig5} and Fig.\ref{fig6}. Similar to
the previous treatments\cite{Buras1,Urban,Herrlich}, we will carry out
the calculation in an arbitrary covariant $\xi$-gauge for the gluon
propagator, where $\xi=0$ represents the Feynman-'t Hooft gauge
and $\xi=1$ the Landau gauge. The ${\rm W}$-propagators is
set in the Feynman-'t Hooft gauge.

\begin{figure}
\setlength{\unitlength}{1mm}
\begin{picture}(140,200)(30,30)
\put(-10,0){\includegraphics{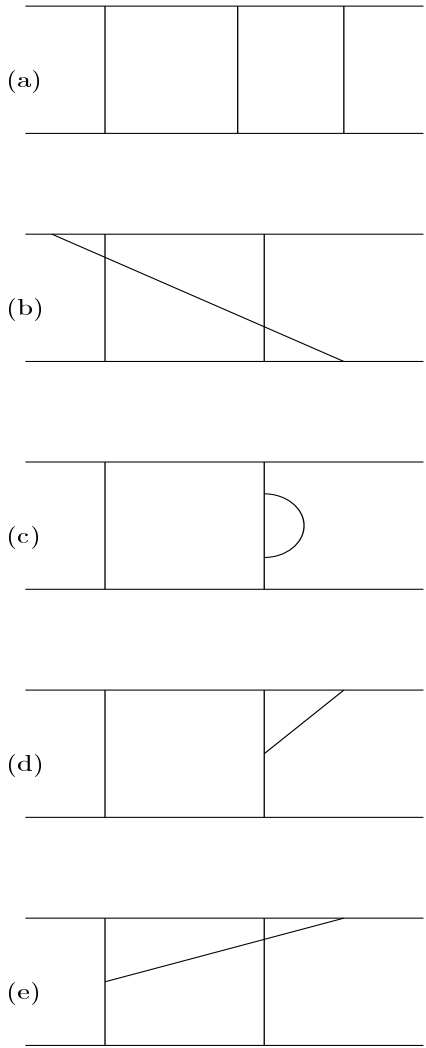}}
\end{picture}
\caption[]{The five topological classes of diagrams appearing in the
NLO-corrections to $B^0-\overline{B}^0$.}
\label{fig7}
\end{figure}

The two-loop Feynman diagrams including all SUSY particles can be
categorized into five distinct topological classes (a),(b),(c),(d) and (e)
in Fig.\ref{fig7}. Fig.\ref{fig7}(c) and Fig.\ref{fig7}(d) are the
self energy- and vertex-insertion diagrams respectively,
whereas the other three classes are of complicated topological structures.

Fig.\ref{fig4}(a, c, g), Fig.\ref{fig5}(a, c, g) and Fig.\ref{fig6}
(a, b, c, d) belong to the topological class shown in Fig.\ref{fig7}(a);
Fig.\ref{fig4}(b) and
Fig.\ref{fig5}(b) belong to the topological class in Fig.\ref{fig7}(b);
Fig.\ref{fig4}(f) and
Fig.\ref{fig5}(f) belong to the topological class in Fig.\ref{fig7}(e);
Fig.\ref{fig4}(d),
Fig.\ref{fig5}(d) and Fig.\ref{fig6}(e, f) belong to the topological
class in Fig.\ref{fig7}(c);
Fig.\ref{fig4}(e), Fig.\ref{fig5}(e) and Fig.\ref{fig6}(g, h, i, j)
belong to the topological class in Fig.\ref{fig7}(d).
The double penguin diagrams Fig.\ref{fig4}(h)
and Fig.\ref{fig5}(h) do not contribute for vanishing external momenta.

To obtain the physical quantities, we have to deal with ultraviolet
divergence. The divergence stems from  diagrams Fig.\ref{fig4}(d, e),
Fig.\ref{fig5}(d, e) and Fig.\ref{fig6}(e, f, g, h, i, j).
In this case we employ dimensional regularization
\cite{Bardeen,Herrlich2} and we carry out the renormalization
in the $\overline{{\rm MS}}$-scheme\cite{Bardeen, Buras&w}.

For an effective Hamiltonian, all internal particles must be
integrated out, namely, a condition that there exists
at least one internal particle with $m_{int}\gg m_{ext}$ where
$m_{int}$ and $m_{ext}$ refer to the masses of the internal and
all external particles (bosons or fermions) respectively, is implied.
In the case for $B^0-\overline B^0$ or $K^0-\overline K^0$ mixing, $m_b$ and
$m_d$ should be set as zero in the resultant effective theory. At
the 0-th order, e.g. when calculating the box diagrams, there is
no problem in the limit of $m_b\sim m_d=0$. However, when the
QCD corrections are taken into account, $b-$ and $d-$quark lines
become internal in diagrams Fig.\ref{fig4}(a, b, c) and
Fig.\ref{fig5}(a, b, c), then under the
limit $m_b\sim m_d=0$ an infrared divergence emerges.
The divergence is artificial
and can be eliminated in the full theory.
The natural way to
handle this problem is keeping all internal-line masses to be non-zero at
denominator of the propagators.

As well known, we need to achieve the effective Hamiltonian at
lower energies and the QCD-corrected box diagrams would determine
the boundary condition of RGE for the running of the Wilson
coefficients. Therefore, the infrared divergence must be properly
eliminated. In next section, following the standard procedures given in
literature to build a matching between the full theory and the effective one
at the scale$-\mu$, we can get rid of the troublesome infrared
divergence.

In the THDM sector of MSSM,   the calculation is standard
and  consistent with the previous work, but because of the
large masses of gluino and squarks, we need to take a more
general treatment for calculating the contribution of the
two-loop diagrams which include gluino. In the following
part, we will illustrate how to compute the loop integral
and separate ultraviolet divergence in one example. In (d)
of Fig.\ref{fig7}, We have an integral as:
\begin{equation}
I_{(d),2}^{a}=\int\frac{d^{D}k}{(2\pi)^{D}}\frac{d^{D}q}{(2\pi)^{D}}
\frac{k^4}{(k^{2}-m_{1}^{2})(k^{2}-m_{2}^{2})(k^{2}-m_{3}^{2})
(k^{2}-m_{4}^{2})((k+q)^{2}-m_{5}^{2})(q^{2}-m_{6}^{2})
(q^{2}-m_{7}^{2})}
\label{integral1}
\end{equation}
where the $m_{i}$ $i=1,\cdots ,7$ are the internal line
(bosons or fermions) masses. The above integral can be decomposed as
\begin{equation}
I_{(d),2}^{a}=I_{(d),2}^{a,1}+(m_{3}^{2}+m_{4}^{2})I_{(d),1}^{a}-
m_{3}^{2}m_{4}^{2}I_{(d),0},
\label{integral2}
\end{equation}
with
\begin{eqnarray}
&&I_{(d),2}^{a,1}=\int\frac{d^{D}k}{(2\pi)^{D}}\frac{d^{D}q}{(2\pi)^{D}}
\frac{1}{(k^{2}-m_{1}^{2})(k^{2}-m_{2}^{2})(((k+q)^{2}-m_{5}^{2})
(q^{2}-m_{6}^{2})(q^{2}-m_{7}^{2})}, \nonumber \\
&&I_{(d),1}^{a}=\int\frac{d^{D}k}{(2\pi)^{D}}\frac{d^{D}q}{(2\pi)^{D}}
\frac{k^2}{(k^{2}-m_{1}^{2})(k^{2}-m_{2}^{2})(k^{2}-m_{3}^{2})
(k^{2}-m_{4}^{2})((k+q)^{2}-m_{5}^{2})(q^{2}-m_{6}^{2})
(q^{2}-m_{7}^{2})},\nonumber \\
&&I_{(d),0}=\int\frac{d^{D}k}{(2\pi)^{D}}\frac{d^{D}q}{(2\pi)^{D}}
\frac{1}{(k^{2}-m_{1}^{2})(k^{2}-m_{2}^{2})(k^{2}-m_{3}^{2})
(k^{2}-m_{4}^{2})((k+q)^{2}-m_{5}^{2})(q^{2}-m_{6}^{2})
(q^{2}-m_{7}^{2})}.\nonumber\\
\label{integral3}
\end{eqnarray}
Now, we calculate the loop integral $I_{D,2}^{a,1}$ step by
step. After the Wick rotation, it is written as:
\begin{eqnarray}
&&I_{(d),2}^{a,1}=\int\frac{d^{D}k}{(2\pi)^{D}}\frac{d^{D}q}{(2\pi)^{D}}
\frac{1}{(k^{2}+m_{1}^{2})(k^{2}+m_{2}^{2})(((k+q)^{2}+m_{5}^{2})
(q^{2}+m_{6}^{2})(q^{2}+m_{7}^{2})}\nonumber \\
&&\hspace{1.2cm}=\frac{1}{(m_{1}^{2}-m_{2}^{2})(m_{6}^{2}-m_{7}^{2})}
\int\frac{d^{D}k}{(2\pi)^{D}}\Bigg(\frac{m_{1}^{2}}{k^{2}(k^{2}+m_{1}^{2})}
-\frac{m_{2}^{2}}{k^{2}(k^{2}+m_{2}^{2})}\Bigg) \nonumber \\
&&\hspace{1.5cm}\frac{1}{(k+q)^{2}+m_{5}^{2}}\Bigg(\frac{1}
{q^{2}+m_{6}^{2}}-\frac{1}{q^{2}+m_{7}^{2}}\Bigg) \nonumber \\
&&\hspace{1.2cm}=\frac{1}{(m_{1}^{2}-m_{2}^{2})(m_{6}^{2}-m_{7}^{2})}
\frac{B(\frac{D}{2},\varepsilon)B(2\varepsilon,1-\varepsilon)}{
\Gamma^{2}(\frac{D}{2})(4\pi)^{D}}\Bigg(\frac{m_{1}^{2}}{m_{1}^{
4\varepsilon}}\int dx x^{-\varepsilon}
(1-x)^{\varepsilon} \nonumber \\
&&\hspace{1.5cm}\bigg(F(\varepsilon,2\varepsilon;1+\varepsilon;
1-\frac{x_{51}}{x}-\frac{x_{61}}{1-x})-
F(\varepsilon,2\varepsilon;1+\varepsilon;1-\frac{x_{51}}{x}-
\frac{x_{71}}{1-x})\bigg)-(m_{1}^{2} \rightarrow m_{2}^{2})
\Bigg)
\label{integral31}
\end{eqnarray}
with $x_{ij}=\frac{m_i^2}{m_j^2}$. $F(\alpha,\beta;\gamma;t)$
is the hypergeometric function
\cite{Wang&Guo} and $\varepsilon=2-\frac{D}{2}$. To derive
Eq.\ref{integral31}, we employ formula\cite{table}
\begin{equation}
\int\limits_{0}^{\infty}dt t^{\lambda -1}(1+t)^{-\mu+\nu}
(t+\beta)^{-\nu} =B(\mu-\lambda, \lambda)F(\nu,\mu-\lambda;
\mu;1-\beta)
\label{formu1}
\end{equation}
with $Re\mu > Re\lambda >0$. Using the definition of the hypergeometric
function\cite{Wang&Guo,table}
\begin{eqnarray}
&&F(\alpha,\beta;\gamma;z)=1+\frac{\alpha\cdot\beta}{\gamma\cdot 1}z+
\frac{\alpha (\alpha +1)\beta (\beta +1)}{\gamma (\gamma+1)\cdot 1
\cdot 2}z^{2} \nonumber \\
&&\hspace{2cm}+\frac{\alpha (\alpha +1)(\alpha +2)\beta
(\beta +1)(\beta +2)}{\gamma (\gamma+1)(\gamma +2)
\cdot 1\cdot 2\cdot 3}z^{3} + \cdots,
\label{F-def}
\end{eqnarray}
we have
\begin{eqnarray}
&&F(\varepsilon,2\varepsilon;1+\varepsilon;z)=1+\frac{2\varepsilon^{2}}
{1\cdot 1}z+\frac{2\varepsilon^{2}\cdot 1 \cdot 1}{(1\cdot 2)
(1\cdot 2)}z^{2}+\frac{2\varepsilon^{2}(1\cdot 2)(1\cdot 2)}
{(1\cdot 2\cdot 3)(1\cdot 2\cdot 3)}z^{3}\nonumber \\
&&\hspace{2cm}+\cdots +\frac{2\varepsilon^{2} (n-1)!(n-1)!}{
n! n!}z^{n}+\cdots\nonumber \\
&&\hspace{1.5cm}=1+2\varepsilon^{2}\int\limits_{0}^{2}dz\bigg(1+
\frac{1\cdot 1}{2\cdot 1}z+\frac{(1\cdot 2)(1\cdot 2)}{2\cdot 3
\hspace{0.25cm}2!}z^{2}
+\cdots +\frac{(n-1)!(n-1)!}{n!(n-1)!}z^{n-1}+\cdots\bigg)+\cdots
\nonumber \\
&&\hspace{1.5cm}=1+2\varepsilon^{2}\int\limits_{0}^{z}dzF(1,1;2;z)
+\cdots \nonumber \\
&&\hspace{1.5cm}=1+2\varepsilon^{2}L_{i_{2}}(z)
+\cdots\;\; .
\label{formu2}
\end{eqnarray}
This is the key formula to proceed our computation and
$L_{i_{2}}(z)$ is the dilogarithm function, which is
defined as
\begin{equation}
L_{i_{2}}(z)=-\int\limits_{0}^{z}dt\frac{\ln(1-t)}{t}=
\sum\limits_{n=1}^{\infty}\frac{z^{n}}{n^{2}},\hspace{1cm} |z| <1.
\label{dilog}
\end{equation}
Using Eq.\ref{formu2}, we have
\begin{eqnarray}
&&I_{(d),2}^{a,1}=\frac{1}{4\pi^{4}}\frac{1}{(m_{1}^{2}-m_{2}^{2})(
m_{6}^{2}-m_{7}^{2})}\Bigg(m_{1}^{2}\bigg({\cal S}L_{i_{2}}(x_{51},
x_{61})-{\cal S}L_{i_{2}}(x_{51},x_{71})\bigg)\nonumber \\
&&\hspace{1.5cm}-m_{2}^{2}
\bigg({\cal S}L_{i_{2}}(x_{51},x_{61})-{\cal S}L_{i_{2}}
(x_{51},x_{71})\bigg)\Bigg)
\label{exp1}
\end{eqnarray}
and ${\cal S}L_{i_{2}}(a,b)$ is
$$ {\cal S}L_{i_{2}}(a,b)=\int_{0}^{1}dt L_{i_{2}}(1-
\frac{a}{t}-\frac{b}{1-t}), $$
which is a continuous and analytic
function\cite{Hilbert}, whose general expression can be found
in  Appendix.\ref{slifun}. In the above example, $I_{(d),2}^{a,1}$
does not contain  divergence. Now, let us look at  another part
that contains ultraviolet divergence. In the same
diagram Fig.\ref{fig7}(d), there exists $I_{(d),2}^d$ which is
ultraviolet divergent, the corresponding integral is
\begin{eqnarray}
&&I_{(d),2}^{d}=\int\frac{d^{D}k}{(2\pi)^{D}}\frac{d^{D}q}{(2\pi)^{D}}
\frac{k^2 q^2}{(k^{2}-m_{1}^{2})(k^{2}-m_{2}^{2})(k^{2}-m_{3}^{2})
(k^{2}-m_{4}^{2})((k+q)^{2}-m_{5}^{2})(q^{2}-m_{6}^{2})
(q^{2}-m_{7}^{2})}\nonumber \\
&&\hspace{1.2cm}=I_{(d),2}^{d,1}+m_{7}^{2}I_{(d),1}^{a}+m_{4}^{2}
I_{(d),1}^{b}-m_{4}^{2}m_{7}^{2}I_{(d),0}.
\label{integral4}
\end{eqnarray}
The explicit forms of $I_{(d),2}^{d,1}$, $I_{(d),1}^{a,1}$, $I_{(d),1}^{b,1}$,
$I_{(d),2}^{c,1}$, $I_{(d),0}$, are given in Appendix.\ref{tloop}.
For  convenience, we calculate only one of them as an example to
display how to deal with them and obtain corresponding result. In
the above expression, the form of $I_{(d),2}^{d,1}$ is
\begin{eqnarray}
&&I_{(d),2}^{d,1}=\int\frac{d^{D}k}{(2\pi)^{D}}\frac{d^{D}q}{(2\pi)^{D}}
\frac{1}{(k^{2}-m_{1}^{2})(k^{2}-m_{2}^{2})(k^{2}-m_{3}^{2})
((k+q)^{2}-m_{5}^{2})(q^{2}-m_{6}^{2})}.
\label{fd2d1}
\end{eqnarray}
Explicitly, we have a solution
\begin{eqnarray}
&&I_{(d),2}^{d,1}=\frac{1}{(4\pi)^4}\sum\limits_{i=1}^{3}
\frac{m_{i}^{2}}{\prod\limits_{j\neq i}(m_{j}^{2}-m_{i}^{2})}
\bigg(\Big(\frac{1}{\varepsilon}-\gamma_{E}+\ln 4\pi\Big)\ln x_{i{\rm w}}
+\Big(3-\gamma_{E}+\ln 4\pi\Big)\ln x_{i{\rm w}} \nonumber \\
&&\hspace{1.5cm}-\ln^{2}x_{i{\rm w}}-{\cal
S}L_{i_{2}}(x_{5i},x_{6i})\bigg).
\label{integral5}
\end{eqnarray}
Generally,
in the self-energy
(class Fig.\ref{fig7}(c)) and vertex (class Fig.\ref{fig7}(d))
insertion diagrams, there is ultraviolet
divergence which needs to be renormalized; in the other topological
classes (Fig.\ref{fig7}(a, b, e)),  no ultraviolet divergence exists.
Certain renormalization procedures can eliminate the ultraviolet
divergence, here we employ the $\overline{\rm MS}$ (the modified minimal
subtraction scheme) to do the job.

Now let us turn to possible infrared divergence which may occur in
the integrations.

We expand the two-loop results with respect to $m_b,m_d$ to
order ${\cal O}(m_{b,d})$ and then
let the masses of the down-type
quarks (d, s and b) approach to zero.
Explicitly,
${\cal S}L_{i_{2}}(a,b)$ is written as
\begin{eqnarray}
&&{\cal S}L_{i_{2}}(a,b)=\bigg(3-\frac{\pi^{2}}{6}-\ln a\ln(1-a)
+a\ln a\ln\frac{a-1}{a}-aL_{i_{2}}(\frac{1}{a})-L_{i_{2}}(a)\bigg)
\nonumber \\
&&\hspace{2.5cm}+\frac{b}{a-1}\bigg(a\Big(\frac{
-\pi^2}{6}+\ln(a(1-a))\ln\frac{a-1}{a}
+L_{i_{2}}(\frac{a}{a-1})-L_{i_{2}}(\frac{1}{a})+L_{i_{2}}(\frac{1}{
1-a})\Big)\nonumber \\
&&\hspace{2.5cm}-\Big(\ln a\ln b-L_{i_{2}}(a)-
\frac{1}{2}\ln^{2}a-\ln(a(1-a))\Big)\bigg)+{\cal O}(b^2)
\label{ser1}
\end{eqnarray}
with $b\rightarrow 0$.
Obviously, Eq.(\ref{ser1}) indicates that as
$m_b\sim m_d=0$, $I_{(a),2}^{a,b,c,d,e,f},$ $I_{(a),1}^{a,b,c}$
and $I_{(b),2}^{a,b,c,d,e,f},$ $I_{(b),1}^{a,b,c}$
would blow up, however the superficial
infrared divergence is benign as long as we retain the masses of
the down-type quarks to be non-zero.

As discussed above, the QCD correction to
the effective Hamiltonian of Eq.\ref{eff0} is given as follows
\begin{equation}
\Delta H_{eff}=\frac{G_{F}^{2}}{4\pi^{2}}m_{\rm w}^{2}
\frac{\alpha_{s}}{4\pi}\sum\limits_{i,j}\lambda_{i}\lambda_{j}
U_{i,j},
\label{qcd-correction}
\end{equation}
where
\begin{eqnarray}
&&U_{i,j}=\sum\limits_{k}^{8}\phi_{k}{\cal O}_{k},
\label{uij}
\end{eqnarray}
with ${\cal O}_{k}$ being defined in Eq.\ref{ope} and $\phi_{k}$
are written as
\begin{equation}
\phi_{k}=\phi_{k}^{g}+\phi^{\tilde{g}}_{k}.
\end{equation}
$\phi_{k}^{g}$ (k=1, 2, $\cdots$, 8) are the contributions of
gluon and $\phi_{k}^{\tilde{g}}$ come from gluino corrections.
$\phi_{k}^{g}$ have been derived and are of following forms
\begin{eqnarray}
&&\phi_{1}^{g}=L_{i,j}^{1}-\xi\Big(\frac{10}{3}S_{1}+\frac{1}{6}S_{4}
+\frac{1}{6}S_{7}\Big)-\frac{10}{3}
(1-\xi)\frac{x_{d{\rm w}}\ln x_{d{\rm w}}-x_{b{\rm w}}\ln x_{b{\rm w}}}{x_{d{\rm w}}-x_{b{\rm w}}}
S_{1}\nonumber \\
&&\hspace{1.2cm}+\big(\frac{4}{3}-\frac{1}{3}\xi\big)
\ln x_{d{\rm w}}x_{b{\rm w}}S_{1}
+\frac{8}{3}(1-\xi)\ln x_{\mu}S_{1}-(4-\xi)S_2\frac{m_{b}m_{d}}{
2(m_{d}^{2}-m_{b}^{2})}\ln \frac{x_{d{\rm w}}}{x_{b{\rm w}}}\nonumber \\
&&\hspace{1.2cm}+2\ln x_{\mu}(\nabla_xS_{1})
, \nonumber \\
&&\phi_{2}^{g}=L_{i,j}^{2}-\frac{11}{3}\xi S_{2}
+(-\frac{25}{3}+\frac{10}{3}\xi)S_{2}
\frac{x_{d{\rm w}}\ln x_{d{\rm w}}-x_{b{\rm w}}\ln x_{b{\rm w}}}{
x_{d{\rm w}}-x_{b{\rm w}}}
+\frac{1}{2}(2-\xi)S_2\ln x_{b{\rm w}}x_{d{\rm w}}
\nonumber\\
&&\hspace{1.2cm}
+(4-\xi)\Big(\frac{5}{6}S_{1}-\frac{5}{12}S_{4}
+\frac{5}{6}S_{6}-\frac{5}{12}S_{7}\Big)
\frac{m_{b}m_{d}}{m_{d}^{2}
-m_{b}^{2}}\ln\frac{x_{d{\rm w}}}
{x_{b{\rm w}}}
\nonumber \\
&&\hspace{1.2cm}+\frac{8}{3}(1-\xi)\ln x_{\mu}S_{2}
+2\ln x_{\mu}(\nabla_xS_{2})
, \nonumber \\
&&\phi_{3}^{g}=L_{i,j}^{3}+\frac{22}{3}\xi S_2
+(\frac{50}{3}-\frac{20}{3}\xi)
\frac{x_{d{\rm w}}\ln x_{d{\rm w}}-x_{b{\rm w}}\ln x_{b{\rm w}}}{
x_{d{\rm w}}-x_{b{\rm w}}}
+(2-\xi)S_2
\ln x_{d{\rm w}}x_{b{\rm w}}
\nonumber \\
&&\hspace{1.2cm}
-(4-\xi)\Big(\frac{5}{3}S_{1}-\frac{5}{6}S_{4}
+\frac{5}{3}S_{6}-\frac{5}{6}S_{7}\Big)
\frac{m_{b}m_{d}}{m_{d}^{2}-m_{b}^{2}}\ln\frac{x_{d{\rm w}}}
{x_{b{\rm w}}}
\nonumber \\
&&\hspace{1.2cm}
-\frac{16}{3}(1-\xi)\ln x_{\mu}S_{2}
-4\ln x_{\mu}(\nabla_xS_{2})
, \nonumber \\
&&\phi_{4}^{g}=L_{i,j}^{4}+\xi\big(\frac{1}{3}S_{1}-
\frac{10}{3}S_{4}-\frac{1}{3}S_{6}\big)+
(-\frac{4}{3}+\frac{10}{3}\xi)S_4
\frac{x_{d{\rm w}}\ln x_{d{\rm w}}-x_{b{\rm w}}\ln x_{b{\rm w}}}{
x_{d{\rm w}}-x_{b{\rm w}}}
\nonumber \\
&&\hspace{1.2cm}+\frac{2-\xi}{3}S_4
\ln x_{d{\rm w}}x_{b{\rm w}}
+\frac{5}{6}(4-\xi)S_2\frac{m_{b}m_{d}}{m_{d}^{2}-
m_{b}^{2}}\ln\frac{x_{d{\rm w}}}{x_{b{\rm w}}}
\nonumber \\
&&\hspace{1.2cm}+\frac{8}{3}(1-\xi)
\ln x_{\mu}S_{4}
+2\ln x_{\mu}(\nabla_xS_{4})
, \nonumber \\
&&\phi_{5}^{g}=L_{i,j}^{5}+\xi\big(\frac{1}{12}S_{1}-
\frac{5}{6}S_{4}-\frac{1}{12}S_{6}\big)+
(-\frac{1}{3}+\frac{5}{6}\xi)S_4
\frac{x_{d{\rm w}}\ln x_{d{\rm w}}-x_{b{\rm w}}\ln x_{b{\rm w}}}{
x_{d{\rm w}}-x_{b{\rm w}}}
\nonumber \\
&&\hspace{1.2cm}+\frac{2-\xi}{12}S_4
\ln x_{d{\rm w}}x_{b{\rm w}}
+\frac{5}{24}(4-\xi)S_2\frac{m_{b}m_{d}}{m_{d}^{2}-
m_{b}^{2}}\ln\frac{x_{d{\rm w}}}{x_{b{\rm w}}}
\nonumber \\
&&\hspace{1.2cm}+\frac{2}{3}(1-\xi)
\ln x_{\mu}S_{4}
+\frac{1}{2}\ln x_{\mu}(\nabla_xS_{4})
, \nonumber \\
&&\phi_{6}^{g}=L_{i,j}^{6}-\xi\Big(\frac{10}{3}S_{6}+\frac{1}{6}S_{4}
+\frac{1}{6}S_{7}\Big)-\frac{10}{3}
(1-\xi)\frac{x_{d{\rm w}}\ln x_{d{\rm w}}-x_{b{\rm w}}\ln x_{b{\rm w}}}{x_{d{\rm w}}-x_{b{\rm w}}}
S_{6}\nonumber \\
&&\hspace{1.2cm}+\big(\frac{4}{3}-\frac{1}{3}\xi\big)
\ln x_{d{\rm w}}x_{b{\rm w}}S_{6}
+\frac{8}{3}(1-\xi)\ln x_{\mu}S_{6}-(4-\xi)S_2\frac{m_{b}m_{d}}{
2(m_{d}^{2}-m_{b}^{2})}\ln \frac{x_{d{\rm w}}}{x_{b{\rm w}}}\nonumber \\
&&\hspace{1.2cm}+2\ln x_{\mu}(\nabla_xS_{6})
, \nonumber \\
&&\phi_{7}^{g}=L_{i,j}^{7}+\xi\big(\frac{1}{3}S_{6}-
\frac{10}{3}S_{7}-\frac{1}{3}S_{1}\big)+
(-\frac{4}{3}+\frac{10}{3}\xi)S_7
\frac{x_{d{\rm w}}\ln x_{d{\rm w}}-x_{b{\rm w}}\ln x_{b{\rm w}}}{
x_{d{\rm w}}-x_{b{\rm w}}}
\nonumber \\
&&\hspace{1.2cm}+\frac{2-\xi}{3}S_7
\ln x_{d{\rm w}}x_{b{\rm w}}
+\frac{5}{6}(4-\xi)S_2\frac{m_{b}m_{d}}{m_{d}^{2}-
m_{b}^{2}}\ln\frac{x_{d{\rm w}}}{x_{b{\rm w}}}
\nonumber \\
&&\hspace{1.2cm}+\frac{8}{3}(1-\xi)
\ln x_{\mu}S_{7}
+2\ln x_{\mu}(\nabla_xS_{7})
, \nonumber \\
&&\phi_{8}^{g}=L_{i,j}^{8}+\xi\big(\frac{1}{12}S_{6}-
\frac{5}{6}S_{7}-\frac{1}{12}S_{1}\big)+
(-\frac{1}{3}+\frac{5}{6}\xi)S_7
\frac{x_{d{\rm w}}\ln x_{d{\rm w}}-x_{b{\rm w}}\ln x_{b{\rm w}}}{
x_{d{\rm w}}-x_{b{\rm w}}}
\nonumber \\
&&\hspace{1.2cm}+\frac{2-\xi}{12}S_7
\ln x_{d{\rm w}}x_{b{\rm w}}
+\frac{5}{24}(4-\xi)S_2\frac{m_{b}m_{d}}{m_{d}^{2}-
m_{b}^{2}}\ln\frac{x_{d{\rm w}}}{x_{b{\rm w}}}
\nonumber \\
&&\hspace{1.2cm}+\frac{2}{3}(1-\xi)
\ln x_{\mu}S_{7}
+\frac{1}{2}\ln x_{\mu}(\nabla_xS_{7}),
\label{coe1}
\end{eqnarray}
where $x_{\mu}=\frac{\mu^{2}}{m_{{\rm {\rm w}}}^{2}}$ and $\mu$ is the scale
at which the heavy degrees of freedom are integrated out.

It is noted that there are $\log x_i$ terms in the expressions.
In fact, the situation for the Feynman  diagrams including
the vertex and self-energy insertions is more subtle, because
these kinds of diagrams are logarithmically divergent. When the
masses of the inner loop (vertex loop or self-energy)
are much greater than that of the outer loop, \footnote{
Here  "inner loop" refers to the loop for the inserted self-energy and vertex
corrections,
whereas "outer loop" is for the loop part outside the
"inner loop". These notations are taken to distinguish "inner" and
"outer" from
"internal" and "external" quantities in the loop evaluation to avoid
possible ambiguities.}
 logarithmic
divergence $\log{m_i^2\over m_e^2}$ may emerge where $m_i$ is the
mass of the particles in the inner loop (vertex correction or
self-energy) and $m_e$ is the mass of particles in the outer loop
\cite{Collins}.

Here, we have defined a new symbol
\begin{eqnarray}
&&\nabla_{x}=3x_{i{\rm w}}\frac{\partial}
{\partial x_{i{\rm w}}}+3x_{j{\rm w}}\frac{\partial}{\partial x_{j{\rm w}}}
2x_{H_{k}^{-}{\rm w}}\frac{\partial}{\partial x_{H_{k}^{-}{\rm w}}}+
+2x_{H_{l}^{-}{\rm w}}\frac{\partial}{\partial x_{H_{l}^{-}{\rm w}}}+
3x_{\tilde{\kappa}_{k}^{-}{\rm w}}\frac{\partial}{\partial
x_{\tilde{\kappa}_{k}^{-}{\rm w}}}\nonumber \\
&&\hspace{1.2cm}+
3x_{\tilde{\kappa}_{l}^{-}{\rm w}}\frac{\partial}{\partial
x_{\tilde{\kappa}_{l}^{-}{\rm w}}}+
2x_{\tilde{\tiny U^{i}}_{m}{\rm w}}\frac{\partial}{\partial
x_{\tilde{\tiny U^{i}}_{m}{\rm w}}}+
2x_{\tilde{\tiny U^{i}}_{n}{\rm w}}\frac{\partial}{\partial
x_{\tilde{\tiny U^{i}}_{n}{\rm w}}}.
\end{eqnarray}
One should note that
all the masses entering the functions are the masses evaluated at the
$\mu$ scale. $L_{i,j}^{a}$ ($a=1,\cdot,8$) are complicated functions
of inner line masses, which are collected in the appendix.
When we derive the above results, the Fierz-transformation
is used to organize the emerging operators into the form of
color-singlet current$\otimes$color-singlet current.
As for the diagrams contain the
ultraviolet divergence, we have taken the ${\cal O}(\varepsilon)$
contributions into account seriously. The results depend on
the gauge parameter, and are infrared-divergent
as $m_{b,d}\rightarrow 0$. However, the infrared divergence and gauge-dependence
vanish after we match the full and effective sides of the theory and the explicit
procedure of the matching is shown in next subsection.

\subsection{Wilson coefficient function of ${\cal O}_{i}$}

The effective Hamiltonian to  order $O(\alpha_{s})$ is given as
\begin{eqnarray}
&&H_{eff}=H_{eff}^{0}+\Delta H_{eff},
\label{heff1}
\end{eqnarray}
where $H_{eff}^0$ is the pure box contribution and
$\Delta H_{eff}$ is the Hamiltonian resulted in by
the SUSY-QCD corrections.
To obtain the Wilson coefficients in Eq.(\ref{heff1}), one needs
to properly handle the matching condition between the full theory
and effective one.

\begin{figure}
\setlength{\unitlength}{1mm}
\begin{picture}(140,200)(30,30)
\put(-10,0){\includegraphics{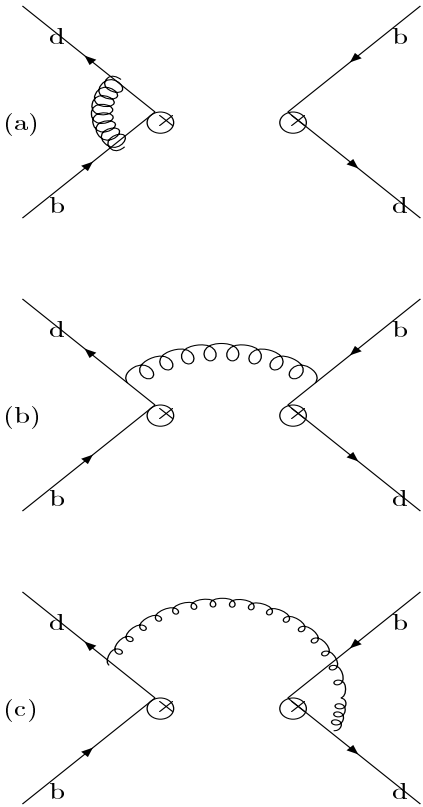}}
\end{picture}
\caption[]{Classes of diagrams in the effective theory contributing
to $Q_i$ up to order $\alpha_s$.}
\label{fig8}
\end{figure}

As stated above, the Hamiltonian contains the infrared divergence
and gauge dependence. In order to obtain physics results,
we need to match the effective theory to the full theory. Before doing this,
we evaluate the matrix elements of the physical operators
${\cal O}_{i}$($i=1,\dots, 8$) up to order $O(\alpha_{s})$ using
the same regularization, renormalization and gauge prescriptions
employed above. The one-loop diagrams which are responsible for the
corrections to the operators
${\cal O}_{i}$ are given in Fig.\ref{fig8}, the results are
\begin{equation}
{\cal O}_{i}={\cal O}^{(0)}_{i}+\frac{\alpha_{s}}{4\pi^{2}}\sum
\limits_{j}\Big(C_{F}r_{ij}^{(1)}{\cal O}^{(1)}_{j}+T^{a}\otimes T^{a}
r_{ij}^{(8)}\tilde{{\cal O}}^{(8)}_{j}\Big)
\label{rdef}
\end{equation}
with
\begin{eqnarray}
&&r_{11}^{(1)}=-3+2(1-\xi)\Big(1+\ln x_{\mu}-\frac{x_{d{\rm w}}\ln x_{d{\rm w}}-
x_{b{\rm w}}\ln x_{b{\rm w}}}{x_{d{\rm w}}-x_{b{\rm w}}}\Big),\nonumber \\
&&r_{12}^{(1)}=(4-\xi)\frac{m_{b}m_{d}}{m_{d}^{2}-m_{b}^{2}}
\ln\frac{x_{d{\rm w}}}{x_{b{\rm w}}},\nonumber \\
&&r_{11}^{(8)}=-5-(4-\xi)\big(2\ln x_{\mu}-\ln x_{d{\rm w}}x_{b{\rm w}}\big)+
2(1-\xi)\Big(1+\ln x_{\mu}-\frac{x_{d{\rm w}}\ln x_{d{\rm w}}-
x_{b{\rm w}}\ln x_{b{\rm w}}}{x_{d{\rm w}}-x_{b{\rm w}}}\Big),\nonumber \\
&&r_{13}^{(8)}=-2(4-\xi)\frac{m_{b}m_{d}}{m_{d}^{2}-m_{b}^{2}}
\ln\frac{x_{d{\rm w}}}{x_{b{\rm w}}},\nonumber \\
&&r_{14}^{(8)}=r_{17}^{(8)}=-(4-\xi);\nonumber \\
&&r_{15}^{(8)}=r_{18}^{(8)}=\frac{1}{4}(4-\xi);\nonumber \\
&&r_{21}^{(1)}=\frac{1}{2}(4-\xi)\frac{m_{b}m_{d}}{m_{d}^{2}-m_{b}^{2}}
\ln\frac{x_{d{\rm w}}}{x_{b{\rm w}}},\nonumber \\
&&r_{22}^{(1)}=-3+2(1-\xi)\Big(1+\ln x_{\mu}-\frac{x_{d{\rm w}}\ln x_{d{\rm w}}-
x_{b{\rm w}}\ln x_{b{\rm w}}}{x_{d{\rm w}}-x_{b{\rm w}}}\Big),\nonumber \\
&&r_{26}^{(1)}=\frac{1}{2}(4-\xi)\frac{m_{b}m_{d}}{m_{d}^{2}-m_{b}^{2}}
\ln\frac{x_{d{\rm w}}}{x_{b{\rm w}}},\nonumber \\
&&r_{22}^{(8)}=(4-\xi)\Big(\ln x_{d{\rm w}}x_{b{\rm w}}-2\frac{x_{d{\rm w}}\ln x_{d{\rm w}}-
x_{b{\rm w}}\ln x_{b{\rm w}}}{x_{d{\rm w}}-x_{b{\rm w}}}\Big)-8-2\xi,\nonumber \\
&&r_{23}^{(8)}=-2(4-\xi),\nonumber \\
&&r_{24}^{(8)}=r_{27}^{(8)}=-(4-\xi)\frac{m_{b}m_{d}}{m_{d}^{2}-m_{b}^{2}}
\ln\frac{x_{d{\rm w}}}{x_{b{\rm w}}},\nonumber \\
&&r_{25}^{(8)}=r_{28}^{(8)}=\frac{4-\xi}{4}\frac{m_{b}m_{d}}
{m_{d}^{2}-m_{b}^{2}}\ln\frac{x_{d{\rm w}}}{x_{b{\rm w}}},\nonumber \\
&&r_{33}^{(1)}=4-2\xi+2(4-\xi)\Big(\ln x_{\mu}-\frac{x_{d{\rm w}}\ln x_{d{\rm w}}-
x_{b{\rm w}}\ln x_{b{\rm w}}}{x_{d{\rm w}}-x_{b{\rm w}}}\Big),\nonumber \\
&&r_{34}^{(1)}=r_{37}^{(1)}=-(4-\xi)\frac{m_{b}m_{d}}{m_{d}^{2}-m_{b}^{2}}
\ln\frac{x_{d{\rm w}}}{x_{b{\rm w}}},\nonumber \\
&&r_{31}^{(8)}=\frac{1}{4}(4-\xi)\frac{m_{b}m_{d}}{m_{d}^{2}-m_{b}^{2}}
\ln\frac{x_{d{\rm w}}}{x_{b{\rm w}}},\nonumber \\
&&r_{32}^{(8)}=-\frac{1}{2}(4-\xi),\nonumber \\
&&r_{33}^{(8)}=-\frac{5}{2}-2\xi+2(1-\xi)\Big(\ln x_{d{\rm w}}x_{b{\rm w}}-
\frac{x_{d{\rm w}}\ln x_{d{\rm w}}-x_{b{\rm w}}\ln x_{b{\rm w}}}{x_{d{\rm w}}-x_{b{\rm w}}}\Big)
,\nonumber \\
&&r_{36}^{(8)}=\frac{1}{4}(4-\xi)\frac{m_{b}m_{d}}{m_{d}^{2}-m_{b}^{2}}
\ln\frac{x_{d{\rm w}}}{x_{b{\rm w}}},\nonumber \\
&&r_{43}^{(1)}=-2(4-\xi)\frac{m_{b}m_{d}}{m_{d}^{2}-m_{b}^{2}}
\ln\frac{x_{d{\rm w}}}{x_{b{\rm w}}},\nonumber \\
&&r_{44}^{(1)}=4-2\xi+2(4-\xi)\Big(\ln x_{\mu}-\frac{x_{d{\rm w}}\ln x_{d{\rm w}}-
x_{b{\rm w}}\ln x_{b{\rm w}}}{x_{d{\rm w}}-x_{b{\rm w}}}\Big),\nonumber \\
&&r_{41}^{(8)}=-\frac{1}{4}(4-\xi),\nonumber \\
&&r_{42}^{(8)}=-\frac{1}{2}(4-\xi)\frac{m_{b}m_{d}}{m_{d}^{2}-m_{b}^{2}}
\ln\frac{x_{d{\rm w}}}{x_{b{\rm w}}},\nonumber \\
&&r_{44}^{(8)}=(1-\xi)\Big(2+\ln x_{d{\rm w}}x_{b{\rm w}}-2
\frac{x_{d{\rm w}}\ln x_{d{\rm w}}-x_{b{\rm w}}\ln x_{b{\rm w}}}{x_{d{\rm w}}-x_{b{\rm w}}}\Big)
,\nonumber \\
&&r_{45}^{(8)}=\frac{3}{4}-\ln x_{\mu}+\frac{1}{4}\ln x_{b{\rm w}}x_{d{\rm w}}+
\frac{1}{2}\frac{x_{d{\rm w}}\ln x_{d{\rm w}}-x_{b{\rm w}}\ln x_{b{\rm w}}}{x_{d{\rm w}}-x_{b{\rm w}}}
,\nonumber \\
&&r_{46}^{(8)}=-\frac{1}{4}(4-\xi),\nonumber \\
&&r_{55}^{(1)}=2\xi\Big(\frac{x_{d{\rm w}}\ln x_{d{\rm w}}-
x_{b{\rm w}}\ln x_{b{\rm w}}}{x_{d{\rm w}}-x_{b{\rm w}}}-\ln x_{\mu}-1\Big), \nonumber \\
&&r_{51}^{(8)}=3(4-\xi),\nonumber \\
&&r_{52}^{(8)}=-24(4-\xi)\frac{m_{b}m_{d}}{m_{d}^{2}-m_{b}^{2}}
\ln\frac{x_{d{\rm w}}}{x_{b{\rm w}}},\nonumber \\
&&r_{54}^{(8)}=-32+48\ln x_{\mu}-12\ln x_{d{\rm w}}x_{b{\rm w}}
-24\frac{x_{d{\rm w}}\ln x_{d{\rm w}}-x_{b{\rm w}}\ln x_{b{\rm w}}}{x_{d{\rm w}}-x_{b{\rm w}}},
\nonumber \\
&&r_{55}^{(8)}=2(3+\xi)\frac{x_{d{\rm w}}\ln x_{d{\rm w}}-x_{b{\rm w}}\ln x_{b{\rm w}}}{x_{d{\rm w}}-x_{b{\rm w}}}-(3+\xi)\ln x_{d{\rm w}}x_{b{\rm w}}-2(1+\xi),\nonumber \\
&&r_{66}^{(1)}=-3+2(1-\xi)\Big(1+\ln x_{\mu}-\frac{x_{d{\rm w}}\ln x_{d{\rm w}}-
x_{b{\rm w}}\ln x_{b{\rm w}}}{x_{d{\rm w}}-x_{b{\rm w}}}\Big),\nonumber \\
&&r_{62}^{(1)}=(4-\xi)\frac{m_{b}m_{d}}{m_{d}^{2}-m_{b}^{2}}
\ln\frac{x_{d{\rm w}}}{x_{b{\rm w}}},\nonumber \\
&&r_{66}^{(8)}=-5-(4-\xi)\big(2\ln x_{\mu}-\ln x_{d{\rm w}}x_{b{\rm w}}\big)+
2(1-\xi)\Big(1+\ln x_{\mu}-\frac{x_{d{\rm w}}\ln x_{d{\rm w}}-
x_{b{\rm w}}\ln x_{b{\rm w}}}{x_{d{\rm w}}-x_{b{\rm w}}}\Big),\nonumber \\
&&r_{63}^{(8)}=-2(4-\xi)\frac{m_{b}m_{d}}{m_{d}^{2}-m_{b}^{2}}
\ln\frac{x_{d{\rm w}}}{x_{b{\rm w}}},\nonumber \\
&&r_{64}^{(8)}=r_{67}^{(8)}=-(4-\xi);\nonumber \\
&&r_{65}^{(8)}=r_{68}^{(8)}=\frac{1}{4}(4-\xi),\nonumber \\
&&r_{73}^{(1)}=-2(4-\xi)\frac{m_{b}m_{d}}{m_{d}^{2}-m_{b}^{2}}
\ln\frac{x_{d{\rm w}}}{x_{b{\rm w}}},\nonumber \\
&&r_{77}^{(1)}=4-2\xi+2(4-\xi)\Big(\ln x_{\mu}-\frac{x_{d{\rm w}}\ln x_{d{\rm w}}-
x_{b{\rm w}}\ln x_{b{\rm w}}}{x_{d{\rm w}}-x_{b{\rm w}}}\Big),\nonumber \\
&&r_{76}^{(8)}=-\frac{1}{4}(4-\xi),\nonumber \\
&&r_{72}^{(8)}=-\frac{1}{2}(4-\xi)\frac{m_{b}m_{d}}{m_{d}^{2}-m_{b}^{2}}
\ln\frac{x_{d{\rm w}}}{x_{b{\rm w}}},\nonumber \\
&&r_{77}^{(8)}=(1-\xi)\Big(2+\ln x_{d{\rm w}}x_{b{\rm w}}-2
\frac{x_{d{\rm w}}\ln x_{d{\rm w}}-x_{b{\rm w}}\ln x_{b{\rm w}}}{x_{d{\rm w}}-x_{b{\rm w}}}\Big)
,\nonumber \\
&&r_{78}^{(8)}=\frac{3}{4}-\ln x_{\mu}+\frac{1}{4}\ln x_{b{\rm w}}x_{d{\rm w}}+
\frac{1}{2}\frac{x_{d{\rm w}}\ln x_{d{\rm w}}-x_{b{\rm w}}\ln x_{b{\rm w}}}{x_{d{\rm w}}-x_{b{\rm w}}}
,\nonumber \\
&&r_{71}^{(8)}=-\frac{1}{4}(4-\xi),\nonumber \\
&&r_{88}^{(1)}=2\xi\Big(\frac{x_{d{\rm w}}\ln x_{d{\rm w}}-
x_{b{\rm w}}\ln x_{b{\rm w}}}{x_{d{\rm w}}-x_{b{\rm w}}}-\ln x_{\mu}-1\Big), \nonumber \\
&&r_{86}^{(8)}=3(4-\xi),\nonumber \\
&&r_{82}^{(8)}=-24(4-\xi)\frac{m_{b}m_{d}}{m_{d}^{2}-m_{b}^{2}}
\ln\frac{x_{d{\rm w}}}{x_{b{\rm w}}},\nonumber \\
&&r_{87}^{(8)}=-32+48\ln x_{\mu}-12\ln x_{d{\rm w}}x_{b{\rm w}}
-24\frac{x_{d{\rm w}}\ln x_{d{\rm w}}-x_{b{\rm w}}\ln x_{b{\rm w}}}{x_{d{\rm w}}-x_{b{\rm w}}},
\nonumber \\
&&r_{88}^{(8)}=2(3+\xi)\frac{x_{d{\rm w}}\ln x_{d{\rm w}}-x_{b{\rm w}}\ln x_{b{\rm w}}}{x_{d{\rm w}}-x_{b{\rm w}}}-(3+\xi)\ln x_{d{\rm w}}x_{b{\rm w}}-2(1+\xi).
\label{r-matr}
\end{eqnarray}
The other elements of $r^{(1,8)}$ are zero. At the scale $\mu$
where matching between the full Hamiltonian and the effective one
is made, the matching condition
can be written as
\begin{eqnarray}
&&H_{eff}=H_{eff}^{0}+\Delta H_{eff}\nonumber \\
&&\hspace{1.0cm}\equiv H_{full}=\frac{G_{F}^{2}}{4\pi^{2}}\lambda_{i}\lambda_{j}
\Big(\vec{{\cal O}}^{(0)T}\cdot \big[\vec{S}+\frac{\alpha_{s}}{
4\pi}\vec{\phi}\big]\Big)\nonumber \\
&&\hspace{1.0cm}=\frac{G_{F}^{2}}{4\pi^{2}}\lambda_{i}\lambda_{j}
\vec{{\cal O}}^{T}(\mu)\cdot\vec{C}(\mu)
\label{heff2}
\end{eqnarray}
where $\vec{{\cal O}}^{(0)}$ are the tree-level operators,
but $\vec{{\cal O}}(\mu)$ are the QCD-modified operators
and $\vec C(\mu)$ are the corresponding coefficients. From
the Eq.\ref{rdef} and Eq.\ref{r-matr}, we obtain
\begin{equation}
\vec{{\cal O}}(\mu)=\Big(1+\frac{\alpha_{s}}{4\pi}\hat{r}\Big)
\vec{{\cal O}}^{(0)},
\label{r1}
\end{equation}
where matrix $\hat r$ can be obtained from $r^{(1)}$, $r^{(8)}$ and  is read
as
\begin{eqnarray}
&&\hat{r}=\frac{1}{2}\Big[C_Fr^{(1)}+\frac{1}{2}r^{(8)}\cdot {\cal F}
-\frac{1}{6}r^{(8)}\Big]\cdot \Big(\hat{I}+{\cal F}\Big),
\label{rhat}
\end{eqnarray}
where $\hat{I}$ denotes the unit matrix and ${\cal F}$ is
the Fierz transformation matrix in the basis Eq.\ref{ope}:
\begin{equation}
{\cal F}=\left(
\begin{array}{cccccccc}
1&0&0&0&0&0&0&0\\
0&0&-2&0&0&0&0&0\\
0&-\frac{1}{2}&0&0&0&0&0&0\\
0&0&0&-\frac{1}{2}&\frac{1}{8}&0&0&0\\
0&0&0&6&\frac{1}{2}&0&0&0\\
0&0&0&0&0&1&0&0\\
0&0&0&0&0&0&-\frac{1}{2}&\frac{1}{8}\\
0&0&0&0&0&0&6&\frac{1}{2}\\
\end{array}\right)
\label{fierz}
\end{equation}
The coefficients $\vec{C}(\mu)$ are obtained by comparing Eq.\ref{heff2}
with Eq.\ref{r1}\cite{ciuchini}:
\begin{equation}
\vec{C}(\mu)=\vec{S}+\frac{\alpha_{s}}{4\pi}\Big(
\vec{\phi}-\hat{r}^{T}\vec{S}\Big)
\label{wi1}
\end{equation}
where $\vec{C}(\mu)$ are given as
\begin{eqnarray}
&&C_{1}(\mu)=S_1+\frac{\alpha_{s}}{4\pi}\Big[
\phi_{i,j}^{\tilde{g},1}+L_{i,j}^{1}+C_{F}\Big(S_{1}+
2\ln x_{\mu}S_{1}+2\ln x_{\mu}\nabla_xS_{1}\Big)\nonumber \\
&&\hspace{1.2cm}+C_{A}\Big(\big(3+6\ln x_{\mu}\big)S_{1}+
4\big(S_{4}+S_{5}\big)-\big(S_{7}+S_{8}\big)\Big)\Big]
, \nonumber \\
&&C_{2}(\mu)=S_2+\frac{\alpha_{s}}{4\pi}\Big[
\phi_{i,j}^{\tilde{g},2}+L_{i,j}^{2}+C_{F}\Big(
-\frac{1}{2}S_2-4\ln x_{\mu}S_{2} +2\ln x_{\mu}\nabla_xS_{2}\Big)
\nonumber \\
&&\hspace{1.2cm}+\frac{13}{12}S_{2}\Big]
,\nonumber \\
&&C_{3}(\mu)=S_3+\frac{\alpha_{s}}{4\pi}\Big[
\phi_{i,j}^{\tilde{g},3}+L_{i,j}^{3}+C_{F}\Big(
S_2+8\ln x_{\mu}S_{2}+2\ln x_{\mu}\nabla_x
S_{3}\Big)\nonumber \\
&&\hspace{1.2cm}-\frac{13}{6}S_{2}\Big]
,\nonumber \\
&&C_{4}(\mu)=S_4+\frac{\alpha_{s}}{4\pi}\Big[
\phi_{i,j}^{\tilde{g},4}+L_{i,j}^{4}+C_{F}\Big(
-12S_{4}-24\ln x_{\mu}S_{4}+2\ln x_{\mu}\nabla_x
S_{4}\Big)\nonumber \\
&&\hspace{1.2cm}-\frac{143}{12}S_{1}+\frac{191}{6}S_{4}
+\frac{31}{16}S_5+\frac{1}{12}S_{6}-24\ln x_{\mu}S_{4}
+\frac{1}{12}\ln x_{\mu}S_{5}\Big]
,\nonumber \\
&&C_{5}(\mu)=S_5+\frac{\alpha_{s}}{4\pi}\Big[
\phi_{i,j}^{\tilde{g},5}+L_{i,j}^{5}+
C_F\Big(-12S_{5}-24\ln x_{\mu}S_{5}+2\ln x_{\mu}\nabla_x
S_{5}\Big)\nonumber \\
&&\hspace{1.2cm}-\frac{143}{48}S_{1}+\frac{191}{24}S_{4}
+\frac{31}{64}S_5+\frac{1}{48}S_{6}-6\ln x_{\mu}S_{4}
+\frac{1}{48}\ln x_{\mu}S_{5}\Big]
,\nonumber \\
&&C_{6}(\mu)=S_6+\frac{\alpha_{s}}{4\pi}\Big[
\phi_{i,j}^{\tilde{g},6}+L_{i,j}^{6}\Big(S_{6}+
2\ln x_{\mu}S_{6}+2\ln x_{\mu}\nabla_xS_{6}\Big)\nonumber \\
&&\hspace{1.2cm}+C_{A}\Big(\big(3+6\ln x_{\mu}\big)S_{6}+
4\big(S_{7}+S_{8}\big)-\big(S_{4}+S_{5}\big)\Big)\Big]
, \nonumber \\
&&C_{7}(\mu)=S_7+\frac{\alpha_{s}}{4\pi}\Big[
\phi_{i,j}^{\tilde{g},7}+L_{i,j}^{7}+C_{F}\Big(
-12S_{7}-24\ln x_{\mu}S_{7}+2\ln x_{\mu}\nabla_x
S_{7}\Big)\nonumber \\
&&\hspace{1.2cm}-\frac{143}{12}S_{6}+\frac{191}{6}S_{7}
+\frac{31}{16}S_8+\frac{1}{12}S_{1}-24\ln x_{\mu}S_{7}
+\frac{1}{12}\ln x_{\mu}S_{8}\Big]
,\nonumber \\
&&C_{8}(\mu)=S_8+\frac{\alpha_{s}}{4\pi}\Big[
\phi_{i,j}^{\tilde{g},8}+L_{i,j}^{8}+
C_F\Big(-12S_{8}-24\ln x_{\mu}S_{8}+2\ln x_{\mu}\nabla_x
S_{8}\Big)\nonumber \\
&&\hspace{1.2cm}-\frac{143}{48}S_{6}+\frac{191}{24}S_{7}
+\frac{31}{64}S_8+\frac{1}{48}S_{1}-6\ln x_{\mu}S_{7}
+\frac{1}{48}\ln x_{\mu}S_{8}\Big],
\end{eqnarray}
with $C_{F}=\frac{4}{3}$, $C_{A}=\frac{1}{3}$, and $S_1$
through $S_8$ are defined in Eq.(7). Hence, at this stage
we have the expressions for the Wilson- coefficients at the
matching scale $\mu$, which do not suffer from the infrared
divergence under limit of $m_b\sim m_d=0$, and this is
consistent with the requirements for the effective
Hamiltonian. The next step is to perform an evolution down
to lower scales. The renormalization group equation for the
Wilson coefficients $\vec{{\cal C}}$ reads
\begin{equation}
\Big[\mu\frac{\partial}{\partial \mu}+\beta(\alpha_{s})\frac{\partial}
{\partial \alpha_{s}}-\frac{\hat{\gamma}^{T}}{2}
\Big]\vec{C}(\mu,\alpha_{s})=0
\label{reg}
\end{equation}
where $\hat{\gamma}$ is the anomalous-dimension matrix and
$\beta(\alpha_{s})$ is the usual $\beta$ function. The
solution of the Eq.\ref{reg} is discussed in \cite{ciuchini}
and we only cite the result here. Through the renormalization-group
evolution matrix $\hat{\rm W}(m,\mu)$, the vectors $\vec{C}(m)$
can be written as
\begin{equation}
\vec{C}(m)=\hat{\rm W}(m,\mu)\vec{C}(\mu)
\label{solution}
\end{equation}
with
\begin{equation}
\hat{\rm W}(m,\mu)=\Big(1+\frac{\alpha_{s}(m)}{4\pi}\hat{J}(m)\Big)
\hat{U(m,\mu)}\Big(1+\frac{\alpha_{s}(\mu)}{4\pi}\hat{J}(\mu)\Big)^{-1},
\label{W-matr}
\end{equation}
where $\hat{U}$ is the leading-order evolution matrix
\begin{equation}
\hat{U}(m,\mu)=\Big[\frac{\alpha_{s}(\mu)}{\alpha_{s}(m)}\Big]^{\hat{
\gamma}^{0T}/2\beta_{0}}
\label{U-matr}
\end{equation}
and the matrix $\hat{J}$ is given in \cite{ciuchini}. To
obtain the above formulae, the computer algebra system
MATHEMATICA 4.0\cite{math4} and MATHEMATICA-based package
FeynArts\cite{feyn} are used. The package
TRACER\cite{tracer} is used to evaluate the spinor
structure.

The main purpose of this work is investigating the gluino
corrections to the $B^0-\overline{B}^0$ mixing in the
supersymmetric scenario with minimal flavor violation.
Before proceeding our discussion, we would analyze the
gluino corrections to $\Delta H_{eff}$ first. In order to
understand the point thoroughly, we neglect the mixing
between the right- and left- squarks. In the case, ${\cal
Z}_{\tilde{Q}^{i}}^{12}={\cal Z}_{\tilde{Q}^{i}}^{21} =0$,
${\cal Z}_{\tilde{Q}^{i}}^{11}={\cal Z}_{\tilde{
Q}^{i}}^{22}=1$ and
$m_{\tilde{Q}^{i}_{1}}=m_{\tilde{Q}^{i}_{R}}$,
$m_{\tilde{Q}^{i}_{2}}=m_{\tilde{Q}^{i}_{L}}$. In the
computation of corrections from gluinos to
$B^0-\overline{B}^0$ mixing, the following terms will
appear in the coefficients of $Q_{1}(\mu)$
(Fig.\ref{fig6}(g,h))
\begin{eqnarray}
&&C_{1}(\mu)\propto -i\frac{G_F^2}{4\pi^2}m_{\rm w}^2\frac{\alpha_{s}}{4\pi}\lambda_{t}
\lambda_{t}^{*}\Big\{2C_F\sum\limits_{\alpha\beta}
{\cal Z}_{\tilde{D}^3}^{1\alpha}
{\cal Z}_{\tilde{D}^3}^{1\alpha *}{\cal Z}_{\tilde{U}^3}^{1\beta}
{\cal Z}_{\tilde{U}^3}^{1\beta *}\Big[F_{D}^{2e}-F_{D}^{2a}
\nonumber \\
&&\hspace{1.0cm}
-F_{D}^{2d}\Big](x_1,x_2,x_3,x_4,x_{\tilde{D}^3_{\alpha}},x_{\tilde{g}{\rm w}},
x_{\tilde{U}^3_{\beta}})\Big\},
\label{large0}
\end{eqnarray}
where $x_i\; (i=1,2)$ represent  $x_{u^I{\rm w}}$ (I=1, 2, 3)
and $x_{k}\; (k=3,4)$
represent  $x_{H^-_{l}{\rm w}}\; (l=1,2)$. When  $x_{\tilde{g}} \gg
x_{\tilde{D}^3_{\alpha}}, x_{\tilde{U}^3_{\beta}}$, we have
\begin{eqnarray}
&&C_{1}(\mu)\propto i\frac{G_F^2}{4\pi^2}m_{\rm w}^2\frac{\alpha_{s}}{4\pi}\lambda_{t}
\lambda_{t}^{*}\Big[2C_F\sum\limits_{i=1}^4\frac{x_{i{\rm w}}^2\ln x_{i{\rm w}}}
{\prod\limits_{j\neq i}(x_{j{\rm w}}-x_{i{\rm w}})}\Big]\ln x_{\tilde{g}{\rm w}}
\nonumber \\
&&\hspace{1.0cm}=i\frac{G_F^2}{4\pi^2}m_{\rm w}^2\frac{\alpha_{s}}{4\pi}
\lambda_{t}\lambda_{t}^{*}\Big[\frac{2x_{t{\rm w}}\ln x_{t{\rm w}}}{(x_{t{\rm w}}-1)^3}
-\frac{1+x_{t{\rm w}}}{(x_{t{\rm w}}-1)^2}\Big]\ln x_{\tilde{g}{\rm w}}
\label{large}
\end{eqnarray}
Here, we have presumed $\tan\beta\sim 1$ and the
contributions to other $C_{i}(\mu)$ ($i=2,\cdots,8$) are
suppressed by the small Yukawa couplings $h_{b}$, $h_{d}$.
In Eq.\ref{large}, we have set $x_1=x_2=x_{t{\rm w}}$ and
$x_3=x_4=1$ (this choice corresponds to exchanging
$W$-boson and top quark in the outer loop). Similar
analysis can be performed in calculating the contributions
of Fig.\ref{fig5}(e,f) (self-insertion diagrams), and we
will find the amplitude growing with $\ln x_{\tilde{g}{\rm
w}}$ when $m_{\tilde{g}}\gg m_{\tilde{U}^3_{1}}$. A similar
conclusion is derived in the SM, where the one-loop
radiation corrections to mass of the W-boson is increasing
with $\ln m_{h}$ ($m_{h}$ is the mass of the standard
Higgs)\cite{group}. When $\tan\beta \gg 1$, the corrections
to the coefficients $C_{i}(\mu)$ ($i=2, \cdots,8$) must be
taken into account seriously, because those terms cannot
cancel each other among themselves and are enhanced
strongly when the mass of gluino $m_{\tilde{g}} $ is much
greater than $m_{\tilde{U}^3_{1}}$. If we consider the
mixing between the left- and right-squarks, the expressions
would be very complicated and we present them in the
appendix. We will further discuss the gluino corrections in
the section of numerical results. However, for illustration
of the physics picture, neglecting such mixing would not
bring up any confusion.

\subsection{Hadronic Matrix Elements}

To numerically evaluate the  $B^{0}-\overline{B}^{0}$ ($K^{0}-\overline{K}^{0}$)
mixing, besides
the low-energy effective $\Delta B=2$ Lagrangian,
one needs to properly calculate the hadronic matrix elements
of the various operators in Eq.\ref{heff1}.
Estimation of such hadronic matrix elements
is notoriously difficult, and is generally accompanied by large
uncertainties due to long-distance, non-perturbative QCD effects.
Fortunately, although the same case holds in the current context, there are
two factors which mitigate those hadronic uncertainties
in our ensuring phenomenological analysis:
\begin{itemize}
\item The supersymmetric contributions to $\overline{B}^{0}-
B^0$ and $\overline{K}^{0}-K^0$ mixing in the MSSM with minimal flavor
violation give rise to the same operator ${\cal O}_{1}$
that exists in the standard model. This makes comparison
of the supersymmetry and standard model contributions relatively
straightforward.
\item For the $\overline{B}^{0}-B^0$ system, the vacuum
saturation
approximation employed below is believed to be a good approximation.
This belief is supported by the lattice Monte Carlo estimates
which give
$B_{B}\simeq 1$\cite{gavela,allton}.
\end{itemize}

We begin by restating the conventional result for the operator
${\cal O}_{1}$:
\begin{equation}
<K^{0}|{\cal O}_{1}|\overline{K}^{0}>=\frac{1}{3}f_{k}^{2}m_{K}^{2}
B_{K}^{1},
\label{via1}
\end{equation}
where $f_{K}\simeq 165$MeV is the K-meson decay constant and
$B_{K}^{1}=1$ corresponds to the "vacuum saturation" result.
Various estimates of this matrix element place $B_{K}^{1}$ in
the range of $0.3\sim 1$\cite{s1}, with a value $B_{K}^{1} \sim
0.7$ is favored by the lattice gauge results\cite{gavela,allton}.
Matrix elements of the other hadronic operators ${\cal O}_{i}$
$(i=2,\cdots,8)$ can be written as
\begin{eqnarray}
&&<K^{0}|{\cal O}_{2}|\overline{K}^{0}>=-\bigg[\frac{1}{4}-\frac{1}{6}
\bigg(\frac{m_{K}}{m_{s}+m_{d}}\bigg)^{2}\bigg]m_{K}^2f_{K}^{2}B_{K}^{2}
,\nonumber \\
&&<K^{0}|{\cal O}_{3}|\overline{K}^{0}>=\bigg[\frac{1}{24}-\frac{1}{4}
\bigg(\frac{m_{K}}{m_{s}+m_{d}}\bigg)^{2}\bigg]m_{K}^2f_{K}^{2}B_{K}^{3}
,\nonumber \\
&&<K^{0}|{\cal O}_{4}|\overline{K}^{0}>=\frac{5}{24}\bigg(\frac{m_{K}}
{m_{s}+m_{d}}\bigg)^{2}m_{K}^2f_{K}^{2}B_{K}^{4}
,\nonumber \\
&&<K^{0}|{\cal O}_{5}|\overline{K}^{0}>=\frac{1}{4}\bigg(\frac{m_{K}}
{m_{s}+m_{d}}\bigg)^{2}m_{K}^2f_{K}^{2}B_{K}^{5}
,\nonumber \\
&&<K^{0}|{\cal O}_{6}|\overline{K}^{0}>=\frac{1}{3}f_{k}^{2}m_{K}^{2}B_{K}^{6}
,\nonumber \\
&&<K^{0}|{\cal O}_{7}|\overline{K}^{0}>=\frac{5}{24}\bigg(\frac{m_{K}}
{m_{s}+m_{d}}\bigg)^{2}m_{K}^2f_{K}^{2}B_{K}^{7}
,\nonumber \\
&&<K^{0}|{\cal O}_{8}|\overline{K}^{0}>=\frac{1}{4}\bigg(\frac{m_{K}}
{m_{s}+m_{d}}\bigg)^{2}m_{K}^2f_{K}^{2}B_{K}^{8}.
\label{via2}
\end{eqnarray}
Similarly, the factors $B_{K}^{i}$ ($i=2,\cdots,8$) are
associated with each of the matrix elements in Eq.\ref{via2}.

The corresponding results for the $\Delta B=2$ matrix elements
are simplified by the fact that the current algebra enhancement factor
$\frac{m_{B}}{m_{b}+m_{d}}\simeq 1$ is sufficiently accurate to
present experimental tolerance. Thus we have
\begin{eqnarray}
&&<B^{0}|{\cal O}_{1}|\overline{B}^{0}>=\frac{1}{3}f_{B}^{2}m_{B}^{2}
,\nonumber \\
&&<B^{0}|{\cal O}_{2}|\overline{B}^{0}>=-\frac{1}{12}f_{B}^{2}m_{B}^2
,\nonumber \\
&&<B^{0}|{\cal O}_{3}|\overline{B}^{0}>=-\frac{5}{24}f_{B}^{2}m_{B}^2
,\nonumber \\
&&<B^{0}|{\cal O}_{4}|\overline{B}^{0}>=\frac{5}{24}f_{B}^{2}m_{B}^2
,\nonumber \\
&&<B^{0}|{\cal O}_{5}|\overline{B}^{0}>=\frac{1}{4}f_{B}^{2}m_{B}^2
,\nonumber \\
&&<B^{0}|{\cal O}_{6}|\overline{B}^{0}>=\frac{1}{3}f_{B}^{2}m_{B}^{2}
,\nonumber \\
&&<B^{0}|{\cal O}_{7}|\overline{B}^{0}>=\frac{5}{24}f_{B}^{2}m_{B}^2
,\nonumber \\
&&<B^{0}|{\cal O}_{8}|\overline{B}^{0}>=\frac{1}{4}f_{B}^{2}m_{B}^2
,\nonumber \\
\label{via3}
\end{eqnarray}
However, the potential benefits gained by setting $B_{B}^{i}\simeq 1$
($i=1,\cdots,8$) for the $\overline{B}^{0}-B^{0}$ matrix elements
of Eq.\ref{via3} are partially offset by our ignorance of $f_{B}$.

\section{Numerical result}

In this section, we will give the numerical discussions and compare
our results with experimental data. Before presenting the numerical
results, we list the input parameters that are used in our discussions.
For the CKM matrix elements, we use the Wolfenstein-parametrization
with parameters $A$, $\lambda$, $\rho$, $\eta$. The SM-parameters
are set as: $G_{F}=1.166\times 10^{-5}{\rm GeV}^{-2}$, $\alpha_{s}
(m_{\rm w})=0.12$, $\alpha_{s}(m_{b})=0.22$, $\alpha_{s}(m_{c})=0.34$,
$A=0.80$, $\lambda=0.22$, $m_b(m_{\rm w})=4.5$GeV, $m_c(m_{\rm w})=1.3$GeV,
$m_t(m_{\rm w})=167$GeV, $f_{B}=0.2$GeV, $f_{K}=0.167$GeV.
For parameters $\rho$, $\eta$, we have $\rho=0.36$, $\eta=0$.
The factor $B_{K}^{i}$ are chosen as $B_K^1=B_K^2=B_K^3=B_K^4=
B_K^5=B_K^6=B_K^7=B_K^8=0.7$. Using above parameters, the
SM-predictions
on $\Delta m_{B}$ and $\Delta m_{K}$ are
$$\Delta m_{B}({\rm SM})=2.18\times 10^{-13}{\rm GeV},
\Delta m_{K}({\rm SM})=2.89\times 10^{-15}{\rm GeV}.$$
At present, the experimental results are
$$\Delta m_{B}=(3.10\pm 0.1)\times 10^{-13}{\rm GeV},
\Delta m_{K}=(3.491\pm 0.009)\times 10^{-15}{\rm GeV}.$$
For the supersymmetric model with minimal flavor violation,
the free parameters to be input are
chosen as follows: $\tan\beta=\frac{\upsilon_2}{\upsilon_1}$,
$m_{H^-}$, $m_{\kappa^-_\lambda}$, $m_{\tilde{U}^i_\alpha}$, $m_{\tilde{B}_\alpha}$,
$m_{\tilde{D}_\alpha}$ ($\alpha, \lambda$=1, 2) and the mixing matrix
\begin{equation}
{\cal Z}_{\tilde{U}^I}=\left(
\begin{array}{cc}
\cos\xi_{\tilde{U}^I} & \sin\xi_{\tilde{U}^I} \\
-\sin\xi_{\tilde{U}^I} & \cos\xi_{\tilde{U}^I}
\end{array}\right),
\end{equation}
\begin{equation}
{\cal Z}_{\tilde{B}}=\left(
\begin{array}{cc}
\cos\zeta_{\tilde{B}} & \sin\zeta_{\tilde{B}} \\
-\sin\zeta_{\tilde{B}} & \cos\zeta_{\tilde{B}}
\end{array}\right),
\end{equation}
\begin{equation}
{\cal Z}_{\tilde{D}}=\left(
\begin{array}{cc}
\cos\zeta_{\tilde{D}} & \sin\zeta_{\tilde{D}} \\
-\sin\zeta_{\tilde{D}} & \cos\zeta_{\tilde{D}}
\end{array}\right).
\end{equation}
As for the mixing matrices of charginos ${\cal Z}_{\pm}$,
they can be fixed by the values of $\tan\beta$ and
$m_{\kappa^-_i}$. In the numerical calculation, we assume
that only one scalar quark is light and other heavy scalar
quarks are taken as $m_{\tilde{D}_1}=4.5{\rm TeV}$,
$m_{\tilde{B}_1}=4.7{\rm TeV}$, $m_{\tilde{D}_2}=4.6{\rm
TeV}$, $m_{\tilde{B}_2}=4.8{\rm TeV}$,
$m_{\tilde{U}^1_1}=4.1{\rm TeV}$,
$m_{\tilde{U}^1_2}=4.9{\rm TeV}$,
$m_{\tilde{U}^2_1}=4.05{\rm TeV}$,
$m_{\tilde{U}^2_2}=4.95{\rm TeV}$ and
$m_{\tilde{U}^3_2}=2.1{\rm TeV}$. For the heavy chargino,
we set $m_{\chi^-_2}=2.2{\rm TeV}$. In order to suppress
the number of free parameters, we assume the mixing angles
to be equal  $\xi_{\tilde{U}^I}=
\zeta_{\tilde{B}}=\zeta_{\tilde{D}}$ and focus on small
value of $\tan\xi_{\tilde{U}^I}$.\footnote{As free
parameters, they can vary in the range $-\frac{\pi}{4}\leq
\xi_{\tilde{U}^I}, \zeta_{\tilde{B}}, \zeta_{\tilde{D}}\leq
\frac{\pi}{4}$}

We obtain the dependence of  $\Delta m_{B}$ on the
lighter scalar top quark mass with $m_{\chi_1^-}=110{\rm GeV}$,
$m_{\tilde{g}}=300{\rm GeV}$, $\tan\xi_{\tilde{U}^I}=
\tan\zeta_{\tilde{D}}=\tan\zeta_{\tilde{B}}=0$ and $\tan\beta=1,
5,30$. We find that as the lighter scalar top mass is
greater than 300 GeV,
the  dependence is very mild, namely $\Delta
m_B$ almost does not change at all as $m_{\tilde{U}^3_1}$ increases
further and
results with and without
the gluino contributions only deviate by a constant of about
$0.3\sim 1.0\times 10^{-13}$ GeV depending on $\tan\beta$ value.

The dependence of $\Delta m_{B}$ on the
lighter chargino mass is similar to that on the lighter stop mass.
With $m_{\tilde{U}^3_1}=150{\rm GeV}$,
$m_{\tilde{g}}=300{\rm GeV}$, $\tan\xi_{\tilde{U}^I}=
\tan\zeta_{\tilde{D}}=\tan\zeta_{\tilde{B}}=0$ and $\tan\beta=1,
5,30$, as the chargino mass is greater than 200 GeV,
the  dependence is very mild, namely $\Delta
m_B$ almost does not change at all as $m_{\chi_1^-}$ increases
further and
the results with and without the gluino contributions only
deviate by a constant of $0.2\sim 0.8\times 10^{-13}$ GeV depending on
the $\tan\beta$ value .

\begin{figure}
\setlength{\unitlength}{1mm}
\begin{picture}(140,200)(30,30)
\put(30,0){\includegraphics{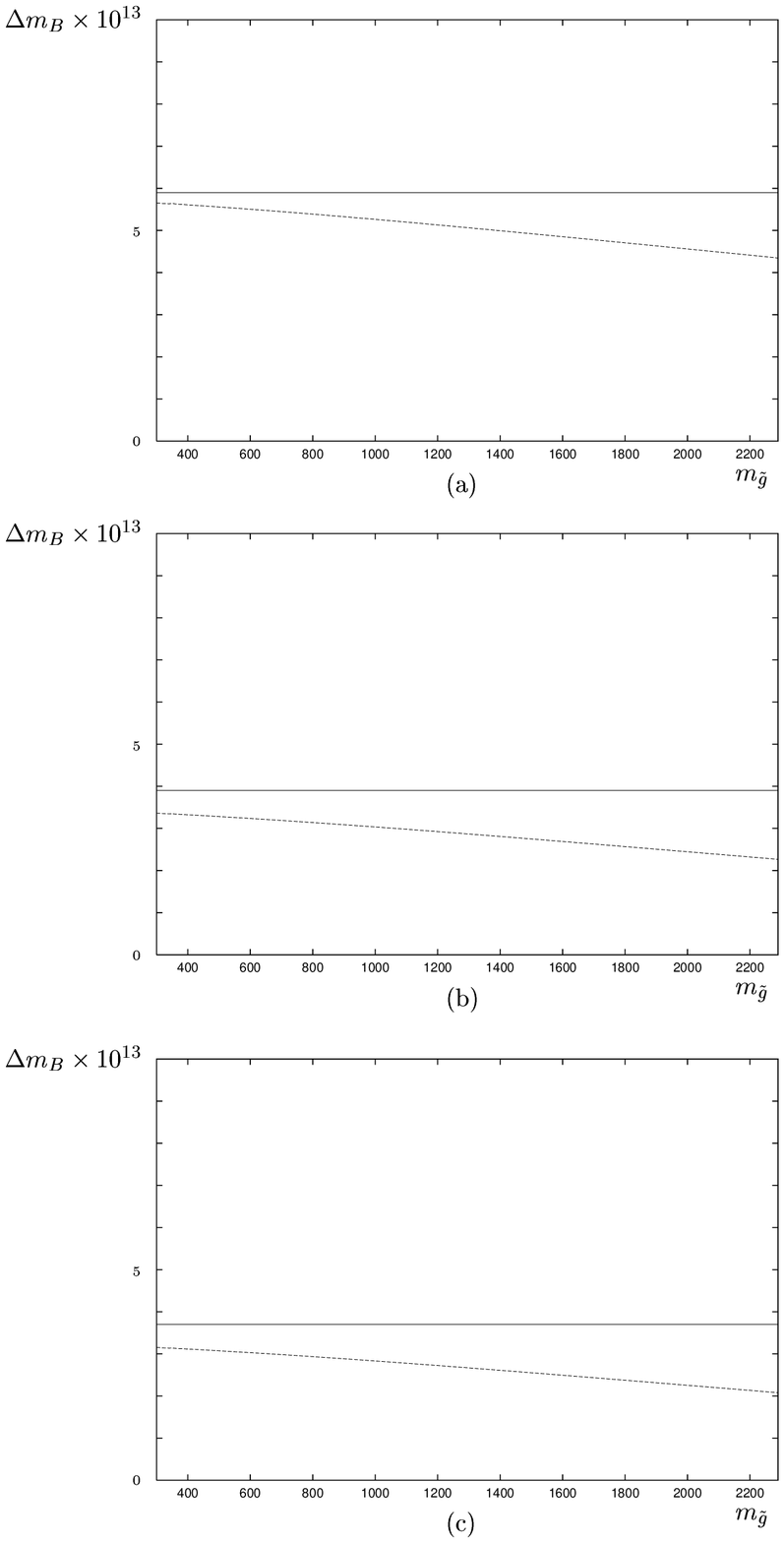}}
\end{picture}
\caption[]{The $\Delta m_{B}$ versus the gluino mass with
$\tan\xi_{\tilde{U}^I}=\tan\zeta_{\tilde{B}}=
\tan\zeta_{\tilde{D}}=0$. and (a)$\tan\beta=1$, (b)$\tan\beta=5$,
(c)$\tan\beta=30$. The dot-line corresponds to the results including the
gluino-corrections and solid-line corresponds to that without the
gluino-corrections. The other parameters are taken as in the text.}
\label{fig9}
\end{figure}

In Fig.\ref{fig9}, we plot the dependence of $\Delta m_B$ on gluino mass
$m_{\tilde{g}}$ with $m_{\chi_1^-}=110$ GeV,
$m_{\tilde{U}_1^3}=150$ GeV,
$\tan\xi_{\tilde{U}^I}=\tan\zeta_{\tilde{D}}=\tan\zeta_{\tilde{B}}=0$
and $\tan\beta=1,5,30$. We find that $\Delta
m_B$ more sensitively depends on the gluino mass. Obviously,
the results including NLO corrections from gluino are
closer to the data than that without gluino contributions. It is
also noted that as $\tan\beta\sim 1$, the data favors heavier
gluino, i.e. $m_{\tilde{g}}$ is greater than a few TeV's. But for
$\tan\beta\geq 5$, the data favors $m_{\tilde{g}}\sim 400\sim 600$ GeV
and the dependence of $\Delta m_B$ is no longer sensitive to
$\tan\beta$.

\begin{figure}
\setlength{\unitlength}{1mm}
\begin{picture}(140,200)(30,30)
\put(30,0){\includegraphics{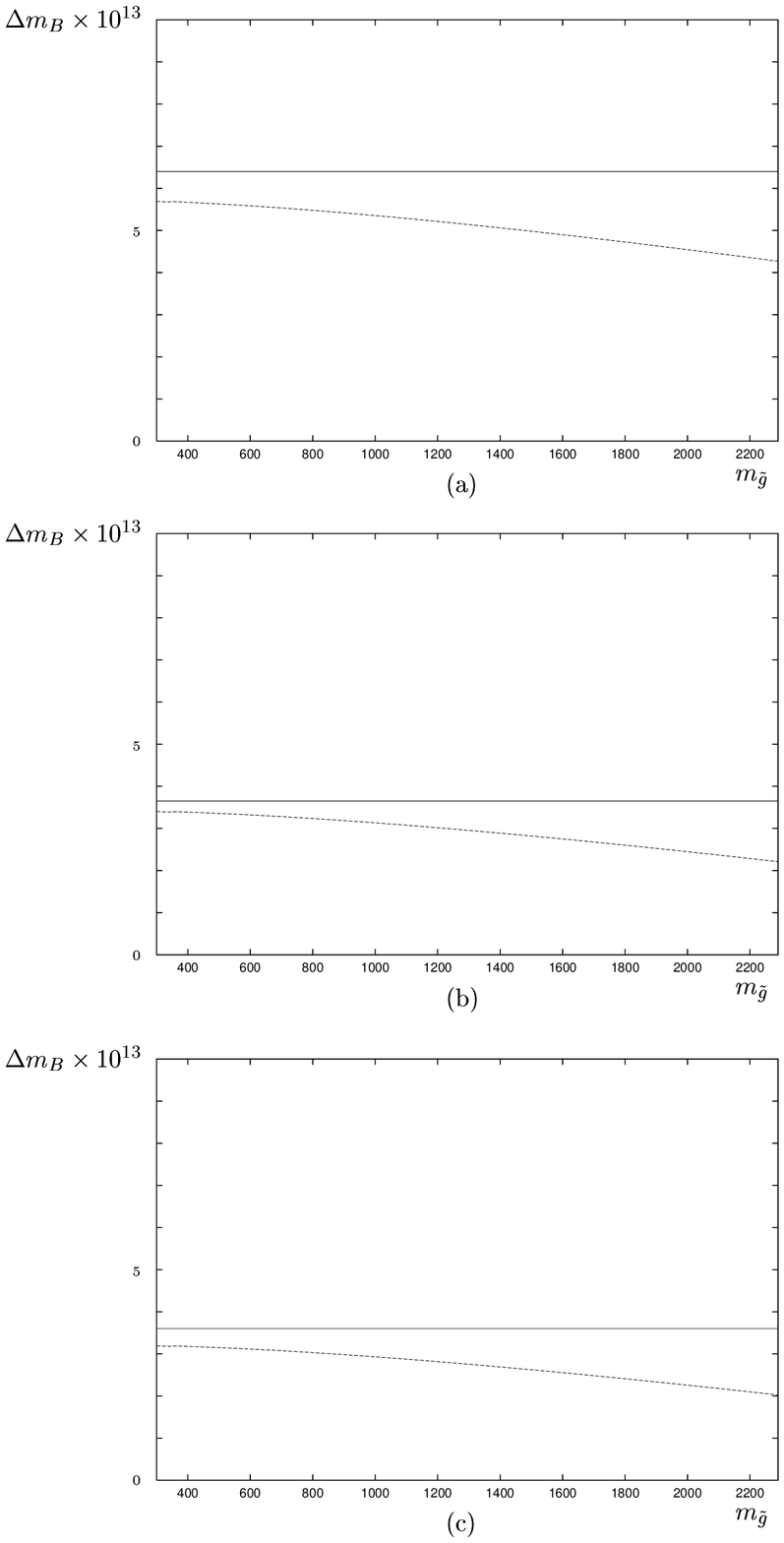}}
\end{picture}
\caption[]{The $\Delta m_{B}$ versus the gluino mass with
$\tan\xi_{\tilde{U}^I}=\tan\zeta_{\tilde{B}}=
\tan\zeta_{\tilde{D}}=0.1$. and (a)$\tan\beta=1$, (b)$\tan\beta=5$,
(c)$\tan\beta=30$. The dot-line corresponds to the results
including the
gluino-corrections and solid-line corresponds to that without the
gluino-corrections. The other parameters are taken as in the text.}
\label{fig10}
\end{figure}

For the case that the mixing between left- and right- squarks is
non-zero, we take $\tan\xi_{\tilde{U}^I}=\tan\zeta_{\tilde{D}}=
\tan\zeta_{\tilde{B}}=0.1$ and plot  $\Delta m_{B}$ versus
$m_{\tilde{g}}$ in Fig.\ref{fig10}. The situation is very
similar to the discussions given above.

\begin{figure}
\setlength{\unitlength}{1mm}
\begin{picture}(140,200)(30,30)
\put(30,0){\includegraphics{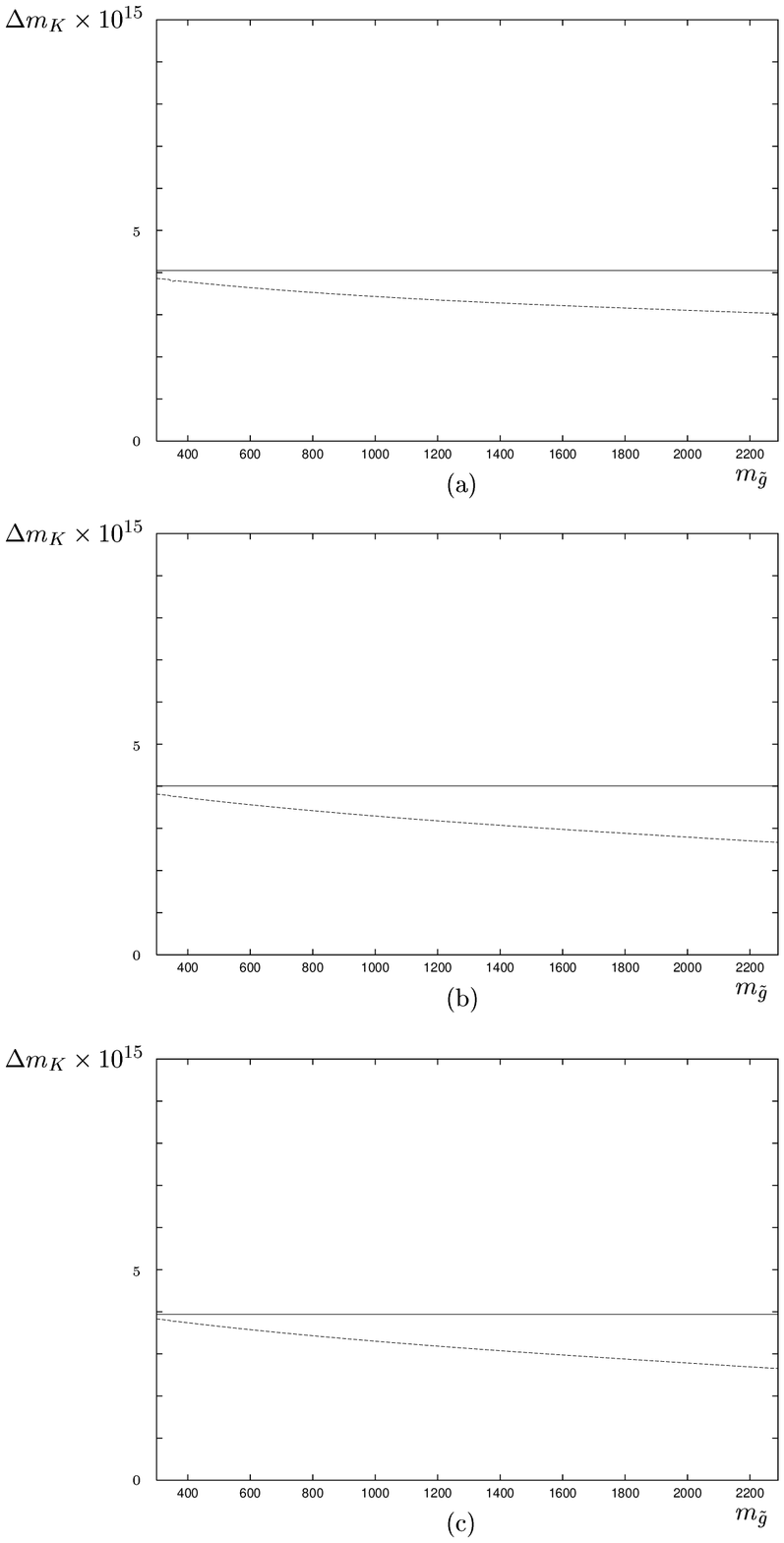}}
\end{picture}
\caption[]{The $\Delta m_{K}$ versus the gluino mass with (a)
$\tan\beta=1.5$, (b)$\tan\beta=5$, (c)$\tan\beta=30$, where
$\tan\xi_{\tilde{U}^I}=\tan\zeta_{\tilde{B}}=\tan\zeta_{\tilde{D}}=0$.
 The dot-line corresponds to the results including the
gluino-corrections and solid-line corresponds to that without the
gluino-corrections. The other parameters are taken as in the text.}
\label{fig11}
\end{figure}

Now, we turn to the $K^0-\overline{K}^{0}$ mixing. In Fig.\ref{fig11},
we plot the $\Delta m_{K}$ versus the mass
of gluino with other parameters being set as
$m_{\chi_1^-}=413{\rm GeV}$, $m_{\tilde{U}^3_1}=150{\rm GeV}$,
$\tan\xi_{\tilde{U}^I}=\tan\zeta_{\tilde{D}}=\tan\zeta_{\tilde{B}}=0$
and $\tan\beta=1.5,5,30$ respectively. From these figures, we find that
$\Delta m_{K}$ is modified obviously when $m_{\tilde{g}}$ varies.

It is noted that we assumed $m_{\tilde{B}_1}\gg m_{\tilde{U}^3_1}$
in the above numerical computations, at present a possibility
$m_{\tilde{B}_1}\sim m_{\tilde{U}^3_1}$ is widely considered. We
have re-calculated the resultant dependence of $\Delta m_B$ and
$\Delta m_K$ on the gluino mass with $m_{\tilde{B}_1}
=m_{\tilde{U}^3_1}=150$GeV
as input. Our numerical results show that for smaller gluino mass
of about 300 GeV, the changes from that with larger $m_{\tilde{B}_1}$
are very small and completely negligible. When the gluino mass
turns larger, we find that the curves drop a bit faster.
Concretely, as gluino mass reaches a region of about 3 TeV, the value
of $\Delta m_B$ is about 1.5\% smaller than that with
$m_{\tilde{B}_1}=4.7$ TeV, and $\Delta m_K$ is only suppressed by
a factor of less than 1\%.

From the above numerical analysis, we find that the gluino corrections
cannot be neglected even when the gluino mass is very heavy. In the
general case, the gluino mediated corrections depend on the choice
of the parameter space and must be taken into account seriously.

\section{Conclusions}

We analyze the gluino mediated corrections to
$B^0-\overline{B}^0$ mixing systematically up to the Next-to-Leading Order
in the supersymmetric extension of standard model with minimal flavor violation.
In the general case, the gluino contributions are evident and cannot
be neglected in the NLO QCD corrections. Our technique can be used in
other rare B processes such as $b\rightarrow s\gamma$, $b\rightarrow s g$,
$b\rightarrow sZ$ and $b\rightarrow se^+e^-$ in the total supersymmetric
calculations. After the systematic analysis on the B- and K- systems,
we can expect to extract
some constraints on the supersymmetric parameter space.

\vspace{1.0cm}
\noindent {\Large{\bf Acknowledgments}}

This work is partially supported by the National Natural Science
Foundation of China.
We also thank Prof. C.-H. Chang for helpful discussions.
\appendix
{\Large{\bf Appendix}}

\section{The functions in the one-loop
calculations\label{oloop}}

The functions in the one loop integrals are
\begin{eqnarray}
&&f_a(x_1,x_2,x_3,x_4)=\sum\limits_{i=1}^{4}\frac{x_i^2\ln x_i}
{\prod\limits_{j\neq i}(x_j-x_i)},\nonumber \\
&&f_b(x_1,x_2,x_3,x_4)=\sum\limits_{i=1}^{4}\frac{x_i\ln x_i}
{\prod\limits_{j\neq i}(x_j-x_i)},
\label{fab}
\end{eqnarray}
when $x_3=x_4=1$, they turn back to
\begin{eqnarray}
&&f_a(x_1,x_2,1,1)=\Big(\frac{x_1^2\ln x_1}{(x_2-x_1)(1-x_1)^2}+
\frac{x_2^2\ln x_2}{(x_1-x_2)(1-x_2)^2}\nonumber \\
&&\hspace{2.5cm}+\frac{1}{(1-x_1)(1-x_2)},\nonumber \\
&&f_b(x_1,x_2,1,1)=\Big(\frac{x_1\ln x_1}{(x_2-x_1)(1-x_1)^2}+
\frac{x_2\ln x_2}{(x_1-x_2)(1-x_2)^2}\nonumber \\
&&\hspace{2.5cm}+\frac{1}{(1-x_1)(1-x_2)}.\nonumber \\
\end{eqnarray}

\section{The integrand functions of two loop\label{tloop}}

In this appendix, we give some necessary integrals that are
used in the context. The five topological diagrams are
drawn in Fig.\ref{fig7}, the integrand functions are
defined as
\begin{eqnarray}
&&I_{(i),0}(m_{1}^{2},m_{2}^{2},m_{3}^{2},m_{4}^{2},
m_{5}^{2},m_{6}^{2},m_{7}^{2})=\int\frac{d^{D}k}{
(2\pi)^{D}}\frac{d^{D}q}{2\pi)^{D}}\frac{1}{A_{(i)}}\; ,
\nonumber \\
&&I_{(i),1}^{a}(m_{1}^{2},m_{2}^{2},m_{3}^{2},m_{4}^{2},
m_{5}^{2},m_{6}^{2},m_{7}^{2})=\int\frac{d^{D}k}{
(2\pi)^{D}}\frac{d^{D}q}{2\pi)^{D}}\frac{k^{2}}{A_{(i)}}\; ,
\nonumber \\
&&I_{(i),1}^{b}(m_{1}^{2},m_{2}^{2},m_{3}^{2},m_{4}^{2},
m_{5}^{2},m_{6}^{2},m_{7}^{2})=\int\frac{d^{D}k}{
(2\pi)^{D}}\frac{d^{D}q}{2\pi)^{D}}\frac{q^{2}}{A_{(i)}}\; ,
\nonumber \\
&&I_{(i),1}^{c}(m_{1}^{2},m_{2}^{2},m_{3}^{2},m_{4}^{2},
m_{5}^{2},m_{6}^{2},m_{7}^{2})=\int\frac{d^{D}k}{
(2\pi)^{D}}\frac{d^{D}q}{2\pi)^{D}}\frac{(k+q)^{2}}
{A_{(i)}}\; ,
\nonumber \\
&&I_{(i),2}^{a}(m_{1}^{2},m_{2}^{2},m_{3}^{2},m_{4}^{2},
m_{5}^{2},m_{6}^{2},m_{7}^{2})=\int\frac{d^{D}k}{
(2\pi)^{D}}\frac{d^{D}q}{2\pi)^{D}}\frac{k^{4}}{A_{(i)}}\; ,
\nonumber \\
&&I_{(i),2}^{b}(m_{1}^{2},m_{2}^{2},m_{3}^{2},m_{4}^{2},
m_{5}^{2},m_{6}^{2},m_{7}^{2})=\int\frac{d^{D}k}{
(2\pi)^{D}}\frac{d^{D}q}{2\pi)^{D}}\frac{q^{4}}{A_{(i)}}\; ,
\nonumber \\
&&I_{(i),2}^{c}(m_{1}^{2},m_{2}^{2},m_{3}^{2},m_{4}^{2},
m_{5}^{2},m_{6}^{2},m_{7}^{2})=\int\frac{d^{D}k}{
(2\pi)^{D}}\frac{d^{D}q}{2\pi)^{D}}\frac{(k+q)^{4}}
{A_{(i)}}\; ,
\nonumber \\
&&I_{(i),2}^{d}(m_{1}^{2},m_{2}^{2},m_{3}^{2},m_{4}^{2},
m_{5}^{2},m_{6}^{2},m_{7}^{2})=\int\frac{d^{D}k}{
(2\pi)^{D}}\frac{d^{D}q}{2\pi)^{D}}\frac{k^{2}q^{2}}
{A_{(i)}}\; ,
\nonumber \\
&&I_{(i),2}^{e}(m_{1}^{2},m_{2}^{2},m_{3}^{2},m_{4}^{2},
m_{5}^{2},m_{6}^{2},m_{7}^{2})=\int\frac{d^{D}k}{
(2\pi)^{D}}\frac{d^{D}q}{2\pi)^{D}}\frac{k^{2}(k+q)^{2}}
{A_{(i)}}\; ,
\nonumber \\
&&I_{(i),2}^{f}(m_{1}^{2},m_{2}^{2},m_{3}^{2},m_{4}^{2},
m_{5}^{2},m_{6}^{2},m_{7}^{2})=\int\frac{d^{D}k}{
(2\pi)^{D}}\frac{d^{D}q}{2\pi)^{D}}\frac{(k+q)^{2}q^{2}}
{A_{(i)}}\; ,
\label{def-integral}
\end{eqnarray}
where the definitions of $A_{(i)}$ are
\begin{eqnarray}
&&A_{(a)}=(k^{2}-m_{1}^{2})(k^{2}-m_{2}^{2})(k^{2}-m_{3}^{
2})((k+q)^{2}-m_{4}^{2})(q^{2}-m_{5}^{2})(q^{2}-m_{6}^{2})
(q^{2}-m_{7}^{2}), \nonumber \\
&&A_{(b)}=(k^{2}-m_{1}^{2})(k^{2}-m_{2}^{2})
((k+q)^{2}-m_{3}^{
2})((k+q)^{2}-m_{4}^{2})(q^{2}-m_{5}^{2})(q^{2}-m_{6}^{2})
(q^{2}-m_{7}^{2}), \nonumber \\
&&A_{(c)}=(k^{2}-m_{1}^{2})(k^{2}-m_{2}^{2})
(k^{2}-m_{3}^{
2})(k^{2}-m_{4}^{2})(k^{2}-m_{5}^{2})((k+q)^{2}-m_{6}^{2})
(q^{2}-m_{7}^{2}), \nonumber \\
&&A_{(d)}=(k^{2}-m_{1}^{2})(k^{2}-m_{2}^{2})
(k^{2}-m_{3}^{
2})(k^{2}-m_{4}^{2})((k+q)^{2}-m_{5}^{2})(q^{2}-m_{6}^{2})
(q^{2}-m_{7}^{2}), \nonumber \\
&&A_{(e)}=(k^{2}-m_{1}^{2})(k^{2}-m_{2}^{2})
(k^{2}-m_{3}^{
2})((k+q)^{2}-m_{4}^{2})((k+q)^{2}-m_{5}^{2})
(q^{2}-m_{6}^{2})
(q^{2}-m_{7}^{2}).
\label{def-number}
\end{eqnarray}
Here, $i=a,b,c,d,e$ are the indices of the diagrams in Fig.\ref{fig7}

The loop integrals for diagram A are decomposed as
\begin{eqnarray}
&&I_{(a),2}^{a}=I_{(a),2}^{a,1}+(m_2^2+m_3^2)I_{(a),1}^{a}-m_2^2
m_3^2I_{(a),0}\; ,\nonumber \\
&&I_{(a),2}^{b}=I_{(a),2}^{b,1}+(m_6^2+m_7^2)I_{(a),1}^{b}-m_6^2
m_7^2I_{(a),0}\; ,\nonumber \\
&&I_{(a),2}^{c}=I_{(a),2}^{c,1}+m_4^2I_{(a),1}^{c}\; ,\nonumber \\
&&I_{(a),2}^{d}=I_{(a),2}^{d,1}+m_3^2I_{(a),1}^{b}+m_7^2I_{(a),1}^{a}
-m_3^2m_7^2I_{(a),0}\; ,\nonumber \\
&&I_{(a),2}^{e}=I_{(a),2}^{e,1}+m_3^2I_{(a),1}^{c}+m_4^2I_{(a),1}^{a}
-m_3^2m_4^2I_{(a),0}\; ,\nonumber \\
&&I_{(a),2}^{f}=I_{(a),2}^{f,1}+m_4^2I_{(a),1}^{b}+m_7^2I_{(a),1}^{c}
-m_4^2m_7^2I_{(a),0}\; ,\nonumber \\
&&I_{(a),1}^{a}=I_{(a),1}^{a,1}+m_3^2I_{(a),0}\; ,\nonumber \\
&&I_{(a),1}^{b}=I_{(a),1}^{b,1}+m_7^2I_{(a),0}\; ,\nonumber \\
&&I_{(a),1}^{c}=I_{(a),1}^{c,1}+m_4^2I_{(a),0}
\label{de1}
\end{eqnarray}
with
\begin{eqnarray}
&&I_{(a),2}^{a,1}=\frac{1}{(4\pi)^4}\sum\limits_{\rho=5}^{7}
\frac{m_{\rho}^2}{\prod\limits_{\sigma\neq \rho}(m_{\sigma}^2
-m_{\rho}^2)}\bigg(\ln x_{\rho\mu}\Big(\frac{1}{\varepsilon}-
\gamma_{E}+\ln(4\pi)\Big)-{\cal S}L_{i_2}(x_{1\rho},x_{4\rho})
\nonumber \\
&&\hspace{1.2cm}+\Big(3-\gamma_{E}+\ln(4\pi)\Big)\ln x_{\rho\mu}
-\ln^2x_{\rho\mu}\bigg)\; ,
\nonumber \\
&&I_{(a),2}^{b,1}=\frac{1}{(4\pi)^4}\sum\limits_{\rho=1}^{3}
\frac{m_{\rho}^2}{\prod\limits_{\sigma\neq \rho}(m_{\sigma}^2
-m_{\rho}^2)}\bigg(\ln x_{\rho\mu}\Big(\frac{1}{\varepsilon}-
\gamma_{E}+\ln(4\pi)\Big)-{\cal S}L_{i_2}(x_{4\rho},x_{5\rho})
\nonumber \\
&&\hspace{1.2cm}+\Big(3-\gamma_{E}+\ln(4\pi)\Big)\ln x_{\rho\mu}
-\ln^2x_{\rho\mu}\bigg)\; ,
\nonumber \\
&&I_{(a),2}^{c,1}=\frac{1}{(4\pi)^4}\sum\limits_{\rho_1=1}^{3}
\sum\limits_{\rho_2=5}^{7}\frac{m_{\rho_1}^4m_{\rho_2}^2+m_{\rho_1}^2
m_{\rho_2}^4}{\prod\limits_{\sigma_1\neq \rho_1}(m_{\sigma_1}^2-
m_{\rho_1}^2)\prod\limits_{\sigma_2\neq \rho_2}(m_{\sigma_2}^2-
m_{\rho_2}^2)}\bigg(\Big(\frac{1}{\varepsilon}-\gamma_{E}+\ln(4\pi)\Big)
\ln(x_{\rho_1\mu}x_{\rho_2\mu})\nonumber \\
&&\hspace{1.2cm}+\Big(2-\gamma_{E}+\ln 4\pi\Big)
\ln(x_{\rho_1\mu}x_{\rho_2\mu})-\frac{1}{2}\ln^2(x_{\rho_1\mu}
x_{\rho_2\mu})\bigg)\; ,
\nonumber \\
&&I_{(a),2}^{d,1}=\frac{1}{m_{5}^2-m_{6}^2}\frac{1}{(4\pi)^2}
\sum\limits_{\rho=1}^{2}\frac{m_{\rho}^2}{\prod\limits_{\sigma\neq \rho}
(m_{\sigma}^2-m_{\rho}^2)}\Big({\cal S}L_{i_2}(x_{4\rho},x_{5\rho})
-{\cal S}L_{i_2}(x_{4\rho},x_{6\rho})\Big)\; ,
\nonumber \\
&&I_{(a),2}^{e,1}=\frac{1}{(4\pi)^4}\sum\limits_{\rho_1=1}^{2}
\sum\limits_{\rho_2=5}^{7}\frac{m_{\rho_1}^2m_{\rho_2}^2}
{\prod\limits_{\sigma_1\neq \rho_1}(m_{\sigma_1}^2-
m_{\rho_1}^2)\prod\limits_{\sigma_2\neq \rho_2}(m_{\sigma_2}^2-
m_{\rho_2}^2)}\bigg(-\Big(\frac{1}{\varepsilon}-\gamma_{E}+\ln(4\pi)\Big)
\ln(x_{\rho_1\mu}x_{\rho_2\mu})\nonumber \\
&&\hspace{1.2cm}-\Big(2-\gamma_{E}+\ln 4\pi\Big)
\ln(x_{\rho_1\mu}x_{\rho_2\mu})+\frac{1}{2}\ln^2(x_{\rho_1\mu}
x_{\rho_2\mu})\bigg)\; ,
\nonumber \\
&&I_{(a),2}^{f,1}=\frac{1}{(4\pi)^4}\sum\limits_{\rho_1=1}^{3}
\sum\limits_{\rho_2=5}^{6}\frac{m_{\rho_1}^2m_{\rho_2}^2}
{\prod\limits_{\sigma_1\neq \rho_1}(m_{\sigma_1}^2-
m_{\rho_1}^2)\prod\limits_{\sigma_2\neq \rho_2}(m_{\sigma_2}^2-
m_{\rho_2}^2)}\bigg(-\Big(\frac{1}{\varepsilon}-\gamma_{E}+\ln(4\pi)\Big)
\ln(x_{\rho_1\mu}x_{\rho_2\mu})\nonumber \\
&&\hspace{1.2cm}-\Big(2-\gamma_{E}+\ln 4\pi\Big)
\ln(x_{\rho_1\mu}x_{\rho_2\mu})+\frac{1}{2}\ln^2(x_{\rho_1\mu}
x_{\rho_2\mu})\bigg)\; ,
\nonumber \\
&&I_{(a),1}^{a,1}=-\frac{1}{m_{1}^2-m_{2}^2}\frac{1}{(4\pi)^4}
\sum\limits_{\rho=5}^{7}\frac{1}{\prod\limits_{\sigma\neq \rho}
(m_{\sigma}^2-m_{\rho}^2)}\Big(m_{1}^2{\cal S}L_{i_2}(x_{41},
x_{\rho 1})-m_{2}^2{\cal S}L_{i_2}(x_{42},x_{\rho 2})\Big)\; ,
\nonumber \\
&&I_{(a),1}^{b,1}=-\frac{1}{m_{5}^2-m_{6}^2}\frac{1}{(4\pi)^4}
\sum\limits_{\rho=1}^{3}\frac{m_{\rho}^2}{\prod\limits_{\sigma\neq \rho}
(m_{\sigma}^2-m_{\rho}^2)}\Big({\cal S}L_{i_2}(x_{4\rho},x_{5\rho})
-{\cal S}L_{i_2}(x_{4\rho},x_{6\rho})\Big)\; ,
\nonumber \\
&&I_{(a),1}^{c,1}=\frac{1}{2(4\pi)^4}\sum\limits_{\rho_1=1}^{3}
\sum\limits_{\rho_2=5}^{7}\frac{m_{\rho_1}^2m_{\rho_2}^2}
{\prod\limits_{\sigma_1\neq \rho_1}(m_{\sigma_1}^2-
m_{\rho_1}^2)\prod\limits_{\sigma_2\neq \rho_2}(m_{\sigma_2}^2-
m_{\rho_2}^2)}\ln^2(x_{\rho_1\mu}x_{\rho_2\mu})\; ,
\nonumber \\
&&I_{(a),0}=-\frac{1}{(4\pi)^4}\sum\limits_{\rho_1=1}^{3}
\sum\limits_{\rho_2=5}^{7}\frac{m_{\rho_1}^2}
{\prod\limits_{\sigma_1\neq \rho_1}(m_{\sigma_1}^2-
m_{\rho_1}^2)\prod\limits_{\sigma_2\neq \rho_2}(m_{\sigma_2}^2-
m_{\rho_2}^2)}{\cal S}L_{i_2}(x_{4\rho_1},x_{\rho_2\rho_1})\; .
\label{ia}
\end{eqnarray}
For the diagrams of class B, the loop integrals are decomposed as
\begin{eqnarray}
&&I_{(b),2}^{a}=I_{(b),2}^{a,1}+(m_1^2+m_2^2)I_{(b),1}^{a}-m_1^2
m_2^2I_{(b),0}\; ,\nonumber \\
&&I_{(b),2}^{b}=I_{(b),2}^{b,1}+(m_6^2+m_7^2)I_{(b),1}^{b}-m_6^2
m_7^2I_{(b),0}\; ,\nonumber \\
&&I_{(b),2}^{c}=I_{(b),2}^{c,1}+(m_3^2+m_4^2)I_{(b),1}^{c}-m_3^2
m_4^2I_{(b),0}\; ,\nonumber \\
&&I_{(b),2}^{d}=I_{(b),2}^{d,1}+m_2^2I_{(b),1}^{b}+m_7^2I_{(b),1}^{a}
-m_2^2m_7^2I_{(b),0}\; ,\nonumber \\
&&I_{(b),2}^{e}=I_{(b),2}^{e,1}+m_2^2I_{(b),1}^{c}+m_4^2I_{(b),1}^{a}
-m_2^2m_4^2I_{(b),0}\; ,\nonumber \\
&&I_{(b),2}^{f}=I_{(b),2}^{f,1}+m_4^2I_{(b),1}^{b}+m_7^2I_{(b),1}^{c}
-m_4^2m_7^2I_{(b),0}\; ,\nonumber \\
&&I_{(b),1}^{a}=I_{(b),1}^{a,1}+m_2^2I_{(b),0}\; ,\nonumber \\
&&I_{(b),1}^{b}=I_{(b),1}^{b,1}+m_7^2I_{(b),0}\; ,\nonumber \\
&&I_{(b),1}^{c}=I_{(b),1}^{c,1}+m_4^2I_{(b),0}
\label{de2}
\end{eqnarray}
and
\begin{eqnarray}
&&I_{(b),2}^{a,1}=\frac{1}{(4\pi)^4}\sum\limits_{\rho=5}^{7}
\frac{m_{\rho}^2}{\prod\limits_{\sigma\neq \rho}
(m_{\sigma}^2-m_{\rho}^2)}\bigg(\Big(\frac{1}{\varepsilon}-
\gamma_{E}+\ln(4\pi)\Big)\ln x_{\rho\mu}+\Big(2-\gamma_{E}+\ln 4\pi\Big)
\ln x_{\rho\mu}\nonumber \\
&&\hspace{1.2cm}-\frac{m_{3}^2}{2(m_3^2-m_4^2)}\ln^2(x_{3\mu}x_{\rho\mu})
+\frac{m_{4}^2}{2(m_3^2-m_4^2)}\ln^2(x_{4\mu}x_{\rho\mu})\bigg)
\; ,\nonumber \\
&&I_{(b),2}^{b,1}=-\frac{1}{(4\pi)^4}\frac{1}{(m_1^2-m_2^2)(m_3^2-m_4^2)}
\bigg(m_1^2\Big({\cal S}L_{i_2}(x_{31},x_{51})-{\cal S}L_{i_2}(x_{41}
,x_{51})\Big)\nonumber \\
&&\hspace{1.2cm}-m_2^2\Big({\cal S}L_{i_2}(x_{32},x_{52})-{\cal S}L_{i_2}
(x_{42},x_{52})\Big)\bigg)
\; ,\nonumber \\
&&I_{(b),2}^{c,1}=\frac{1}{(4\pi)^4}\sum\limits_{\rho=5}^{7}
\frac{m_{\rho}^2}{\prod\limits_{\sigma\neq \rho}
(m_{\sigma}^2-m_{\rho}^2)}\bigg(\Big(\frac{1}{\varepsilon}-
\gamma_{E}+\ln(4\pi)\Big)\ln x_{\rho\mu}+\Big(2-\gamma_{E}+\ln 4\pi\Big)
\ln x_{\rho\mu}\nonumber \\
&&\hspace{1.2cm}-\frac{m_{1}^2}{2(m_1^2-m_2^2)}\ln^2(x_{1\mu}x_{\rho\mu})
+\frac{m_{2}^2}{2(m_1^2-m_2^2)}\ln^2(x_{2\mu}x_{\rho\mu})\bigg)
\; ,\nonumber \\
&&I_{(b),2}^{d,1}=-\frac{1}{(4\pi)^4}\frac{1}{(m_3^2-m_4^2)(m_5^2-m_6^2)}
\bigg(m_5^2\Big({\cal S}L_{i_2}(x_{15},x_{35})-{\cal S}L_{i_2}(x_{15}
,x_{45})\Big)\nonumber \\
&&\hspace{1.2cm}-m_6^2\Big({\cal S}L_{i_2}(x_{16},x_{36})-{\cal S}L_{i_2}
(x_{16},x_{46})\Big)\bigg)
\; ,\nonumber \\
&&I_{(b),2}^{e,1}=\frac{1}{(4\pi)^4}\sum\limits_{\rho=5}^{7}
\frac{m_{\rho}^2}{\prod\limits_{\sigma\neq \rho}
(m_{\sigma}^2-m_{\rho}^2)}\bigg(\Big(\frac{1}{\varepsilon}-
\gamma_{E}+\ln(4\pi)\Big)\ln x_{\rho\mu}+\Big(3-\gamma_{E}+\ln 4\pi\Big)
\ln x_{\rho\mu}\nonumber \\
&&\hspace{1.2cm}-\ln^2x_{\rho\mu}-{\cal S}L_{i_2}(x_{1\rho},x_{3\rho})
\bigg)
\; ,\nonumber \\
&&I_{(b),2}^{f,1}=-\frac{1}{(4\pi)^4}\frac{1}{(m_1^2-m_2^2)(m_5^2-m_6^2)}
\bigg(m_1^2\Big({\cal S}L_{i_2}(x_{31},x_{51})-{\cal S}L_{i_2}(x_{31},
x_{61})\Big)\nonumber \\
&&\hspace{1.2cm}-m_2^2\Big({\cal S}L_{i_2}(x_{32},x_{52})-{\cal S}L_{i_2}
(x_{32},x_{62})\Big)\bigg)
\; ,\nonumber \\
&&I_{(b),1}^{a,1}=-\frac{1}{(4\pi)^4}\frac{1}{m_3^2-m_4^2}\sum\limits_{\rho=5}^{7}
\frac{m_{\rho}^2}{\prod\limits_{\sigma\neq \rho}(m_{\sigma}^2-m_{\rho}^2)}
\Big({\cal S}L_{i_2}(x_{1\rho},x_{3\rho})-{\cal S}L_{i_2}(x_{1\rho}
,x_{4\rho})\Big)
\; ,\nonumber \\
&&I_{(b),1}^{b,1}=-\frac{1}{(4\pi)^4}\frac{1}{(m_1^2-m_2^2)(m_3^2-m_4^2)
(m_5^2-m_6^2)}\bigg(m_1^2\Big({\cal S}L_{i_2}(x_{31},x_{51})-
{\cal S}L_{i_2}(x_{31},x_{61})\nonumber \\
&&\hspace{1.2cm}-{\cal S}L_{i_2}(x_{41},x_{51})+
{\cal S}L_{i_2}(x_{41},x_{61})\Big)-
m_2^2\Big({\cal S}L_{i_2}(x_{32},x_{52})-
{\cal S}L_{i_2}(x_{32},x_{62})\nonumber \\
&&\hspace{1.2cm}-{\cal S}L_{i_2}(x_{42},x_{52})+
{\cal S}L_{i_2}(x_{42},x_{62})\Big)\bigg)
\; ,\nonumber \\
&&I_{(b),1}^{c,1}=\frac{1}{(4\pi)^4}\frac{1}{m_1^2-m_2^2}
\sum\limits_{\rho=5}^{7}\frac{1}{\prod\limits_{\sigma\neq \rho}
(m_{\sigma}^2-m_{\rho}^2)}\Big(m_1^2{\cal S}L_{i_2}(x_{31},x_{\rho 1})
-m_2^2{\cal S}L_{i_2}(x_{32},x_{\rho 2})\Big)
\; ,\nonumber \\
&&I_{(b),0}=-\frac{1}{(4\pi)^4}\frac{1}{(m_1^2-m_2^2)(m_3^2-m_4^2)}
\sum\limits_{\rho=5}^{7}\frac{1}{\prod\limits_{\sigma\neq
\rho}(m_{\sigma}^2-m_{\rho}^2)}\bigg(m_1^2\Big({\cal S}L_{i_2}(x_{31},
x_{\rho 1})-{\cal S}L_{i_2}(x_{41},x_{\rho 1})\Big)\nonumber \\
&&\hspace{1.2cm}-
m_2^2\Big({\cal S}L_{i_2}(x_{32},x_{\rho 2})-{\cal S}L_{i_2}(x_{42},
x_{\rho 2})\Big)\bigg)\; .
\label{ib}
\end{eqnarray}
The reduced formulae for the self energy insertion diagrams
(class C) are:
\begin{eqnarray}
&&I_{(c),2}^{a}=I_{(c),2}^{a,1}+(m_4^2+m_5^2)I_{(c),1}^{a}-m_4^2
m_5^2I_{(c),0}\; ,\nonumber \\
&&I_{(c),2}^{b}=I_{(c),2}^{b,1}+m_7^2I_{(c),1}^{b}\; ,\nonumber \\
&&I_{(c),2}^{c}=I_{(c),2}^{c,1}+m_6^2I_{(c),1}^{c}\; ,\nonumber \\
&&I_{(c),2}^{d}=I_{(c),2}^{d,1}+m_5^2I_{(c),1}^{b}+m_7^2I_{(c),1}^{a}
-m_5^2m_7^2I_{(c),0}\; ,\nonumber \\
&&I_{(c),2}^{e}=I_{(c),2}^{e,1}+m_5^2I_{(c),1}^{c}+m_6^2I_{(c),1}^{a}
-m_5^2m_6^2I_{(c),0}\; ,\nonumber \\
&&I_{(c),2}^{f}=I_{(c),2}^{f,1}+m_6^2I_{(c),1}^{c}+m_7^2I_{(c),1}^{b}
-m_6^2m_7^2I_{(c),0}\; ,\nonumber \\
&&I_{(c),1}^{a}=I_{(c),1}^{a,1}+m_5^2I_{(c),0}\; ,\nonumber \\
&&I_{(c),1}^{b}=I_{(c),1}^{b,1}+m_7^2I_{(c),0}\; ,\nonumber \\
&&I_{(c),1}^{c}=I_{(c),1}^{c,1}+m_6^2I_{(c),0}
\label{de3}
\end{eqnarray}
and
\begin{eqnarray}
&&I_{(c),2}^{a,1}=\frac{1}{(4\pi)^4}\sum\limits_{\rho=1}^{3}
\frac{m_{\rho}^2}{\prod\limits_{\sigma\neq \rho}
(m_{\sigma}^2-m_{\rho}^2)}\bigg(\Big(\frac{1}{\varepsilon}-
\gamma_{E}+\ln(4\pi)\Big)\ln x_{\rho\mu}+\Big(3-\gamma_{E}+\ln 4\pi\Big)
\ln x_{\rho\mu}\nonumber \\
&&\hspace{1.2cm}-\ln^2x_{\rho\mu}-{\cal S}L_{i_2}(x_{6\rho},x_{7\rho})
\bigg)\; ,\nonumber \\
&&I_{(c),2}^{b,1}=\frac{1}{(4\pi)^4}\sum\limits_{\rho=1}^{5}
\frac{m_{\rho}^4m_6^2+m_{\rho}^2m_{6}^4}{\prod\limits_{\sigma\neq \rho}
(m_{\sigma}^2-m_{\rho}^2)}\bigg(\Big(\frac{1}{\varepsilon}-
\gamma_{E}+\ln(4\pi)\Big)\ln(x_{\rho\mu}x_{6\mu})
+\Big(2-\gamma_{E}+\ln 4\pi\Big)\ln(x_{\rho\mu}x_{6\mu})\nonumber \\
&&\hspace{1.2cm}-\frac{1}{2}\ln^2(x_{\rho\mu}x_{6\mu})\bigg)\; ,\nonumber \\
&&I_{(c),2}^{c,1}=\frac{1}{(4\pi)^4}\sum\limits_{\rho=1}^{5}
\frac{m_{\rho}^4m_7^2+m_{\rho}^2m_{7}^4}{\prod\limits_{\sigma\neq \rho}
(m_{\sigma}^2-m_{\rho}^2)}\bigg(\Big(\frac{1}{\varepsilon}-
\gamma_{E}+\ln(4\pi)\Big)\ln(x_{\rho\mu}x_{7\mu})
+\Big(2-\gamma_{E}+\ln 4\pi\Big)\ln(x_{\rho\mu}x_{7\mu})\nonumber \\
&&\hspace{1.2cm}-\frac{1}{2}\ln^2(x_{\rho\mu}x_{7\mu})\bigg)\; ,\nonumber \\
&&I_{(c),2}^{d,1}=-\frac{1}{(4\pi)^4}\sum\limits_{\rho=1}^{4}
\frac{m_{\rho}^2m_6^2}{\prod\limits_{\sigma\neq \rho}
(m_{\sigma}^2-m_{\rho}^2)}\bigg(\Big(\frac{1}{\varepsilon}-
\gamma_{E}+\ln(4\pi)\Big)\ln(x_{\rho\mu}x_{6\mu})
+\Big(2-\gamma_{E}+\ln 4\pi\Big)\ln(x_{\rho\mu}x_{6\mu})\nonumber \\
&&\hspace{1.2cm}-\frac{1}{2}\ln^2(x_{\rho\mu}x_{6\mu})\bigg)\; ,\nonumber \\
&&I_{(c),2}^{e,1}=-\frac{1}{(4\pi)^4}\sum\limits_{\rho=1}^{4}
\frac{m_{\rho}^2m_7^2}{\prod\limits_{\sigma\neq \rho}
(m_{\sigma}^2-m_{\rho}^2)}\bigg(\Big(\frac{1}{\varepsilon}-
\gamma_{E}+\ln(4\pi)\Big)\ln(x_{\rho\mu}x_{7\mu})
+\Big(2-\gamma_{E}+\ln 4\pi\Big)\ln(x_{\rho\mu}x_{7\mu})\nonumber \\
&&\hspace{1.2cm}-\frac{1}{2}\ln^2(x_{\rho\mu}x_{7\mu})\bigg)\; ,\nonumber \\
&&I_{(c),2}^{f,1}=0\; ,\nonumber \\
&&I_{(c),1}^{a,1}=-\frac{1}{(4\pi)^4}\sum\limits_{\rho=1}^{4}
\frac{m_{\rho}^2}{\prod\limits_{\sigma\neq \rho}(m_{\sigma}^2-m_{\rho}^2)}
\bigg(\Big(\frac{1}{\varepsilon}-
\gamma_{E}+\ln(4\pi)\Big)\ln x_{\rho\mu}+\Big(3-\gamma_{E}+\ln 4\pi\Big)
\ln x_{\rho\mu}\nonumber \\
&&\hspace{1.2cm}-\ln^2x_{\rho\mu}-{\cal S}L_{i_2}(x_{6\rho},x_{7\rho})
\bigg)\; ,\nonumber \\
&&I_{(c),1}^{b,1}=\frac{1}{(4\pi)^4}\sum\limits_{\rho=1}^{5}
\frac{m_{\rho}^2m_6^2}{\prod\limits_{\sigma\neq \rho}
(m_{\sigma}^2-m_{\rho}^2)}\bigg(\Big(\frac{1}{\varepsilon}-
\gamma_{E}+\ln(4\pi)\Big)\ln(x_{\rho\mu}x_{6\mu})
+\Big(2-\gamma_{E}+\ln 4\pi\Big)\ln(x_{\rho\mu}x_{6\mu})\nonumber \\
&&\hspace{1.2cm}-\frac{1}{2}\ln^2(x_{\rho\mu}x_{6\mu})\bigg)\; ,\nonumber \\
&&I_{(c),1}^{c,1}=\frac{1}{(4\pi)^4}\sum\limits_{\rho=1}^{5}
\frac{m_{\rho}^2m_7^2}{\prod\limits_{\sigma\neq \rho}
(m_{\sigma}^2-m_{\rho}^2)}\bigg(\Big(\frac{1}{\varepsilon}-
\gamma_{E}+\ln(4\pi)\Big)\ln(x_{\rho\mu}x_{7\mu})
+\Big(2-\gamma_{E}+\ln 4\pi\Big)\ln(x_{\rho\mu}x_{7\mu})\nonumber \\
&&\hspace{1.2cm}-\frac{1}{2}\ln^2(x_{\rho\mu}x_{7\mu})\bigg)\; ,\nonumber \\
&&I_{(c),0}=\frac{1}{(4\pi)^4}\sum\limits_{\rho=1}^{5}
\frac{m_{\rho}^2}{\prod\limits_{\sigma\neq \rho}(m_{\sigma}^2-m_{\rho}^2)}
\bigg(\Big(\frac{1}{\varepsilon}-
\gamma_{E}+\ln(4\pi)\Big)\ln x_{\rho\mu}+\Big(3-\gamma_{E}+\ln 4\pi\Big)
\ln x_{\rho\mu}\nonumber \\
&&\hspace{1.2cm}-\ln^2x_{\rho\mu}-{\cal S}L_{i_2}(x_{6\rho},x_{7\rho})
\bigg)\; .
\label{ic}
\end{eqnarray}
In a similar way, the formulae for vertex insertion
diagrams (class D) are decomposed into
\begin{eqnarray}
&&I_{(d),2}^{a}=I_{(d),2}^{a,1}+(m_3^2+m_4^2)I_{(d),1}^{a}-m_3^2
m_4^2I_{(d),0}\; ,\nonumber \\
&&I_{(d),2}^{b}=I_{(d),2}^{b,1}+(m_6^2+m_7^2)I_{(d),1}^{b}
-m_6^2m_7^2I_{(d),0}\; ,\nonumber \\
&&I_{(d),2}^{c}=I_{(d),2}^{c,1}+m_5^2I_{(d),1}^{c}\; ,\nonumber \\
&&I_{(d),2}^{d}=I_{(d),2}^{d,1}+m_4^2I_{(d),1}^{b}+m_7^2I_{(d),1}^{a}
-m_4^2m_7^2I_{(d),0}\; ,\nonumber \\
&&I_{(d),2}^{e}=I_{(d),2}^{e,1}+m_4^2I_{(d),1}^{c}+m_5^2I_{(d),1}^{a}
-m_4^2m_5^2I_{(d),0}\; ,\nonumber \\
&&I_{(d),2}^{f}=I_{(d),2}^{f,1}+m_5^2I_{(d),1}^{b}+m_7^2I_{(d),1}^{c}
-m_5^2m_7^2I_{(d),0}\; ,\nonumber \\
&&I_{(d),1}^{a}=I_{(d),1}^{a,1}+m_4^2I_{(d),0}\; ,\nonumber \\
&&I_{(d),1}^{b}=I_{(d),1}^{b,1}+m_7^2I_{(d),0}\; ,\nonumber \\
&&I_{(d),1}^{c}=I_{(d),1}^{c,1}+m_5^2I_{(d),0}
\label{de4}
\end{eqnarray}
with
\begin{eqnarray}
&&I_{(d),2}^{a,1}=\frac{1}{(4\pi)^4}\frac{1}{(m_1^2-m_2^2)(m_6^2-m_7^2)}
\bigg(m_1^2\Big({\cal S}L_{i_2}(x_{51},x_{61})-{\cal S}L_{i_2}(x_{51}
,x_{71})\Big)\nonumber \\
&&\hspace{1.2cm}-m_2^2\Big({\cal S}L_{i_2}(x_{52},x_{62})-{\cal
S}L_{i_2}(x_{52},x_{72})\Big)\bigg)\; ,\nonumber \\
&&I_{(d),2}^{b,1}=-\frac{1}{(4\pi)^4}\sum\limits_{\rho=1}^{4}
\frac{m_{\rho}^2m_5^2}{\prod\limits_{\sigma\neq \rho}
(m_{\sigma}^2-m_{\rho}^2)}\bigg(\Big(\frac{1}{\varepsilon}-
\gamma_{E}+\ln(4\pi)\Big)\ln(x_{\rho\mu}x_{5\mu})
+\Big(2-\gamma_{E}+\ln 4\pi\Big)\ln(x_{\rho\mu}x_{5\mu})\nonumber \\
&&\hspace{1.2cm}+\frac{1}{2}\ln^2(x_{\rho\mu}x_{5\mu})\bigg)\; ,\nonumber \\
&&I_{(d),2}^{c,1}=-\frac{1}{(4\pi)^4}\frac{1}{m_6^2-m_7^2}
\sum\limits_{\rho=1}^{4}\frac{1}{\prod\limits_{\sigma\neq \rho}
(m_{\sigma}^2-m_{\rho}^2)}\bigg(\Big(m_{\rho}^2m_6^2(m_{\rho}^2+m_6^2)
-m_{\rho}^2m_7^2(m_{\rho}^2+m_7^2)\Big)\Big(\frac{1}{\varepsilon}-
\gamma_{E}\nonumber \\
&&\hspace{1.2cm}+\ln(4\pi)\Big)\ln x_{\rho\mu}
+m_{\rho}^2m_6^2(m_{\rho}^2+m_6^2)\Big((2-\gamma_{E}+\ln 4\pi)\ln(x_{\rho\mu}x_{6\mu})-\frac{1}{2}\ln^2(x_{\rho\mu}x_{6\mu})\Big)
\nonumber \\
&&\hspace{1.2cm}-m_{\rho}^2m_7^2(m_{\rho}^2+m_7^2)\Big((2-\gamma_{E}+\ln 4\pi)\ln(x_{\rho\mu}x_{7\mu})-\frac{1}{2}\ln^2(x_{\rho\mu}x_{7\mu})\Big)
\bigg)\; ,\nonumber \\
&&I_{(d),2}^{d,1}=\frac{1}{(4\pi)^4}
\sum\limits_{\rho=1}^{3}\frac{m_{\rho}^2}{\prod\limits_{\sigma\neq \rho}
(m_{\sigma}^2-m_{\rho}^2)}\bigg(\Big(\frac{1}{\varepsilon}-
\gamma_{E}+\ln(4\pi)\Big)\ln x_{\rho\mu}+\Big(3-\gamma_{E}+\ln 4\pi\Big)
\ln x_{\rho\mu}\nonumber \\
&&\hspace{1.2cm}-\ln^2x_{\rho\mu}-{\cal S}L_{i_2}(x_{5\rho},x_{6\rho})
\bigg)\; ,\nonumber \\
&&I_{(d),2}^{e,1}=\frac{1}{(4\pi)^4}
\sum\limits_{\rho=1}^{3}\frac{m_{\rho}^2}{\prod\limits_{\sigma\neq \rho}
(m_{\sigma}^2-m_{\rho}^2)}\bigg(\Big(\frac{1}{\varepsilon}-
\gamma_{E}+\ln(4\pi)\Big)\ln x_{\rho\mu}+\Big(2-\gamma_{E}+\ln 4\pi\Big)
\ln x_{\rho\mu}\nonumber \\
&&\hspace{1.2cm}-\frac{m_6^2\ln^2(x_{6\mu}x_{\rho\mu})-m_7^2
\ln^2(x_{7\mu}x_{\rho\mu})}{2(m_6^2-m_7^2)}\bigg)
\; ,\nonumber \\
&&I_{(d),2}^{f,1}=-\frac{1}{(4\pi)^4}\sum\limits_{\rho=1}^{3}
\frac{m_6^2m_{\rho}^2}{\prod\limits_{\sigma\neq \rho}
(m_{\sigma}^2-m_{\rho}^2)}\bigg(\Big(\frac{1}{\varepsilon}-
\gamma_{E}+\ln(4\pi)\Big)\ln(x_{6\mu}x_{\rho\mu})+\Big(2-\gamma_{E}+
\ln 4\pi\Big)\ln(x_{6\mu}x_{\rho\mu})\nonumber \\
&&\hspace{1.2cm}+\frac{1}{2}\ln^2(x_{6\mu}x_{\rho\mu})\bigg)
\; ,\nonumber \\
&&I_{(d),1}^{a,1}=-\frac{1}{(4\pi)^4}\frac{1}{m_6^2-m_7^2}
\sum\limits_{\rho=1}^{3}\frac{m_{\rho}^2}{\prod\limits_{\sigma\neq \rho}
(m_{\sigma}^2-m_{\rho}^2)}\Big({\cal S}L_{i_2}(x_{5\rho},x_{6\rho})
-{\cal S}L_{i_2}(x_{5\rho},x_{7\rho})\Big)\; ,\nonumber \\
&&I_{(d),1}^{b,1}=-\frac{1}{(4\pi)^4}
\sum\limits_{\rho=1}^{4}\frac{m_{\rho}^2}{\prod\limits_{\sigma\neq \rho}
(m_{\sigma}^2-m_{\rho}^2)}\bigg(\Big(\frac{1}{\varepsilon}-
\gamma_{E}+\ln(4\pi)\Big)\ln x_{\rho\mu}+\Big(3-\gamma_E+\ln 4\pi\Big)
\ln x_{\rho\mu}-\ln^2x_{\rho\mu}\bigg)\; ,\nonumber \\
&&I_{(d),1}^{c,1}=-\frac{1}{(4\pi)^4}
\sum\limits_{\rho=1}^{4}\frac{m_{\rho}^2}{\prod\limits_{\sigma\neq \rho}
(m_{\sigma}^2-m_{\rho}^2)}\bigg(\Big(\frac{1}{\varepsilon}-
\gamma_{E}+\ln(4\pi)\Big)\ln x_{\rho\mu}+\Big(2-\gamma_E+\ln 4\pi\Big)
\ln x_{\rho\mu}\nonumber \\
&&\hspace{1.2cm}-\frac{m_6^2\ln^2(x_{6\mu}x_{\rho\mu})-m_7^2
\ln^2(x_{7\mu}x_{\rho\mu})}{2(m_6^2-m_7^2)}\bigg)\; ,\nonumber \\
&&I_{(d),0}=\frac{1}{(4\pi)^4}\frac{1}{m_6^2-m_7^2}
\sum\limits_{\rho=1}^{4}\frac{m_{\rho}^2}{\prod\limits_{\sigma\neq \rho}
(m_{\sigma}^2-m_{\rho}^2)}\Big({\cal S}L_{i_2}(x_{5\rho},x_{6\rho})
-{\cal S}L_{i_2}(x_{5\rho},x_{7\rho})\Big)\; .
\label{id}
\end{eqnarray}
For the topological class E, the formulae are written as
\begin{eqnarray}
&&I_{(e),2}^{a}=I_{(e),2}^{a,1}+(m_2^2+m_3^2)I_{(e),1}^{a}-m_2^2
m_3^2I_{(e),0}\; ,\nonumber \\
&&I_{(e),2}^{b}=I_{(e),2}^{b,1}+(m_6^2+m_7^2)I_{(e),1}^{b}
-m_6^2m_7^2I_{(e),0}\; ,\nonumber \\
&&I_{(e),2}^{c}=I_{(e),2}^{c,1}+(m_4^2+m_5^2)I_{(e),1}^{c}
-m_4^2m_5^2I_{(e),0}\; ,\nonumber \\
&&I_{(e),2}^{d}=I_{(e),2}^{d,1}+m_3^2I_{(e),1}^{b}+m_7^2I_{(e),1}^{a}
-m_3^2m_7^2I_{(e),0}\; ,\nonumber \\
&&I_{(e),2}^{e}=I_{(e),2}^{e,1}+m_3^2I_{(e),1}^{c}+m_5^2I_{(e),1}^{a}
-m_3^2m_5^2I_{(e),0}\; ,\nonumber \\
&&I_{(e),2}^{f}=I_{(e),2}^{f,1}+m_5^2I_{(e),1}^{c}+m_7^2I_{(e),1}^{b}
-m_5^2m_7^2I_{(e),0}\; ,\nonumber \\
&&I_{(e),1}^{a}=I_{(e),1}^{a,1}+m_3^2I_{(e),0}\; ,\nonumber \\
&&I_{(e),1}^{b}=I_{(e),1}^{b,1}+m_7^2I_{(e),0}\; ,\nonumber \\
&&I_{(e),1}^{c}=I_{(e),1}^{c,1}+m_5^2I_{(e),0}
\label{de5}
\end{eqnarray}
and
\begin{eqnarray}
&&I_{(e),2}^{a,1}=-\frac{1}{(4\pi)^4}\frac{1}{(m_4^2-m_5^2)(m_6^2-m_7^2)}
\bigg(m_6^2\Big({\cal S}L_{i_2}(x_{16},x_{46})-
{\cal S}L_{i_2}(x_{16},x_{56})\Big)\nonumber \\
&&\hspace{1.2cm}-m_7^2\Big({\cal S}L_{i_2}(x_{17},x_{47})-
{\cal S}L_{i_2}(x_{17},x_{57})\Big)\bigg)\; ,\nonumber \\
&&I_{(e),2}^{b,1}=\frac{1}{(4\pi)^4}
\sum\limits_{\rho=1}^{3}\frac{m_{\rho}^2}{\prod\limits_{\sigma\neq \rho}
(m_{\sigma}^2-m_{\rho}^2)}\bigg(\Big(\frac{1}{\varepsilon}-
\gamma_{E}+\ln(4\pi)\Big)\ln x_{\rho\mu}+\Big(2-\gamma_E+\ln 4\pi\Big)
\ln x_{\rho\mu}\nonumber \\
&&\hspace{1.2cm}-\frac{m_4^2\ln^2(x_{4\mu}x_{\rho\mu})-
m_5^2\ln^2(x_{5\mu}x_{\rho\mu})}{2(m_4^2-m_5^2)}\bigg)\; ,\nonumber \\
&&I_{(e),2}^{c,1}=\frac{1}{(4\pi)^4}
\sum\limits_{\rho=1}^{3}\frac{m_{\rho}^2}{\prod\limits_{\sigma\neq \rho}
(m_{\sigma}^2-m_{\rho}^2)}\bigg(\Big(\frac{1}{\varepsilon}-
\gamma_{E}+\ln(4\pi)\Big)\ln x_{\rho\mu}+\Big(2-\gamma_E+\ln 4\pi\Big)
\ln x_{\rho\mu}\nonumber \\
&&\hspace{1.2cm}-\frac{m_6^2\ln^2(x_{6\mu}x_{\rho\mu})-
m_7^2\ln^2(x_{7\mu}x_{\rho\mu})}{2(m_6^2-m_7^2)}\bigg)\; ,\nonumber \\
&&I_{(e),2}^{d,1}=-\frac{1}{(4\pi)^4}\frac{1}{(m_1^2-m_2^2)(m_4^2-m_5^2)}
\bigg(m_1^2\Big({\cal S}L_{i_2}(x_{41},x_{61})-
{\cal S}L_{i_2}(x_{51},x_{61})\Big)\nonumber \\
&&\hspace{1.2cm}-m_2^2\Big({\cal S}L_{i_2}(x_{42},x_{62})-
{\cal S}L_{i_2}(x_{52},x_{62})\Big)\bigg)\; ,\nonumber \\
&&I_{(e),2}^{e,1}=-\frac{1}{(4\pi)^4}\frac{1}{(m_1^2-m_2^2)(m_6^2-m_7^2)}
\bigg(m_1^2\Big({\cal S}L_{i_2}(x_{41},x_{61})-
{\cal S}L_{i_2}(x_{41},x_{71})\Big)\nonumber \\
&&\hspace{1.2cm}-m_2^2\Big({\cal S}L_{i_2}(x_{42},x_{62})-
{\cal S}L_{i_2}(x_{42},x_{72})\Big)\bigg)\; ,\nonumber \\
&&I_{(e),2}^{f,1}=\frac{1}{(4\pi)^4}
\sum\limits_{\rho=1}^{3}\frac{m_{\rho}^2}{\prod\limits_{\sigma\neq \rho}
(m_{\sigma}^2-m_{\rho}^2)}\bigg(\Big(\frac{1}{\varepsilon}-
\gamma_{E}+\ln(4\pi)\Big)\ln x_{\rho\mu}+\Big(3-\gamma_E+\ln 4\pi\Big)
\ln x_{\rho\mu}\nonumber \\
&&\hspace{1.2cm}-\ln^2x_{\rho\mu}-{\cal S}L_{i_2}(x_{4\rho},x_{6\rho})
\bigg)\; ,\nonumber \\
&&I_{(e),1}^{a,1}=-\frac{1}{(4\pi)^4}\frac{1}{(m_1^2-m_2^2)(m_4^2-m_5^2)
(m_6^2-m_7^2)}\bigg(m_1^2\Big({\cal S}L_{i_2}(x_{41},x_{61})-
{\cal S}L_{i_2}(x_{41},x_{71})\nonumber \\
&&\hspace{1.2cm}-{\cal S}L_{i_2}(x_{51},x_{61})+
{\cal S}L_{i_2}(x_{51},x_{71})\Big)-m_2^2\Big({\cal S}L_{i_2}(x_{42}
,x_{62})-{\cal S}L_{i_2}(x_{42},x_{72})\nonumber \\
&&\hspace{1.2cm}-{\cal S}L_{i_2}(x_{52},x_{62})+
{\cal S}L_{i_2}(x_{52},x_{72})\Big)\bigg)\; ,\nonumber \\
&&I_{(e),1}^{b,1}=-\frac{1}{(4\pi)^4}\frac{1}{(m_4^2-m_5^2)}
\sum\limits_{\rho=1}^{3}\frac{m_{\rho}^2}{\prod\limits_{\sigma\neq \rho}
(m_{\sigma}^2-m_{\rho}^2)}\Big({\cal S}L_{i_2}(x_{4\rho},x_{6\rho})-
{\cal S}L_{i_2}(x_{5\rho},x_{6\rho})\Big)\; ,\nonumber \\
&&I_{(e),1}^{c,1}=-\frac{1}{(4\pi)^4}\frac{1}{(m_6^2-m_7^2)}
\sum\limits_{\rho=1}^{3}\frac{m_{\rho}^2}{\prod\limits_{\sigma\neq \rho}
(m_{\sigma}^2-m_{\rho}^2)}\Big({\cal S}L_{i_2}(x_{4\rho},x_{6\rho})-
{\cal S}L_{i_2}(x_{4\rho},x_{7\rho})\Big)\; ,\nonumber \\
&&I_{(e),0}=-\frac{1}{(4\pi)^4}\frac{1}{(m_4^2-m_5^2)(m_6^2-m_7^2)}
\sum\limits_{\rho=1}^{3}\frac{m_{\rho}^2}{\prod\limits_{\sigma\neq \rho}
(m_{\sigma}^2-m_{\rho}^2)}\Big({\cal S}L_{i_2}(x_{4\rho},x_{6\rho})
-{\cal S}L_{i_2}(x_{4\rho},x_{7\rho})\nonumber \\
&&\hspace{1.2cm}-{\cal S}L_{i_2}(x_{5\rho},x_{6\rho})
+{\cal S}L_{i_2}(x_{5\rho},x_{7\rho})\Big)\; .
\label{ie}
\end{eqnarray}
When we compute the amplitude of $\Delta B=2$ processes, the
class C and class D contain the ultraviolet divergence and we adopt the $\overline{MS}$-
scheme to remove them. After the above step, the functions that appear
in the effective Hamiltonian are
\begin{eqnarray}
&&F_{(a),2}^{a,1}=\sum\limits_{\rho=5}^{7}
\frac{x_{\rho\mu}}{\prod\limits_{\sigma\neq \rho}(x_{\sigma\mu}
-x_{\rho\mu})}\bigg(-{\cal S}L_{i_2}(x_{1\rho},x_{4\rho})
+\ln x_{\rho\mu}-\ln^2x_{\rho\mu}\bigg)\; ,
\nonumber \\
&&F_{(a),2}^{b,1}=\sum\limits_{\rho=1}^{3}
\frac{x_{\rho\mu}}{\prod\limits_{\sigma\neq \rho}(x_{\sigma\mu}
-x_{\rho\mu})}\bigg(-{\cal S}L_{i_2}(x_{4\rho},x_{5\rho})
+\ln x_{\rho\mu}-\ln^2x_{\rho\mu}\bigg)\; ,
\nonumber \\
&&F_{(a),2}^{c,1}=-\frac{1}{2}\sum\limits_{\rho_1=1}^{3}
\sum\limits_{\rho_2=5}^{7}\frac{x_{\rho_1\mu}^2x_{\rho_2\mu}+x_{\rho_1\mu}
x_{\rho_2\mu}^2}{\prod\limits_{\sigma_1\neq \rho_1}(x_{\sigma_1\mu}-
x_{\rho_1\mu})\prod\limits_{\sigma_2\neq \rho_2}(x_{\sigma_2\mu}-
x_{\rho_2\mu})}\ln^2(x_{\rho_1\mu}x_{\rho_2\mu})\; ,
\nonumber \\
&&F_{(a),2}^{d,1}=\frac{1}{x_{5\mu}-x_{6\mu}}
\sum\limits_{\rho=1}^{2}\frac{x_{\rho\mu}}{\prod\limits_{\sigma\neq \rho}
(x_{\sigma\mu}-x_{\rho\mu})}\Big({\cal S}L_{i_2}(x_{4\rho},x_{5\rho})
-{\cal S}L_{i_2}(x_{4\rho},x_{6\rho})\Big)\; ,
\nonumber \\
&&F_{(a),2}^{e,1}=\frac{1}{2}\sum\limits_{\rho_1=1}^{2}
\sum\limits_{\rho_2=5}^{7}\frac{x_{\rho_1\mu}x_{\rho_2\mu}}
{\prod\limits_{\sigma_1\neq \rho_1}(x_{\sigma_1\mu}-
x_{\rho_1\mu})\prod\limits_{\sigma_2\neq \rho_2}(x_{\sigma_2\mu}-
x_{\rho_2\mu})}\ln^2(x_{\rho_1\mu}x_{\rho_2\mu})\; ,
\nonumber \\
&&F_{(a),2}^{f,1}=\frac{1}{2}\sum\limits_{\rho_1=1}^{3}
\sum\limits_{\rho_2=5}^{6}\frac{x_{\rho_1\mu}x_{\rho_2\mu}}
{\prod\limits_{\sigma_1\neq \rho_1}(x_{\sigma_1\mu}-
x_{\rho_1\mu})\prod\limits_{\sigma_2\neq \rho_2}(x_{\sigma_2\mu}-
x_{\rho_2\mu})}\ln^2(x_{\rho_1\mu}x_{\rho_2\mu})\; ,
\nonumber \\
&&F_{(a),1}^{a,1}=-\frac{1}{x_{1\mu}-x_{2\mu}}
\sum\limits_{\rho=5}^{7}\frac{1}{\prod\limits_{\sigma\neq \rho}
(x_{\sigma\mu}-x_{\rho\mu})}\Big(x_{1\mu}{\cal S}L_{i_2}(x_{41},
x_{\rho 1})-x_{2\mu}{\cal S}L_{i_2}(x_{42},x_{\rho 2})\Big)\; ,
\nonumber \\
&&F_{(a),1}^{b,1}=-\frac{1}{x_{5\mu}-x_{6\mu}}
\sum\limits_{\rho=1}^{3}\frac{x_{\rho\mu}}{\prod\limits_{\sigma\neq \rho}
(x_{\sigma\mu}-x_{\rho\mu})}\Big({\cal S}L_{i_2}(x_{4\rho},x_{5\rho})
-{\cal S}L_{i_2}(x_{4\rho},x_{6\rho})\Big)\; ,
\nonumber \\
&&F_{(a),1}^{c,1}=\frac{1}{2}\sum\limits_{\rho_1=1}^{3}
\sum\limits_{\rho_2=5}^{7}\frac{x_{\rho_1\mu}x_{\rho_2\mu}}
{\prod\limits_{\sigma_1\neq \rho_1}(x_{\sigma_1\mu}-
x_{\rho_1\mu})\prod\limits_{\sigma_2\neq \rho_2}(x_{\sigma_2\mu}-
x_{\rho_2\mu})}\ln^2(x_{\rho_1\mu}x_{\rho_2\mu})\; ,
\nonumber \\
&&F_{(a),0}=-\sum\limits_{\rho_1=1}^{3}
\sum\limits_{\rho_2=5}^{7}\frac{x_{\rho_1\mu}}
{\prod\limits_{\sigma_1\neq \rho_1}(x_{\sigma_1\mu}-
x_{\rho_1\mu})\prod\limits_{\sigma_2\neq \rho_2}(x_{\sigma_2\mu}-
x_{\rho_2\mu})}{\cal S}L_{i_2}(x_{4\rho_1},x_{\rho_2\rho_1})\; ,
\label{ia1}
\end{eqnarray}
and
\begin{eqnarray}
&&F_{(b),2}^{a,1}=\sum\limits_{\rho=5}^{7}
\frac{x_{\rho\mu}}{\prod\limits_{\sigma\neq \rho}
(x_{\sigma\mu}-x_{\rho\mu})}\bigg(
-\frac{x_{3\mu}}{2(x_{3\mu}-x_{4\mu})}\ln^2(x_{3\mu}x_{\rho\mu})
+\frac{x_{4\mu}}{2(x_{3\mu}-x_{4\mu})}\ln^2(x_{4\mu}x_{\rho\mu})\bigg)
\; ,\nonumber \\
&&F_{(b),2}^{b,1}=-\frac{1}{(x_{1\mu}-x_{2\mu})(x_{3\mu}-x_{4\mu})}
\bigg(x_{1\mu}\Big({\cal S}L_{i_2}(x_{31},x_{51})-{\cal S}L_{i_2}(x_{41}
,x_{51})\Big)\nonumber \\
&&\hspace{1.2cm}-x_{2\mu}\Big({\cal S}L_{i_2}(x_{32},x_{52})-{\cal S}L_{i_2}
(x_{42},x_{52})\Big)\bigg)
\; ,\nonumber \\
&&F_{(b),2}^{c,1}=\sum\limits_{\rho=5}^{7}
\frac{x_{\rho\mu}}{\prod\limits_{\sigma\neq \rho}
(x_{\sigma\mu}-x_{\rho\mu})}\bigg(
-\frac{x_{1\mu}}{2(x_{1\mu}-x_{2\mu})}\ln^2(x_{1\mu}x_{\rho\mu})
+\frac{x_{2\mu}}{2(x_{1\mu}-x_{2\mu})}\ln^2(x_{2\mu}x_{\rho\mu})\bigg)
\; ,\nonumber \\
&&F_{(b),2}^{d,1}=-\frac{1}{(x_{3\mu}-x_{4\mu})(x_{5\mu}-x_{6\mu})}
\bigg(x_{5\mu}\Big({\cal S}L_{i_2}(x_{15},x_{35})-{\cal S}L_{i_2}(x_{15}
,x_{45})\Big)\nonumber \\
&&\hspace{1.2cm}-x_{6\mu}\Big({\cal S}L_{i_2}(x_{16},x_{36})-{\cal S}L_{i_2}
(x_{16},x_{46})\Big)\bigg)
\; ,\nonumber \\
&&F_{(b),2}^{e,1}=\sum\limits_{\rho=5}^{7}
\frac{x_{\rho\mu}}{\prod\limits_{\sigma\neq \rho}
(x_{\sigma\mu}-x_{\rho\mu})}\bigg(\ln x_{\rho\mu}
-\ln^2x_{\rho\mu}-{\cal S}L_{i_2}(x_{1\rho},x_{3\rho})
\bigg)\; ,
\nonumber \\
&&F_{(b),2}^{f,1}=-\frac{1}{(x_{1\mu}-x_{2\mu})(x_{5\mu}-x_{6\mu})}
\bigg(x_{1\mu}\Big({\cal S}L_{i_2}(x_{31},x_{51})-{\cal S}L_{i_2}(x_{31},
x_{61})\Big)\nonumber \\
&&\hspace{1.2cm}-x_{2\mu}\Big({\cal S}L_{i_2}(x_{32},x_{52})-{\cal S}L_{i_2}
(x_{32},x_{62})\Big)\bigg)\; ,
\nonumber \\
&&F_{(b),1}^{a,1}=-\frac{1}{x_{3\mu}-x_{4\mu}}\sum\limits_{\rho=5}^{7}
\frac{x_{\rho\mu}}{\prod\limits_{\sigma\neq \rho}(x_{\sigma\mu}-x_{\rho\mu})}
\Big({\cal S}L_{i_2}(x_{1\rho},x_{3\rho})-{\cal S}L_{i_2}(x_{1\rho}
,x_{4\rho})\Big)\; ,
\nonumber \\
&&F_{(b),1}^{b,1}=-\frac{1}{(x_{1\mu}-x_{2\mu})(x_{3\mu}-x_{4\mu})
(x_{5\mu}-x_{6\mu})}\bigg(x_{1\mu}\Big({\cal S}L_{i_2}(x_{31},x_{51})-
{\cal S}L_{i_2}(x_{31},x_{61})\nonumber \\
&&\hspace{1.2cm}-{\cal S}L_{i_2}(x_{41},x_{51})+
{\cal S}L_{i_2}(x_{41},x_{61})\Big)-
x_{2\mu}\Big({\cal S}L_{i_2}(x_{32},x_{52})-
{\cal S}L_{i_2}(x_{32},x_{62})\nonumber \\
&&\hspace{1.2cm}-{\cal S}L_{i_2}(x_{42},x_{52})+
{\cal S}L_{i_2}(x_{42},x_{62})\Big)\bigg)\; ,
\nonumber \\
&&F_{(b),1}^{c,1}=\frac{1}{x_{1\mu}-x_{2\mu}}
\sum\limits_{\rho=5}^{7}\frac{1}{\prod\limits_{\sigma\neq \rho}
(x_{\sigma\mu}-x_{\rho\mu})}\Big(x_{1\mu}{\cal S}L_{i_2}(x_{31},x_{\rho 1})
-x_{2\mu}{\cal S}L_{i_2}(x_{32},x_{\rho 2})\Big)\; ,
\nonumber \\
&&F_{(b),0}=-\frac{1}{(x_{1\mu}-x_{2\mu})(x_{3\mu}-x_{4\mu})}
\sum\limits_{\rho=5}^{7}\frac{1}{\prod\limits_{\sigma\neq
\rho}(x_{\sigma\mu}-x_{\rho\mu})}\bigg(x_{1\mu}\Big({\cal S}L_{i_2}(x_{31},
x_{\rho 1})-{\cal S}L_{i_2}(x_{41},x_{\rho 1})\Big)\nonumber \\
&&\hspace{1.2cm}-
x_{2\mu}\Big({\cal S}L_{i_2}(x_{32},x_{\rho 2})-{\cal S}L_{i_2}(x_{42},
x_{\rho 2})\Big)\bigg)\; ,
\label{ib1}
\end{eqnarray}
\begin{eqnarray}
&&F_{(c),2}^{a,1}=\sum\limits_{\rho=1}^{3}
\frac{x_{\rho\mu}}{\prod\limits_{\sigma\neq \rho}
(x_{\sigma\mu}-x_{\rho\mu})}\bigg(\ln x_{\rho\mu}
-\ln^2x_{\rho\mu}-{\cal S}L_{i_2}(x_{6\rho},x_{7\rho})
\bigg)\; ,\nonumber \\
&&F_{(c),2}^{b,1}=-\frac{1}{2}\sum\limits_{\rho=1}^{5}
\frac{x_{\rho\mu}^2x_{6\mu}+x_{\rho\mu}x_{6\mu}^2}{\prod\limits_{\sigma\neq \rho}(x_{\sigma\mu}-x_{\rho\mu})}\ln^2(x_{\rho\mu}x_{6\mu})
\; ,\nonumber \\
&&F_{(c),2}^{c,1}=-\frac{1}{2}\sum\limits_{\rho=1}^{5}
\frac{x_{\rho\mu}^2x_{7\mu}+x_{\rho\mu}x_{7\mu}^2}{\prod\limits_{\sigma\neq \rho}
(x_{\sigma\mu}-x_{\rho\mu})}\ln^2(x_{\rho\mu}x_{7\mu})
\; ,\nonumber \\
&&F_{(c),2}^{d,1}=\frac{1}{2}\sum\limits_{\rho=1}^{4}
\frac{x_{\rho\mu}x_{6\mu}}{\prod\limits_{\sigma\neq \rho}
(x_{\sigma\mu}-x_{\rho\mu})}\ln^2(x_{\rho\mu}x_{6\mu})
\; ,\nonumber \\
&&F_{(c),2}^{e,1}=\frac{1}{2}\sum\limits_{\rho=1}^{4}
\frac{x_{\rho\mu}x_{7\mu}}{\prod\limits_{\sigma\neq \rho}
(x_{\sigma\mu}-x_{\rho\mu})}\ln^2(x_{\rho\mu}x_{7\mu})\; ,\nonumber \\
&&F_{(c),2}^{f,1}=0\; ,\nonumber \\
&&F_{(c),1}^{a,1}=-\sum\limits_{\rho=1}^{4}
\frac{x_{\rho\mu}}{\prod\limits_{\sigma\neq \rho}(x_{\sigma\mu}-x_{\rho\mu})}\Big(\ln x_{\rho\mu}-\ln^2x_{\rho\mu}-
{\cal S}L_{i_2}(x_{6\rho},x_{7\rho})\Big)\; ,\nonumber \\
&&F_{(c),1}^{b,1}=-\frac{1}{2(4\pi)^4}\sum\limits_{\rho=1}^{5}
\frac{x_{\rho\mu}x_{6\mu}}{\prod\limits_{\sigma\neq \rho}
(x_{\sigma\mu}-x_{\rho\mu})}\ln^2(x_{\rho\mu}x_{6\mu})
\; ,\nonumber \\
&&F_{(c),1}^{c,1}=-\frac{1}{2}\sum\limits_{\rho=1}^{5}
\frac{x_{\rho\mu}x_{7\mu}}{\prod\limits_{\sigma\neq \rho}
(x_{\sigma\mu}-x_{\rho\mu})}\ln^2(x_{\rho\mu}x_{7\mu})\; ,\nonumber \\
&&F_{(c),0}=\sum\limits_{\rho=1}^{5}
\frac{x_{\rho\mu}}{\prod\limits_{\sigma\neq \rho}(x_{\sigma\mu}-x_{\rho\mu})}\Big(\ln x_{\rho\mu}-\ln^2x_{\rho\mu}-{\cal S}L_{i_2}(x_{6\rho},x_{7\rho})\Big)\; ,
\label{ic1}
\end{eqnarray}
\begin{eqnarray}
&&F_{(d),2}^{a,1}=\frac{1}{(x_{1\mu}-x_{2\mu})(x_{6\mu}-x_{7\mu})}
\bigg(x_{1\mu}\Big({\cal S}L_{i_2}(x_{51},x_{61})-{\cal S}L_{i_2}(x_{51}
,x_{71})\Big)\nonumber \\
&&\hspace{1.2cm}-x_{2\mu}\Big({\cal S}L_{i_2}(x_{52},x_{62})-{\cal
S}L_{i_2}(x_{52},x_{72})\Big)\bigg)\; ,\nonumber \\
&&F_{(d),2}^{b,1}=-\frac{1}{2(4\pi)^4}\sum\limits_{\rho=1}^{4}
\frac{x_{\rho\mu}x_{5\mu}}{\prod\limits_{\sigma\neq \rho}
(x_{\sigma\mu}-x_{\rho\mu})}\ln^2(x_{\rho\mu}x_{5\mu})\; ,\nonumber \\
&&F_{(d),2}^{c,1}=\frac{1}{2}\frac{1}{x_{6\mu}-x_{7\mu}}
\sum\limits_{\rho=1}^{4}\frac{1}{\prod\limits_{\sigma\neq \rho}
(x_{\sigma\mu}-x_{\rho\mu})}\Big(
x_{\rho\mu}x_{6\mu}(x_{\rho\mu}+x_{6\mu})
\ln^2(x_{\rho\mu}x_{6\mu})\nonumber \\
&&\hspace{1.2cm}-x_{\rho\mu}x_{7\mu}(x_{\rho\mu}+x_{7\mu})
\ln^2(x_{\rho\mu}x_{7\mu})\Big)\; ,
\nonumber \\
&&F_{(d),2}^{d,1}=
\sum\limits_{\rho=1}^{3}\frac{x_{\rho\mu}}{
\prod\limits_{\sigma\neq \rho}(x_{\sigma\mu}-x_{\rho\mu})}\bigg(\ln x_{\rho\mu}
-\ln^2x_{\rho\mu}-{\cal S}L_{i_2}(x_{5\rho},x_{6\rho})
\bigg)\; ,\nonumber \\
&&F_{(d),2}^{e,1}=-
\sum\limits_{\rho=1}^{3}\frac{x_{\rho\mu}}{\prod\limits_
{\sigma\neq \rho}(x_{\sigma\mu}-x_{\rho\mu})}
\frac{x_{6\mu}\ln^2(x_{6\mu}x_{\rho\mu})-x_{7\mu}
\ln^2(x_{7\mu}x_{\rho\mu})}{2(x_{6\mu}-x_{7\mu})}\; ,
\nonumber \\
&&F_{(d),2}^{f,1}=-\frac{1}{2}\sum\limits_{\rho=1}^{3}
\frac{x_{6\mu}x_{\rho\mu}}{\prod\limits_{\sigma\neq \rho}
(x_{\sigma\mu}-x_{\rho\mu})}\ln^2(x_{6\mu}x_{\rho\mu})\; ,
\nonumber \\
&&F_{(d),1}^{a,1}=-\frac{1}{x_{6\mu}-x_{7\mu}}
\sum\limits_{\rho=1}^{3}\frac{x_{\rho\mu}}{\prod\limits_{\sigma\neq \rho}
(x_{\sigma\mu}-x_{\rho\mu})}\Big({\cal S}L_{i_2}(x_{5\rho},x_{6\rho})
-{\cal S}L_{i_2}(x_{5\rho},x_{7\rho})\Big)\; ,\nonumber \\
&&F_{(d),1}^{b,1}=-
\sum\limits_{\rho=1}^{4}\frac{x_{\rho\mu}}{\prod\limits_{\sigma\neq \rho}
(x_{\sigma\mu}-x_{\rho\mu})}\Big(\ln x_{\rho\mu}-\ln^2x_{\rho\mu}\Big)
\; ,\nonumber \\
&&F_{(d),1}^{c,1}=\frac{1}{2}
\sum\limits_{\rho=1}^{4}\frac{x_{\rho\mu}}{\prod\limits_
{\sigma\neq \rho}(x_{\sigma\mu}-x_{\rho\mu})}
\frac{x_{6\mu}\ln^2(x_{6\mu}x_{\rho\mu})-x_{7\mu}
\ln^2(x_{7\mu}x_{\rho\mu})}{(x_{6\mu}-x_{7\mu})}\; ,\nonumber \\
&&F_{(d),0}=\frac{1}{x_{6\mu}-x_{7\mu}}
\sum\limits_{\rho=1}^{4}\frac{x_{\rho\mu}}{\prod\limits_{\sigma\neq \rho}
(x_{\sigma\mu}-x_{\rho\mu})}\Big({\cal S}L_{i_2}(x_{5\rho},x_{6\rho})
-{\cal S}L_{i_2}(x_{5\rho},x_{7\rho})\Big)\; ,
\label{id1}
\end{eqnarray}
\begin{eqnarray}
&&F_{(e),2}^{a,1}=-\frac{1}{(x_{4\mu}-x_{5\mu})(x_{6\mu}-x_{7\mu})}
\bigg(x_{6\mu}\Big({\cal S}L_{i_2}(x_{16},x_{46})-
{\cal S}L_{i_2}(x_{16},x_{56})\Big)\nonumber \\
&&\hspace{1.2cm}-x_{7\mu}\Big({\cal S}L_{i_2}(x_{17},x_{47})-
{\cal S}L_{i_2}(x_{17},x_{57})\Big)\bigg)\; ,\nonumber \\
&&F_{(e),2}^{b,1}=-\frac{1}{2}
\sum\limits_{\rho=1}^{3}\frac{x_{\rho\mu}}{\prod\limits_
{\sigma\neq \rho}(x_{\sigma\mu}-x_{\rho\mu})}
\frac{x_{4\mu}\ln^2(x_{4\mu}x_{\rho\mu})-
x_{5\mu}\ln^2(x_{5\mu}x_{\rho\mu})}{(x_{4\mu}-x_{5\mu})}
\; ,\nonumber \\
&&F_{(e),2}^{c,1}=-\frac{1}{2}
\sum\limits_{\rho=1}^{3}\frac{x_{\rho\mu}}{\prod\limits_
{\sigma\neq \rho}(x_{\sigma\mu}-x_{\rho\mu})}
\frac{x_{6\mu}\ln^2(x_{6\mu}x_{\rho\mu})-
x_{7\mu}\ln^2(x_{7\mu}x_{\rho\mu})}{(x_{6\mu}-x_{7\mu})}
\; ,\nonumber \\
&&F_{(e),2}^{d,1}=-\frac{1}{(x_{1\mu}-x_{2\mu})(x_{4\mu}-x_{5\mu})}
\bigg(x_{1\mu}\Big({\cal S}L_{i_2}(x_{41},x_{61})-
{\cal S}L_{i_2}(x_{51},x_{61})\Big)\nonumber \\
&&\hspace{1.2cm}-x_{2\mu}\Big({\cal S}L_{i_2}(x_{42},x_{62})-
{\cal S}L_{i_2}(x_{52},x_{62})\Big)\bigg)\; ,\nonumber \\
&&F_{(e),2}^{e,1}=-\frac{1}{(x_{1\mu}-x_{2\mu})(x_{6\mu}-x_{7\mu})}
\bigg(x_{1\mu}\Big({\cal S}L_{i_2}(x_{41},x_{61})-
{\cal S}L_{i_2}(x_{41},x_{71})\Big)\nonumber \\
&&\hspace{1.2cm}-x_{2\mu}\Big({\cal S}L_{i_2}(x_{42},x_{62})-
{\cal S}L_{i_2}(x_{42},x_{72})\Big)\bigg)\; ,\nonumber \\
&&F_{(e),2}^{f,1}=
\sum\limits_{\rho=1}^{3}\frac{x_{\rho\mu}}{\prod\limits_{\sigma\neq \rho}
(x_{\sigma\mu}-x_{\rho\mu})}\bigg(\ln x_{\rho\mu}
-\ln^2x_{\rho\mu}-{\cal S}L_{i_2}(x_{4\rho},x_{6\rho})
\bigg)\; ,\nonumber \\
&&F_{(e),1}^{a,1}=-\frac{1}{(x_{1\mu}-x_{2\mu})(x_{4\mu}-x_{5\mu})
(x_{6\mu}-x_{7\mu})}\bigg(x_{1\mu}\Big({\cal S}L_{i_2}(x_{41},x_{61})-
{\cal S}L_{i_2}(x_{41},x_{71})\nonumber \\
&&\hspace{1.2cm}-{\cal S}L_{i_2}(x_{51},x_{61})+
{\cal S}L_{i_2}(x_{51},x_{71})\Big)-x_{2\mu}\Big({\cal S}L_{i_2}(x_{42}
,x_{62})-{\cal S}L_{i_2}(x_{42},x_{72})\nonumber \\
&&\hspace{1.2cm}-{\cal S}L_{i_2}(x_{52},x_{62})+
{\cal S}L_{i_2}(x_{52},x_{72})\Big)\bigg)\; ,\nonumber \\
&&F_{(e),1}^{b,1}=-\frac{1}{(x_{4\mu}-x_{5\mu})}
\sum\limits_{\rho=1}^{3}\frac{x_{\rho\mu}}{\prod\limits_{\sigma\neq \rho}
(x_{\sigma\mu}-x_{\rho\mu})}\Big({\cal S}L_{i_2}(x_{4\rho},x_{6\rho})-
{\cal S}L_{i_2}(x_{5\rho},x_{6\rho})\Big)\; ,\nonumber \\
&&F_{(e),1}^{c,1}=-\frac{1}{(x_{6\mu}-x_{7\mu})}
\sum\limits_{\rho=1}^{3}\frac{x_{\rho\mu}}{\prod\limits_{\sigma\neq \rho}
(x_{\sigma\mu}-x_{\rho\mu})}\Big({\cal S}L_{i_2}(x_{4\rho},x_{6\rho})-
{\cal S}L_{i_2}(x_{4\rho},x_{7\rho})\Big)\; ,\nonumber \\
&&F_{(e),0}=-\frac{1}{(x_{4\mu}-x_{5\mu})(x_{6\mu}-x_{7\mu})}
\sum\limits_{\rho=1}^{3}\frac{x_{\rho\mu}}{\prod\limits_{\sigma\neq \rho}
(x_{\sigma\mu}-x_{\rho\mu})}\Big({\cal S}L_{i_2}(x_{4\rho},x_{6\rho})
-{\cal S}L_{i_2}(x_{4\rho},x_{7\rho})\nonumber \\
&&\hspace{1.2cm}-{\cal S}L_{i_2}(x_{5\rho},x_{6\rho})
+{\cal S}L_{i_2}(x_{5\rho},x_{7\rho})\Big)
\label{ie1}\; .
\end{eqnarray}
The reduced formulae are similar to $I_{(i)}$, but with replacements
of $m_i^2 \rightarrow x_{i\mu}$, $I_{(i)}\rightarrow F_{(i)}$.

\section{The expressions of ${\cal S}L_{i_2}(a,b)$, $\AE(a,b)$,
${\cal R}_{i_2}(a)$ and $\Upsilon(a)$\label{slifun}}

We use the new symbols $y_a$, $y_b$ to represent the two roots of
equation $a(1-x)+bx-x(1-x)=0$, and the functions ${\cal S}L_{i_2}(a,b)$
can be written as
\begin{eqnarray}
&&{\cal S}L_{i_2}(a,b)=14+\frac{(b-1)\ln^2b}{2(b-a)}+\frac{(a-1)\ln^2a}
{2(a-b)}+\frac{1}{y_a-y_b}\bigg(\Big(2y_a+(b-a-1)\Big)
L_{i_2}(\frac{1}{1-y_a})\nonumber \\
&&\hspace{1.2cm}-\Big(2y_b+(b-a-1)\Big)L_{i_2}(\frac{1}{1-y_b})
\bigg)-\bigg(\ln b\ln(1-y_b)-y_a\Big(\ln(a+(b-a)y_a)\ln\frac{y_a-1}{y_a}
\nonumber \\
&&\hspace{1.2cm}
+L_{i_2}(\frac{(a-b)y_a}{(a-b)y_a-a})-L_{i_2}(\frac{(a-b)(y_a-1)}
{(a-b)y_a-a})\Big)-\Big((2y_a-1)\ln\frac{y_a-1}{y_a}-y_aL_{i_2}(\frac{1}
{y_a})\nonumber \\
&&\hspace{1.2cm}
+(y_a-1)L_{i_2}(\frac{1}{1-y_a})-\ln(y_a(y_a-1))\Big)+
(y_a\rightarrow y_b)\bigg).
\label{sli2}
\end{eqnarray}
The expression of $\AE(a,b)$ is
\begin{eqnarray}
&&\AE(a,b)=\frac{1}{2(b-a)}\Big(b(\ln b+1)^2-a(\ln a+1)^2\Big)
-\frac{2b}{a-b}\ln a-\frac{2a}{b-a}\ln b \nonumber \\
&&\hspace{1.2cm}
-\frac{a}{b-a}\Big(\ln\frac{b}{a}\ln\frac{a-b}{a}
+L_{i_2}(\frac{b}{a})\Big)-\frac{b}{a-b}\Big(\ln\frac{a}{b}
\ln\frac{b-a}{b}+L_{i_2}(\frac{a}{b})\Big).
\label{AE}
\end{eqnarray}
It is easy to note that when $a=b$, $\AE(a,b)=2\ln a+\frac{1}{2}\ln^2a$.
The definition of function $\Upsilon(a)$ is
\begin{eqnarray}
&&\Upsilon(a)=-\Big(\ln a\ln(1-a)+L_{i_2}(a)\Big)+a\Big(\ln a\ln(1-
\frac{1}{a})-L_{i_2}(\frac{1}{a})\Big).
\label{Upsi}
\end{eqnarray}
The expression of ${\cal R}_{i_2}$ is written as
\begin{eqnarray}
&&{\cal R}_{i_2}(a)=\ln a\ln(1-a)+L_{i_2}(a)-\frac{1}{2}\ln^2a
-\frac{\ln^2a}{1-a}\; .
\label{rli2}
\end{eqnarray}
\section{The coefficients at Next-to-leading Order}
\subsection{The gluon corrections}
We present here the gluon corrections $L_{i,j}^{\alpha}$
$(\alpha=1,\cdots ,8)$ as follows
\begin{eqnarray}
&&L_{i,j}^{\alpha}=WW_{\alpha}+2WH_{\alpha}+HH_{\alpha}+cc_{\alpha}
\end{eqnarray}
with
\begin{eqnarray}
&&WW_{1}=\frac{1}{3(-1+x_{i{\rm w}})^3(x_{i{\rm w}}-x_{j{\rm w}})^2(-1+x_{j{\rm w}})^2}
\bigg(\Big(2x_{i{\rm w}}\big(4-27x_{i{\rm w}}+19x_{i{\rm w}}^2-5x_{i{\rm w}}^4+9x_{i{\rm w}}^4\big)
\nonumber \\
&&\hspace{1.0cm}-2x_{j{\rm w}}\big(4-30x_{i{\rm w}}-39x_{i{\rm w}}^2+17x_{i{\rm w}}^3+
39x_{i{\rm w}}^4+x_{i{\rm w}}^5\big)-2x_{j{\rm w}}^2\big(3+79x_{i{\rm w}}+49x_{i{\rm w}}^2
-91x_{i{\rm w}}^3-40x_{i{\rm w}}^4\big)\nonumber \\
&&\hspace{1.0cm}+2x_{j{\rm w}}^3\big(21+101x_{i{\rm w}}-37x_{i{\rm w}}^2-89x_{i{\rm w}}^3
+4x_{i{\rm w}}^4\big)-4x_{j{\rm w}}^4\big(15+20x_{i{\rm w}}-32x_{i{\rm w}}^2+2x_{i{\rm w}}^3\big)
\nonumber \\
&&\hspace{1.0cm}+32x_{j{\rm w}}^5(1-x_{i{\rm w}})\Big)\big(L_{i_2}(\frac{x_{j{\rm w}}}{
x_{i{\rm w}}})+\ln\frac{x_{j{\rm w}}}{x_{i{\rm w}}}\ln(1-\frac{x_{j{\rm w}}}{x_{i{\rm w}}})\big)
\nonumber \\
&&\hspace{1.0cm}+\Big(-2x_{i{\rm w}}(4+3x_{i{\rm w}}-41x_{i{\rm w}}^2+65x_{i{\rm w}}^3
-39x_{i{\rm w}}^4+8x_{i{\rm w}}^5)+2x_{j{\rm w}}(4+30x_{i{\rm w}}-155x_{i{\rm w}}^2+265x_{i{\rm w}}^3
\nonumber \\
&&\hspace{1.0cm}-221x_{i{\rm w}}^4+93x_{i{\rm w}}^5-16x_{i{\rm w}}^6)-2x_{j{\rm w}}^2(27-
99x_{i{\rm w}}+171x_{i{\rm w}}^2-181x_{i{\rm w}}^3+110x_{i{\rm w}}^4-28x_{i{\rm w}}^5)
\nonumber \\
&&\hspace{1.0cm}+2x_{j{\rm w}}^3(15-29x_{i{\rm w}}+x_{i{\rm w}}^2+25x_{i{\rm w}}^3
-12x_{i{\rm w}}^4)\Big)\big(L_{i_2}(\frac{x_{i{\rm w}}}{
x_{j{\rm w}}})+\ln\frac{x_{i{\rm w}}}{x_{j{\rm w}}}\ln(1-\frac{x_{i{\rm w}}}{x_{j{\rm w}}})\big)
\nonumber \\
&&\hspace{1.0cm}+\Big(4x_{i{\rm w}}^2(13-53x_{i{\rm w}}+55x_{i{\rm w}}^2-19x_{i{\rm w}}^3
+4x_{i{\rm w}}^4)-4x_{i{\rm w}}x_{j{\rm w}}(20-77x_{i{\rm w}}+77x_{j{\rm w}}^2-59x_{i{\rm w}}^3+47x_{i{\rm w}}^4
-8x_{i{\rm w}}^5)\nonumber \\
&&\hspace{1.0cm}+4x_{j{\rm w}}^2(7-21x_{i{\rm w}}+29x_{i{\rm w}}^2-85x_{i{\rm w}}^3+84x_{i{\rm w}}^4
-14x_{i{\rm w}}^5)-4x_{j{\rm w}}^3(3+7x_{i{\rm w}}-45x_{i{\rm w}}^2+41x_{i{\rm w}}^3\nonumber \\
&&\hspace{1.0cm}
-6x_{i{\rm w}}^4)\Big)
\big(L_{i_2}(x_{i{\rm w}})+\ln x_{i{\rm w}}\ln(1-x_{i{\rm w}})\big)
\nonumber \\
&&\hspace{1.0cm}+\Big(4x_{i{\rm w}}^2(3+10x_{i{\rm w}}-13x_{i{\rm w}}^2)-4x_{i{\rm w}}x_{j{\rm w}}(
12+10x_{i{\rm w}}+16x_{i{\rm w}}^2-38x_{i{\rm w}}^3)+4x_{j{\rm w}}^2(9+16x_{i{\rm w}}+46x_{i{\rm w}}^2
\nonumber \\
&&\hspace{1.0cm}
-48x_{i{\rm w}}^3-23x_{i{\rm w}}^4)-8x_{j{\rm w}}^3(8+16x_{i{\rm w}}+x_{i{\rm w}}^2-26x_{i{\rm w}}^3
+x_{i{\rm w}}^4)+4x_{j{\rm w}}^4(15+20x_{i{\rm w}}-37x_{i{\rm w}}^2
+2x_{i{\rm w}}^3)\nonumber \\
&&\hspace{1.0cm}
-32x_{j{\rm w}}^5(1-x_{i{\rm w}})\Big)\big(L_{i_2}(x_{j{\rm w}})
+\ln x_{j{\rm w}}\ln(1-x_{j{\rm w}})\big)
\nonumber \\
&&\hspace{1.0cm}+\Big(8x_{i{\rm w}}(3+x_{i{\rm w}}-104x_{i{\rm w}}^2+98x_{i{\rm w}}^3-46x_{i{\rm w}}^4)
+8x_{j{\rm w}}(3-29x_{i{\rm w}}+209x_{i{\rm w}}^2-110x_{i{\rm w}}^2+25x_{i{\rm w}}^4+46x_{i{\rm w}}^5)
\nonumber \\
&&\hspace{1.0cm}
-8x_{j{\rm w}}^2(2+51x_{i{\rm w}}+69x_{i{\rm w}}^2-86x_{i{\rm w}}^3+108x_{i{\rm w}}^4)+
8x_{j{\rm w}}^3(30-41x_{i{\rm w}}+3x_{i{\rm w}}^2+50x_{i{\rm w}}^3+6x_{i{\rm w}}^4)
\nonumber \\
&&\hspace{1.0cm}
-32x_{j{\rm w}}^4(4-13x_{i{\rm w}}+9x_{i{\rm w}}^2)\Big)+\Big(8x_{i{\rm w}}^2(47-80x_{i{\rm w}}
-42x_{i{\rm w}}^2-x_{i{\rm w}}^3)+8x_{i{\rm w}}x_{j{\rm w}}(-47+13x_{i{\rm w}}+191x_{i{\rm w}}^2
\nonumber \\
&&\hspace{1.0cm}
+70x_{i{\rm w}}^3+x_{i{\rm w}}^4)+8x_{i{\rm w}}x_{j{\rm w}}^2(105-211x_{i{\rm w}}-94x_{i{\rm w}}^2
-26x_{i{\rm w}}^3)-48x_{i{\rm w}}x_{j{\rm w}}^4(1-x_{i{\rm w}})\Big)\ln x_{i{\rm w}}
\nonumber \\
&&\hspace{1.0cm}+\Big(x_{i{\rm w}}(4+175x_{i{\rm w}}+38x_{i{\rm w}}^2+57x_{i{\rm w}}^3
-34x_{i{\rm w}}^4)+2x_{j{\rm w}}(2-179x_{i{\rm w}}-116x_{i{\rm w}}^2-69x_{i{\rm w}}^3
-31x_{i{\rm w}}^4+33x_{i{\rm w}}^5)\nonumber \\
&&\hspace{1.0cm}-x_{j{\rm w}}^2(5-751x_{i{\rm w}}+88x_{i{\rm w}}^2-125x_{i{\rm w}}^3
+31x_{i{\rm w}}^4+32x_{i{\rm w}}^5)-x_{j{\rm w}}^3(13+453x_{i{\rm w}}-219x_{i{\rm w}}^2+29x_{i{\rm w}}^3
-36x_{i{\rm w}}^4)\nonumber \\
&&\hspace{1.0cm}
+2x_{j{\rm w}}^4(15+20x_{i{\rm w}}-37x_{i{\rm w}}^2+2x_{i{\rm w}}^3)-16x_{j{\rm w}}^5(
1-x_{i{\rm w}})\Big)\ln^2 x_{i{\rm w}}
\nonumber \\
&&\hspace{1.0cm}+\Big(-16x_{i{\rm w}}^2(4-9x_{i{\rm w}}+6x_{i{\rm w}}^2-x_{i{\rm w}}^3)+8x_{i{\rm w}}
x_{j{\rm w}}(13-27x_{i{\rm w}}+15x_{i{\rm w}}^2-x_{i{\rm w}}^3)-8x_{j{\rm w}}^2(5+29x_{i{\rm w}}\nonumber \\
&&\hspace{1.0cm}-117x_{i{\rm w}}^2
+127x_{i{\rm w}}^3-44x_{i{\rm w}}^4)-96x_{i{\rm w}}x_{j{\rm w}}^3(1-x_{i{\rm w}})^2-48x_{j{\rm w}}^4
(1-x_{i{\rm w}})^2\Big)\ln x_{j{\rm w}}\nonumber \\
&&\hspace{1.0cm}+\Big(-2x_{i{\rm w}}(4+15x_{i{\rm w}}-41x_{i{\rm w}}^2+53x_{i{\rm w}}^3
-39x_{i{\rm w}}^4+8x_{i{\rm w}}^5)+2x_{j{\rm w}}(-4+110x_{i{\rm w}}-189x_{i{\rm w}}^2+145x_{i{\rm w}}^3
\nonumber \\
&&\hspace{1.0cm}
-79x_{i{\rm w}}^4+33x_{i{\rm w}}^5-16x_{i{\rm w}}^6)+2x_{j{\rm w}}^2(5-137x_{i{\rm w}}+119x_{i{\rm w}}^2
+101x_{i{\rm w}}^3-148x_{i{\rm w}}^4+60x_{i{\rm w}}^5)\nonumber \\
&&\hspace{1.0cm}
+2x_{j{\rm w}}^3(13+123x_{i{\rm w}}-237x_{i{\rm w}}^2
+149x_{i{\rm w}}^3-48x_{i{\rm w}}^4)-4x_{j{\rm w}}^4(15-17x_{i{\rm w}}+2x_{i{\rm w}}^2)\nonumber \\
&&\hspace{1.0cm}
+32x_{j{\rm w}}^5(1-x_{i{\rm w}})\Big)\ln x_{i{\rm w}}\ln x_{j{\rm w}}\nonumber \\
&&\hspace{1.0cm}+\Big(x_{i{\rm w}}(4+3x_{i{\rm w}}-41x_{i{\rm w}}^2+65x_{i{\rm w}}^3-39x_{i{\rm w}}^4)
+8x_{i{\rm w}}^5)+x_{j{\rm w}}(4+18x_{i{\rm w}}+39x_{i{\rm w}}^2-285x_{i{\rm w}}^2+377x_{i{\rm w}}^3
\nonumber \\
&&\hspace{1.0cm}
-169x_{i{\rm w}}^4+16x_{i{\rm w}}^5)+x_{j{\rm w}}^2(-33+12x_{i{\rm w}}+230x_{i{\rm w}}^2-424x_{i{\rm w}}^3
+275x_{i{\rm w}}^4-60x_{i{\rm w}}^5)\nonumber \\
&&\hspace{1.0cm}
+x_{j{\rm w}}^3(-10-26x_{i{\rm w}}+134x_{i{\rm w}}^2-150x_{i{\rm w}}^3
+52x_{i{\rm w}}^4)\Big)\ln^2 x_{j{\rm w}}+(x_{i{\rm w}}\leftrightarrow x_{j{\rm w}})\bigg)
\; ,
\label{ww1}
\end{eqnarray}
\begin{eqnarray}
&&WH_{1}=\frac{({\cal Z}_{H}^{2k})^2}{\sin^2\beta}x_{i{\rm w}}x_{j{\rm w}}\bigg(
\Big(\frac{16}{(-1+x_{i{\rm w}})(x_{H^-_k{\rm w}}-x_{i{\rm w}})}-\frac{16}{(-1+x_{i{\rm w}})
(x_{H^-_k{\rm w}}-x_{j{\rm w}})}\Big)\Big(L_{i_2}(\frac{x_{j{\rm w}}}{x_{i{\rm w}}})
\nonumber \\
&&\hspace{1.0cm}+\ln\frac{x_{j{\rm w}}}{x_{i{\rm w}}}\ln(1-\frac{x_{j{\rm w}}}{x_{i{\rm w}}})\Big)
\nonumber \\
&&\hspace{1.0cm}+\Big(\frac{32}{3(x_{i{\rm w}}-1)(x_{j{\rm w}}-1)}+\frac{32}
{3(x_{i{\rm w}}-1)^2(x_{j{\rm w}}-1)}-\frac{32x_{H^-_k{\rm w}}}{3(x_{i{\rm w}}-1)^2(x_{j{\rm w}}-1)
(x_{H^-_k{\rm w}}-1)}\nonumber \\
&&\hspace{1.0cm}+\frac{32x_{H^-_k{\rm w}}}{3(x_{i{\rm w}}-1)(x_{j{\rm w}}-1)(x_{H^-_k{\rm w}}-1)}
+\frac{4}{3(x_{i{\rm w}}-1)(x_{j{\rm w}}-1)(x_{i{\rm w}}-x_{j{\rm w}})}\Big)\Big(L_{i_2}(x_{i{\rm w}})
+\ln x_{i{\rm w}}\ln(1-x_{i{\rm w}})\Big)\nonumber \\
&&\hspace{1.0cm}-\Big(\frac{16}{(-1+x_{i{\rm w}})(x_{H^-_k{\rm w}}-1)}+
\frac{16}{(x_{i{\rm w}}-1)(x_{H^-_k{\rm w}}-x_{j{\rm w}})}+\frac{4}
{3(x_{i{\rm w}}-1)(x_{j{\rm w}}-1)(x_{i{\rm w}}-x_{j{\rm w}})}\Big)\nonumber \\
&&\hspace{1.0cm}\Big(L_{i_2}(x_{j{\rm w}})
+\ln x_{j{\rm w}}\ln(1-x_{j{\rm w}})\Big)-\frac{16}{3(x_{H^-_k{\rm w}}-1)
(x_{H^-_k{\rm w}}-x_{i{\rm w}})}\Big(L_{i_2}(\frac{x_{j{\rm w}}}{x_{H^-_k{\rm w}}})+
\ln\frac{x_{j{\rm w}}}{x_{H^-_k{\rm w}}}\ln(1-\frac{x_{j{\rm w}}}{x_{H^-_k{\rm w}}})\Big)
\nonumber \\
&&\hspace{1.0cm}+\frac{32x_{H^-_k{\rm w}}^2(1+x_{H^-_k{\rm w}}-x_{i{\rm w}})}
{3(-1+x_{H^-_k{\rm w}})(x_{H^-_k{\rm w}}-x_{i{\rm w}})^2(x_{H^-_k{\rm w}}-x_{j{\rm w}})}
\Big(L_{i_2}(\frac{x_{i{\rm w}}}{x_{H^-_k{\rm w}}})+
\ln\frac{x_{i{\rm w}}}{x_{H^-_k{\rm w}}}\ln(1-\frac{x_{i{\rm w}}}{x_{H^-_k{\rm w}}})\Big)
\nonumber \\
&&\hspace{1.0cm}-\Big(\frac{4x_{H^-_k{\rm w}}^2(1+x_{H^-_k{\rm w}})}
{3(x_{H^-_k{\rm w}}-x_{i{\rm w}})(x_{i{\rm w}}-1)(x_{H^-_k{\rm w}}-x_{j{\rm w}})(x_{i{\rm w}}-x_{j{\rm w}})}
+\frac{4x_{H^-_k{\rm w}}}{3(x_{H^-_k{\rm w}}-x_{i{\rm w}})(x_{H^-_k{\rm w}}-x_{j{\rm w}})
(x_{i{\rm w}}-x_{j{\rm w}})}\nonumber \\
&&\hspace{1.0cm}+\frac{4}{3(x_{i{\rm w}}-1)(x_{i{\rm w}}-x_{j{\rm w}})}-
\frac{16}{3(x_{i{\rm w}}-1)(x_{H^-_k{\rm w}}-x_{j{\rm w}})}\Big)
\Big(L_{i_2}(\frac{x_{H^-_k{\rm w}}}{x_{i{\rm w}}})+
\ln\frac{x_{H^-_k{\rm w}}}{x_{i{\rm w}}}\ln(1-\frac{x_{H^-_k{\rm w}}}{x_{i{\rm w}}})\Big)
\nonumber \\
&&\hspace{1.0cm}+\Big(\frac{4x_{H^-_k{\rm w}}^2(1+x_{H^-_k{\rm w}})}
{3(x_{H^-_k{\rm w}}-x_{i{\rm w}})(x_{j{\rm w}}-1)(x_{H^-_k{\rm w}}-x_{j{\rm w}})(x_{i{\rm w}}-x_{j{\rm w}})}
+\frac{4x_{H^-_k{\rm w}}}{3(x_{H^-_k{\rm w}}-x_{i{\rm w}})(x_{H^-_k{\rm w}}-x_{j{\rm w}})
(x_{i{\rm w}}-x_{j{\rm w}})}\nonumber \\
&&\hspace{1.0cm}-\frac{4}{3(x_{i{\rm w}}-1)(x_{i{\rm w}}-x_{j{\rm w}})}\Big)
\Big(L_{i_2}(\frac{x_{H^-_k{\rm w}}}{x_{j{\rm w}}})+
\ln\frac{x_{H^-_k{\rm w}}}{x_{j{\rm w}}}\ln(1-\frac{x_{H^-_k{\rm w}}}{x_{j{\rm w}}})\Big)
\nonumber \\
&&\hspace{1.0cm}+\Big(\frac{4x_{H^-_k{\rm w}}(1-x_{H^-_k{\rm w}}^2)}
{3(x_{H^-_k{\rm w}}-x_{i{\rm w}})(x_{H^-_k{\rm w}}-x_{j{\rm w}})(x_{i{\rm w}}-1)(x_{j{\rm w}}-1)}
-\frac{4}{3(x_{i{\rm w}}-1)(x_{j{\rm w}}-1)}\nonumber \\
&&\hspace{1.0cm}+\frac{16}{3(x_{i{\rm w}}-1)(x_{H^-_k{\rm w}}-x_{j{\rm w}})}\Big)
\Big(L_{i_2}(x_{H^-_k{\rm w}})+\ln x_{H^-_k{\rm w}}\ln(1-x_{H^-_k{\rm w}})\Big)
\nonumber \\
&&\hspace{1.0cm}-\frac{128+32\ln x_{i{\rm w}}+30\ln^2 x_{i{\rm w}}+4\ln x_{H^-_k{\rm w}}
\ln x_{i{\rm w}}}{3(x_{i{\rm w}}-1)(x_{i{\rm w}}-x_{j{\rm w}})}+\frac{8\ln^2(\frac{x_{H^-_k{\rm w}}}
{x_{j{\rm w}}})}
{3(x_{H^-_k{\rm w}}-1)(x_{H^-_k{\rm w}}-x_{i{\rm w}})}\nonumber \\
&&\hspace{1.0cm}+\frac{2\ln^2 x_{H^-_k{\rm w}}-16\ln^2x_{i{\rm w}}}{3(x_{i{\rm w}}-1)
(x_{j{\rm w}}-1)}-\frac{16(\ln^2x_{H^-_k{\rm w}}-\ln x_{i{\rm w}}\ln x_{H^-_k{\rm w}}+
\ln x_{i{\rm w}}\ln x_{j{\rm w}})}{3(x_{i{\rm w}}-1)(x_{H^-_k{\rm w}}-x_{j{\rm w}})}\nonumber \\
&&\hspace{1.0cm}+\frac{8\ln^2\frac{x_{i{\rm w}}}{x_{j{\rm w}}}}{3(x_{i{\rm w}}-1)
(x_{H^-_k{\rm w}}-1)}+\frac{4\ln x_{H^-_k{\rm w}}\ln x_{j{\rm w}}-2\ln^2 x_{j{\rm w}}}
{3(x_{i{\rm w}}-x_{j{\rm w}})(x_{j{\rm w}}-1)}+\frac{32\ln^2 x_{i{\rm w}}}{3(x_{i{\rm w}}-1)^2
(x_{j{\rm w}}-1)}\nonumber \\
&&\hspace{1.0cm}+\frac{8\ln^2 x_{j{\rm w}}}{(x_{i{\rm w}}-1)(x_{i{\rm w}}-x_{H^-_k{\rm w}})}
+\frac{2x_{H^-_k{\rm w}}^2(1+x_{H^-_k{\rm w}})\ln^2(\frac{x_{H^-_k{\rm w}}}{x_{i{\rm w}}})}
{3(x_{H^-_k{\rm w}}-x_{i{\rm w}})(x_{i{\rm w}}-1)(x_{H^-_k{\rm w}}-x_{j{\rm w}})(x_{i{\rm w}}-x_{j{\rm w}})}
\nonumber \\
&&\hspace{1.0cm}+\frac{x_{H^-_k{\rm w}}(2x_{H^-_k{\rm w}}+1)(x_{H^-_k{\rm w}}-1)\ln^2
x_{H^-_k{\rm w}}}{3(x_{H^-_k{\rm w}}-x_{i{\rm w}})(x_{i{\rm w}}-1)(x_{H^-_k{\rm w}}-x_{j{\rm w}})(x_{j{\rm w}}-1)}
-\frac{4x_{j{\rm w}}\ln^2\frac{x_{j{\rm w}}}{x_{H^-_k{\rm w}}}}{3(x_{H^-_k{\rm w}}-x_{j{\rm w}})
(x_{j{\rm w}}-1)(x_{j{\rm w}}-x_{i{\rm w}})}\nonumber \\
&&\hspace{1.0cm}-\frac{2x_{H^-_k{\rm w}}^2(1+x_{H^-_k{\rm w}})\ln^2\frac{x_{j{\rm w}}}
{x_{H^-_k{\rm w}}}}{3(x_{H^-_k{\rm w}}-x_{i{\rm w}})(x_{i{\rm w}}-x_{j{\rm w}})(x_{H^-_k{\rm w}}-x_{j{\rm w}})
(x_{j{\rm w}}-1)}-\frac{4(\ln^2x_{H^-_k{\rm w}}-4x_{H^-_k{\rm w}}\ln x_{i{\rm w}}+4\ln^2
x_{j{\rm w}})}{3(x_{H^-_k{\rm w}}-1)(x_{i{\rm w}}-1)(x_{j{\rm w}}-1)}
\nonumber \\
&&\hspace{1.0cm}+\frac{32x_{H^-_k{\rm w}}(\ln^2 x_{i{\rm w}}+3\ln x_{i{\rm w}}+4)
-4x_{i{\rm w}}\ln^2\frac{x_{H^-_k{\rm w}}}{x_{i{\rm w}}}-16x_{i{\rm w}}\ln^2\frac{x_{j{\rm w}}}{x_{i{\rm w}}}}
{3(x_{H^-_k{\rm w}}-x_{i{\rm w}})(x_{i{\rm w}}-1)(x_{i{\rm w}}-x_{j{\rm w}})}
\nonumber \\
&&\hspace{1.0cm}-\frac{2x_{H^-_k{\rm w}}(2\ln x_{H^-_k{\rm w}}-\ln(x_{i{\rm w}}x_{j{\rm w}}))
\ln\frac{x_{i{\rm w}}}{x_{j{\rm w}}}}
{3(x_{H^-_k{\rm w}}-x_{i{\rm w}})(x_{H^-_k{\rm w}}-x_{j{\rm w}})(x_{i{\rm w}}-x_{j{\rm w}})}-\frac{16\ln^2
x_{j{\rm w}}}{3(x_{i{\rm w}}-1)(x_{j{\rm w}}-1)(x_{j{\rm w}}-x_{H^-_k{\rm w}})}
\nonumber \\
&&\hspace{1.0cm}-\frac{16x_{H^-_k{\rm w}}^2\ln^2\frac{x_{H^-_k{\rm w}}}{x_{i{\rm w}}}
+16x_{H^-_k{\rm w}}(7\ln x_{H^-_k{\rm w}}+2\ln x_{H^-_k{\rm w}}\ln x_{j{\rm w}}-\ln^2 x_{j{\rm w}})}
{3(x_{H^-_k{\rm w}}-1)(x_{H^-_k{\rm w}}-x_{i{\rm w}})(x_{H^-_k{\rm w}}-x_{j{\rm w}})}
\nonumber \\
&&\hspace{1.0cm}+\frac{16x_{i{\rm w}}\ln^2\frac{x_{i{\rm w}}}{x_{j{\rm w}}}}{3(x_{i{\rm w}}-1)
(x_{H^-_k{\rm w}}-x_{j{\rm w}})(x_{i{\rm w}}-x_{j{\rm w}})}-\frac{32x_{H^-_k{\rm w}}\ln^2 x_{i{\rm w}}}
{3(x_{i{\rm w}}-1)^2(x_{H^-_k{\rm w}}-1)(x_{j{\rm w}}-1)}\nonumber \\
&&\hspace{1.0cm}+\frac{16x_{i{\rm w}}\ln^2\frac{x_{H^-_k{\rm w}}}{x_{i{\rm w}}}}
{3(x_{H^-_k{\rm w}}-x_{i{\rm w}})(x_{i{\rm w}}-1)(x_{H^-_k{\rm w}}-x_{j{\rm w}})}
+\frac{16\ln^2x_{H^-_k{\rm w}}}{3(x_{H^-_k{\rm w}}-1)(x_{i{\rm w}}-1)(x_{H^-_k{\rm w}}-x_{j{\rm w}})}
\nonumber \\
&&\hspace{1.0cm}-\frac{32x_{H^-_k{\rm w}}^2(\ln x_{H^-_k{\rm w}}+x_{H^-_k{\rm w}}
\ln^2x_{H^-_k{\rm w}}+x_{H^-_k{\rm w}}\ln^2\frac{x_{H^-_k{\rm w}}}{x_{i{\rm w}}}}
{3(x_{H^-_k{\rm w}}-1)(x_{H^-_k{\rm w}}-x_{i{\rm w}})^2(x_{H^-_k{\rm w}}-x_{j{\rm w}})}+(x_{i{\rm w}}
\leftrightarrow x_{j{\rm w}})\bigg)\nonumber \\
&&\hspace{1.0cm}+({\cal Z}_{H}^{1k})^2h_bh_d\sqrt{x_{i{\rm w}}x_{j{\rm w}}}\bigg(\Big(
\frac{8}{3(x_{j{\rm w}}-x_{H^-_k{\rm w}})(x_{j{\rm w}}-x_{i{\rm w}})}-\frac{8(1-x_{j{\rm w}})}{3(
x_{i{\rm w}}-1)(x_{i{\rm w}}-x_{j{\rm w}})(x_{j{\rm w}}-x_{H^-_k{\rm w}})}\Big)\nonumber \\
&&\hspace{1.0cm}\Big(L_{i_2}(
\frac{x_{j{\rm w}}}{x_{i{\rm w}}})-\ln\frac{x_{j{\rm w}}}{x_{i{\rm w}}}\ln(1-\frac{x_{i{\rm w}}}
{x_{j{\rm w}}})\Big)+\frac{8(1-x_{j{\rm w}})}{3(x_{i{\rm w}}-1)(x_{j{\rm w}}-1)(x_{H^-_k{\rm w}}-
x_{j{\rm w}})}\Big(L_{i_2}(x_{j{\rm w}})+\ln x_{j{\rm w}}\ln(1-x_{j{\rm w}})\Big)
\nonumber \\
&&\hspace{1.0cm}+\Big(\frac{8(1-x_{j{\rm w}})}{3(x_{i{\rm w}}-1)(x_{i{\rm w}}-x_{j{\rm w}})
(x_{j{\rm w}}-x_{H^-_k{\rm w}})}+\frac{8}{3(x_{i{\rm w}}-x_{j{\rm w}})(x_{j{\rm w}}-x_{H^-_k{\rm w}})}
\nonumber \\
&&\hspace{1.0cm}
+\frac{8}{3(x_{i{\rm w}}-1)(x_{i{\rm w}}-x_{j{\rm w}})}\Big)\Big(L_{i_2}(\frac{x_{H^-_k{\rm w}}}
{x_{i{\rm w}}})+\ln \frac{x_{H^-_k{\rm w}}}{x_{i{\rm w}}}\ln(1-\frac{x_{H^-_k{\rm w}}}{x_{i{\rm w}}})
\Big)\nonumber \\
&&\hspace{1.0cm}-\Big(\frac{8(1-x_{j{\rm w}})}{3(x_{j{\rm w}}-1)(x_{i{\rm w}}-x_{j{\rm w}})
(x_{j{\rm w}}-x_{H^-_k{\rm w}})}+\frac{8}{3(x_{i{\rm w}}-x_{j{\rm w}})(x_{j{\rm w}}-x_{H^-_k{\rm w}})}
\nonumber \\
&&\hspace{1.0cm}
+\frac{8}{3(x_{i{\rm w}}-1)(x_{i{\rm w}}-x_{j{\rm w}})}\Big)\Big(L_{i_2}(\frac{x_{H^-_k{\rm w}}}
{x_{j{\rm w}}})+\ln \frac{x_{H^-_k{\rm w}}}{x_{j{\rm w}}}\ln(1-\frac{x_{H^-_k{\rm w}}}{x_{j{\rm w}}})
\Big)\nonumber \\
&&\hspace{1.0cm}+\Big(\frac{8(x_{j{\rm w}}-1)}{3(x_{i{\rm w}}-1)(x_{j{\rm w}}-1)(x_{H^-_k{\rm w}}-
x_{j{\rm w}})}+\frac{8}{3(x_{i{\rm w}}-1)(x_{j{\rm w}}-1)}\Big)\Big(L_{i_2}(x_{H^-_k{\rm w}})
+\ln x_{H^-_k{\rm w}}\ln(1-x_{H^-_k{\rm w}})\Big)\nonumber \\
&&\hspace{1.0cm}
+\frac{8x_{i{\rm w}}(x_{j{\rm w}}-1)\ln^2\frac{x_{H^-_k{\rm w}}}{x_{i{\rm w}}}}{3(x_{H^-_k{\rm w}}
-x_{i{\rm w}})(x_{i{\rm w}}-1)(x_{H^-_k{\rm w}}-x_{j{\rm w}})(x_{i{\rm w}}-x_{j{\rm w}})}
+\frac{8(1-x_{j{\rm w}})\ln^2 x_{H^-_k{\rm w}}}{3(x_{H^-_k{\rm w}}-1)(x_{i{\rm w}}-1)(x_{j{\rm w}}-1)
(x_{j{\rm w}}-x_{H^-_k{\rm w}})}\nonumber \\
&&\hspace{1.0cm}
-\frac{8x_{j{\rm w}}(1-x_{j{\rm w}})\ln^2\frac{x_{H^-_k{\rm w}}}{x_{j{\rm w}}}}{3(x_{H^-_k{\rm w}}-
x_{j{\rm w}})^2(x_{j{\rm w}}-1)(x_{j{\rm w}}-x_{i{\rm w}})}+\frac{8(1-x_{j{\rm w}})}{3(x_{H^-_k{\rm w}}-
x_{j{\rm w}})(x_{i{\rm w}}-1)(x_{j{\rm w}}-1)^2}\nonumber \\
&&\hspace{1.0cm}+\frac{8x_{i{\rm w}}\ln^2\frac{x_{H^-_k{\rm w}}}{x_{i{\rm w}}}}
{3(x_{H^-_k{\rm w}}-x_{i{\rm w}})(x_{i{\rm w}}-1)(x_{i{\rm w}}-x_{j{\rm w}})}
-\frac{4(1+x_{j{\rm w}})\ln^2\frac{x_{H^-_k{\rm w}}}{x_{j{\rm w}}}}{3(x_{H^-_k{\rm w}}-x_{j{\rm w}})
(x_{i{\rm w}}-x_{j{\rm w}})(x_{j{\rm w}}-1)}
\nonumber \\
&&\hspace{1.0cm}
+\frac{8x_{i{\rm w}}(x_{j{\rm w}}-1)\ln^2\frac{x_{i{\rm w}}}{x_{j{\rm w}}}}{3(x_{H^-_k{\rm w}}
-x_{j{\rm w}})(x_{i{\rm w}}-1)(x_{i{\rm w}}-x_{j{\rm w}})^2}
-\frac{8x_{i{\rm w}}\ln^2\frac{x_{H^-_k{\rm w}}}{x_{i{\rm w}}}}{3(x_{H^-_k{\rm w}}-x_{i{\rm w}})
(x_{H^-_k{\rm w}}-x_{j{\rm w}})(x_{i{\rm w}}-x_{j{\rm w}})}
\nonumber \\
&&\hspace{1.0cm}+\frac{8\ln^2x_{H^-_k{\rm w}}}{3(x_{H^-_k{\rm w}}-1)(x_{i{\rm w}}-1)(
x_{j{\rm w}}-1)}-\frac{4(x_{j{\rm w}}-1)\ln(\frac{x_{H^-_k{\rm w}}}{x_{j{\rm w}}})
\ln(\frac{x_{H^-_k{\rm w}}x_{j{\rm w}}}{x_{i{\rm w}}^2})}
{3x_{i{\rm w}}-1)(x_{H^-_k{\rm w}}-x_{j{\rm w}})(x_{i{\rm w}}-x_{j{\rm w}})}
\nonumber \\
&&\hspace{1.0cm}
-\frac{4(\ln^2x_{H^-_k{\rm w}}-\ln^2x_{j{\rm w}})}{3(x_{i{\rm w}}-1)(x_{H^-_k{\rm w}}
-x_{j{\rm w}})}+\frac{8x_{j{\rm w}}\ln^2(\frac{x_{H^-_k{\rm w}}}{x_{j{\rm w}}})}{3(x_{H^-_k{\rm w}}
-x_{j{\rm w}})^2(x_{i{\rm w}}-x_{j{\rm w}})}+\frac{4\ln^2x_{H^-_k{\rm w}}}{3(x_{i{\rm w}}-1)(1-x_{j{\rm w}})}
\nonumber \\
&&\hspace{1.0cm}-\frac{4\ln^2\frac{x_{H^-_k{\rm w}}}{x_{i{\rm w}}}}
{3(x_{i{\rm w}}-1)(x_{i{\rm w}}-x_{j{\rm w}})}+\frac{8(\ln x_{H^-_k{\rm w}}\ln x_{i{\rm w}}-
\ln x_{H^-_k{\rm w}}\ln x_{j{\rm w}}-\ln x_{i{\rm w}}\ln x_{j{\rm w}}+\ln^2x_{j{\rm w}})}
{3(x_{i{\rm w}}-x_{j{\rm w}})(x_{j{\rm w}}-x_{H^-_k{\rm w}})}\nonumber \\
&&\hspace{1.0cm}+\frac{8x_{i{\rm w}}(\ln^2\frac{x_{H^-_k{\rm w}}}{x_{i{\rm w}}})}
{3(x_{i{\rm w}}-x_{j{\rm w}})^2(x_{j{\rm w}}-x_{H^-_k{\rm w}})}-\frac{4\ln^2(\frac{x_{H^-_k{\rm w}}}
{x_{j{\rm w}}})}{3(x_{j{\rm w}}-1)(x_{j{\rm w}}-x_{i{\rm w}})}+(x_{i{\rm w}}\leftrightarrow x_{j{\rm w}})
\bigg)\; ,
\label{wh1}
\end{eqnarray}

\begin{eqnarray}
&&\hspace{-12cm}WH_{3}=-2WH_{2}\; ,
\label{wh3}
\end{eqnarray}
\begin{eqnarray}
&&HH_{1}=-\bigg(\frac{2}{3\sin^4\beta}x_{i{\rm w}}^2x_{j{\rm w}}^2({\cal Z}_{H}^{2k})^2
({\cal Z}_{H}^{2l})^2F_{A}^{0}-\frac{1}{6\sin^4\beta}x_{i{\rm w}}x_{j{\rm w}}(x_{i{\rm w}}
+x_{j{\rm w}})({\cal Z}_{H}^{2k})^2({\cal Z}_{H}^{2l})^2\Big(F_{A}^{1a}+F_{A}^{1b}
\nonumber \\
&&\hspace{1.0cm}-F_{A}^{1c}\Big)+\frac{1}{12\sin^3\beta}(h_{d}+h_{b})x_{i{\rm w}}^{1\over 2}
x_{j{\rm w}}({\cal Z}_{H}^{2k})^2{\cal Z}_{H}^{1l}{\cal Z}_{H}^{2l}\Big(F_{A}^{2a}
+F_{A}^{2b}+F_{A}^{2c}+2F_{A}^{2d}-2F_{A}^{2e}\nonumber \\
&&\hspace{1.0cm}-2F_{A}^{2f}\Big)\bigg)
(x_{i{\rm w}},x_{j{\rm w}},x_{H^-_l{\rm w}},0,x_{i{\rm w}},x_{j{\rm w}},x_{H^-_k{\rm w}})
\nonumber \\
&&\hspace{1.0cm}+\frac{4}{3\sin^4\beta}x_{i{\rm w}}x_{j{\rm w}}({\cal Z}_{H}^{2k})^2
({\cal Z}_{H}^{2l})^2\bigg(-\sum\limits_{\sigma=u^i,H^-_k,H^-_l}\Big(
\frac{x_{\sigma}\ln^2 x_{\sigma}}{\prod\limits_{\rho\neq \sigma}(x_{\rho}-
x_{\sigma})}+(x_{H^-_k{\rm w}}+x_{j{\rm w}})\frac{{\cal R}_{i_2}(\frac{x_{j{\rm w}}}{x_{\sigma}})}
{\prod\limits_{\rho\neq \sigma}(x_{\rho}-x_{\sigma})}\Big)
\nonumber \\
&&\hspace{1.0cm}-\frac{{\cal R}_{i_2}(\frac{x_{j{\rm w}}}{x_{i{\rm w}}})
-{\cal R}_{i_2}(\frac{x_{j{\rm w}}}{x_{H^-_l{\rm w}}})}
{(x_{H^-_k{\rm w}}-x_{H^-_l{\rm w}})(x_{i{\rm w}}-x_{H^-_l{\rm w}})}\bigg)
\nonumber \\
&&\hspace{1.0cm}-\frac{x_{i{\rm w}}x_{j{\rm w}}}
{3\sin^4\beta}({\cal Z}_{H}^{2k})^2({\cal Z}_{H}^{2l})^2\bigg(
\sum\limits_{\sigma=u^i,u^j,H^-_l}\Big(
-\frac{3x_{\sigma}\ln^2 x_{\sigma}}{2\prod\limits_{\rho\neq \sigma}(x_{\rho}-
x_{\sigma})}+2(x_{H^-_k{\rm w}}\nonumber \\
&&\hspace{1.0cm}
+x_{H^-_l{\rm w}})\frac{{\cal R}_{i_2}(\frac{x_{H^-_k{\rm w}}}{x_{\sigma}})}
{\prod\limits_{\rho\neq \sigma}(x_{\rho}-x_{\sigma})}+
\frac{x_{\sigma}(\ln x_{\sigma}-\ln^2 x_{\sigma} -
\Upsilon(\frac{x_{H^-_k{\rm w}}}{x_{\sigma}}))}
{\prod\limits_{\rho\neq \sigma}(x_{\rho}-x_{\sigma})}\Big)
+2\frac{{\cal R}_{i_2}(\frac{x_{H^-_k{\rm w}}}{x_{i{\rm w}}})-
{\cal R}_{i_2}(\frac{x_{H^-_k{\rm w}}}{x_{j{\rm w}}})}
{-x_{i{\rm w}}+x_{j{\rm w}}}\bigg)
\nonumber \\
&&\hspace{1.0cm}-\frac{32}{3\sin^2\beta}x_{i{\rm w}}x_{j{\rm w}}(h_b^2+h_d^2){\cal Z}_{H}^{1k}
{\cal Z}_{H}^{2k}{\cal Z}_{H}^{1l}{\cal Z}_{H}^{2l}\bigg(
\sum\limits_{\sigma=u^j,H^-_k,H^-_l}
\frac{x_{\sigma}(\ln x_{\sigma}-2\ln^2 x_{\sigma}
-\Upsilon(\frac{x_{i{\rm w}}}{x_{\sigma}}))}
{2(x_{i{\rm w}}-x_{\sigma})^2\prod\limits_{\rho\neq \sigma}(x_{\rho}-x_{\sigma})}
\nonumber \\
&&\hspace{1.0cm}+\frac{(-1+3\ln x_{i{\rm w}}+2\ln^2 x_{i{\rm w}})}
{2(x_{H^-_k{\rm w}}-x_{i{\rm w}})(x_{H^-_l{\rm w}}-x_{i{\rm w}})(x_{j{\rm w}}-x_{i{\rm w}})}+
\frac{x_{i{\rm w}}(-\ln x_{i{\rm w}}+2\ln^2 x_{i{\rm w}})}
{2(x_{H^-_k{\rm w}}-x_{i{\rm w}})(x_{H^-_l{\rm w}}-x_{i{\rm w}})(x_{j{\rm w}}-x_{i{\rm w}})^2}
\nonumber \\
&&\hspace{1.0cm}+
\frac{x_{i{\rm w}}(-\ln x_{i{\rm w}}+2\ln^2 x_{i{\rm w}})}
{2(x_{H^-_k{\rm w}}-x_{i{\rm w}})(x_{H^-_l{\rm w}}-x_{i{\rm w}})^2(x_{j{\rm w}}-x_{i{\rm w}})}
+\frac{x_{i{\rm w}}(-\ln x_{i{\rm w}}+2\ln^2 x_{i{\rm w}})}
{2(x_{H^-_k{\rm w}}-x_{i{\rm w}})^2(x_{H^-_l{\rm w}}-x_{i{\rm w}})(x_{j{\rm w}}-x_{i{\rm w}})}
\bigg)\nonumber \\
&&\hspace{1.0cm}-\frac{32}{3\sin^4\beta}x_{i{\rm w}}^2x_{j{\rm w}}
({\cal Z}_{H}^{2k})^2({\cal Z}_{H}^{2l})^2\bigg(
\sum\limits_{\sigma=u^j,H^-_k,H^-_l}
\frac{x_{H^-_k{\rm w}}x_{\sigma}(-\ln x_{\sigma}-\ln^2 x_{\sigma}
-\Upsilon(\frac{x_{i{\rm w}}}{x_{\sigma}}))}
{2(x_{i{\rm w}}-x_{\sigma})^2\prod\limits_{\rho\neq \sigma}(x_{\rho}-x_{\sigma})}
\nonumber \\
&&\hspace{1.0cm}+\frac{x_{H^-_k{\rm w}}(1-3\ln x_{i{\rm w}}+\ln^2 x_{i{\rm w}})}
{(x_{H^-_k{\rm w}}-x_{i{\rm w}})(x_{H^-_l{\rm w}}-x_{i{\rm w}})(x_{j{\rm w}}-x_{i{\rm w}})}+
\frac{x_{H^-_k{\rm w}}x_{i{\rm w}}(\ln x_{i{\rm w}}+\ln^2 x_{i{\rm w}})}
{2(x_{H^-_k{\rm w}}-x_{i{\rm w}})(x_{H^-_l{\rm w}}-x_{i{\rm w}})(x_{j{\rm w}}-x_{i{\rm w}})^2}
\nonumber \\
&&\hspace{1.0cm}+
\frac{x_{H^-_k{\rm w}}x_{i{\rm w}}(\ln x_{i{\rm w}}+\ln^2 x_{i{\rm w}})}
{2(x_{H^-_k{\rm w}}-x_{i{\rm w}})(x_{H^-_l{\rm w}}-x_{i{\rm w}})^2(x_{j{\rm w}}-x_{i{\rm w}})}
+\frac{x_{H^-_k{\rm w}}x_{i{\rm w}}(\ln x_{i{\rm w}}+\ln^2 x_{i{\rm w}})}
{2(x_{H^-_k{\rm w}}-x_{i{\rm w}})^2(x_{H^-_l{\rm w}}-x_{i{\rm w}})(x_{j{\rm w}}-x_{i{\rm w}})}
\nonumber \\
&&\hspace{1.0cm}+\Big(\frac{\big(x_{H^-_l{\rm w}}(\ln x_{H^-_l{\rm w}}+\ln^2 x_{H^-_l{\rm w}}
+\Upsilon(\frac{x_{i{\rm w}}}{x_{H^-_l{\rm w}}}))
\big)}{(-x_{H^-_l{\rm w}}+x_{i{\rm w}})^2(-x_{H^-_l{\rm w}}+x_{j{\rm w}})}
+(x_{H^-_l{\rm w}}\rightarrow x_{j{\rm w}})\Big)\nonumber \\
&&\hspace{1.0cm}-\Big(\frac{1+3\ln x_{i{\rm w}}+ln^2 x_{i{\rm w}}}
{(x_{H^-_l{\rm w}}-x_{i{\rm w}})(x_{j{\rm w}}-x_{i{\rm w}})}
+\frac{x_{i{\rm w}}(\ln x_{i{\rm w}}+\ln^2 x_{i{\rm w}})}
{(x_{H^-_l{\rm w}}-x_{i{\rm w}})(x_{j{\rm w}}-x_{i{\rm w}})^2}
\nonumber \\
&&\hspace{1.0cm}+\frac{x_{i{\rm w}}(\ln x_{i{\rm w}}+\ln^2 x_{i{\rm w}})}
{(x_{H^-_l{\rm w}}-x_{i{\rm w}})^2(x_{j{\rm w}}-x_{i{\rm w}})}\Big)\bigg)
+\frac{16}{3\sin^2\beta}(h_b^2+h_{d}^2)x_{i{\rm w}}x_{j{\rm w}}
{\cal Z}_{H}^{1k}{\cal Z}_{H}^{2k}{\cal Z}_{H}^{1l}{\cal Z}_{H}^{2l}
\bigg(\nonumber \\
&&\hspace{1.0cm}\sum\limits_{\sigma=u^j,H^-_k,H^-_l}
\frac{x_{\sigma}(2x_{i{\rm w}}\ln x_{\sigma}-2(x_{H^-_k{\rm w}}+x_{i{\rm w}})\ln^2 x_{\sigma}
-x_{i{\rm w}}\ln^2 (x_{i{\rm w}}x_{\sigma})-2(x_{H^-_k{\rm w}}-x_{i{\rm w}})
\Upsilon(\frac{x_{i{\rm w}}}{x_{\sigma}}))}
{2(x_{i{\rm w}}-x_{\sigma})^2\prod\limits_{\rho\neq \sigma}(x_{\rho}-x_{\sigma})}
\nonumber \\
&&\hspace{1.0cm}+\frac{\big(-x_{i{\rm w}}+(x_{H^-_k{\rm w}}+4x_{i{\rm w}})\ln x_{i{\rm w}}
+(x_{H^-_k{\rm w}}+5x_{i{\rm w}})\ln^2 x_{i{\rm w}}\big)}
{(x_{H^-_k{\rm w}}-x_{i{\rm w}})(x_{H^-_l{\rm w}}-x_{i{\rm w}})(x_{j{\rm w}}-x_{i{\rm w}})}
\nonumber \\
&&\hspace{1.0cm}+\frac{-x_{i{\rm w}}^2\ln x_{i{\rm w}}+(3x_{i{\rm w}}
+x_{H^-_k{\rm w}})x_{i{\rm w}}\ln^2 x_{i{\rm w}}}
{(x_{H^-_k{\rm w}}-x_{i{\rm w}})(x_{H^-_l{\rm w}}-x_{i{\rm w}})(x_{j{\rm w}}-x_{i{\rm w}})^2}
+\frac{-x_{i{\rm w}}^2\ln x_{i{\rm w}}+(3x_{i{\rm w}}
+x_{H^-_k{\rm w}})x_{i{\rm w}}\ln^2 x_{i{\rm w}}}
{(x_{H^-_k{\rm w}}-x_{i{\rm w}})(x_{H^-_l{\rm w}}-x_{i{\rm w}})^2(x_{j{\rm w}}-x_{i{\rm w}})}\nonumber \\
&&\hspace{1.0cm}+\frac{-x_{i{\rm w}}^2\ln x_{i{\rm w}}+(3x_{i{\rm w}}
+x_{H^-_k{\rm w}})x_{i{\rm w}}\ln^2 x_{i{\rm w}}}
{(x_{H^-_k{\rm w}}-x_{i{\rm w}})^2(x_{H^-_l{\rm w}}-x_{i{\rm w}})(x_{j{\rm w}}-x_{i{\rm w}})}
+\Big(\frac{x_{H^-_l{\rm w}}(\ln^2 x_{H^-_l{\rm w}}
+\Upsilon(\frac{x_{i{\rm w}}}{x_{H^-_l{\rm w}}}))}
{(-x_{H^-_l{\rm w}}+x_{i{\rm w}})^2(-x_{H^-_l{\rm w}}+x_{j{\rm w}})}
+(x_{H^-_l{\rm w}}\leftrightarrow x_{j{\rm w}})\Big)\nonumber \\
&&\hspace{1.0cm}-\frac{2\ln x_{i{\rm w}}+\ln^2 x_{i{\rm w}}}
{(x_{H^-_l{\rm w}}-x_{i{\rm w}})(x_{j{\rm w}}-x_{i{\rm w}})}-
\frac{x_{i{\rm w}}\ln^2 x_{i{\rm w}}}{(x_{H^-_l{\rm w}}-x_{i{\rm w}})^2(x_{j{\rm w}}-x_{i{\rm w}})}\bigg)
+\frac{4}{3\sin^4\beta}x_{i{\rm w}}^2x_{j{\rm w}}({\cal Z}_{H}^{2k})^2
({\cal Z}_{H}^{2l})^2\bigg(\nonumber \\
&&\hspace{1.0cm}\sum\limits_{\sigma=u^j,H^-_k,H^-_l}
\frac{x_{\sigma}(-(3x_{H^-_k{\rm w}}-2x_{i{\rm w}})\ln x_{\sigma}-2(x_{H^-_k{\rm w}}+x_{i{\rm w}})
(\ln^2 x_{\sigma}+\Upsilon(\frac{x_{i{\rm w}}}{x_{H^-_k{\rm w}}})))}
{2(x_{i{\rm w}}-x_{\sigma})^2\prod\limits_{\rho\neq \sigma}(x_{\rho}-x_{\sigma})}
\nonumber \\
&&\hspace{1.0cm}+\frac{\big(3x_{H^-_k{\rm w}}-2x_{i{\rm w}}+(7x_{H^-_k{\rm w}}-11x_{i{\rm w}})\ln x_{i{\rm w}}
+2(x_{H^-_k{\rm w}}+5x_{i{\rm w}})\ln^2 x_{i{\rm w}}\big)}
{2(x_{H^-_k{\rm w}}-x_{i{\rm w}})(x_{H^-_l{\rm w}}-x_{i{\rm w}})(x_{j{\rm w}}-x_{i{\rm w}})}
\nonumber \\
&&\hspace{1.0cm}+\frac{x_{i{\rm w}}(3x_{H^-_k{\rm w}}-2x_{i{\rm w}})\ln x_{\sigma}
+2x_{i{\rm w}}(x_{H^-_k{\rm w}}+3x_{i{\rm w}})\ln^2 x_{i{\rm w}}}
{2(x_{H^-_k{\rm w}}-x_{i{\rm w}})(x_{H^-_l{\rm w}}-x_{i{\rm w}})(x_{j{\rm w}}-x_{i{\rm w}})^2}
\nonumber \\
&&\hspace{1.0cm}+\frac{x_{i{\rm w}}(3x_{H^-_k{\rm w}}-2x_{i{\rm w}})\ln x_{\sigma}
+2x_{i{\rm w}}(x_{H^-_k{\rm w}}+3x_{i{\rm w}})\ln^2 x_{i{\rm w}}}
{2(x_{H^-_k{\rm w}}-x_{i{\rm w}})(x_{H^-_l{\rm w}}-x_{i{\rm w}})^2(x_{j{\rm w}}-x_{i{\rm w}})}
\nonumber \\
&&\hspace{1.0cm}+\frac{x_{i{\rm w}}(3x_{H^-_k{\rm w}}-2x_{i{\rm w}})\ln x_{\sigma}
+2x_{i{\rm w}}(x_{H^-_k{\rm w}}+3x_{i{\rm w}})\ln^2 x_{i{\rm w}}}
{2(x_{H^-_k{\rm w}}-x_{i{\rm w}})^2(x_{H^-_l{\rm w}}-x_{i{\rm w}})(x_{j{\rm w}}-x_{i{\rm w}})}
\nonumber \\
&&\hspace{1.0cm}+\Big(\frac{-\frac{3}{2}x_{i{\rm w}}\ln x_{i{\rm w}}+x_{i{\rm w}}\ln^2 x_{i{\rm w}}
+x_{H^-_l{\rm w}}\Upsilon(\frac{x_{i{\rm w}}}{x_{H^-_l{\rm w}}})}
{(-x_{H^-_l{\rm w}}+x_{i{\rm w}})^2(-x_{H^-_l{\rm w}}+x_{j{\rm w}})}
+(x_{H^-_l{\rm w}}\leftrightarrow x_{j{\rm w}})\Big)\nonumber \\
&&\hspace{1.0cm}+\frac{-3-7\ln x_{i{\rm w}}-2\ln^2 x_{i{\rm w}}}
{2(-x_{H^-_l{\rm w}}+x_{i{\rm w}})(-x_{i{\rm w}}+x_{j{\rm w}})}
+\frac{3x_{H^-_l{\rm w}}\ln x_{H^-_l{\rm w}}+2x_{H^-_l{\rm w}}\ln^2 x_{H^-_l{\rm w}}}
{2(-x_{H^-_l{\rm w}}+x_{i{\rm w}})^2(-x_{i{\rm w}}+x_{j{\rm w}})}\bigg)
\nonumber \\
&&\hspace{1.0cm}+\frac{4}{3\sin^2\beta}x_{i{\rm w}}x_{j{\rm w}}({\cal Z}_{H}^{2k})^2
({\cal Z}_{H}^{2l})^2\bigg(\sum\limits_{\sigma=u^j,H^-_k,H^-_l}\Big(
-\frac{x_{H^-_k{\rm w}}^2x_{i{\rm w}}x_{\sigma}\ln^2 (x_{\sigma}x_{i{\rm w}})}
{2(x_{i{\rm w}}-x_{\sigma})^2\prod\limits_{\rho\neq \sigma}(x_{\rho}-x_{\sigma})}
\nonumber \\
&&\hspace{1.0cm}+
\frac{x_{\sigma}\big((-3x_{H^-_k{\rm w}}^2+5x_{H^-_k{\rm w}}x_{i{\rm w}}-3x_{i{\rm w}}^2)\ln x_{\sigma}
-2(x_{H^-_k{\rm w}}^2+x_{i{\rm w}}^2)(\ln^2 x_{\sigma}+\Upsilon(\frac
{x_{i{\rm w}}}{x_{\sigma}}))\big)}
{2(x_{i{\rm w}}-x_{\sigma})^2\prod\limits_{\rho\neq \sigma}(x_{\rho}-x_{\sigma})}
\Big)
\nonumber \\
&&\hspace{1.0cm}+\frac{3x_{H^-_k{\rm w}}^2-5x_{H^-_k{\rm w}}x_{i{\rm w}}+3x_{i{\rm w}}^2+
(7x_{H^-_k{\rm w}}^2-2x_{H^-_k{\rm w}}x_{i{\rm w}}+7x_{i{\rm w}}^2)\ln x_{i{\rm w}}}
{2(x_{H^-_k{\rm w}}-x_{i{\rm w}})(x_{H^-_l{\rm w}}-x_{i{\rm w}})(x_{j{\rm w}}-x_{i{\rm w}})}
\nonumber \\
&&\hspace{1.0cm}+\frac{(x_{H^-_k{\rm w}}^2
+4x_{H^-_k{\rm w}}x_{i{\rm w}}+x_{i{\rm w}}^2)\ln^2 x_{i{\rm w}}}
{(x_{H^-_k{\rm w}}-x_{i{\rm w}})(x_{H^-_l{\rm w}}-x_{i{\rm w}})(x_{j{\rm w}}-x_{i{\rm w}})}
\nonumber \\
&&\hspace{1.0cm}+\frac{(3x_{H^-_k{\rm w}}^2x_{i{\rm w}}-5x_{H^-_k{\rm w}}x_{i{\rm w}}^2
+3x_{i{\rm w}}^3)\ln x_{i{\rm w}}+2(x_{H^-_k{\rm w}}^2x_{i{\rm w}}+x_{i{\rm w}}^3+x_{H^-_k{\rm w}}x_{i{\rm w}}^2)
\ln^2 x_{i{\rm w}}}
{(x_{H^-_k{\rm w}}-x_{i{\rm w}})(x_{H^-_l{\rm w}}-x_{i{\rm w}})(x_{j{\rm w}}-x_{i{\rm w}})^2}
\nonumber \\
&&\hspace{1.0cm}+\frac{(3x_{H^-_k{\rm w}}^2x_{i{\rm w}}-5x_{H^-_k{\rm w}}x_{i{\rm w}}^2
+3x_{i{\rm w}}^3)\ln x_{i{\rm w}}+2(x_{H^-_k{\rm w}}^2x_{i{\rm w}}+x_{i{\rm w}}^3+x_{H^-_k{\rm w}}x_{i{\rm w}}^2)
\ln^2 x_{i{\rm w}}}
{(x_{H^-_k{\rm w}}-x_{i{\rm w}})(x_{H^-_l{\rm w}}-x_{i{\rm w}})^2(x_{j{\rm w}}-x_{i{\rm w}})}
\nonumber \\
&&\hspace{1.0cm}+\frac{(3x_{H^-_k{\rm w}}^2x_{i{\rm w}}-5x_{H^-_k{\rm w}}x_{i{\rm w}}^2
+3x_{i{\rm w}}^3)\ln x_{i{\rm w}}+2(x_{H^-_k{\rm w}}^2x_{i{\rm w}}+x_{i{\rm w}}^3+x_{H^-_k{\rm w}}x_{i{\rm w}}^2)
\ln^2 x_{i{\rm w}}}
{(x_{H^-_k{\rm w}}-x_{i{\rm w}})^2(x_{H^-_l{\rm w}}-x_{i{\rm w}})(x_{j{\rm w}}-x_{i{\rm w}})}
\nonumber \\
&&\hspace{1.0cm}+\Big(\frac{(3x_{H^-_k{\rm w}}-5x_{i{\rm w}})x_{H^-_l{\rm w}}\ln x_{H^-_l{\rm w}}
+2(x_{H^-_k{\rm w}}+x_{H^-_l{\rm w}})x_{H^-_l{\rm w}}\ln^2 x_{H^-_l{\rm w}}-x_{H^-_l{\rm w}}x_{i{\rm w}}\ln^2
(x_{H^-_l{\rm w}}x_{i{\rm w}})}
{2(-x_{H^-_l{\rm w}}+x_{i{\rm w}})^2(-x_{H^-_l{\rm w}}+x_{j{\rm w}})}
\nonumber \\
&&\hspace{1.0cm}+\frac{x_{H^-_l{\rm w}}(x_{H^-_k{\rm w}}+x_{H^-_l{\rm w}})\Upsilon(
\frac{x_{i{\rm w}}}{x_{H^-_l{\rm w}}})}
{(-x_{H^-_l{\rm w}}+x_{i{\rm w}})^2(-x_{H^-_l{\rm w}}+x_{j{\rm w}})}+
(x_{H^-_l{\rm w}}\leftrightarrow x_{j{\rm w}})\Big)
\nonumber \\
&&\hspace{1.0cm}+\frac{-(x_{H^-_k{\rm w}}+x_{H^-_l{\rm w}})(3+7\ln x_{i{\rm w}}+\ln^2 x_{i{\rm w}}
)+x_{i{\rm w}}(5+9\ln x_{i{\rm w}}+4\ln^2 x_{i{\rm w}})}
{2(-x_{H^-_l{\rm w}}+x_{i{\rm w}})(-x_{i{\rm w}}+x_{j{\rm w}})}
\nonumber \\
&&\hspace{1.0cm}+\frac{(x_{H^-_k{\rm w}}+x_{H^-_l{\rm w}})x_{i{\rm w}}(-3\ln x_{i{\rm w}}
-2\ln^2 x_{i{\rm w}})+x_{i{\rm w}}^2(5\ln x_{i{\rm w}}+4\ln^2 x_{i{\rm w}})}
{2(-x_{H^-_l{\rm w}}+x_{i{\rm w}})(-x_{i{\rm w}}+x_{j{\rm w}})^2}
\nonumber \\
&&\hspace{1.0cm}+\frac{(x_{H^-_k{\rm w}}+x_{H^-_l{\rm w}})x_{i{\rm w}}(-3\ln x_{i{\rm w}}
-2\ln^2 x_{i{\rm w}})+x_{i{\rm w}}^2(5\ln x_{i{\rm w}}+4\ln^2 x_{i{\rm w}})}
{2(-x_{H^-_l{\rm w}}+x_{i{\rm w}})^2(-x_{i{\rm w}}+x_{j{\rm w}})}
\nonumber \\
&&\hspace{1.0cm}+\frac{3+7\ln x_{i{\rm w}}+\ln^2 x_{i{\rm w}}}{2(x_{j{\rm w}}-x_{i{\rm w}})}
\nonumber \\
&&\hspace{1.0cm}+\frac{\Big(x_{i{\rm w}}(3\ln x_{i{\rm w}}+2\ln^2 x_{i{\rm w}})+
(x_{i{\rm w}}\leftrightarrow x_{j{\rm w}})\Big)-2x_{j{\rm w}}\Upsilon(\frac{x_{i{\rm w}}}{x_{j{\rm w}}})}
{2(x_{j{\rm w}}-x_{i{\rm w}})^2}
\nonumber \\
&&\hspace{1.0cm}+\frac{3x_{H^-_l{\rm w}}^2\ln x_{H^-_l{\rm w}}}
{2(-x_{H^-_l{\rm w}}+x_{i{\rm w}})^2(-x_{H^-_l{\rm w}}+x_{j{\rm w}})}
+\frac{3x_{H^-_l{\rm w}}x_{j{\rm w}}\ln x_{j{\rm w}}}{2(-x_{j{\rm w}}+x_{i{\rm w}})^2(x_{H^-_l{\rm w}}-x_{j{\rm w}})}
\bigg)\nonumber \\
&&\hspace{1.0cm}+\frac{8}{3\sin^4\beta}x_{i{\rm w}}^2x_{j{\rm w}}({\cal Z}_{H}^{2k})^2
({\cal Z}_{H}^{2l})^2\bigg(-\sum\limits_{\sigma=u^i,u^j,H^-_l}
\frac{x_{\sigma}{\cal R}_{i_2}(\frac{x_{i{\rm w}}}{x_{\sigma}})}
{\prod\limits_{\rho\neq \sigma}(x_{\rho}-x_{\sigma})}
\nonumber \\
&&\hspace{1.0cm}\sum\limits_{\sigma=u^i,u^j,H^-_k,H^-_l}
\frac{x_{\sigma}(-2\ln x_{\sigma}+\ln^2 x_{\sigma}
+2(x_{H^-_k{\rm w}}-x_{i{\rm w}}){\cal R}_{i_2}(\frac{x_{i{\rm w}}}{x_{\sigma}})+
2\Upsilon(\frac{x_{i{\rm w}}}{x_{\sigma}}))}
{2\prod\limits_{\rho\neq \sigma}(x_{\rho}-x_{\sigma})}
\nonumber \\
&&\hspace{1.0cm}+\frac{x_{H^-_l{\rm w}}(5\ln x_{H^-_l{\rm w}} +\ln^2 x_{H^-_l{\rm w}})}
{2(x_{H^-_k{\rm w}}-x_{H^-_l{\rm w}})(x_{i{\rm w}}-x_{H^-_l{\rm w}})(x_{j{\rm w}}-x_{H^-_l{\rm w}})}
-\frac{x_{H^-_l{\rm w}}(5\ln x_{H^-_l{\rm w}} +\ln^2 x_{H^-_l{\rm w}})}
{2(x_{H^-_k{\rm w}}-x_{H^-_l{\rm w}})(x_{i{\rm w}}-x_{H^-_l{\rm w}})^2}\bigg)
\nonumber \\
&&\hspace{1.0cm}-\frac{16}{3\sin^4\beta}x_{i{\rm w}}x_{j{\rm w}}({\cal Z}_{H}^{2k})^2
({\cal Z}_{H}^{2l})^2\bigg(
\nonumber \\
&&\hspace{1.0cm}\sum\limits_{\sigma=u^i,u^j,H^-_l}
\frac{x_{\sigma}(3\ln x_{\sigma}+3\ln^2 x_{\sigma}
-2(x_{H^-_k{\rm w}}-x_{i{\rm w}}+x_{\sigma}){\cal R}_{i_2}(\frac{x_{i{\rm w}}}{x_{\sigma}})
+2\Upsilon(\frac{x_{i{\rm w}}}{x_{\sigma}}))}
{2\prod\limits_{\rho\neq \sigma}(x_{\rho}-x_{\sigma})}
\nonumber \\
&&\hspace{1.0cm}+\sum\limits_{\sigma=u^i,u^j,H^-_k,H^-_l}
\frac{x_{H^-_k{\rm w}}x_{\sigma}(-6\ln x_{\sigma}-3\ln^2 x_{\sigma}
+2(x_{H^-_k{\rm w}}-x_{i{\rm w}}){\cal R}_{i_2}(\frac{x_{i{\rm w}}}{x_{\sigma}})-
2\Upsilon(\frac{x_{i{\rm w}}}{x_{\sigma}}))}
{2\prod\limits_{\rho\neq \sigma}(x_{\rho}-x_{\sigma})}
\nonumber \\
&&\hspace{1.0cm}+\frac{x_{H^-_k{\rm w}}x_{H^-_l{\rm w}}(-\ln x_{H^-_l{\rm w}} +\ln^2 x_{H^-_l{\rm w}})}
{2(x_{H^-_k{\rm w}}-x_{H^-_l{\rm w}})(x_{i{\rm w}}-x_{H^-_l{\rm w}})(x_{j{\rm w}}-x_{H^-_l{\rm w}})}
-\frac{x_{H^-_k{\rm w}}x_{H^-_l{\rm w}}(4\ln x_{H^-_l{\rm w}} +\ln^2 x_{H^-_l{\rm w}})}
{2(x_{H^-_k{\rm w}}-x_{H^-_l{\rm w}})(x_{i{\rm w}}-x_{H^-_l{\rm w}})^2}\bigg)
\nonumber \\
&&\hspace{1.0cm}+\frac{1}{3\sin^4\beta}x_{i{\rm w}}x_{j{\rm w}}^2({\cal Z}_{H}^{2k})^2
({\cal Z}_{H}^{2l})^2\bigg(-\sum\limits_{\sigma=u^i,u^j,H^-_l}
\Big(\frac{{\cal R}_{i_2}(\frac{x_{H^-_k{\rm w}}}{x_{\sigma}})}
{\prod\limits_{\rho\neq \sigma}(x_{\rho}-x_{\sigma})}
\nonumber \\
&&\hspace{1.0cm}+\frac{(x_{H^-_l{\rm w}}-x_{j{\rm w}})
({\cal R}_{i_2}(\frac{x_{H^-_k{\rm w}}}{x_{\sigma}})-
{\cal R}_{i_2}(\frac{x_{j{\rm w}}}{x_{\sigma}}))+x_{\sigma}
(\Upsilon(\frac{x_{H^-_k{\rm w}}}{x_{\sigma}})-
\Upsilon(\frac{x_{j{\rm w}}}{x_{\sigma}}))}
{(x_{H^-_k{\rm w}}-x_{j{\rm w}})\prod\limits_{\rho\neq \sigma}(x_{\rho}-x_{\sigma})}\Big)
\nonumber \\
&&\hspace{1.0cm}+\frac{{\cal R}_{i_2}(\frac{x_{H^-_k{\rm w}}}{x_{i{\rm w}}})
-{\cal R}_{i_2}(\frac{x_{H^-_k{\rm w}}}{x_{j{\rm w}}})-
{\cal R}_{i_2}(\frac{x_{j{\rm w}}}{x_{i{\rm w}}})}
{(x_{H^-_k{\rm w}}-x_{j{\rm w}})(x_{i{\rm w}}-x_{j{\rm w}})}\bigg)
\nonumber \\
&&\hspace{1.0cm}-\frac{1}{3\sin^4\beta}x_{i{\rm w}}x_{j{\rm w}}({\cal Z}_{H}^{2k})^2
({\cal Z}_{H}^{2l})^2\bigg(
\sum\limits_{\sigma=u^i,u^j,H^-_l}
\Big(-\frac{x_{\sigma}(x_{H^-_k{\rm w}}\ln^2(x_{H^-_k{\rm w}}x_{\sigma})-
x_{j{\rm w}}\ln (x_{j{\rm w}}x_{\sigma}))}
{(x_{H^-_k{\rm w}}-x_{j{\rm w}})\prod\limits_{\rho\neq \sigma}(x_{\rho}-x_{\sigma})}
\nonumber \\
&&\hspace{1.0cm}-\frac{x_{\sigma}\ln^2 x_{\sigma}-2(x_{j{\rm w}}+x_{H^-_k{\rm w}}){\cal R}_
{i_2}(\frac{x_{H^-_k{\rm w}}}{x_{\sigma}})-
4x_{\sigma}\Upsilon(\frac{x_{H^-_k{\rm w}}}{x_{\sigma}})}
{\prod\limits_{\rho\neq \sigma}(x_{\rho}-x_{\sigma})}
\nonumber \\
&&\hspace{1.0cm}+\frac{(2x_{j{\rm w}}^2-6x_{H^-_l{\rm w}}^2)({\cal R}_
{i_2}(\frac{x_{H^-_k{\rm w}}}{x_{\sigma}})-{\cal R}_
{i_2}(\frac{x_{j{\rm w}}}{x_{\sigma}}))+8x_{H^-_l{\rm w}}x_{\sigma}
(\Upsilon(\frac{x_{H^-_k{\rm w}}}{x_{\sigma}})
-\Upsilon(\frac{x_{j{\rm w}}}{x_{\sigma}}))}
{(x_{H^-_k{\rm w}}-x_{j{\rm w}})\prod\limits_{\rho\neq \sigma}(x_{\rho}-x_{\sigma})}
\nonumber \\
&&\hspace{1.0cm}-\frac{6(x_{H^-_l{\rm w}}+x_{j{\rm w}})(
{\cal R}_{i_2}(\frac{x_{H^-_k{\rm w}}}{x_{i{\rm w}}})
-{\cal R}_{i_2}(\frac{x_{H^-_k{\rm w}}}{x_{j{\rm w}}})-
{\cal R}_{i_2}(\frac{x_{j{\rm w}}}{x_{i{\rm w}}}))}
{(x_{H^-_k{\rm w}}-x_{j{\rm w}})(x_{i{\rm w}}-x_{j{\rm w}})}
\nonumber \\
&&\hspace{1.0cm}-\frac{8(x_{i{\rm w}}\Upsilon(
\frac{x_{H^-_k{\rm w}}}{x_{i{\rm w}}})-x_{j{\rm w}}\Upsilon(\frac{x_{H^-_k{\rm w}}}{x_{j{\rm w}}}))}
{(x_{H^-_k{\rm w}}-x_{j{\rm w}})(x_{i{\rm w}}-x_{j{\rm w}})}+6\frac{\AE(x_{i{\rm w}},x_{H^-_k{\rm w}})
-\AE(x_{i{\rm w}},x_{j{\rm w}})}{x_{H^-_k{\rm w}}-x_{j{\rm w}}}\bigg)
\nonumber \\
&&\hspace{1.0cm}+\frac{4}{3\sin^4\beta}x_{i{\rm w}}x_{j{\rm w}}({\cal Z}_{H}^{2k})^2
({\cal Z}_{H}^{2l})^2\bigg(\frac{\AE(x_{i{\rm w}},x_{H^-_k{\rm w}})
-\AE(x_{i{\rm w}},x_{j{\rm w}})}{x_{H^-_k{\rm w}}-x_{j{\rm w}}}-
\frac{{\cal R}_{i_2}(\frac{x_{j{\rm w}}}{x_{H^-_l{\rm w}}})
-{\cal R}_{i_2}(\frac{x_{j{\rm w}}}{x_{i{\rm w}}})}
{x_{H^-_l{\rm w}}-x_{i{\rm w}}}\nonumber \\
&&\hspace{1.0cm}-\frac{(x_{H^-_k{\rm w}}+x_{H^-_l{\rm w}})
(-{\cal R}_{i_2}(\frac{x_{H^-_k{\rm w}}}{x_{H^-_l{\rm w}}})
-{\cal R}_{i_2}(\frac{x_{H^-_k{\rm w}}}{x_{i{\rm w}}})
-{\cal R}_{i_2}(\frac{x_{j{\rm w}}}{x_{H^-_l{\rm w}}})+
{\cal R}_{i_2}(\frac{x_{j{\rm w}}}{x_{i{\rm w}}}))}
{(x_{H^-_l{\rm w}}-x_{i{\rm w}})(x_{H^-_k{\rm w}}-x_{j{\rm w}})}
\nonumber \\
&&\hspace{1.0cm}+\frac{(x_{i{\rm w}}-x_{H^-_l{\rm w}})
(-\Upsilon(\frac{x_{H^-_k{\rm w}}}{x_{H^-_l{\rm w}}})
+\Upsilon(\frac{x_{j{\rm w}}}{x_{H^-_l{\rm w}}})+(x_{H^-_l{\rm w}}\rightarrow x_{i{\rm w}}))}
{(x_{H^-_l{\rm w}}-x_{i{\rm w}})(x_{H^-_k{\rm w}}-x_{j{\rm w}})}\bigg)+(x_{i{\rm w}}\leftrightarrow x_{j{\rm w}})\; ,
\label{hh1}
\end{eqnarray}
\begin{eqnarray}
&&HH_{2}=-\frac{1}{12\sin^2\beta}h_{b}h_{d}
{\cal Z}_{H}^{1k}{\cal Z}_{H}^{2k}{\cal Z}_{H}^{1l}{\cal Z}_{H}^{2l}
\bigg((x_{i{\rm w}}^2x_{j{\rm w}}+x_{i{\rm w}}x_{j{\rm w}}^2)F_{A}^{0}
+(x_{i{\rm w}}+x_{j{\rm w}})^2\Big(F_{A}^{1a}+F_A^{1b}-F_A^{1c}\Big)
\nonumber \\
&&\hspace{1.0cm}+(x_{i{\rm w}}+x_{j{\rm w}})
F_{A}^{2d}\bigg)(x_{i{\rm w}},x_{j{\rm w}},x_{H^-_l{\rm w}},0,x_{i{\rm w}},x_{j{\rm w}},x_{H^-_k{\rm w}})
\nonumber \\
&&\hspace{1.0cm}+\frac{1}{3\sin^2\beta}(x_{i{\rm w}}+x_{j{\rm w}})h_{b}h_{d}
{\cal Z}_{H}^{1k}{\cal Z}_{H}^{1l}{\cal Z}_{H}^{2k}{\cal Z}_{H}^{2l}
\bigg[-\sum\limits_{\sigma=u^i,u^j,H^-_l}
\frac{x_{H^-_l{\rm w}}{\cal R}_{i_2}(\frac{x_{H^-_k{\rm w}}}{x_{\sigma}})}
{\prod\limits_{\rho\neq \sigma}(x_{\rho}-x_{\sigma})}
-\frac{{\cal R}_{i_2}(\frac{x_{H^-_k{\rm w}}}{x_{i{\rm w}}})-
{\cal R}_{i_2}(\frac{x_{H^-_k{\rm w}}}{x_{j{\rm w}}})}{-x_{i{\rm w}}+x_{j{\rm w}}}
\nonumber \\
&&\hspace{1.0cm}+2\bigg(\frac{{\cal R}_{i_2}(\frac{x_{j{\rm w}}}{x_{i{\rm w}}})
-{\cal R}_{i_2}(\frac{x_{j{\rm w}}}{x_{H^-_l{\rm w}}})}
{(x_{H^-_k{\rm w}}-x_{H^-_l{\rm w}})(x_{i{\rm w}}-x_{H^-_l{\rm w}})}
+\sum\limits_{\sigma=u^i,H^-_k,H^-_l}\Big(
\frac{x_{\sigma}\ln^2 x_{\sigma}}{\prod\limits_{\rho\neq \sigma}(x_{\rho}-
x_{\sigma})}+(x_{H^-_k{\rm w}}+x_{j{\rm w}})\frac{{\cal R}_{i_2}(\frac{x_{j{\rm w}}}{x_{\sigma}})}
{\prod\limits_{\rho\neq \sigma}(x_{\rho}-x_{\sigma})}\Big)\bigg)
\nonumber \\
&&\hspace{1.0cm}
+\sum\limits_{\sigma=u^j,H^-_k,H^-_l}\Big(
-\frac{x_{H^-_l{\rm w}}x_{i{\rm w}}\ln^2(x_{H^-_l{\rm w}}x_{\sigma})}
{(x_{i{\rm w}}-x_{\sigma})^2\prod\limits_{\rho\neq \sigma}
(x_{\rho}-x_{\sigma})}\nonumber \\
&&\hspace{1.0cm}
+\frac{x_{\sigma}\big((x_{H^-_l{\rm w}}^2+x_{H^-_l{\rm w}}x_{i{\rm w}}+x_{i{\rm w}}^2)
\ln x_{\sigma}-2(x_{H^-_l{\rm w}}^2+x_{i{\rm w}}^2)(\ln^2 x_{\sigma}
+\Upsilon(\frac{x_{i{\rm w}}}{x_{\sigma}}))\big)}
{(x_{i{\rm w}}-x_{\sigma})^2\prod\limits_{\rho\neq \sigma}(x_{\rho}-x_{\sigma})}\Big)
\nonumber \\
&&\hspace{1.0cm}+\frac{x_{H^-_k{\rm w}}x_{i{\rm w}}(-1+2\ln x_{i{\rm w}}+8\ln^2 x_{i{\rm w}})+
(x_{H^-_k{\rm w}}^2+x_{i{\rm w}}^2)(-1+3\ln x_{i{\rm w}}+2\ln^2 x_{i{\rm w}})}
{(x_{H^-_k{\rm w}}-x_{i{\rm w}})(x_{H^-_l{\rm w}}-x_{i{\rm w}})(x_{j{\rm w}}-x_{i{\rm w}})}
\nonumber \\
&&\hspace{1.0cm}+\frac{(x_{H^-_k{\rm w}}^2+x_{i{\rm w}}^2)x_{i{\rm w}}(-\ln x_{i{\rm w}}
+2\ln^2 x_{i{\rm w}})+3x_{H^-_k{\rm w}}x_{i{\rm w}}^2\ln^2 x_{i{\rm w}})}
{(x_{H^-_k{\rm w}}-x_{i{\rm w}})(x_{H^-_l{\rm w}}-x_{i{\rm w}})(x_{j{\rm w}}-x_{i{\rm w}})^2}
\nonumber \\
&&\hspace{1.0cm}+\frac{(x_{H^-_k{\rm w}}^2+x_{i{\rm w}}^2)x_{i{\rm w}}(-\ln x_{i{\rm w}}
+2\ln^2 x_{i{\rm w}})+3x_{H^-_k{\rm w}}x_{i{\rm w}}^2\ln^2 x_{i{\rm w}})}
{(x_{H^-_k{\rm w}}-x_{i{\rm w}})(x_{H^-_l{\rm w}}-x_{i{\rm w}})^2(x_{j{\rm w}}-x_{i{\rm w}})}
\nonumber \\
&&\hspace{1.0cm}+\frac{(x_{H^-_k{\rm w}}^2+x_{i{\rm w}}^2)x_{i{\rm w}}(-\ln x_{i{\rm w}}
+2\ln^2 x_{i{\rm w}})+3x_{H^-_k{\rm w}}x_{i{\rm w}}^2\ln^2 x_{i{\rm w}})}
{(x_{H^-_k{\rm w}}-x_{i{\rm w}})^2(x_{H^-_l{\rm w}}-x_{i{\rm w}})(x_{j{\rm w}}-x_{i{\rm w}})}
\nonumber \\
&&\hspace{1.0cm}+\frac{(x_{H^-_k{\rm w}}+x_{H^-_l{\rm w}})(1-3\ln x_{i{\rm w}}
-2\ln^2 x_{i{\rm w}})+x_{i{\rm w}}(10\ln x_{i{\rm w}}+4\ln^2 x_{i{\rm w}})}
{(x_{H^-_l{\rm w}}-x_{i{\rm w}})(x_{j{\rm w}}-x_{i{\rm w}})}
\nonumber \\
&&\hspace{1.0cm}+\Big(\frac{(x_{H^-_k{\rm w}}+x_{H^-_l{\rm w}})x_{H^-_l{\rm w}}
(-\ln x_{H^-_l{\rm w}}+2\ln^2x_{H^-_l{\rm w}}+2\Upsilon(\frac{x_{i{\rm w}}}{x_{H^-_l{\rm w}}}))}
{(-x_{H^-_l{\rm w}}+x_{i{\rm w}})^2(x_{j{\rm w}}-x_{H^-_l{\rm w}})}
\nonumber \\
&&\hspace{1.0cm}+\frac{x_{H^-_l{\rm w}}x_{i{\rm w}}(\ln x_{H^-_l{\rm w}}+\ln^2(x_{H^-_l{\rm w}}x_{i{\rm w}}))}
{(-x_{H^-_l{\rm w}}+x_{i{\rm w}})^2(x_{j{\rm w}}-x_{H^-_l{\rm w}})}+
(x_{H^-_l{\rm w}}\leftrightarrow x_{j{\rm w}})\Big)
\nonumber \\
&&\hspace{1.0cm}+\frac{x_{i{\rm w}}(x_{H^-_k{\rm w}}+x_{H^-_l{\rm w}})(\ln x_{i{\rm w}}
-2\ln^2 x_{i{\rm w}})+3x_{i{\rm w}}^2\ln^2 x_{i{\rm w}}}
{(-x_{H^-_l{\rm w}}+x_{i{\rm w}})^2(x_{j{\rm w}}-x_{H^-_l{\rm w}})}
\nonumber \\
&&\hspace{1.0cm}+\frac{x_{i{\rm w}}(x_{H^-_k{\rm w}}+x_{H^-_l{\rm w}})(\ln x_{i{\rm w}}
-2\ln^2 x_{i{\rm w}})+3x_{i{\rm w}}^2\ln^2 x_{i{\rm w}}}
{(-x_{j{\rm w}}+x_{i{\rm w}})^2(-x_{j{\rm w}}+x_{H^-_l{\rm w}})}
\nonumber \\
&&\hspace{1.0cm}+\frac{-1+3\ln x_{i{\rm w}}+2\ln^2 x_{i{\rm w}}}{2(x_{j{\rm w}}-x_{i{\rm w}})}
+\frac{-x_{i{\rm w}}\ln x_{i{\rm w}}+2x_{i{\rm w}}\ln^2x_{i{\rm w}}+x_{j{\rm w}}(\ln x_{j{\rm w}}
-2\ln^2 x_{j{\rm w}})}{(x_{j{\rm w}}-x_{i{\rm w}})^2}
\nonumber \\
&&\hspace{1.0cm}
+\frac{21(\AE(x_{i{\rm w}},x_{H^-_k{\rm w}})
-\AE(x_{i{\rm w}},x_{j{\rm w}}))}{16(x_{H^-_k{\rm w}}-x_{j{\rm w}})}
\nonumber \\
&&\hspace{1.0cm}
-4\bigg(\sum\limits_{\sigma=u^i,u^j,H^-_k,H^-_l}
\frac{x_{H^-_k{\rm w}}x_{\sigma}(2\ln x_{\sigma}-3\ln^2 x_{\sigma}
+2(x_{H^-_k{\rm w}}-x_{i{\rm w}}){\cal R}_{i_2}(\frac{x_{i{\rm w}}}{x_{\sigma}})
+\Upsilon(\frac{x_{i{\rm w}}}{x_{\sigma}}))}
{\prod\limits_{\rho\neq \sigma}(x_{\rho}-x_{\sigma})}
\nonumber \\
&&\hspace{1.0cm}-\sum\limits_{\sigma=u^i,u^j,H^-_l}
\frac{x_{\sigma}(2\ln x_{\sigma}-3\ln^2 x_{\sigma}
+2(x_{H^-_k{\rm w}}-x_{i{\rm w}}){\cal R}_{i_2}(\frac{x_{i{\rm w}}}{x_{\sigma}})
+\Upsilon(\frac{x_{i{\rm w}}}{x_{\sigma}}))}
{\prod\limits_{\rho\neq \sigma}(x_{\rho}-x_{\sigma})}
\nonumber \\
&&\hspace{1.0cm}
-\frac{2{\cal R}_{i_2}(\frac{x_{i{\rm w}}}{x_{j{\rm w}}})}
{x_{i{\rm w}}-x_{j{\rm w}}}+\frac{x_{H^-_k{\rm w}}x_{H^-_l{\rm w}}
(x_{i{\rm w}}-x_{j{\rm w}})\ln^2 x_{H^-_l{\rm w}}}{(x_{H^-_k{\rm w}}-x_{H^-_l{\rm w}})
(x_{i{\rm w}}-x_{H^-_l{\rm w}})^2(x_{j{\rm w}}-x_{H^-_l{\rm w}})}\bigg)
\nonumber \\
&&\hspace{1.0cm}
-\frac{x_{H^-_l{\rm w}}(x_{H^-_l{\rm w}}-x_{j{\rm w}})}
{x_{H^-_k{\rm w}}-x_{j{\rm w}}}\sum\limits_{\sigma=u^i,u^j,H^-_l}
\frac{{\cal R}_{i_2}(\frac{x_{H^-_k{\rm w}}}{x_{\sigma}})-
{\cal R}_{i_2}(\frac{x_{j{\rm w}}}{x_{\sigma}})}
{\prod\limits_{\rho\neq \sigma}(x_{\rho}-x_{\sigma})}
+\frac{x_{i{\rm w}}\Upsilon(\frac{x_{H^-_k{\rm w}}}{x_{i{\rm w}}})
-x_{j{\rm w}}\Upsilon(\frac{x_{H^-_k{\rm w}}}{x_{j{\rm w}}})}
{(x_{H^-_k{\rm w}}-x_{j{\rm w}})(x_{i{\rm w}}-x_{j{\rm w}})}
\nonumber \\
&&\hspace{1.0cm}
+\frac{x_{H^-_l{\rm w}}}
{x_{H^-_k{\rm w}}-x_{j{\rm w}}}\sum\limits_{\sigma=u^i,u^j,H^-_l}
\frac{x_{\sigma}(\Upsilon(\frac{x_{H^-_k{\rm w}}}{x_{\sigma}})-
\Upsilon(\frac{x_{j{\rm w}}}{x_{\sigma}}))}
{\prod\limits_{\rho\neq \sigma}(x_{\rho}-x_{\sigma})}
-\frac{x_{H^-_l{\rm w}}(
{\cal R}_{i_2}(\frac{x_{H^-_k{\rm w}}}{x_{i{\rm w}}})-
{\cal R}_{i_2}(\frac{x_{H^-_k{\rm w}}}{x_{j{\rm w}}})-
{\cal R}_{i_2}(\frac{x_{j{\rm w}}}{x_{i{\rm w}}}))}
{(x_{H^-_k{\rm w}}-x_{j{\rm w}})^2(x_{i{\rm w}}-x_{j{\rm w}})}
\nonumber \\
&&\hspace{1.0cm}+\frac{5}{16}\bigg(
\frac{{\cal R}_{i_2}(\frac{x_{j{\rm w}}}{x_{H^-_l{\rm w}}})
-{\cal R}_{i_2}(\frac{x_{j{\rm w}}}{x_{i{\rm w}}})}
{x_{i{\rm w}}-x_{H^-_l{\rm w}}}-\frac{(-\Upsilon(\frac{x_{H^-_k{\rm w}}}{x_{H^-_l{\rm w}}})
+\Upsilon(\frac{x_{j{\rm w}}}{x_{H^-_l{\rm w}}})+(x_{H^-_l{\rm w}}\rightarrow x_{i{\rm w}}))}
{(x_{H^-_k{\rm w}}-x_{j{\rm w}})}
\nonumber \\
&&\hspace{1.0cm}-\frac{(x_{H^-_k{\rm w}}+x_{H^-_l{\rm w}})
(-{\cal R}_{i_2}(\frac{x_{H^-_k{\rm w}}}{x_{H^-_l{\rm w}}})
-{\cal R}_{i_2}(\frac{x_{H^-_k{\rm w}}}{x_{i{\rm w}}})
-{\cal R}_{i_2}(\frac{x_{j{\rm w}}}{x_{H^-_l{\rm w}}})+
{\cal R}_{i_2}(\frac{x_{j{\rm w}}}{x_{i{\rm w}}}))}
{(x_{H^-_l{\rm w}}-x_{i{\rm w}})(x_{H^-_k{\rm w}}-x_{j{\rm w}})}\bigg)\bigg]
\nonumber \\
&&\hspace{1.0cm}
-\frac{x_{i{\rm w}}(x_{i{\rm w}}+x_{j{\rm w}})}{3\sin^2\beta}h_bh_d
{\cal Z}_{H}^{1k}{\cal Z}_{H}^{1l}{\cal Z}_{H}^{2k}{\cal Z}_{H}^{2l}
\bigg[\nonumber \\
&&\hspace{1.0cm}
-2\sum\limits_{\sigma=u^j,H^-_k,H^-_l}
\frac{(3x_{H^-_k{\rm w}}-x_{i{\rm w}})x_{\sigma}(\ln x_{\sigma}-\ln^2 x_{\sigma}
-\Upsilon(\frac{x_{i{\rm w}}}{x_{\sigma}}))+x_{H^-_k{\rm w}}\ln x_{\sigma}
+x_{i{\rm w}}\ln^2(x_{H^-_k{\rm w}}x_{i{\rm w}})}
{(x_{i{\rm w}}-x_{\sigma})^2\prod\limits_{\rho\neq \sigma}(x_{\rho}-x_{\sigma})}
\nonumber \\
&&\hspace{1.0cm}+\frac{2(5x_{i{\rm w}}-3x_{H^-_k{\rm w}})(-1+\ln x_{i{\rm w}}+\ln^2 x_{i{\rm w}})
+x_{H^-_k{\rm w}}(1+\ln x_{i{\rm w}})+x_{i{\rm w}}(8-\ln x_{i{\rm w}})}
{(x_{H^-_k{\rm w}}-x_{i{\rm w}})(x_{H^-_l{\rm w}}-x_{i{\rm w}})(x_{j{\rm w}}-x_{i{\rm w}})}
\nonumber \\
&&\hspace{1.0cm}
+\frac{6x_{i{\rm w}}(x_{H^-_k{\rm w}}-x_{i{\rm w}})(\ln x_{i{\rm w}}-\ln^2 x_{i{\rm w}})+x_{i{\rm w}}
(4x_{i{\rm w}}+x_{H^-_k{\rm w}})\ln x_{i{\rm w}}}
{(x_{H^-_k{\rm w}}-x_{i{\rm w}})(x_{H^-_l{\rm w}}-x_{i{\rm w}})(x_{j{\rm w}}-x_{i{\rm w}})^2}
\nonumber \\
&&\hspace{1.0cm}+
\frac{6x_{i{\rm w}}(x_{H^-_k{\rm w}}-x_{i{\rm w}})(\ln x_{i{\rm w}}-\ln^2 x_{i{\rm w}})+x_{i{\rm w}}
(4x_{i{\rm w}}+x_{H^-_k{\rm w}})\ln x_{i{\rm w}}}
{(x_{H^-_k{\rm w}}-x_{i{\rm w}})(x_{H^-_l{\rm w}}-x_{i{\rm w}})^2(x_{j{\rm w}}-x_{i{\rm w}})}
\nonumber \\
&&\hspace{1.0cm}
+\frac{6x_{i{\rm w}}(x_{H^-_k{\rm w}}-x_{i{\rm w}})(\ln x_{i{\rm w}}-\ln^2 x_{i{\rm w}})+x_{i{\rm w}}
(4x_{i{\rm w}}+x_{H^-_k{\rm w}})\ln x_{i{\rm w}}}
{(x_{H^-_k{\rm w}}-x_{i{\rm w}})^2(x_{H^-_l{\rm w}}-x_{i{\rm w}})(x_{j{\rm w}}-x_{i{\rm w}})}
\nonumber \\
&&\hspace{1.0cm}+\Big(\frac{16x_{H^-_l{\rm w}}
\Upsilon(\frac{x_{i{\rm w}}}{x_{H^-_l{\rm w}}})+x_{i{\rm w}}(\ln x_{i{\rm w}}
-2\ln^2x_{i{\rm w}})-x_{H^-_l{\rm w}}(13\ln x_{H^-_l{\rm w}}-14\ln^2x_{H^-_l{\rm w}})
}{(-x_{H^-_l{\rm w}}+x_{i{\rm w}})^2(-x_{H^-_l{\rm w}}+x_{j{\rm w}})}
+(x_{H^-_l{\rm w}}\leftrightarrow x_{j{\rm w}})\Big)\nonumber \\
&&\hspace{1.0cm}+16\Big(\frac{-1-\ln x_{i{\rm w}}-ln^2 x_{i{\rm w}}}
{(x_{H^-_l{\rm w}}-x_{i{\rm w}})(x_{j{\rm w}}-x_{i{\rm w}})}
+\frac{x_{i{\rm w}}(\ln x_{i{\rm w}}-\ln^2 x_{i{\rm w}})}
{(x_{H^-_l{\rm w}}-x_{i{\rm w}})(x_{j{\rm w}}-x_{i{\rm w}})^2}
+\frac{x_{i{\rm w}}(\ln x_{i{\rm w}}-\ln^2 x_{i{\rm w}})}
{(x_{H^-_l{\rm w}}-x_{i{\rm w}})^2(x_{j{\rm w}}-x_{i{\rm w}})}\nonumber \\
&&\hspace{1.0cm}
-\frac{1-3\ln x_{i{\rm w}}-2\ln^2x_{i{\rm w}}}
{(x_{H^-_l{\rm w}}-x_{i{\rm w}})(x_{H^-_l{\rm w}}-x_{j{\rm w}})}\Big)\bigg)
-2\bigg(\sum\limits_{\sigma=u^i,u^j,H^-_l}
\Big(\frac{{\cal R}_{i_2}(\frac{x_{H^-_k{\rm w}}}{x_{\sigma}})}
{\prod\limits_{\rho\neq \sigma}(x_{\rho}-x_{\sigma})}
\nonumber \\
&&\hspace{1.0cm}
-\frac{(x_{H^-_l{\rm w}}-x_{j{\rm w}})
({\cal R}_{i_2}(\frac{x_{H^-_k{\rm w}}}{x_{\sigma}})-
{\cal R}_{i_2}(\frac{x_{j{\rm w}}}{x_{\sigma}}))+x_{\sigma}
(\Upsilon(\frac{x_{H^-_k{\rm w}}}{x_{\sigma}})-
\Upsilon(\frac{x_{j{\rm w}}}{x_{\sigma}}))}
{(x_{H^-_k{\rm w}}-x_{j{\rm w}})\prod\limits_{\rho\neq \sigma}(x_{\rho}-x_{\sigma})}\Big)
\nonumber \\
&&\hspace{1.0cm}-\frac{
{\cal R}_{i_2}(\frac{x_{H^-_k{\rm w}}}{x_{i{\rm w}}})
-{\cal R}_{i_2}(\frac{x_{H^-_k{\rm w}}}{x_{j{\rm w}}})-
{\cal R}_{i_2}(\frac{x_{j{\rm w}}}{x_{i{\rm w}}})}
{(x_{H^-_k{\rm w}}-x_{j{\rm w}})(x_{i{\rm w}}-x_{j{\rm w}})}\bigg)
+\frac{2x_{H^-_l{\rm w}}(\ln x_{H^-_l{\rm w}}+\ln^2x_{H^-_l{\rm w}})}
{(x_{H^-_k{\rm w}}-x_{H^-_k{\rm w}})(-x_{H^-_k{\rm w}}+x_{i{\rm w}})^2}
\nonumber \\
&&\hspace{1.0cm}
+2\sum\limits_{\sigma=u^i,u^j,H^-_k,H^-_l}
\frac{x_{\sigma}(-2\ln x_{\sigma}+\ln^2 x_{\sigma}+2(x_{H^-_k{\rm w}}
-x_{i{\rm w}}){\cal R}_{i_2}(\frac{x_{i{\rm w}}}{x_{\sigma}})+2
\Upsilon(\frac{x_{i{\rm w}}}{x_{\sigma}}))}
{2\prod\limits_{\rho\neq \sigma}(x_{\rho}-x_{\sigma})}\bigg]
\nonumber \\
&&\hspace{1.0cm}
-\frac{1}{3\sin^2\beta}x_{i{\rm w}}x_{j{\rm w}}h_bh_d
(({\cal Z}_{H}^{1k})^2({\cal Z}_{H}^{2l})^2+
({\cal Z}_{H}^{2k})^2({\cal Z}_{H}^{1l})^2)
\bigg[\nonumber \\
&&\hspace{1.0cm}
-8\bigg(\sum\limits_{\sigma=u^j,H^-_k,H^-_l}
\frac{x_{H^-_k{\rm w}}x_{\sigma}(\ln x_{\sigma}-2\ln^2x_{\sigma}
-2\Upsilon(\frac{x_{i{\rm w}}}{x_{\sigma}}))}
{(x_{i{\rm w}}-x_{\sigma})^2\prod\limits_{\rho\neq \sigma}
(x_{\rho}-x_{\sigma})}
\nonumber \\
&&\hspace{1.0cm}
+\Big(\frac{-x_{H^-_l{\rm w}}\ln x_{H^-_l{\rm w}}
+\ln^2 x_{H^-_l{\rm w}}+2\Upsilon(\frac{x_{i{\rm w}}}{x_{H^-_l{\rm w}}})}
{2(-x_{H^-_l{\rm w}}+x_{i{\rm w}})^2(-x_{H^-_l{\rm w}}+x_{j{\rm w}})}
+(x_{H^-_l{\rm w}} \leftrightarrow x_{j{\rm w}})\Big)
\nonumber \\
&&\hspace{1.0cm}+\frac{x_{H^-_k{\rm w}}(-1+3\ln x_{i{\rm w}}+2\ln^2 x_{i{\rm w}}
)}
{(x_{H^-_k{\rm w}}-x_{i{\rm w}})(x_{H^-_l{\rm w}}-x_{i{\rm w}})(x_{j{\rm w}}-x_{i{\rm w}})}
+\frac{x_{H^-_k{\rm w}}x_{i{\rm w}}(-\ln x_{i{\rm w}}+2\ln^2 x_{i{\rm w}}
)}
{(x_{H^-_k{\rm w}}-x_{i{\rm w}})(x_{H^-_l{\rm w}}-x_{i{\rm w}})(x_{j{\rm w}}-x_{i{\rm w}})^2}
\nonumber \\
&&\hspace{1.0cm}+\frac{x_{H^-_k{\rm w}}x_{i{\rm w}}(-\ln x_{i{\rm w}}+2\ln^2 x_{i{\rm w}}
)}
{(x_{H^-_k{\rm w}}-x_{i{\rm w}})(x_{H^-_l{\rm w}}-x_{i{\rm w}})^2(x_{j{\rm w}}-x_{i{\rm w}})}
+\frac{x_{H^-_k{\rm w}}x_{i{\rm w}}(-\ln x_{i{\rm w}}+2\ln^2 x_{i{\rm w}}
)}
{(x_{H^-_k{\rm w}}-x_{i{\rm w}})^2(x_{H^-_l{\rm w}}-x_{i{\rm w}})(x_{j{\rm w}}-x_{i{\rm w}})}
\nonumber \\
&&\hspace{1.0cm}-\frac{x_{i{\rm w}}(\ln x_{i{\rm w}}-2\ln^2 x_{i{\rm w}})}
{(x_{H^-_l{\rm w}}-x_{i{\rm w}})(x_{j{\rm w}}-x_{i{\rm w}})^2}
-\frac{x_{i{\rm w}}(\ln x_{i{\rm w}}-2\ln^2 x_{i{\rm w}})}
{(x_{H^-_l{\rm w}}-x_{i{\rm w}})^2(x_{j{\rm w}}-x_{i{\rm w}})}
-\frac{1-3\ln x_{i{\rm w}}-2\ln^2 x_{i{\rm w}})}
{(x_{H^-_l{\rm w}}-x_{i{\rm w}})(x_{j{\rm w}}-x_{i{\rm w}})}\bigg)
\nonumber \\
&&\hspace{1.0cm}
+\sum\limits_{\sigma=u^i,u^j,H^-_l}
\Big(\frac{{\cal R}_{i_2}(\frac{x_{H^-_k{\rm w}}}{x_{\sigma}})}
{\prod\limits_{\rho\neq \sigma}(x_{\rho}-x_{\sigma})}
+\frac{(x_{H^-_l{\rm w}}-x_{j{\rm w}})
({\cal R}_{i_2}(\frac{x_{H^-_k{\rm w}}}{x_{\sigma}})-
{\cal R}_{i_2}(\frac{x_{j{\rm w}}}{x_{\sigma}}))+x_{\sigma}
(\Upsilon(\frac{x_{H^-_k{\rm w}}}{x_{\sigma}})-
\Upsilon(\frac{x_{j{\rm w}}}{x_{\sigma}}))}
{(x_{H^-_k{\rm w}}-x_{j{\rm w}})\prod\limits_{\rho\neq \sigma}
(x_{\rho}-x_{\sigma})}\Big)
\nonumber \\
&&\hspace{1.0cm}-\frac{
{\cal R}_{i_2}(\frac{x_{H^-_k{\rm w}}}{x_{i{\rm w}}})
-{\cal R}_{i_2}(\frac{x_{H^-_k{\rm w}}}{x_{j{\rm w}}})-
{\cal R}_{i_2}(\frac{x_{j{\rm w}}}{x_{i{\rm w}}})}
{(x_{H^-_k{\rm w}}-x_{j{\rm w}})(x_{i{\rm w}}-x_{j{\rm w}})}\bigg)
-16\sum\limits_{\sigma=u^i,H^-_k,H^-_l}
\frac{{\cal R}_{i_2}(\frac{x_{j{\rm w}}}{x_{\sigma}})}
{\prod\limits_{\rho\neq \sigma}(x_{\rho}-x_{\sigma})}
+48\sum\limits_{\sigma=u^i,u^j,H^-_l}
\frac{x_{\sigma}{\cal R}_{i_2}(\frac{x_{i{\rm w}}}{x_{\sigma}})}
{\prod\limits_{\rho\neq \sigma}(x_{\rho}-x_{\sigma})}
\nonumber \\
&&\hspace{1.0cm}
-8\bigg(\sum\limits_{\sigma=u^i,u^j,H^-_k,H^-_l}
\frac{13x_{\sigma}\ln^2 x_{\sigma}-2x_{\sigma}\ln x_{\sigma}
-6x_{\sigma}(x_{H^-_k{\rm w}}
-x_{i{\rm w}}){\cal R}_{i_2}(\frac{x_{i{\rm w}}}{x_{\sigma}})+10x_{\sigma}
\Upsilon(\frac{x_{i{\rm w}}}{x_{\sigma}})}
{\prod\limits_{\rho\neq \sigma}(x_{\rho}-x_{\sigma})}
\nonumber \\
&&\hspace{1.0cm}+\frac{x_{H^-_l{\rm w}}(2\ln x_{H^-_l{\rm w}}+3\ln^2x_{H^-_l{\rm w}})}
{(x_{H^-_k{\rm w}}-x_{H^-_l{\rm w}})(-x_{H^-_l{\rm w}}+x_{i{\rm w}})(-x_{H^-_l{\rm w}}+x_{j{\rm w}})}
-\frac{x_{H^-_l{\rm w}}(2\ln x_{H^-_l{\rm w}}+3\ln^2x_{H^-_l{\rm w}})}
{(x_{H^-_k{\rm w}}-x_{H^-_l{\rm w}})(-x_{H^-_l{\rm w}}+x_{i{\rm w}})^2}\bigg)
\nonumber \\
&&\hspace{1.0cm}
-16\bigg(\sum\limits_{\sigma=u^j,H^-_k,H^-_l}
\frac{x_{\sigma}(2x_{i{\rm w}}\ln x_{\sigma}-2(x_{H^-_k{\rm w}}
+x_{i{\rm w}})(\ln^2x_{\sigma}+\Upsilon(\frac{x_{i{\rm w}}}{x_{\sigma}}))
-x_{i{\rm w}}\ln^2(x_{i{\rm w}}x_{j{\rm w}}))}
{2(x_{i{\rm w}}-x_{\sigma})^2\prod\limits_{\rho\neq \sigma}
(x_{\rho}-x_{\sigma})}\nonumber \\
&&\hspace{1.0cm}+\frac{-x_{i{\rm w}}+(2x_{H^-_k{\rm w}}+4x_{i{\rm w}})\ln x_{i{\rm w}}+
(x_{H^-_k{\rm w}}+5x_{i{\rm w}})\ln^2 x_{i{\rm w}}}
{(x_{H^-_k{\rm w}}-x_{i{\rm w}})(x_{H^-_l{\rm w}}-x_{i{\rm w}})(x_{j{\rm w}}-x_{i{\rm w}})}
+\frac{x_{i{\rm w}}^2\ln x_{i{\rm w}}+x_{i{\rm w}}(x_{H^-_k{\rm w}}
+3x_{i{\rm w}})\ln^2 x_{i{\rm w}}}
{(x_{H^-_k{\rm w}}-x_{i{\rm w}})(x_{H^-_l{\rm w}}-x_{i{\rm w}})(x_{j{\rm w}}-x_{i{\rm w}})^2}
\nonumber \\
&&\hspace{1.0cm}+\frac{x_{i{\rm w}}^2\ln x_{i{\rm w}}+x_{i{\rm w}}(x_{H^-_k{\rm w}}
+3x_{i{\rm w}})\ln^2 x_{i{\rm w}}}
{(x_{H^-_k{\rm w}}-x_{i{\rm w}})(x_{H^-_l{\rm w}}-x_{i{\rm w}})^2(x_{j{\rm w}}-x_{i{\rm w}})}
+\frac{x_{i{\rm w}}^2\ln x_{i{\rm w}}+x_{i{\rm w}}(x_{H^-_k{\rm w}}
+3x_{i{\rm w}})\ln^2 x_{i{\rm w}}}
{(x_{H^-_k{\rm w}}-x_{i{\rm w}})^2(x_{H^-_l{\rm w}}-x_{i{\rm w}})(x_{j{\rm w}}-x_{i{\rm w}})}
\nonumber \\
&&\hspace{1.0cm}+\Big(\frac{x_{H^-_l{\rm w}}(
ln^2 x_{H^-_l{\rm w}}+\Upsilon(\frac{x_{i{\rm w}}}{x_{H^-_l{\rm w}}}))}
{(-x_{H^-_l{\rm w}}+x_{i{\rm w}})^2(-x_{H^-_l{\rm w}}+x_{j{\rm w}})}
+(x_{H^-_l{\rm w}} \leftrightarrow x_{j{\rm w}})\Big)
-\frac{2\ln x_{i{\rm w}}+\ln^2 x_{i{\rm w}}}
{(-x_{H^-_l{\rm w}}+x_{i{\rm w}})(-x_{i{\rm w}}+x_{j{\rm w}})}\nonumber \\
&&\hspace{1.0cm}-\frac{x_{i{\rm w}}\ln^2 x_{i{\rm w}}}
{(-x_{H^-_l{\rm w}}+x_{i{\rm w}})(-x_{i{\rm w}}+x_{j{\rm w}})^2}
-\frac{x_{i{\rm w}}\ln^2 x_{i{\rm w}}}
{(-x_{H^-_l{\rm w}}+x_{i{\rm w}})^2(-x_{i{\rm w}}+x_{j{\rm w}})}\bigg)\bigg]\; ,
\label{hh2}
\end{eqnarray}
\begin{eqnarray}
HH_3=-2HH_2\; ,
\label{hh3}
\end{eqnarray}
\begin{eqnarray}
&&HH_{4}=\frac{h_b^2+h_d^2}{\sin^2\beta}x_{i{\rm w}}x_{j{\rm w}}
{\cal Z}_{H}^{1k}{\cal Z}_{H}^{1l}{\cal Z}_{H}^{2k}{\cal Z}_{H}^{2l}
\bigg[\Big(-F_{A}^{1a}-F_{A}^{1b}+\frac{13}{32}F_{A}^{1c}\Big)
(x_{i{\rm w}},x_{j{\rm w}},x_{H^-_l{\rm w}},0,x_{i{\rm w}},x_{j{\rm w}},x_{H^-_k{\rm w}})
\nonumber \\
&&\hspace{1.0cm}-\frac{4}{3}
\bigg(-\sum\limits_{\sigma=u^j,H^-_k,H^-_l}
\frac{x_{H^-_k{\rm w}}x_{\sigma}(\ln x_{\sigma}-2\ln^2x_{\sigma}
-2\Upsilon(\frac{x_{i{\rm w}}}{x_{\sigma}}))}
{(x_{i{\rm w}}-x_{\sigma})^2\prod\limits_{\rho\neq \sigma}
(x_{\rho}-x_{\sigma})}
\nonumber \\
&&\hspace{1.0cm}+\Big(\frac{-x_{H^-_l{\rm w}}\ln x_{H^-_l{\rm w}})
+\ln^2 x_{H^-_l{\rm w}}+2\Upsilon(\frac{x_{i{\rm w}}}{x_{H^-_l{\rm w}}})}
{(-x_{H^-_l{\rm w}}+x_{i{\rm w}})^2(-x_{H^-_l{\rm w}}+x_{j{\rm w}})}
+(x_{H^-_l{\rm w}} \leftrightarrow x_{j{\rm w}})\Big)
\nonumber \\
&&\hspace{1.0cm}+\frac{x_{H^-_k{\rm w}}(3\ln x_{i{\rm w}}+2\ln^2 x_{i{\rm w}}
)}
{(x_{H^-_k{\rm w}}-x_{i{\rm w}})(x_{H^-_l{\rm w}}-x_{i{\rm w}})(x_{j{\rm w}}-x_{i{\rm w}})}
+\frac{x_{H^-_k{\rm w}}x_{i{\rm w}}(-\ln x_{i{\rm w}}+2\ln^2 x_{i{\rm w}})}
{(x_{H^-_k{\rm w}}-x_{i{\rm w}})(x_{H^-_l{\rm w}}-x_{i{\rm w}})(x_{j{\rm w}}-x_{i{\rm w}})^2}
\nonumber \\
&&\hspace{1.0cm}+\frac{x_{H^-_k{\rm w}}x_{i{\rm w}}(-\ln x_{i{\rm w}}+2\ln^2 x_{i{\rm w}}
)}
{(x_{H^-_k{\rm w}}-x_{i{\rm w}})(x_{H^-_l{\rm w}}-x_{i{\rm w}})^2(x_{j{\rm w}}-x_{i{\rm w}})}
+\frac{x_{H^-_k{\rm w}}x_{i{\rm w}}(-\ln x_{i{\rm w}}+2\ln^2 x_{i{\rm w}}
)}
{(x_{H^-_k{\rm w}}-x_{i{\rm w}})^2(x_{H^-_l{\rm w}}-x_{i{\rm w}})(x_{j{\rm w}}-x_{i{\rm w}})}
\nonumber \\
&&\hspace{1.0cm}+\frac{x_{i{\rm w}}(\ln x_{i{\rm w}}-2\ln^2 x_{i{\rm w}})}
{(x_{H^-_l{\rm w}}-x_{i{\rm w}})(x_{j{\rm w}}-x_{i{\rm w}})^2}
+\frac{x_{i{\rm w}}(\ln x_{i{\rm w}}-2\ln^2 x_{i{\rm w}})}
{(x_{H^-_l{\rm w}}-x_{i{\rm w}})^2(x_{j{\rm w}}-x_{i{\rm w}})}
+\frac{1-3\ln x_{i{\rm w}}-2\ln^2 x_{i{\rm w}})}
{(x_{H^-_l{\rm w}}-x_{i{\rm w}})(x_{j{\rm w}}-x_{i{\rm w}})}\bigg)\nonumber \\
&&\hspace{1.0cm}-\frac{4}{3}
\bigg(-\sum\limits_{\sigma=u^i,u^j,H^-_l}
\frac{3x_{\sigma}{\cal R}_{i_2}(\frac{x_{i{\rm w}}}{x_{\sigma}})}
{\prod\limits_{\rho\neq \sigma}(x_{\rho}-x_{\sigma})}
+\frac{x_{H^-_l{\rm w}}(2\ln x_{H^-_l{\rm w}}+3\ln^2x_{H^-_l{\rm w}})}
{2(x_{H^-_k{\rm w}}-x_{H^-_l{\rm w}})(-x_{H^-_l{\rm w}}+x_{i{\rm w}})(-x_{H^-_l{\rm w}}+x_{j{\rm w}})}
\nonumber \\
&&\hspace{1.0cm}+\sum\limits_{\sigma=u^i,u^j,H^-_k,H^-_l}
\frac{x_{\sigma}\ln x_{\sigma}-13x_{\sigma}\ln^2 x_{\sigma}
+6x_{\sigma}(x_{H^-_k{\rm w}}-x_{i{\rm w}}){\cal R}_{i_2}(\frac{x_{i{\rm w}}}{x_{\sigma}})
-10x_{\sigma}\Upsilon(\frac{x_{i{\rm w}}}{x_{\sigma}})}
{2\prod\limits_{\rho\neq \sigma}(x_{\rho}-x_{\sigma})}
\nonumber \\
&&\hspace{1.0cm}-\frac{x_{H^-_l{\rm w}}(2\ln x_{H^-_l{\rm w}}+3\ln^2x_{H^-_l{\rm w}})}
{2(x_{H^-_k{\rm w}}-x_{H^-_l{\rm w}})(-x_{H^-_l{\rm w}}+x_{i{\rm w}})^2}\bigg)
+\frac{7}{6}\bigg(\frac{
{\cal R}_{i_2}(\frac{x_{H^-_k{\rm w}}}{x_{i{\rm w}}})
-{\cal R}_{i_2}(\frac{x_{H^-_k{\rm w}}}{x_{j{\rm w}}})-
{\cal R}_{i_2}(\frac{x_{j{\rm w}}}{x_{i{\rm w}}})}
{(x_{H^-_k{\rm w}}-x_{j{\rm w}})(x_{i{\rm w}}-x_{j{\rm w}})}\nonumber \\
&&\hspace{1.0cm}
-\sum\limits_{\sigma=u^i,u^j,H^-_l}
\Big(\frac{{\cal R}_{i_2}(\frac{x_{H^-_k{\rm w}}}{x_{\sigma}})}
{\prod\limits_{\rho\neq \sigma}(x_{\rho}-x_{\sigma})}
+\frac{(x_{H^-_l{\rm w}}-x_{j{\rm w}})
({\cal R}_{i_2}(\frac{x_{H^-_k{\rm w}}}{x_{\sigma}})-
{\cal R}_{i_2}(\frac{x_{j{\rm w}}}{x_{\sigma}}))+x_{\sigma}
(\Upsilon(\frac{x_{H^-_k{\rm w}}}{x_{\sigma}})-
\Upsilon(\frac{x_{j{\rm w}}}{x_{\sigma}}))}
{(x_{H^-_k{\rm w}}-x_{j{\rm w}})\prod\limits_{\rho\neq \sigma}
(x_{\rho}-x_{\sigma})}\Big)\bigg)
\nonumber \\
&&\hspace{1.0cm}
-\frac{1}{2}\frac{{\cal R}_{i_2}(\frac{x_{j{\rm w}}}{x_{i{\rm w}}})-
{\cal R}_{i_2}(\frac{x_{H^-_k{\rm w}}}{x_{H^-_l{\rm w}}})-
{\cal R}_{i_2}(\frac{x_{H^-_k{\rm w}}}{x_{i{\rm w}}})-
{\cal R}_{i_2}(\frac{x_{j{\rm w}}}{x_{H^-_l{\rm w}}})}
{(-x_{H^-_l{\rm w}}+x_{i{\rm w}})(-x_{H^-_k{\rm w}}+x_{j{\rm w}})}
\nonumber \\
&&\hspace{1.0cm}
+\frac{1}{4}\sum\limits_{\sigma=u^i,u^j,H^-_l}
\frac{{\cal R}_{i_2}(\frac{x_{H^-_k{\rm w}}}{x_{\sigma}})}
{\prod\limits_{\rho\neq \sigma}(x_{\rho}-x_{\sigma})}
+(x_{i{\rm w}} \leftrightarrow x_{j{\rm w}})\bigg]\; ,
\label{hh4}
\end{eqnarray}
\begin{eqnarray}
&&HH_5=\frac{1}{4}HH_4\; ,
\label{hh5}
\end{eqnarray}
\begin{eqnarray}
&&cc_{1}=\frac{1}{6}a_+^{(c)} b_-^{(c)} c_+^{(c)} d_-^{(c)}
\Big(F_{A}^{2a}+F_{A}^{2b}+\frac{1}{2}F_{A}^{2c}+2F_{A}^{2d}
-3F_{A}^{2e}-F_{A}^{2f}\Big)
(x_{\kappa_\lambda^- {\rm w}}, x_{\tilde{U}^i_\alpha {\rm w}},
x_{\tilde{U}^j_\beta {\rm w}}, 0,x_{\kappa_\eta^- {\rm w}},
x_{\tilde{U}^i_\alpha {\rm w}}, x_{\tilde{U}^j_\beta {\rm w}})\nonumber \\
&&\hspace{1.0cm}+\frac{4}{3}a_+^{(c)}b_-^{(c)}c_+^{(c)}d_-^{(c)}
\bigg(
\frac{{\cal R}_{i_2}(\frac{x_{\kappa^-_{\lambda}{\rm w}}}
{x_{\tilde{\tiny U}^{i}_{\alpha}{\rm w}}})-
{\cal R}_{i_2}(\frac{x_{\kappa^-_{\lambda}{\rm w}}}
{x_{\tilde{\tiny U}^{j}_{\beta}{\rm w}}})}{x_{\tilde{\tiny U}^{i}_{\alpha}{\rm w}}
-x_{\tilde{\tiny U}^{j}_{\beta}{\rm w}}}-\Big(x_{\kappa^-_{\eta}{\rm w}}-
x_{\kappa^-_{\lambda}{\rm w}}\Big)\sum\limits_{\sigma=\tilde{\tiny U}^{i}_{\alpha},
\tilde{\tiny U}^{i}_{\beta},\kappa^-_{\eta}}
\frac{{\cal R}_{i_2}(\frac{x_{\kappa^-_{\lambda}{\rm w}}}{x_\sigma})}
{\prod\limits_{\rho \neq \sigma}(x_{\rho}-x_{\sigma})}\bigg)
\nonumber \\
&&\hspace{1.0cm}+\frac{1}{3}a_+^{(c)}b_-^{(c)}c_+^{(c)}d_-^{(c)}
\bigg(
-\sum\limits_{\sigma=\tilde{\tiny U}^{i}_{\alpha},
\tilde{\tiny U}^{i}_{\beta},\kappa^-_{\eta}}
\frac{x_{\sigma}(x_{\kappa^-_{\lambda}{\rm w}}\ln^2(x_{\sigma}
x_{\kappa^-_{\lambda}{\rm w}})-x_{\tilde{\tiny U}^{j}_{\beta}{\rm w}}
\ln^2(x_{\sigma}x_{\tilde{\tiny U}^{j}_{\beta}{\rm w}}))}
{2(x_{\kappa^-_{\lambda}{\rm w}}-x_{\tilde{\tiny U}^{j}_{\beta}{\rm w}})
\prod\limits_{\rho \neq \sigma}(x_{\rho}-x_{\sigma})}
\nonumber \\
&&\hspace{1.0cm}+\frac{x_{\tilde{\tiny U}^{j}_{\beta}{\rm w}}^2+
2x_{\kappa^-_{\lambda}{\rm w}}x_{\tilde{\tiny U}^{j}_{\beta}{\rm w}}
-2x_{\kappa^-_{\lambda}{\rm w}}^2}
{x_{\kappa^-_{\lambda}{\rm w}}-x_{\tilde{\tiny U}^{j}_{\beta}{\rm w}}}
\sum\limits_{\sigma=\tilde{\tiny U}^{i}_{\alpha},
\tilde{\tiny U}^{i}_{\beta},\kappa^-_{\eta}}
\frac{{\cal R}_{i_2}(\frac{x_{\kappa^-_{\lambda}{\rm w}}}{x_{\sigma}})
-{\cal R}_{i_2}(\frac{x_{\tilde{\tiny U}^{j}_{\beta}{\rm w}}}{x_\sigma})}
{\prod\limits_{\rho \neq \sigma}(x_{\rho}-x_{\sigma})}
\nonumber \\
&&\hspace{1.0cm}+(x_{\tilde{\tiny U}^{j}_{\beta}{\rm w}}+
2x_{\kappa^-_{\eta}{\rm w}})\sum\limits_{\sigma=\tilde{\tiny U}^{i}_{\alpha},
\tilde{\tiny U}^{i}_{\beta},\kappa^-_{\eta}}
\frac{{\cal R}_{i_2}(\frac{x_{\kappa^-_{\lambda}{\rm w}}}{x_{\sigma}})}
{\prod\limits_{\rho \neq \sigma}(x_{\rho}-x_{\sigma})}
+2\frac{{\cal R}_{i_2}(\frac{x_{\kappa^-_{\eta}{\rm w}}}
{x_{\tilde{\tiny U}^{i}_{\alpha}{\rm w}}})-
{\cal R}_{i_2}(\frac{x_{\kappa^-_{\eta}{\rm w}}}
{x_{\tilde{\tiny U}^{j}_{\beta}{\rm w}}})}
{x_{\tilde{\tiny U}^{i}_{\alpha}{\rm w}}-x_{\tilde{\tiny U}^{j}_{\beta}{\rm w}}}
\nonumber \\
&&\hspace{1.0cm}+2\frac{\AE(x_{\tilde{\tiny U}^{i}_{\alpha}{\rm w}}
,x_{\kappa^-_{\eta}{\rm w}})-\AE(x_{\tilde{\tiny U}^{i}_{\alpha}{\rm w}},
x_{\tilde{\tiny U}^{j}_{\beta}{\rm w}})}
{x_{\kappa^-_{\lambda}{\rm w}}-x_{\tilde{\tiny U}^{i}_{\beta}{\rm w}}}
+\sum\limits_{\sigma=\tilde{\tiny U}^{i}_{\alpha},
\tilde{\tiny U}^{i}_{\beta},\kappa^-_{\eta}}
\frac{x_{\sigma}\Upsilon(\frac{x_{\kappa^-_{\lambda}{\rm w}}}{x_{\sigma}})}
{\prod\limits_{\rho \neq \sigma}(x_{\rho}-x_{\sigma})}
\nonumber \\
&&\hspace{1.0cm}+3\frac{x_{\tilde{\tiny U}^{i}_{\alpha}{\rm w}}
\Upsilon(\frac{x_{\kappa^-_{\lambda}{\rm w}}}
{x_{\tilde{\tiny U}^{j}_{\beta}{\rm w}}})-x_{\tilde{\tiny U}^{j}_{\beta}{\rm w}}
\Upsilon(\frac{x_{\kappa^-_{\lambda}{\rm w}}}{x_{\tilde{\tiny U}^{j}_{\beta}{\rm w}}})}
{(x_{\kappa^-_{\lambda}{\rm w}}-x_{\tilde{\tiny U}^{j}_{\beta}{\rm w}})
(x_{\tilde{\tiny U}^{i}_{\alpha}{\rm w}}-x_{\tilde{\tiny U}^{j}_{\beta}{\rm w}})}
+\sum\limits_{\sigma=\tilde{\tiny U}^{i}_{\alpha},
\tilde{\tiny U}^{i}_{\beta},\kappa^-_{\eta}}
\frac{x_{\kappa^-_{\eta}{\rm w}}x_{\sigma}
(\Upsilon(\frac{x_{\kappa^-_{\lambda}{\rm w}}}
{x_{\sigma}})-\Upsilon(\frac{x_{\tilde{\tiny U}^{j}_{\beta}{\rm w}}}
{x_{\sigma}}))}
{(x_{\kappa^-_{\lambda}{\rm w}}-x_{\tilde{\tiny U}^{j}_{\beta}{\rm w}})
\prod\limits_{\rho \neq \sigma}(x_{\rho}-x_{\sigma})}\bigg)
\nonumber \\
&&\hspace{1.0cm}+\frac{16}{3}a_+^{(c)}b_-^{(c)}c_-^{(c)}d_+^{(c)}
\bigg(-2(x_{\tilde{\tiny U}^{i}_{\alpha}{\rm w}}-
x_{\tilde{\tiny U}^{j}_{\beta}{\rm w}})
\sum\limits_{\sigma=\tilde{\tiny U}^{i}_{\alpha},
\kappa^-_{\eta},\kappa^-_{\lambda}}
\frac{{\cal R}_{i_2}(\frac{x_{\tilde{\tiny U}^{j}_{\beta}{\rm w}}}{x_{\sigma}})}
{\prod\limits_{\rho \neq \sigma}(x_{\rho}-x_{\sigma})}
\nonumber \\
&&\hspace{1.0cm}+\sum\limits_{\sigma=\tilde{\tiny U}^{i}_{\alpha},
\kappa^-_{\eta},\kappa^-_{\lambda}}
\frac{x_{\sigma}(2\ln x_{\sigma}-3\ln^2 x_{\sigma}
-2\Upsilon(\frac{x_{\tilde{\tiny U}^{j}_{\beta}{\rm w}}}{x_{\sigma}}))}
{2\prod\limits_{\rho \neq \sigma}(x_{\rho}-x_{\sigma})}
+2\frac{{\cal R}_{i_2}(\frac{x_{\tilde{\tiny U}^{j}_{\beta}{\rm w}}}
{x_{\kappa^-_{\eta}{\rm w}}})-{\cal R}_{i_2}(\frac{x_{\tilde{\tiny
U}^{j}_{\beta}{\rm w}}}{x_{\kappa^-_{\lambda}{\rm w}}})}
{-x_{\kappa^-_{\eta}{\rm w}}+x_{\kappa^-_{\lambda}{\rm w}}}
\nonumber \\
&&\hspace{1.0cm}-\frac{x_{\kappa^-_{\eta}{\rm w}}\ln^2 x_{\kappa^-_{\eta}{\rm w}}}
{(-x_{\kappa^-_{\eta}{\rm w}}+x_{\kappa^-_{\lambda}{\rm w}})
(-x_{\kappa^-_{\eta}{\rm w}}+x_{\tilde{\tiny U}^{i}_{\alpha}{\rm w}})}
-\frac{x_{\tilde{\tiny U}^{i}_{\alpha}{\rm w}}\ln^2 x_{\tilde{\tiny U}^{i}_{\alpha}{\rm w}}}
{(x_{\kappa^-_{\eta}{\rm w}}-x_{\tilde{\tiny U}^{i}_{\alpha}{\rm w}})
(x_{\kappa^-_{\lambda}{\rm w}}-x_{\tilde{\tiny U}^{i}_{\alpha}{\rm w}})}\bigg)
\nonumber \\
&&\hspace{1.0cm}-\frac{2}{3}a_+^{(c)} b_-^{(c)} c_+^{(c)} d_-^{(c)}\bigg(
-\frac{-\AE(x_{\tilde{U}^i_\alpha {\rm w}}, x_{\kappa_\lambda^- {\rm w}})
+\AE(x_{\tilde{U}^i_\alpha {\rm w}}, x_{\tilde{U}^j_\beta {\rm w}})}{
(-x_{\kappa_\lambda^- {\rm w}} + x_{\tilde{U}^j_\beta {\rm w}})}
+\frac{-{\cal R}_{i_2}(\frac{x_{\tilde{U}^j_\beta {\rm w}}}
{x_{\kappa_\eta^- {\rm w}}})+{\cal R}_{i_2}(\frac{x_{\tilde{U}^j_\beta {\rm w}}}
{x_{\tilde{U}^i_\alpha {\rm w}}})}{-x_{\kappa_\eta^- {\rm w}} + x_{\tilde{U}^i_\alpha {\rm w}}}
\nonumber \\
&&\hspace{1.0cm}+\frac{x_{\kappa_\eta^- {\rm w}}+x_{\kappa_\lambda^- {\rm w}}}
{(-x_{\kappa_\eta^- {\rm w}} + x_{\tilde{U}^i_\alpha {\rm w}})
(-x_{\kappa_\lambda^- {\rm w}} + x_{\tilde{U}^j_\beta {\rm w}})}
\Big(-{\cal R}_{i_2}(\frac{x_{\kappa_\lambda^- {\rm w}}}{x_{\kappa_\eta^- {\rm w}}})
-{\cal R}_{i_2}(\frac{x_{\kappa_\lambda^- {\rm w}}}{x_{\tilde{U}^i_\alpha {\rm w}}})
-{\cal R}_{i_2}(\frac{x_{\tilde{U}^j_\beta {\rm w}}}{x_{\kappa_\eta^- {\rm w}}})
+{\cal R}_{i_2}(\frac{x_{\tilde{U}^j_\beta {\rm w}}}{x_{\tilde{U}^i_\alpha {\rm w}}})
\Big)\nonumber \\
&&\hspace{1.0cm}-\frac{1}{(-x_{\kappa_\eta^- {\rm w}}+ x_{\tilde{U}^i_\alpha {\rm w}})
(-x_{\kappa_\lambda^- {\rm w}} + x_{\tilde{U}^j_\beta {\rm w}})}\Big(x_{\kappa_\eta^- {\rm w}}
(\Upsilon(\frac{x_{\kappa_\lambda^- {\rm w}}}{x_{\kappa_\eta^- {\rm w}}})
-\Upsilon(\frac{x_{\tilde{U}^j_\beta {\rm w}}}{x_{\kappa_\eta^- {\rm w}}}))
-x_{\tilde{U}^i_\alpha {\rm w}} (\Upsilon(\frac{x_{\kappa_\lambda^- {\rm w}}}{
x_{\tilde{U}^i_\alpha {\rm w}}}) -
\Upsilon(\frac{x_{\tilde{U}^j_\beta {\rm w}}}
{x_{\tilde{U}^i_\alpha {\rm w}}}))\Big)\bigg)
\nonumber \\
&&\hspace{1.0cm}+\frac{16}{3}a_+^{(c)} b_-^{(c)} c_+^{(c)} d_-^{(c)}
\bigg(\sum\limits_{\sigma=\tilde{\tiny U}^{i}_{\alpha},
\kappa^-_{\eta},\kappa^-_{\lambda}}\Big(
\frac{x_{\tilde{U}^i_\alpha {\rm w}}^2x_{\sigma}(-2\ln x_{\sigma}-16
\ln^2 x_{\sigma}-3\ln^2(x_{\sigma}x_{\tilde{U}^i_\alpha {\rm w}}))}
{4(x_{\sigma}-x_{\tilde{U}^i_\alpha {\rm w}})^2
\prod\limits_{\rho \neq \sigma}(x_{\rho}-x_{\sigma})}
\nonumber \\
&&\hspace{1.0cm}+\frac{x_{\tilde{U}^i_\alpha {\rm w}}x_{\sigma}(3\ln x_{\sigma}+12
\ln^2 x_{\sigma}-\ln^2(x_{\sigma}x_{\tilde{U}^i_\alpha {\rm w}}))}
{2(x_{\sigma}-x_{\tilde{U}^i_\alpha {\rm w}})
\prod\limits_{\rho \neq \sigma}(x_{\rho}-x_{\sigma})}
-\frac{2x_{\sigma}\ln^2 x_{\sigma}}
{\prod\limits_{\rho \neq \sigma}(x_{\rho}-x_{\sigma})}\Big)
\nonumber \\
&&\hspace{1.0cm}+\frac{x_{\tilde{U}^i_\alpha {\rm w}}^3(
\ln x_{\tilde{U}^i_\alpha {\rm w}}+14\ln^2 x_{\tilde{U}^i_\alpha {\rm w}})}
{2(x_{\kappa^-_{\eta}{\rm w}}-x_{\tilde{U}^i_\alpha {\rm w}})
(x_{\kappa^-_{\lambda}{\rm w}}-x_{\tilde{U}^i_\alpha {\rm w}})
(x_{\tilde{U}^j_\beta {\rm w}}-x_{\tilde{U}^i_\alpha {\rm w}})^2}
+\frac{x_{\tilde{U}^i_\alpha {\rm w}}^2(1+
29\ln x_{\tilde{U}^i_\alpha {\rm w}}+28\ln^2 x_{\tilde{U}^i_\alpha {\rm w}})}
{2(x_{\kappa^-_{\eta}{\rm w}}-x_{\tilde{U}^i_\alpha {\rm w}})
(x_{\kappa^-_{\lambda}{\rm w}}-x_{\tilde{U}^i_\alpha {\rm w}})
(x_{\tilde{U}^j_\beta {\rm w}}-x_{\tilde{U}^i_\alpha {\rm w}})}
\nonumber \\
&&\hspace{1.0cm}+\frac{x_{\tilde{U}^i_\alpha {\rm w}}^3(
\ln x_{\tilde{U}^i_\alpha {\rm w}}+14\ln^2 x_{\tilde{U}^i_\alpha {\rm w}})}
{2(x_{\kappa^-_{\eta}{\rm w}}-x_{\tilde{U}^i_\alpha {\rm w}})
(x_{\kappa^-_{\lambda}{\rm w}}-x_{\tilde{U}^i_\alpha {\rm w}})^2
(x_{\tilde{U}^j_\beta {\rm w}}-x_{\tilde{U}^i_\alpha {\rm w}})}
+\frac{x_{\tilde{U}^i_\alpha {\rm w}}^3(
\ln x_{\tilde{U}^i_\alpha {\rm w}}+14\ln^2 x_{\tilde{U}^i_\alpha {\rm w}})}
{2(x_{\kappa^-_{\eta}{\rm w}}-x_{\tilde{U}^i_\alpha {\rm w}})^2
(x_{\kappa^-_{\lambda}{\rm w}}-x_{\tilde{U}^i_\alpha {\rm w}})
(x_{\tilde{U}^j_\beta {\rm w}}-x_{\tilde{U}^i_\alpha {\rm w}})}
\nonumber \\
&&\hspace{1.0cm}-\frac{x_{\kappa^-_{\eta}{\rm w}}\ln x_{\kappa^-_{\eta}{\rm w}}}
{(x_{\kappa^-_{\lambda}{\rm w}}-x_{\kappa^-_{\eta}{\rm w}})
(x_{\tilde{U}^j_\beta {\rm w}}-x_{\kappa^-_{\eta}{\rm w}})}+
2\frac{x_{\kappa^-_{\lambda}{\rm w}}\ln x_{\kappa^-_{\lambda}{\rm w}}}
{(x_{\kappa^-_{\eta}{\rm w}}-x_{\kappa^-_{\lambda}{\rm w}})
(x_{\tilde{U}^j_\beta {\rm w}}-x_{\kappa^-_{\lambda}{\rm w}})}
-\frac{3x_{\kappa^-_{\lambda}{\rm w}}\ln x_{\kappa^-_{\lambda}{\rm w}}}
{(x_{\kappa^-_{\eta}{\rm w}}-x_{\kappa^-_{\lambda}{\rm w}})^2}\bigg)
\nonumber \\
&&\hspace{1.0cm}+\frac{16}{3}a_+^{(c)} b_-^{(c)} c_+^{(c)} d_-^{(c)}
\bigg(\sum\limits_{\sigma=\tilde{\tiny U}^{i}_{\alpha},
\tilde{\tiny U}^{j}_{\beta},\kappa^-_{\eta},\kappa^-_{\lambda}}
\frac{x_{\kappa^-_{\lambda}{\rm w}}x_{\sigma}(6\ln x_{\sigma}-7\ln^2 x_{\sigma})}
{\prod\limits_{\rho \neq \sigma}(x_{\rho}-x_{\sigma})}
+\sum\limits_{\sigma=\tilde{\tiny U}^{i}_{\alpha},
\kappa^-_{\eta},\kappa^-_{\lambda}}
\frac{x_{\sigma}(3\ln x_{\sigma}+7\ln^2 x_{\sigma})}
{\prod\limits_{\rho \neq \sigma}(x_{\rho}-x_{\sigma})}
\nonumber \\
&&\hspace{1.0cm}
-\frac{x_{\kappa^-_{\eta}{\rm w}}x_{\kappa^-_{\lambda}{\rm w}}(3\ln x_{\kappa^-_{\eta}{\rm w}}
+\ln^2 x_{\kappa^-_{\eta}{\rm w}})}
{2(x_{\kappa^-_{\eta}{\rm w}}-x_{\tilde{U}^i_\alpha {\rm w}})
(x_{\kappa^-_{\eta}{\rm w}}-x_{\tilde{U}^i_\alpha {\rm w}})^2}
+\frac{4x_{\kappa^-_{\eta}{\rm w}}x_{\kappa^-_{\lambda}{\rm w}}
\ln^2 x_{\kappa^-_{\eta}{\rm w}}}
{2(x_{\kappa^-_{\eta}{\rm w}}-x_{\tilde{U}^i_\alpha {\rm w}})
(x_{\kappa^-_{\eta}{\rm w}}-x_{\tilde{U}^i_\alpha {\rm w}})
(x_{\kappa^-_{\eta}{\rm w}}-x_{\tilde{U}^j_\beta {\rm w}})}
\nonumber \\
&&\hspace{1.0cm}+\sum\limits_{\sigma=\tilde{\tiny U}^{i}_{\alpha},
{\tiny U}^{j}_{\beta},\kappa^-_{\eta},\kappa^-_{\lambda}}
\Big(\frac{(x_{\kappa^-_{\lambda}{\rm w}}+x_{\tilde{U}^j_\beta {\rm w}})
x_{\kappa^-_{\lambda}{\rm w}}x_{\sigma}{\cal R}_{i_2}
(\frac{x_{\tilde{U}^j_\beta {\rm w}}}{x_{\sigma}})}
{\prod\limits_{\rho \neq \sigma}(x_{\rho}-x_{\sigma})}
-\frac{3x_{\kappa^-_{\lambda}{\rm w}}x_{\sigma}
\Upsilon(\frac{x_{\tilde{U}^j_\beta {\rm w}}}{x_{\sigma}})}
{\prod\limits_{\rho \neq \sigma}(x_{\rho}-x_{\sigma})}
\Big)\nonumber \\
&&\hspace{1.0cm}+\sum\limits_{\sigma=\tilde{\tiny U}^{i}_{\alpha},
{\tiny U}^{j}_{\beta},\kappa^-_{\eta}}\Big(
\frac{(x_{\kappa^-_{\lambda}{\rm w}}+x_{\tilde{U}^j_\beta {\rm w}}+
x_{\kappa^-_{\lambda}{\rm w}})x_{\sigma}{\cal R}_{i_2}
(\frac{x_{\tilde{U}^j_\beta {\rm w}}}{x_{\sigma}})}
{\prod\limits_{\rho \neq \sigma}(x_{\rho}-x_{\sigma})}
+\frac{3x_{\sigma}\Upsilon(\frac{x_{\tilde{U}^j_\beta {\rm w}}}{x_{\sigma}})}
{\prod\limits_{\rho \neq \sigma}(x_{\rho}-x_{\sigma})}\Big)
+\frac{{\cal R}_{i_2}(\frac{x_{\tilde{U}^j_\beta {\rm w}}}
{x_{\tilde{U}^i_\alpha {\rm w}}})}{x_{\tilde{U}^i_\alpha {\rm w}}
-x_{\tilde{U}^j_\beta {\rm w}}}\bigg)\; ,
\label{cc1}
\end{eqnarray}
\begin{eqnarray}
&&cc_2=\frac{1}{24}(a_-^{(c)} b_+^{(c)} c_+^{(c)} d_-^{(c)}+
a_+^{(c)} b_-^{(c)} c_-^{(c)} d_+^{(c)})\Big(F_{A}^{2a}+F_{A}^{2b}
+\frac{3}{2}F_{A}^{2c}+6F_{A}^{2d}
\nonumber \\
&&\hspace{1.0cm}-\frac{5}{2}F_{A}^{2e}
+\frac{1}{2}F_{A}^{2f}\Big)
(x_{\kappa_\lambda^- {\rm w}}, x_{\tilde{U}^i_\alpha {\rm w}},
x_{\tilde{U}^j_\beta {\rm w}},0, x_{\kappa_\eta^- {\rm w}},
x_{\tilde{U}^i_\alpha {\rm w}}, x_{\tilde{U}^j_\beta {\rm w}})
\nonumber \\
&&\hspace{1.0cm}
+\Big(a_-^{(c)}b_+^{(c)}c_+^{(c)}d_-^{(c)}+
a_+^{(c)}b_-^{(c)}c_-^{(c)}d_+^{(c)}\Big)\bigg[\bigg(
\frac{{\cal R}_{i_2}(\frac{x_{\kappa^-_{\lambda}{\rm w}}}
{x_{\tilde{\tiny U}^{i}_{\alpha}{\rm w}}})-
{\cal R}_{i_2}(\frac{x_{\kappa^-_{\lambda}{\rm w}}}
{x_{\tilde{\tiny U}^{j}_{\beta}{\rm w}}})}{x_{\tilde{\tiny U}^{i}_{\alpha}{\rm w}}
-x_{\tilde{\tiny U}^{j}_{\beta}{\rm w}}}+
\sum\limits_{\sigma=\tilde{\tiny U}^{i}_{\alpha},
\tilde{\tiny U}^{i}_{\beta},\kappa^-_{\eta}}
\frac{4\Big(x_{\sigma}-
x_{\kappa^-_{\lambda}{\rm w}}\Big)
{\cal R}_{i_2}(\frac{x_{\kappa^-_{\lambda}{\rm w}}}{x_\sigma})}
{3\prod\limits_{\rho \neq \sigma}(x_{\rho}-x_{\sigma})}
\nonumber \\
&&\hspace{1.0cm}
+\frac{7}{48}\frac{-x_{\kappa^-_{\eta}{\rm w}}(-
\Upsilon(\frac{x_{\kappa^-_{\lambda}{\rm w}}}
{x_{\kappa^-_{\eta}{\rm w}}})+\Upsilon(\frac{x_{\tilde{\tiny U}^{j}_{\beta}{\rm w}}}
{x_{\kappa^-_{\eta}{\rm w}}}))+x_{\tilde{\tiny U}^{i}_{\alpha}{\rm w}}
(-\Upsilon(\frac{x_{\kappa^-_{\lambda}{\rm w}}}
{x_{\tilde{\tiny U}^{i}_{\alpha}{\rm w}}})+
\Upsilon(\frac{x_{\tilde{\tiny U}^{j}_{\beta}{\rm w}}}
{x_{\tilde{\tiny U}^{i}_{\alpha}{\rm w}}}))}{(x_{\kappa^-_{\eta}{\rm w}}-
x_{\tilde{\tiny U}^{i}_{\alpha}{\rm w}})(x_{\kappa^-_{\lambda}{\rm w}}-
x_{\tilde{\tiny U}^{j}_{\beta}{\rm w}})}
\bigg)
\nonumber \\
&&\hspace{1.0cm}
-\frac{1}{48}\bigg(
-\sum\limits_{\sigma=\tilde{\tiny U}^{i}_{\alpha},
\tilde{\tiny U}^{i}_{\beta},\kappa^-_{\eta}}
\frac{x_{\sigma}(x_{\kappa^-_{\lambda}{\rm w}}\ln^2(x_{\sigma}
x_{\kappa^-_{\lambda}{\rm w}})-x_{\tilde{\tiny U}^{j}_{\beta}{\rm w}}
\ln^2(x_{\sigma}x_{\tilde{\tiny U}^{j}_{\beta}{\rm w}}))}
{2(x_{\kappa^-_{\lambda}{\rm w}}-x_{\tilde{\tiny U}^{j}_{\beta}{\rm w}})
\prod\limits_{\rho \neq \sigma}(x_{\rho}-x_{\sigma})}
\nonumber \\
&&\hspace{1.0cm}+\frac{x_{\tilde{\tiny U}^{j}_{\beta}{\rm w}}^2+
2x_{\kappa^-_{\lambda}{\rm w}}x_{\tilde{\tiny U}^{j}_{\beta}{\rm w}}
-2x_{\kappa^-_{\lambda}{\rm w}}^2}
{x_{\kappa^-_{\lambda}{\rm w}}-x_{\tilde{\tiny U}^{j}_{\beta}{\rm w}}}
\sum\limits_{\sigma=\tilde{\tiny U}^{i}_{\alpha},
\tilde{\tiny U}^{i}_{\beta},\kappa^-_{\eta}}
\frac{{\cal R}_{i_2}(\frac{x_{\kappa^-_{\lambda}{\rm w}}}{x_{\sigma}})
-{\cal R}_{i_2}(\frac{x_{\tilde{\tiny U}^{j}_{\beta}{\rm w}}}{x_\sigma})}
{\prod\limits_{\rho \neq \sigma}(x_{\rho}-x_{\sigma})}
+2\frac{{\cal R}_{i_2}(\frac{x_{\kappa^-_{\eta}{\rm w}}}
{x_{\tilde{\tiny U}^{i}_{\alpha}{\rm w}}})-
{\cal R}_{i_2}(\frac{x_{\kappa^-_{\eta}{\rm w}}}
{x_{\tilde{\tiny U}^{j}_{\beta}{\rm w}}})}
{x_{\tilde{\tiny U}^{i}_{\alpha}{\rm w}}-x_{\tilde{\tiny U}^{j}_{\beta}{\rm w}}}
\nonumber \\
&&\hspace{1.0cm}+(x_{\tilde{\tiny U}^{j}_{\beta}{\rm w}}+
2x_{\kappa^-_{\eta}{\rm w}})\sum\limits_{\sigma=\tilde{\tiny U}^{i}_{\alpha},
\tilde{\tiny U}^{i}_{\beta},\kappa^-_{\eta}}
\frac{{\cal R}_{i_2}(\frac{x_{\kappa^-_{\lambda}{\rm w}}}{x_{\sigma}})}
{\prod\limits_{\rho \neq \sigma}(x_{\rho}-x_{\sigma})}
+2\frac{\AE(x_{\tilde{\tiny U}^{i}_{\alpha}{\rm w}}
,x_{\kappa^-_{\eta}{\rm w}})-\AE(x_{\tilde{\tiny U}^{i}_{\alpha}{\rm w}},
x_{\tilde{\tiny U}^{j}_{\beta}{\rm w}})}
{x_{\kappa^-_{\lambda}{\rm w}}-x_{\tilde{\tiny U}^{i}_{\beta}{\rm w}}}
\nonumber \\
&&\hspace{1.0cm}+3\frac{x_{\tilde{\tiny U}^{i}_{\alpha}{\rm w}}
\Upsilon(\frac{x_{\kappa^-_{\lambda}{\rm w}}}
{x_{\tilde{\tiny U}^{j}_{\beta}{\rm w}}})-x_{\tilde{\tiny U}^{j}_{\beta}{\rm w}}
\Upsilon(\frac{x_{\kappa^-_{\lambda}{\rm w}}}{x_{\tilde{\tiny U}^{j}_{\beta}{\rm w}}})}
{(x_{\kappa^-_{\lambda}{\rm w}}-x_{\tilde{\tiny U}^{j}_{\beta}{\rm w}})
(x_{\tilde{\tiny U}^{i}_{\alpha}{\rm w}}-x_{\tilde{\tiny U}^{j}_{\beta}{\rm w}})}
+\sum\limits_{\sigma=\tilde{\tiny U}^{i}_{\alpha},
\tilde{\tiny U}^{i}_{\beta},\kappa^-_{\eta}}
\frac{x_{\kappa^-_{\eta}{\rm w}}x_{\sigma}
(\Upsilon(\frac{x_{\kappa^-_{\lambda}{\rm w}}}
{x_{\sigma}})-\Upsilon(\frac{x_{\tilde{\tiny U}^{j}_{\beta}{\rm w}}}
{x_{\sigma}}))}
{(x_{\kappa^-_{\lambda}{\rm w}}-x_{\tilde{\tiny U}^{j}_{\beta}{\rm w}})
\prod\limits_{\rho \neq \sigma}(x_{\rho}-x_{\sigma})}
\nonumber \\
&&\hspace{1.0cm}+\sum\limits_{\sigma=\tilde{\tiny U}^{i}_{\alpha},
\tilde{\tiny U}^{i}_{\beta},\kappa^-_{\eta}}
\frac{x_{\sigma}\Upsilon(\frac{x_{\kappa^-_{\lambda}{\rm w}}}{x_{\sigma}})}
{\prod\limits_{\rho \neq \sigma}(x_{\rho}-x_{\sigma})}\bigg)
-\frac{8}{3}\bigg(
\sum\limits_{\sigma=\tilde{\tiny U}^{i}_{\alpha},
\tilde{\tiny U}^{j}_{\beta},\kappa^-_{\eta}}
\frac{x_{\sigma}(-10\ln x_{\sigma}+7\ln^2 x_{\sigma})}
{2\prod\limits_{\rho \neq \sigma}(x_{\rho}-x_{\sigma})}
\nonumber \\
&&\hspace{1.0cm}
+\sum\limits_{\sigma=\tilde{\tiny U}^{i}_{\alpha},
\tilde{\tiny U}^{i}_{\beta},\kappa^-_{\lambda}}
\frac{x_{\sigma}(10\ln x_{\sigma}-7\ln^2 x_{\sigma})}
{2\prod\limits_{\rho \neq \sigma}(x_{\rho}-x_{\sigma})}
+\sum\limits_{\sigma=\tilde{\tiny U}^{i}_{\alpha},
\tilde{\tiny U}^{i}_{\beta},\kappa^-_{\eta},\kappa^-_{\lambda}}
\frac{(x_{\kappa^-_{\lambda}{\rm w}}^2-
x_{\tilde{\tiny U}^{j}_{\beta}{\rm w}}x_{\kappa^-_{\lambda}{\rm w}})
x_{\sigma}{\cal R}_{i_2}
(\frac{x_{\tilde{\tiny U}^{j}_{\beta}{\rm w}}}{x_{\sigma}})}
{\prod\limits_{\rho \neq \sigma}(x_{\rho}-x_{\sigma})}
\nonumber \\
&&\hspace{1.0cm}-\sum\limits_{\sigma=\tilde{\tiny U}^{i}_{\alpha},
\tilde{\tiny U}^{i}_{\beta}}\frac{(x_{\kappa^-_{\eta}{\rm w}}
+x_{\kappa^-_{\lambda}{\rm w}}-x_{\tilde{\tiny U}^{j}_{\beta}{\rm w}})x_{\sigma}
{\cal R}_{i_2}(\frac{x_{\tilde{\tiny U}^{j}_{\beta}{\rm w}}}{x_{\sigma}})}
{\prod\limits_{\rho \neq \sigma}(x_{\rho}-x_{\sigma})}+
\sum\limits_{\sigma=\tilde{\tiny U}^{i}_{\alpha},
\tilde{\tiny U}^{j}_{\beta},\kappa^-_{\eta},\kappa^-_{\lambda}}
\frac{3x_{\kappa^-_{\lambda}{\rm w}}x_{\sigma}
\Upsilon(\frac{x_{\tilde{\tiny U}^{j}_{\beta}{\rm w}}}{x_{\sigma}})}
{\prod\limits_{\rho \neq \sigma}(x_{\rho}-x_{\sigma})}
\nonumber \\
&&\hspace{1.0cm}+\sum\limits_{\sigma=\tilde{\tiny U}^{i}_{\alpha},
\tilde{\tiny U}^{j}_{\beta},\kappa^-_{\eta}}
\frac{3x_{\sigma}
\Upsilon(\frac{x_{\tilde{\tiny U}^{j}_{\beta}{\rm w}}}{x_{\sigma}})}
{\prod\limits_{\rho \neq \sigma}(x_{\rho}-x_{\sigma})}+
\frac{{\cal R}_{i_2}(\frac{x_{\tilde{\tiny U}^{j}_{\beta}{\rm w}}}
{x_{\tilde{\tiny U}^{i}_{\alpha}{\rm w}}})}
{x_{\tilde{\tiny U}^{i}_{\alpha}{\rm w}}-x_{\tilde{\tiny U}^{j}_{\beta}{\rm w}}}
+\frac{x_{\kappa^-_{\eta}{\rm w}}x_{\kappa^-_{\lambda}{\rm w}}(
\ln x_{\kappa^-_{\eta}{\rm w}}-\ln^2 x_{\kappa^-_{\eta}{\rm w}})}
{2(-x_{\kappa^-_{\eta}{\rm w}}+x_{\kappa^-_{\lambda}{\rm w}})(-x_{\kappa^-_{\eta}{\rm w}}
+x_{\tilde{\tiny U}^{i}_{\alpha}{\rm w}})^2}\nonumber \\
&&\hspace{1.0cm}+\frac{x_{\kappa^-_{\eta}{\rm w}}x_{\kappa^-_{\lambda}{\rm w}}(
9\ln x_{\kappa^-_{\eta}{\rm w}}-6\ln^2 x_{\kappa^-_{\eta}{\rm w}})}
{2(-x_{\kappa^-_{\eta}{\rm w}}+x_{\kappa^-_{\lambda}{\rm w}})(-x_{\kappa^-_{\eta}{\rm w}}
+x_{\tilde{\tiny U}^{i}_{\alpha}{\rm w}})(-x_{\kappa^-_{\eta}{\rm w}}
+x_{\tilde{\tiny U}^{j}_{\beta}{\rm w}})}+
\sum\limits_{\sigma=\tilde{\tiny U}^{j}_{\beta},
\kappa^-_{\eta},\kappa^-_{\lambda}}\Big(\frac{x_{\sigma}(3\ln x_{\sigma}
-2\ln^2 x_{\sigma})}
{\prod\limits_{\rho \neq \sigma}(x_{\rho}-x_{\sigma})}
\nonumber \\
&&\hspace{1.0cm}+\frac{x_{\tilde{\tiny U}^{i}_{\alpha}{\rm w}}^2
x_{\sigma}(22\ln x_{\sigma}-16\ln^2 x_{\sigma}+
\ln^2(x_{\tilde{\tiny U}^{i}_{\alpha}{\rm w}}x_{\sigma}))}
{4(-x_{\sigma}+x_{\tilde{\tiny U}^{i}_{\alpha}{\rm w}})^2
\prod\limits_{\rho \neq \sigma}(x_{\rho}-x_{\sigma})}-
\frac{x_{\tilde{\tiny U}^{i}_{\alpha}{\rm w}}^2
x_{\sigma}(17\ln x_{\sigma}-12\ln^2 x_{\sigma}+
\ln^2(x_{\tilde{\tiny U}^{i}_{\alpha}{\rm w}}x_{\sigma}))}
{2(-x_{\sigma}+x_{\tilde{\tiny U}^{i}_{\alpha}{\rm w}})
\prod\limits_{\rho \neq \sigma}(x_{\rho}-x_{\sigma})}\Big)
\nonumber \\
&&\hspace{1.0cm}-\frac{15x_{\tilde{\tiny U}^{i}_{\alpha}{\rm w}}^2
\ln x_{\tilde{\tiny U}^{i}_{\alpha}{\rm w}}-
6x_{\tilde{\tiny U}^{i}_{\alpha}{\rm w}}^2
\ln^2 x_{\tilde{\tiny U}^{i}_{\alpha}{\rm w}}
+11x_{\tilde{\tiny U}^{i}_{\alpha}{\rm w}}^2}{2
(x_{\kappa^-_{\eta}{\rm w}}-x_{\tilde{\tiny U}^{i}_{\alpha}{\rm w}})
(x_{\kappa^-_{\lambda}{\rm w}}-x_{\tilde{\tiny U}^{i}_{\alpha}{\rm w}})
(x_{\tilde{\tiny U}^{j}_{\beta}{\rm w}}-x_{\tilde{\tiny U}^{i}_{\alpha}{\rm w}})}
-\frac{11x_{\tilde{\tiny U}^{i}_{\alpha}{\rm w}}^2
\ln x_{\tilde{\tiny U}^{i}_{\alpha}{\rm w}}-6x_{\tilde{\tiny U}^{i}_{\alpha}{\rm w}}^3
\ln^2 x_{\tilde{\tiny U}^{i}_{\alpha}{\rm w}}}
{2(x_{\kappa^-_{\eta}{\rm w}}-x_{\tilde{\tiny U}^{i}_{\alpha}{\rm w}})
(x_{\kappa^-_{\lambda}{\rm w}}-x_{\tilde{\tiny U}^{i}_{\alpha}{\rm w}})
(x_{\tilde{\tiny U}^{j}_{\beta}{\rm w}}-x_{\tilde{\tiny U}^{i}_{\alpha}{\rm w}})^2}
\nonumber \\
&&\hspace{1.0cm}+\frac{x_{\tilde{\tiny U}^{i}_{\alpha}{\rm w}}^3(6
\ln^2 x_{\tilde{\tiny U}^{i}_{\alpha}{\rm w}}-11\ln x_{\tilde{\tiny U}^{i}_{\alpha}{\rm w}})}
{2(x_{\kappa^-_{\eta}{\rm w}}-x_{\tilde{\tiny U}^{i}_{\alpha}{\rm w}})
(x_{\kappa^-_{\lambda}{\rm w}}-x_{\tilde{\tiny U}^{i}_{\alpha}{\rm w}})^2
(x_{\tilde{\tiny U}^{j}_{\beta}{\rm w}}-x_{\tilde{\tiny U}^{i}_{\alpha}{\rm w}})}
+\frac{x_{\tilde{\tiny U}^{i}_{\alpha}{\rm w}}^3(6
\ln^2 x_{\tilde{\tiny U}^{i}_{\alpha}{\rm w}}-11\ln x_{\tilde{\tiny U}^{i}_{\alpha}{\rm w}})}
{2(x_{\kappa^-_{\eta}{\rm w}}-x_{\tilde{\tiny U}^{i}_{\alpha}{\rm w}})^2
(x_{\kappa^-_{\lambda}{\rm w}}-x_{\tilde{\tiny U}^{i}_{\alpha}{\rm w}})
(x_{\tilde{\tiny U}^{j}_{\beta}{\rm w}}-x_{\tilde{\tiny U}^{i}_{\alpha}{\rm w}})}
\nonumber \\
&&\hspace{1.0cm}+\frac{x_{\kappa^-_{\lambda}{\rm w}}\ln x_{\kappa^-_{\lambda}{\rm w}}}
{(x_{\kappa^-_{\eta}{\rm w}}-x_{\kappa^-_{\lambda}{\rm w}})^2}-
\frac{x_{\kappa^-_{\lambda}{\rm w}}\ln x_{\kappa^-_{\lambda}{\rm w}}}
{(x_{\kappa^-_{\eta}{\rm w}}-x_{\kappa^-_{\lambda}{\rm w}})
(x_{\tilde{\tiny U}^{j}_{\beta}{\rm w}}-x_{\kappa^-_{\lambda}{\rm w}})}
-\sum\limits_{\sigma=\tilde{\tiny U}^{j}_{\beta},
\kappa^-_{\eta},\kappa^-_{\lambda}}\Big(\frac{2x_{\sigma}
\Upsilon(\frac{x_{\tilde{\tiny U}^{i}_{\alpha}{\rm w}}}{x_{\sigma}})}
{\prod\limits_{\rho \neq \sigma}(x_{\rho}-x_{\sigma})}
\nonumber \\
&&\hspace{1.0cm}+
\frac{4x_{\tilde{\tiny U}^{i}_{\alpha}{\rm w}}^4 x_{\sigma}
\Upsilon(\frac{x_{\tilde{\tiny U}^{i}_{\alpha}{\rm w}}}{x_{\sigma}})}
{(-x_{\sigma}+x_{\tilde{\tiny U}^{i}_{\alpha}{\rm w}})^2
\prod\limits_{\rho \neq \sigma}(x_{\rho}-x_{\sigma})}
-\frac{6x_{\tilde{\tiny U}^{i}_{\alpha}{\rm w}}
x_{\sigma}\Upsilon(\frac{x_{\tilde{\tiny U}^{i}_{\alpha}{\rm w}}}
{x_{\sigma}})}
{(-x_{\sigma}+x_{\tilde{\tiny U}^{i}_{\alpha}{\rm w}})
\prod\limits_{\rho \neq \sigma}(x_{\rho}-x_{\sigma})}\Big)
\bigg)\bigg]
\nonumber \\
&&\hspace{1.0cm}
+\Big(a_+^{(c)}b_+^{(c)}c_-^{(c)}d_-^{(c)}
+a_-^{(c)}b_-^{(c)}c_+^{(c)}d_+^{(c)}\Big)\sqrt{x_{\kappa^-_{\eta}{\rm w}}
x_{\kappa^-_{\lambda}{\rm w}}}\bigg[\frac{4}{3}\bigg(
-\sum\limits_{\sigma=\tilde{\tiny U}^{i}_{\alpha},
\tilde{\tiny U}^{i}_{\beta},\kappa^-_{\eta}}\frac{
{\cal R}_{i_2}(\frac{x_{\kappa^-_{\lambda}{\rm w}}}{x_{\sigma}})}
{\prod\limits_{\rho \neq \sigma}(x_{\rho}-x_{\sigma})}\nonumber \\
&&\hspace{1.0cm}+\frac{x_{\kappa^-_{\eta}{\rm w}}-
x_{\tilde{\tiny U}^{j}_{\beta}{\rm w}}}{x_{\kappa^-_{\lambda}{\rm w}}
-x_{\tilde{\tiny U}^{j}_{\beta}{\rm w}}}
\sum\limits_{\sigma=\tilde{\tiny U}^{i}_{\alpha},
\tilde{\tiny U}^{i}_{\beta},\kappa^-_{\eta}}
\frac{{\cal R}_{i_2}(\frac{x_{\kappa^-_{\lambda}{\rm w}}}{x_{\sigma}})
-{\cal R}_{i_2}(\frac{x_{\tilde{\tiny U}^{j}_{\beta}{\rm w}}}{x_\sigma})}
{\prod\limits_{\rho \neq \sigma}(x_{\rho}-x_{\sigma})}
+\frac{
{\cal R}_{i_2}(\frac{x_{\kappa^-_{\lambda}{\rm w}}}
{x_{\tilde{\tiny U}^{i}_{\alpha}{\rm w}}})-
{\cal R}_{i_2}(\frac{x_{\kappa^-_{\lambda}{\rm w}}}
{x_{\tilde{\tiny U}^{j}_{\beta}{\rm w}}})-
{\cal R}_{i_2}(\frac{x_{\tilde{\tiny U}^{j}_{\beta}{\rm w}}}
{x_{\tilde{\tiny U}^{i}_{\alpha}{\rm w}}})}{(x_{\kappa^-_{\lambda}{\rm w}}
-x_{\tilde{\tiny U}^{j}_{\beta}{\rm w}})
(x_{\tilde{\tiny U}^{i}_{\alpha}{\rm w}}-x_{\tilde{\tiny U}^{j}_{\beta}{\rm w}})}
\bigg)\nonumber \\
&&\hspace{1.0cm}
-\frac{{\cal R}_{i_2}(\frac{x_{\kappa^-_{\lambda}{\rm w}}}{x_{\kappa^-_{\eta}{\rm w}}})
+{\cal R}_{i_2}(\frac{x_{\kappa^-_{\lambda}{\rm w}}}{x_{\tilde{\tiny U}^{i}_{\alpha}{\rm w}}})
+{\cal R}_{i_2}(\frac{x_{\tilde{\tiny U}^{i}_{\alpha}{\rm w}}}{x_{\kappa^-_{\eta}{\rm w}}})
-{\cal R}_{i_2}(\frac{x_{\tilde{\tiny U}^{j}_{\beta}{\rm w}}}
{x_{\tilde{\tiny U}^{i}_{\alpha}{\rm w}}})}
{2(x_{\tilde{\tiny U}^{i}_{\alpha}{\rm w}}-x_{\kappa^-_{\eta}{\rm w}})
(x_{\tilde{\tiny U}^{j}_{\beta}{\rm w}}-x_{\kappa^-_{\lambda}{\rm w}})}
+\frac{28}{3}\bigg(\sum\limits_{\sigma=
\tilde{\tiny U}^{j}_{\beta},\kappa^-_{\eta},\kappa^-_{\lambda}}\Big(
\frac{2x_{\tilde{\tiny U}^{i}_{\alpha}{\rm w}}(x_{\sigma}\ln x_{\sigma}
-\ln^2 x_{\sigma})}
{(x_{\tilde{\tiny U}^{i}_{\alpha}{\rm w}}-x_{\sigma})
\prod\limits_{\rho \neq \sigma}(x_{\rho}-x_{\sigma})}\nonumber \\
&&\hspace{1.0cm}-
\frac{x_{\sigma}(2\ln x_{\sigma}-\ln^2 x_{\sigma})}
{\prod\limits_{\rho \neq \sigma}(x_{\rho}-x_{\sigma})}
-\frac{2x_{\sigma}x_{\tilde{\tiny U}^{i}_{\alpha}{\rm w}}
\Upsilon(\frac{x_{\tilde{\tiny U}^{i}_{\alpha}{\rm w}}}{x_{\sigma}})}
{(x_{\tilde{\tiny U}^{i}_{\alpha}{\rm w}}-x_{\sigma})
\prod\limits_{\rho \neq \sigma}(x_{\rho}-x_{\sigma})}
+\frac{x_{\sigma}\Upsilon(\frac{x_{\tilde{\tiny U}^{i}_{\alpha}{\rm w}}}{x_{\sigma}})}
{\prod\limits_{\rho \neq \sigma}(x_{\rho}-x_{\sigma})}\Big)
\nonumber \\
&&\hspace{1.0cm}+\frac{-2x_{\tilde{\tiny U}^{i}_{\alpha}{\rm w}}
+x_{\tilde{\tiny U}^{i}_{\alpha}{\rm w}}\ln x_{\tilde{\tiny U}^{i}_{\alpha}{\rm w}}
+3x_{\tilde{\tiny U}^{i}_{\alpha}{\rm w}}\ln^2 x_{\tilde{\tiny U}^{i}_{\alpha}{\rm w}}}
{(x_{\kappa^-_{\eta}{\rm w}}-x_{\tilde{\tiny U}^{i}_{\alpha}{\rm w}})
(x_{\kappa^-_{\lambda}{\rm w}}-x_{\tilde{\tiny U}^{i}_{\alpha}{\rm w}})
(x_{\tilde{\tiny U}^{j}_{\beta}{\rm w}}-x_{\tilde{\tiny U}^{i}_{\alpha}{\rm w}})}
+\frac{-2x_{\tilde{\tiny U}^{i}_{\alpha}{\rm w}}^2(\ln
x_{\tilde{\tiny U}^{i}_{\alpha}{\rm w}}+\ln^2 x_{\tilde{\tiny U}^{i}_{\alpha}{\rm w}})}
{(x_{\kappa^-_{\eta}{\rm w}}-x_{\tilde{\tiny U}^{i}_{\alpha}{\rm w}})
(x_{\kappa^-_{\lambda}{\rm w}}-x_{\tilde{\tiny U}^{i}_{\alpha}{\rm w}})
(x_{\tilde{\tiny U}^{j}_{\beta}{\rm w}}-x_{\tilde{\tiny U}^{i}_{\alpha}{\rm w}})^2}
\nonumber \\
&&\hspace{1.0cm}+\frac{-2x_{\tilde{\tiny U}^{i}_{\alpha}{\rm w}}^2(\ln
x_{\tilde{\tiny U}^{i}_{\alpha}{\rm w}}+\ln^2 x_{\tilde{\tiny U}^{i}_{\alpha}{\rm w}})}
{(x_{\kappa^-_{\eta}{\rm w}}-x_{\tilde{\tiny U}^{i}_{\alpha}{\rm w}})
(x_{\kappa^-_{\lambda}{\rm w}}-x_{\tilde{\tiny U}^{i}_{\alpha}{\rm w}})^2
(x_{\tilde{\tiny U}^{j}_{\beta}{\rm w}}-x_{\tilde{\tiny U}^{i}_{\alpha}{\rm w}})}
+\frac{-2x_{\tilde{\tiny U}^{i}_{\alpha}{\rm w}}^2(\ln
x_{\tilde{\tiny U}^{i}_{\alpha}{\rm w}}+\ln^2 x_{\tilde{\tiny U}^{i}_{\alpha}{\rm w}})}
{(x_{\kappa^-_{\eta}{\rm w}}-x_{\tilde{\tiny U}^{i}_{\alpha}{\rm w}})^2
(x_{\kappa^-_{\lambda}{\rm w}}-x_{\tilde{\tiny U}^{i}_{\alpha}{\rm w}})
(x_{\tilde{\tiny U}^{j}_{\beta}{\rm w}}-x_{\tilde{\tiny U}^{i}_{\alpha}{\rm w}})}\bigg)
\nonumber \\
&&\hspace{1.0cm}-\frac{16}{3}
\bigg(-\frac{x_{\kappa^-_{\eta}{\rm w}}\ln^2 x_{\kappa^-_{\eta}{\rm w}}}
{2(-x_{\kappa^-_{\eta}{\rm w}}+x_{\kappa^-_{\lambda}{\rm w}})
(-x_{\kappa^-_{\eta}{\rm w}}+x_{\tilde{\tiny U}^{i}_{\alpha}{\rm w}})^2}
+\frac{4x_{\kappa^-_{\eta}{\rm w}}\ln^2 x_{\kappa^-_{\eta}{\rm w}}}
{(-x_{\kappa^-_{\eta}{\rm w}}+x_{\kappa^-_{\lambda}{\rm w}})
(-x_{\kappa^-_{\eta}{\rm w}}+x_{\tilde{\tiny U}^{i}_{\alpha}{\rm w}})
(-x_{\kappa^-_{\eta}{\rm w}}+x_{\tilde{\tiny U}^{j}_{\beta}{\rm w}})}
\nonumber \\
&&\hspace{1.0cm}
+\sum\limits_{\tiny \sigma=\tilde{\tiny U}^{i}_{\alpha}
\tilde{\tiny U}^{j}_{\beta},\kappa^-_{\eta},\kappa^-_{\lambda}}\Big(
\frac{x_{\sigma}(6\ln x_{\sigma}-7\ln^2 x_{\sigma})}
{2\prod\limits_{\rho \neq \sigma}(x_{\rho}-x_{\sigma})}
+\frac{(x_{\kappa^-_{\lambda}{\rm w}}-x_{\tilde{\tiny U}^{j}_{\beta}{\rm w}}
)x_{\sigma}{\cal R}_{i_2}(\frac{\tilde{\tiny U}^{j}_{\beta}}
{x_{\sigma}})}{\prod\limits_{\rho \neq \sigma}(x_{\rho}-x_{\sigma})}
-3\frac{x_{\sigma}\Upsilon(\frac{\tilde{\tiny U}^{j}_{\beta}}
{x_{\sigma}})}{\prod\limits_{\rho \neq \sigma}(x_{\rho}-x_{\sigma})}\Big)
\nonumber \\
&&\hspace{1.0cm}
-\sum\limits_{\tiny \sigma=\tilde{\tiny U}^{i}_{\alpha},
\tilde{\tiny U}^{j}_{\beta},\kappa^-_{\eta}}
\frac{x_{\sigma}{\cal R}_{i_2}(\frac{\tilde{\tiny U}^{j}_{\beta}}
{x_{\sigma}})}{\prod\limits_{\rho \neq \sigma}(x_{\rho}-x_{\sigma})}
\bigg)\bigg]\; ,
\label{cc2}
\end{eqnarray}
\begin{eqnarray}
&&cc_{3}=-2cc_2\; ,
\label{cc3}
\end{eqnarray}
\begin{eqnarray}
&&cc_{4}=a_-^{(c)}b_-^{(c)}c_-^{(c)}d_-^{(c)}\sqrt{
x_{\kappa^-_{\eta}{\rm w}}x_{\kappa^-_{\lambda}{\rm w}}}\bigg(
\frac{4\Big({\cal R}_{i_2}(\frac{x_{\kappa^-_{\lambda}{\rm w}}}
{x_{\kappa^-_{\eta}{\rm w}}})+{\cal R}_{i_2}(\frac{x_{\kappa^-_{\lambda}{\rm w}}}
{x_{\tilde{U}^{i}_{\alpha}{\rm w}}})+{\cal R}_{i_2}(\frac{
x_{\tilde{U}^{j}_{\beta}{\rm w}}}{x_{\kappa^-_{\eta}{\rm w}}})
-{\cal R}_{i_2}(\frac{x_{\tilde{U}^{j}_{\beta}{\rm w}}}
{x_{\tilde{U}^{i}_{\alpha}{\rm w}}})\Big)}
{3(-x_{\kappa^-_{\eta}{\rm w}}+x_{\tilde{U}^{i}_{\alpha}{\rm w}})
(-x_{\kappa^-_{\lambda}{\rm w}}+x_{\tilde{U}^{j}_{\beta}{\rm w}})}
\nonumber \\
&&\hspace{1.0cm}-\frac{2}{3}
\bigg(-\sum\limits_{\sigma=\tilde{\tiny U}^{i}_{\alpha},
\tilde{\tiny U}^{i}_{\beta},\kappa^-_{\eta}}\frac{
{\cal R}_{i_2}(\frac{x_{\kappa^-_{\lambda}{\rm w}}}{x_{\sigma}})}
{\prod\limits_{\rho \neq \sigma}(x_{\rho}-x_{\sigma})}
+\frac{x_{\kappa^-_{\eta}{\rm w}}-
x_{\tilde{\tiny U}^{j}_{\beta}{\rm w}}}{x_{\kappa^-_{\lambda}{\rm w}}
-x_{\tilde{\tiny U}^{j}_{\beta}{\rm w}}}
\sum\limits_{\sigma=\tilde{\tiny U}^{i}_{\alpha},
\tilde{\tiny U}^{i}_{\beta},\kappa^-_{\eta}}
\frac{{\cal R}_{i_2}(\frac{x_{\kappa^-_{\lambda}{\rm w}}}{x_{\sigma}})
-{\cal R}_{i_2}(\frac{x_{\tilde{\tiny U}^{j}_{\beta}{\rm w}}}{x_\sigma})}
{\prod\limits_{\rho \neq \sigma}(x_{\rho}-x_{\sigma})}
\nonumber \\
&&\hspace{1.0cm}+\frac{
{\cal R}_{i_2}(\frac{x_{\kappa^-_{\lambda}{\rm w}}}
{x_{\tilde{\tiny U}^{i}_{\alpha}{\rm w}}})-
{\cal R}_{i_2}(\frac{x_{\kappa^-_{\lambda}{\rm w}}}
{x_{\tilde{\tiny U}^{j}_{\beta}{\rm w}}})-
{\cal R}_{i_2}(\frac{x_{\tilde{\tiny U}^{j}_{\beta}{\rm w}}}
{x_{\tilde{\tiny U}^{i}_{\alpha}{\rm w}}})}{(x_{\kappa^-_{\lambda}{\rm w}}
-x_{\tilde{\tiny U}^{j}_{\beta}{\rm w}})
(x_{\tilde{\tiny U}^{i}_{\alpha}{\rm w}}-x_{\tilde{\tiny U}^{j}_{\beta}{\rm w}})}
\bigg)-\frac{64}{3}
\bigg(-\sum\limits_{\sigma=\tilde{\tiny U}^{j}_{\beta},
\kappa^-_{\eta},\kappa^-_{\lambda}}\frac{2x_{\tilde{\tiny U}^{i}_{\alpha}{\rm w}}
x_{\sigma}(\ln x_{\sigma}-\ln^2 x_{\sigma})}
{(-x_{\sigma}+x_{\tilde{\tiny U}^{i}_{\alpha}{\rm w}})^2
\prod\limits_{\rho \neq \sigma}(x_{\rho}-x_{\sigma})}
\nonumber \\
&&\hspace{1.0cm}+\frac{x_{\tilde{\tiny U}^{i}_{\alpha}{\rm w}}(2-
\ln x_{\tilde{\tiny U}^{i}_{\alpha}{\rm w}}
-3\ln^2 x_{\tilde{\tiny U}^{i}_{\alpha}{\rm w}})}
{(x_{\kappa^-_{\eta}{\rm w}}-x_{\tilde{\tiny U}^{i}_{\alpha}{\rm w}})
(x_{\kappa^-_{\lambda}{\rm w}}-x_{\tilde{\tiny U}^{i}_{\alpha}{\rm w}})
(x_{\tilde{\tiny U}^{j}_{\beta}{\rm w}}-x_{\tilde{\tiny U}^{i}_{\alpha}{\rm w}})}
+\frac{2x_{\tilde{\tiny U}^{i}_{\alpha}{\rm w}}^2(
\ln x_{\tilde{\tiny U}^{i}_{\alpha}{\rm w}}-\ln^2 x_{\tilde{\tiny U}^{i}_{\alpha}{\rm w}})}
{(x_{\kappa^-_{\eta}{\rm w}}-x_{\tilde{\tiny U}^{i}_{\alpha}{\rm w}})
(x_{\kappa^-_{\lambda}{\rm w}}-x_{\tilde{\tiny U}^{i}_{\alpha}{\rm w}})
(x_{\tilde{\tiny U}^{j}_{\beta}{\rm w}}-x_{\tilde{\tiny U}^{i}_{\alpha}{\rm w}})^2}
\nonumber \\
&&\hspace{1.0cm}+\frac{2x_{\tilde{\tiny U}^{i}_{\alpha}{\rm w}}^2(
\ln x_{\tilde{\tiny U}^{i}_{\alpha}{\rm w}}-\ln^2 x_{\tilde{\tiny U}^{i}_{\alpha}{\rm w}})}
{(x_{\kappa^-_{\eta}{\rm w}}-x_{\tilde{\tiny U}^{i}_{\alpha}{\rm w}})
(x_{\kappa^-_{\lambda}{\rm w}}-x_{\tilde{\tiny U}^{i}_{\alpha}{\rm w}})^2
(x_{\tilde{\tiny U}^{j}_{\beta}{\rm w}}-x_{\tilde{\tiny U}^{i}_{\alpha}{\rm w}})}
+\frac{2x_{\tilde{\tiny U}^{i}_{\alpha}{\rm w}}^2(
\ln x_{\tilde{\tiny U}^{i}_{\alpha}{\rm w}}-\ln^2 x_{\tilde{\tiny U}^{i}_{\alpha}{\rm w}})}
{(x_{\kappa^-_{\eta}{\rm w}}-x_{\tilde{\tiny U}^{i}_{\alpha}{\rm w}})^2
(x_{\kappa^-_{\lambda}{\rm w}}-x_{\tilde{\tiny U}^{i}_{\alpha}{\rm w}})
(x_{\tilde{\tiny U}^{j}_{\beta}{\rm w}}-x_{\tilde{\tiny U}^{i}_{\alpha}{\rm w}})}
\nonumber \\
&&\hspace{1.0cm}+\sum\limits_{\sigma=\tilde{\tiny U}^{j}_{\beta},
\kappa^-_{\eta},\kappa^-_{\lambda}}\frac{(x_{\tilde{\tiny U}^{i}_{\alpha}{\rm w}}
+x_{\sigma})x_{\sigma}\Upsilon(\frac{x_{\tilde{\tiny U}^{i}_{\alpha}{\rm w}}}
{x_{\sigma}})}
{(-x_{\sigma}+x_{\tilde{\tiny U}^{i}_{\alpha}{\rm w}})^2
\prod\limits_{\rho \neq \sigma}(x_{\rho}-x_{\sigma})}
\bigg)\nonumber \\
&&\hspace{1.0cm}+\frac{32}{3}\bigg(\sum\limits_{\sigma=
\tilde{\tiny U}^{i}_{\alpha},\tilde{\tiny U}^{j}_{\beta},
\kappa^-_{\eta},\kappa^-_{\lambda}}\frac{x_{\sigma}(-3\ln x_{\sigma}
+7\ln^2 x_{\sigma})}{\prod\limits_{\rho \neq \sigma}(x_{\rho}-x_{\sigma})}
+\frac{x_{\kappa^-_{\eta}{\rm w}}\ln^2 x_{\kappa^-_{\eta}{\rm w}}}
{2(-x_{\kappa^-_{\eta}{\rm w}}+x_{\kappa^-_{\lambda}{\rm w}})
(-x_{\kappa^-_{\eta}{\rm w}}+x_{\tilde{\tiny U}^{i}_{\alpha}{\rm w}})^2}
\nonumber \\
&&\hspace{1.0cm}-\frac{4x_{\kappa^-_{\eta}{\rm w}}\ln^2 x_{\kappa^-_{\eta}{\rm w}}}
{(-x_{\kappa^-_{\eta}{\rm w}}+x_{\kappa^-_{\lambda}{\rm w}})
(-x_{\kappa^-_{\eta}{\rm w}}+x_{\tilde{\tiny U}^{i}_{\alpha}{\rm w}})
(-x_{\kappa^-_{\eta}{\rm w}}+x_{\tilde{\tiny U}^{j}_{\beta}{\rm w}})}
+\sum\limits_{\sigma=
\tilde{\tiny U}^{i}_{\alpha},\tilde{\tiny U}^{j}_{\beta},
\kappa^-_{\eta}}\frac{x_{\sigma}{\cal R}_{i_2}(\frac{
x_{\tilde{\tiny U}^{j}_{\beta}{\rm w}}}{x_{\sigma}})}
{\prod\limits_{\rho \neq \sigma}(x_{\rho}-x_{\sigma})}
\nonumber \\
&&\hspace{1.0cm}+\sum\limits_{\sigma=
\tilde{\tiny U}^{i}_{\alpha},\tilde{\tiny U}^{j}_{\beta},
\kappa^-_{\eta},\kappa^-_{\lambda}}
\frac{x_{\sigma}(-x_{\kappa^-_{\lambda}{\rm w}}+x_{\tilde{\tiny U}^{j}_{\beta}{\rm w}})
{\cal R}_{i_2}(\frac{x_{\tilde{\tiny U}^{j}_{\beta}{\rm w}}}{x_{\sigma}})}
{\prod\limits_{\rho \neq \sigma}(x_{\rho}-x_{\sigma})}
+\sum\limits_{\sigma=
\tilde{\tiny U}^{i}_{\alpha},\tilde{\tiny U}^{j}_{\beta},
\kappa^-_{\eta},\kappa^-_{\lambda}}
\frac{3x_{\sigma}\Upsilon(\frac{x_{\tilde{
\tiny U}^{j}_{\beta}{\rm w}}}{x_{\sigma}})}
{\prod\limits_{\rho \neq \sigma}(x_{\rho}-x_{\sigma})}\bigg)
\bigg)\; ,
\label{cc4}
\end{eqnarray}
\begin{eqnarray}
&&cc_{5}=\frac{1}{4}cc_4\; .
\label{cc5}
\end{eqnarray}
In the above expressions
Eqs.(\ref{cc1},\ref{cc2},\ref{cc3},\ref{cc4},\ref{cc5}), the coupling
constants are defined as
\begin{eqnarray}
&&a_+^{(c)}=\Big(-{\cal Z}_{\tilde{U}^i}^{1\alpha}{\cal Z}_{+}^{1l}+
\frac{m_{u^i}}{\sqrt{2}m_{\rm w}\sin\beta}{\cal Z}_{\tilde{U}^i}^{2\alpha}
{\cal Z}_{+}^{2l}\Big)\; ,\nonumber \\
&&a_-^{(c)}=\frac{h_d}{\sqrt{2}}{\cal Z}_{\tilde{U}^i}^{1\alpha}
{\cal Z}_{-}^{2l}\; ,\nonumber \\
&&b_+^{(c)}=\frac{h_b}{\sqrt{2}}{\cal Z}_{\tilde{U}^j}^{1\beta}
{\cal Z}_{-}^{2l}\; ,\nonumber \\
&&b_-^{(c)}=\Big(-{\cal Z}_{\tilde{U}^j}^{1\beta}{\cal Z}_{+}^{1l}+
\frac{m_{u^j}}{\sqrt{2}m_{\rm w}\sin\beta}{\cal Z}_{\tilde{U}^j}^{2\beta}
{\cal Z}_{+}^{2l}\Big)\; ,\nonumber \\
&&c_+^{(c)}=\Big(-{\cal Z}_{\tilde{U}^j}^{1\beta}{\cal Z}_{+}^{1k}+
\frac{m_{u^j}}{\sqrt{2}m_{\rm w}\sin\beta}{\cal Z}_{\tilde{U}^j}^{2\beta}
{\cal Z}_{+}^{2k}\Big)\; ,\nonumber \\
&&c_-^{(c)}=\frac{h_d}{\sqrt{2}}{\cal Z}_{\tilde{U}^j}^{1\beta}
{\cal Z}_{-}^{2k}\; ,\nonumber \\
&&d_+^{(c)}=\frac{h_b}{\sqrt{2}}{\cal Z}_{\tilde{U}^i}^{1\alpha}
{\cal Z}_{-}^{2k}\; ,\nonumber \\
&&d_-^{(c)}=\Big(-{\cal Z}_{\tilde{U}^i}^{1\alpha}{\cal Z}_{+}^{1k}+
\frac{m_{u^i}}{\sqrt{2}m_{\rm w}\sin\beta}{\cal Z}_{\tilde{U}^i}^{2\alpha}
{\cal Z}_{+}^{2k}\Big)\; .\nonumber \\
\end{eqnarray}
\subsection{The corrections due to gluino contributions}
The corrections caused by gluino can be written as
\begin{eqnarray}
&&\phi^{\tilde{g}}_{\alpha}=\phi^{ww\tilde{g}}_{\alpha}+
2\phi^{wh\tilde{g}}_{\alpha}+\phi^{hh\tilde{g}}_{\alpha}+
\phi^{sw\tilde{g}}_{\alpha}+\phi^{sh\tilde{g}}_{\alpha}+
\phi^{p\tilde{g}}_{\alpha}
\end{eqnarray}
with
\begin{eqnarray}
&&\phi^{ww\tilde{g}}_{1}={\cal Z}_{\tilde{D}^3}^{1\gamma}{\cal Z}_{\tilde{D}^1}^{1\delta}
{\cal Z}_{\tilde{U}^i}^{1\alpha}{\cal Z}_{\tilde{U}^i}^{1\alpha}
\Big(\frac{4}{3}F_{A}^{2a}+\frac{16}{3}F_{A}^{2b}-4F_{A}^{2d}-\frac{4}{3}F_{A}^{2e}-\frac{16}{3}F_{A}^{2f}\Big)
(x_{j{\rm w}}, 1, 1, x_{\tilde{\tiny U}^{i}_{\alpha}{\rm w}},
x_{\tilde{D}^1_\delta {\rm w}}, x_{\tilde{D}^3_\gamma {\rm w}},
x_{\tilde{g}{\rm w}})\nonumber \\
&&\hspace{1.2cm}+\frac{16}{3}({\cal Z}_{\tilde{U}^i}^{1\alpha})^2)
\Big(F_{C}^{2a}+F_{C}^{2d}-F_{C}^{2e}\Big)
(x_{j{\rm w}}, x_{i{\rm w}}, x_{i{\rm w}}, 1, 1, x_{\tilde{U}^i_\alpha {\rm w}}, x_{\tilde{g}{\rm w}})
\nonumber \\
&&\hspace{1.2cm}-\frac{16}{3}{\cal Z}_{\tilde{D}^3}^{1\delta}
{\cal Z}_{\tilde{U}^i}^{1\alpha}{\cal Z}_{\tilde{D}^3}^{1\delta}
{\cal Z}_{\tilde{U}^i}^{1\alpha}\Big(F_{D}^{2e}-F_{D}^{2a}-
F_{D}^{2d}\Big)
(x_{i{\rm w}}, x_{j{\rm w}}, 1, 1, x_{\tilde{U}^i_\alpha {\rm w}},
x_{\tilde{D}^1_\delta {\rm w}}, x_{\tilde{g}{\rm w}})\nonumber \\
&&\hspace{1.2cm}-\frac{64}{3}\sqrt{x_{\tilde{g}{\rm w}}x_{i{\rm w}}}
{\cal Z}_{\tilde{U}^i}^{1\alpha}{\cal Z}_{\tilde{U}^i}^{2\alpha}
F_{C}^{1a}(x_{j{\rm w}}, x_{i{\rm w}}, x_{i{\rm w}}, 1, 1, x_{\tilde{U}^i_\alpha {\rm w}},
x_{\tilde{g}{\rm w}}) \nonumber \\
&&\hspace{1.2cm}+\frac{32}{3}\sqrt{x_{\tilde{g}{\rm w}}x_{i{\rm w}}}
{\cal Z}_{\tilde{D}^3}^{1\delta}{\cal Z}_{\tilde{D}^3}^{1\delta}
{\cal Z}_{\tilde{U}^i}^{1\alpha}{\cal Z}_{\tilde{U}^i}^{2\alpha}
\Big(F_{D}^{1b}-F_{D}^{1c}\Big)(x_{i{\rm w}}, x_{j{\rm w}}, 1, 1,
x_{\tilde{U}^i_\alpha {\rm w}},
x_{\tilde{D}^1_\delta {\rm w}}, x_{\tilde{g}{\rm w}})\nonumber \\
&&\hspace{1.2cm}+\frac{16}{3}x_{i{\rm w}}
({\cal Z}_{\tilde{U}^i}^{2\alpha})^2\Big(F_{C}^{1a}+F_{C}^{1b}
-F_{C}^{1c}\Big)(x_{j{\rm w}}, x_{i{\rm w}}, x_{i{\rm w}}, 1, 1,
x_{\tilde{U}^i_\alpha {\rm w}}, x_{\tilde{g}{\rm w}})\nonumber \\
&&\hspace{1.2cm}
+\frac{56}{3}\Big(\frac{x_{i{\rm w}}\ln x_{i{\rm w}}}{(-x_{i{\rm w}} + x_{j{\rm w}})^2}
+\frac{x_{i{\rm w}}\ln x_{i{\rm w}}}{(1 - x_{i{\rm w}})^2 (-x_{i{\rm w}} + x_{j{\rm w}})^2}
-\frac{2x_{i{\rm w}}\ln x_{i{\rm w}}}{(1 - x_{i{\rm w}}) (-x_{i{\rm w}} + x_{j{\rm w}})^2}
 + \frac{\ln x_{i{\rm w}}}{-x_{i{\rm w}} + x_{j{\rm w}}} \nonumber \\
&&\hspace{1.2cm}+
\frac{\ln x_{i{\rm w}}}{(1 - x_{i{\rm w}})^2 (-x_{i{\rm w}} + x_{j{\rm w}})}
+ \frac{2x_{i{\rm w}} \ln x_{i{\rm w}}}{(1 - x_{i{\rm w}})^3 (-x_{i{\rm w}} + x_{j{\rm w}})}
-\frac{2x_{i{\rm w}}\ln x_{i{\rm w}}}{(1 - x_{i{\rm w}})^2 (-x_{i{\rm w}} + x_{j{\rm w}})}
\nonumber \\
&&\hspace{1.2cm}
-\frac{2\ln x_{i{\rm w}}}{(1 - x_{i{\rm w}}) (-x_{i{\rm w}} + x_{j{\rm w}})}\Big)
-\frac{64}{3}\frac{x_{i{\rm w}}^2 \ln x_{i{\rm w}}}{
(1 - x_{i{\rm w}})^2 (-x_{i{\rm w}} + x_{j{\rm w}})}+\frac{16 x_{i{\rm w}}^2 \ln x_{i{\rm w}}}
{(1 - x_{i{\rm w}})^2 (-x_{i{\rm w}} + x_{j{\rm w}})}\nonumber \\
&&\hspace{1.2cm}-\frac{56}{3}\frac{x_{j{\rm w}}^3 \ln x_{j{\rm w}}}
{(1 - x_{j{\rm w}})^2 (x_{i{\rm w}} - x_{j{\rm w}})^2} + \frac{64}{3}
\frac{x_{j{\rm w}}^2 \ln x_{j{\rm w}}}{(1 - x_{j{\rm w}})^2 (x_{i{\rm w}} - x_{j{\rm w}})}
+\frac{16 x_{j{\rm w}}^2 \ln x_{j{\rm w}}}{(1 - x_{j{\rm w}})^2 (x_{i{\rm w}} - x_{j{\rm w}})}
+(i\leftrightarrow j)\; ,
\label{chww1}
\end{eqnarray}
\begin{eqnarray}
&&\phi^{ww\tilde{g}}_2={\cal Z}_{\tilde{D}^3}^{2\gamma}{\cal Z}_{\tilde{D}^1}^{2\delta}
{\cal Z}_{\tilde{U}^i}^{1\alpha}{\cal Z}_{\tilde{U}^i}^{1\alpha}
\Big(\frac{2}{3}F_{A}^{2a}+\frac{8}{3}F_{A}^{2b}-2F_{A}^{2d}-\frac{2}{3}F_{A}^{2e}
-\frac{8}{3}F_{A}^{2f}\Big)
(x_{j{\rm w}}, 1, 1, x_{\tilde{\tiny U}^{i}_{\alpha}{\rm w}},
x_{\tilde{D}^1_\delta {\rm w}}, x_{\tilde{D}^3_\gamma {\rm w}},
x_{\tilde{g}{\rm w}})\nonumber \\
&&\hspace{1.2cm}
+(i\leftrightarrow j)\; ,
\label{chww2}
\end{eqnarray}
\begin{eqnarray}
&&\phi^{ww\tilde{g}}_3=-2\phi^{ww\tilde{g}}_2\; ,
\label{chww3}
\end{eqnarray}
\begin{eqnarray}
&&\phi^{wh\tilde{g}}_{1}=-\frac{64}{3\sin^2\beta}({\cal Z}_{H}^{2k})^{2}\frac{x_{i{\rm w}}^{\frac{5}{2}}
x_{j{\rm w}}^{\frac{3}{2}}}{(1 - x_{i{\rm w}})(x_{H^{-}_{k}{\rm w}}-x_{i{\rm w}})(-x_{i{\rm w}} + x_{j{\rm w}})}
\nonumber \\
&&\hspace{1.2cm}-\frac{16}{3\sin^2\beta}(x_{i{\rm w}} x_{j{\rm w}})^{\frac{3}{2}}
({\cal Z}_{H}^{2k})^{2}({\cal Z}_{\tilde{U}^i}^{1\alpha})^2
\Big(F_{C}^{1a}+F_{C}^{1b}-F_{C}^{1c}\Big)
(x_{j{\rm w}}, x_{i{\rm w}}, x_{i{\rm w}}, x_{H^{-}_{k}{\rm w}}, 1,
x_{\tilde{U}^i_\alpha {\rm w}}, x_{\tilde{g}{\rm w}})\nonumber \\
&&\hspace{1.2cm}+\frac{32}{3 \sin^2\beta}({\cal Z}_{H}^{2k})^{2}
{\cal Z}_{\tilde{U}^i}^{1\alpha}{\cal Z}_{\tilde{U}^i}^{2\alpha}
x_{i{\rm w}}^2 x_{j{\rm w}} \sqrt{x_{\tilde{g}{\rm w}} x_{j{\rm w}}}
F_{C}^{0}(x_{j{\rm w}}, x_{i{\rm w}}, x_{i{\rm w}}, x_{H^{-}_{k}{\rm w}}, 1,
x_{\tilde{U}^i_\alpha {\rm w}}, x_{\tilde{g}{\rm w}})
\nonumber \\
&&\hspace{1.2cm}+\frac{32}{3\sin^2\beta}({\cal Z}_{H}^{2k})^{2}
{\cal Z}_{\tilde{U}^i}^{1\alpha}{\cal Z}_{\tilde{U}^i}^{2\alpha}
x_{i{\rm w}}x_{j{\rm w}}\sqrt{x_{\tilde{g}{\rm w}} x_{j{\rm w}}}
F_{C}^{1a}(x_{j{\rm w}}, x_{i{\rm w}}, x_{i{\rm w}}, x_{H^{-}_{k}{\rm w}},
1, x_{\tilde{U}^i_\alpha {\rm w}}, x_{\tilde{g}{\rm w}})\nonumber \\
&&\hspace{1.2cm}-\frac{16}{3\sin^2\beta}({\cal Z}_{H}^{2k})^{2}
({\cal Z}_{\tilde{U}^i}^{2\alpha})^2(x_{i{\rm w}} x_{j{\rm w}})^{\frac{3}{2}}
\Big(F_{C}^{1a}+F_{C}^{1b}-F_{C}^{1c}\Big)
(x_{j{\rm w}}, x_{i{\rm w}}, x_{i{\rm w}}, x_{H^{-}_{k}{\rm w}},
1, x_{\tilde{U}^i_\alpha {\rm w}}, x_{\tilde{g}{\rm w}})\nonumber \\
&&\hspace{1.2cm}-\frac{8}{3\sin^2\beta}(x_{i{\rm w}} x_{j{\rm w}})^{\frac{3}{2}}
({\cal Z}_{H}^{2k})^{2}\bigg({\cal Z}_{\tilde{D}^1}^{1\delta}
{\cal Z}_{\tilde{U}^i}^{1\alpha}+{\cal Z}_{\tilde{D}^1}^{1\delta}
{\cal Z}_{\tilde{U}^i}^{2\alpha}
\bigg){\cal Z}_{\tilde{D}^1}^{1\delta}
{\cal Z}_{\tilde{U}^j}^{1\beta}\Big(F_{D}^{1a}-
F_{D}^{1b}\nonumber \\
&&\hspace{1.2cm}-F_{D}^{1c}\Big)
(x_{i{\rm w}}, x_{j{\rm w}}, x_{H^{-}_{k}{\rm w}}, 1,
x_{\tilde{U}^i_\alpha {\rm w}}, x_{\tilde{D}^1_\delta {\rm w}}, x_{\tilde{g}{\rm w}})
\nonumber \\
&&\hspace{1.2cm}+\frac{64}{3\sin^2\beta}
(x_{i{\rm w}}x_{j{\rm w}})^{\frac{3}{2}}({\cal Z}_{H}^{2k})^{2}
\frac{x_{H^{-}_{k}{\rm w}}^2\ln x_{H^{-}_{k}{\rm w}}}
{(1 - x_{H^{-}_{k}{\rm w}}) (-x_{H^{-}_{k}{\rm w}} + x_{i{\rm w}})^2
(-x_{H^{-}_{k}{\rm w}} + x_{j{\rm w}})}\nonumber \\
&&\hspace{1.2cm}+({\cal Z}_{H}^{2k})^{2}
(x_{i{\rm w}} x_{j{\rm w}})^{\frac{3}{2}}\frac{8x_{H^{-}_{k}{\rm w}}
\ln x_{H^{-}_{k}{\rm w}}}{\sin^2\beta(1 - x_{H^{-}_{k}{\rm w}})
(-x_{H^{-}_{k}{\rm w}} + x_{i{\rm w}})(-x_{H^{-}_{k}{\rm w}} + x_{j{\rm w}})}\nonumber \\
&&\hspace{1.2cm}+\frac{4}{3}h_{d} {\cal E}^{ib}
\sqrt{x_{i{\rm w}}}{\cal Z}_{H}^{1j}
{\cal Z}_{\tilde{D}^3}^{1\gamma}{\cal Z}_{\tilde{D}^1}^{1\delta}
\Big(F_{A}^{1a}-F_{A}^{1b}-F_{A}^{1c}\Big)
(x_{i{\rm w}}, x_{H^{-}_{j}{\rm w}}, 1, x_{\tilde{U}^i_\alpha {\rm w}},
x_{\tilde{D}^3_\gamma {\rm w}}, x_{\tilde{D}^1_\delta {\rm w}},
x_{\tilde{g}{\rm w}})\nonumber \\
&&\hspace{1.2cm}
-\frac{64}{3\sin^2\beta}({\cal Z}_{H}^{2k})^2
\Big(-x_{H^{-}_{k}{\rm w}}x_{i{\rm w}}^3 x_{j{\rm w}}+x_{i{\rm w}}^5 x_{j{\rm w}}+2x_{H^{-}_{k}{\rm w}}
x_{i{\rm w}}^2 x_{j{\rm w}}^2 - x_{i{\rm w}}^3 x_{j{\rm w}}^2-x_{H^{-}_{k}{\rm w}}x_{i{\rm w}}^3 x_{j{\rm w}}^2
\Big)\nonumber \\
&&\hspace{1.2cm}
\frac{\sqrt{x_{i{\rm w}} x_{j{\rm w}}}\ln x_{i{\rm w}}}
{(1 - x_{i{\rm w}})^2 (x_{H^{-}_{k}{\rm w}} - x_{i{\rm w}})^2 (-x_{i{\rm w}} + x_{j{\rm w}})^2}
+\frac{8}{\sin^2\beta}({\cal Z}_{H}^{2k})^{2}\Big(
\frac{8x_{i{\rm w}}^{3\over 2}x_{j{\rm w}}^{7\over 2}\ln x_{j{\rm w}}}{
(1 - x_{j{\rm w}})(x_{H^{-}_{k}{\rm w}} - x_{j{\rm w}})(x_{i{\rm w}} - x_{j{\rm w}})^2}
\nonumber \\
&&\hspace{1.2cm}
+\frac{x_{i{\rm w}}^{\frac{5}{2}} x_{j{\rm w}}^{\frac{3}{2}}\ln x_{i{\rm w}}}{
( (1 - x_{i{\rm w}}) (x_{H^{-}_{k}{\rm w}} - x_{i{\rm w}})(x_{i{\rm w}} - x_{j{\rm w}})}+
\frac{x_{i{\rm w}}^{3\over 2} x_{j{\rm w}}^{5\over 2}\ln x_{j{\rm w}}}{(1 - x_{j{\rm w}})
(x_{H^{-}_{k}{\rm w}} - x_{j{\rm w}})(x_{i{\rm w}} - x_{j{\rm w}})}
\Big)
\nonumber \\
&&\hspace{1.2cm}+(i\leftrightarrow j)\; ,
\label{chwh1}
\end{eqnarray}
\begin{eqnarray}
&&\phi^{wh\tilde{g}}_{2}=-\frac{4}{3}h_{b}h_{d}
\Big(\sqrt{x_{i{\rm w}} x_{j{\rm w}}}{\cal Z}_{\tilde{D}^1}^{1\delta}
{\cal Z}_{\tilde{U}^i}^{1\alpha}+ \sqrt{x_{\tilde{g}{\rm w}} x_{j{\rm w}}}
{\cal Z}_{\tilde{D}^1}^{1\delta}
{\cal Z}_{\tilde{U}^i}^{2\alpha}\Big)
({\cal Z}_{H}^{1k})^{2}
{\cal Z}_{\tilde{D}^1}^{1\delta}{\cal Z}_{\tilde{U}^j}^{1\beta}
\nonumber \\
&&\hspace{1.2cm}
\Big(
F_{D}^{1a}-3F_{D}^{1b}-F_{D}^{1c}\Big)
(x_{i{\rm w}}, x_{j{\rm w}}, x_{H^{-}_{k}{\rm w}}, 1,
x_{\tilde{U}^i_\alpha {\rm w}}, x_{\tilde{D}^1_\delta {\rm w}}, x_{\tilde{g}{\rm w}})
\nonumber \\
&&\hspace{1.2cm}-\frac{4x_{i{\rm w}} x_{j{\rm w}}}{3\sin^2\beta}\Big(
\sqrt{x_{\tilde{g}{\rm w}} x_{j{\rm w}}}{\cal Z}_{\tilde{D}^1}^{2\delta}
{\cal Z}_{\tilde{U}^i}^{1\alpha}+\sqrt{x_{i{\rm w}} x_{j{\rm w}}}
{\cal Z}_{\tilde{D}^1}^{2\delta}
{\cal Z}_{\tilde{U}^i}^{2\alpha}\Big)({\cal Z}_{H}^{2k})^{2}
{\cal Z}_{\tilde{D}^1}^{1\delta}{\cal Z}_{\tilde{U}^j}^{1\beta}
\nonumber \\
&&\hspace{1.2cm}
\Big(F_{D}^{1a}-3F_{D}^{1b}-F_{D}^{1c}\Big)
(x_{i{\rm w}}, x_{j{\rm w}}, x_{H^{-}_{k}{\rm w}}, 1,
x_{\tilde{U}^i_\alpha {\rm w}}, x_{\tilde{D}^1_\delta {\rm w}},
x_{\tilde{g}{\rm w}})
\nonumber \\
&&\hspace{1.2cm}+ \frac{2}{3}h_{d}\sqrt{x_{i{\rm w}}}{\cal Z}_{H}^{1j}
{\cal Z}_{\tilde{D}^3}^{2\gamma}{\cal Z}_{\tilde{D}^1}^{2\delta}
{\cal E}^{ib}\Big(F_{A}^{1a}-3F_{A}^{1b}-F_{A}^{1c}\Big)
(x_{i{\rm w}}, x_{H^{-}_{j}{\rm w}}, 1, x_{\tilde{U}^i_\alpha {\rm w}},
x_{\tilde{D}^3_\gamma {\rm w}}, x_{\tilde{D}^1_\delta {\rm w}},
x_{\tilde{g}{\rm w}})\nonumber \\
&&\hspace{1.2cm}-\frac{4}{3}h_{b}h_{d}({\cal Z}_{H}^{1k})^{2}\Big(
\frac{x_{H^{-}_{k}{\rm w}} \sqrt{x_{i{\rm w}} x_{j{\rm w}}}\ln x_{H^{-}_{k}{\rm w}}}
{(1 - x_{H^{-}_{k}{\rm w}})(-x_{H^{-}_{k}{\rm w}} + x_{i{\rm w}})
(-x_{H^{-}_{k}{\rm w}} + x_{j{\rm w}})}
+\frac{x_{i{\rm w}}\sqrt{x_{i{\rm w}}x_{j{\rm w}}}\ln x_{i{\rm w}}}{(1 - x_{i{\rm w}})
(x_{H^{-}_{k}{\rm w}} - x_{i{\rm w}})(-x_{i{\rm w}} + x_{j{\rm w}})}
\nonumber \\
&&\hspace{1.2cm}+\frac{x_{j{\rm w}} \sqrt{x_{i{\rm w}} x_{j{\rm w}}}\ln x_{j{\rm w}}}{(1 - x_{j{\rm w}})
(x_{H^{-}_{k}{\rm w}} - x_{j{\rm w}})(x_{i{\rm w}} - x_{j{\rm w}})}\Big)
-\frac{4}{3\sin^2\beta}({\cal Z}_{H}^{2k})^{2}\Big(
\frac{x_{H^{-}_{k}{\rm w}}x_{i{\rm w}}x_{j{\rm w}}\sqrt{x_{\tilde{g}{\rm w}} x_{j{\rm w}}}
 \ln x_{H^{-}_{k}{\rm w}}}{(1 - x_{H^{-}_{k}{\rm w}})
(-x_{H^{-}_{k}{\rm w}} + x_{i{\rm w}})(-x_{H^{-}_{k}{\rm w}} + x_{j{\rm w}})}
\nonumber \\
&&\hspace{1.2cm}+
\frac{x_{i{\rm w}}^2 x_{j{\rm w}} \sqrt{x_{\tilde{g}{\rm w}} x_{j{\rm w}}}\ln x_{i{\rm w}}}
{(1 - x_{i{\rm w}})(x_{H^{-}_{k}{\rm w}} - x_{i{\rm w}})(-x_{i{\rm w}} + x_{j{\rm w}})}
+\frac{x_{i{\rm w}}x_{j{\rm w}}^2 \sqrt{x_{\tilde{g}{\rm w}}x_{j{\rm w}}}\ln x_{j{\rm w}}}{
(1 - x_{j{\rm w}})(x_{H^{-}_{k}{\rm w}} - x_{j{\rm w}})(x_{i{\rm w}} - x_{j{\rm w}})}\Big)
\nonumber \\
&&\hspace{1.2cm}+\frac{4}{3}h_{b}h_{d}({\cal Z}_{H}^{1k})^{2}
\Big(({\cal Z}_{\tilde{U}^i}^{1\alpha})^2+({\cal Z}_{\tilde{U}^i}^{1\alpha})^2
\Big)\Big(F_{C}^{2a}+F_{C}^{2d}-F_{C}^{2e}\Big)
(x_{j{\rm w}}, x_{i{\rm w}}, x_{i{\rm w}}, x_{H^{-}_{k}{\rm w}}, 1,
x_{\tilde{U}^i_\alpha {\rm w}}, x_{\tilde{g}{\rm w}})\nonumber \\
&&\hspace{1.2cm}-\frac{16}{3}h_{b}h_{d}\sqrt{x_{\tilde{g}{\rm w}}x_{i{\rm w}}}
({\cal Z}_{H}^{1k})^{2}{\cal Z}_{\tilde{U}^i}^{1\alpha}
{\cal Z}_{\tilde{U}^i}^{2\alpha}F_{C}^{1a}(x_{j{\rm w}}, x_{i{\rm w}}, x_{i{\rm w}},
x_{H^{-}_{k}{\rm w}}, 1, x_{\tilde{U}^i_\alpha {\rm w}}, x_{\tilde{g}{\rm w}})\nonumber \\
&&\hspace{1.2cm}-\frac{8}{3\sin\beta}{\cal Z}_{H}^{1k}
{\cal Z}_{H}^{2k}\Big(h_{d}x_{i{\rm w}}{\cal Z}_{\tilde{D}^1}^{1\delta}
{\cal Z}_{\tilde{U}^i}^{1\alpha}+h_{b}x_{j{\rm w}}{\cal Z}_{\tilde{D}^1}^{2\delta}
{\cal Z}_{\tilde{U}^i}^{2\alpha}\Big)
{\cal Z}_{\tilde{D}^1}^{1\delta}{\cal Z}_{\tilde{U}^j}^{1\beta}
\nonumber \\
&&\hspace{1.2cm}\Big(F_{D}^{2b}+F_{D}^{2d}-2F_{D}^{2f}\Big)
(x_{i{\rm w}}, x_{j{\rm w}}, x_{H^{-}_{k}{\rm w}}, 1,
x_{\tilde{U}^i_\alpha {\rm w}}, x_{\tilde{D}^1_\delta {\rm w}}, x_{\tilde{g}{\rm w}})
\nonumber \\
&&\hspace{1.2cm}+\frac{16}{3\sin\beta}\sqrt{x_{\tilde{g}{\rm w}} x_{i{\rm w}}}
{\cal Z}_{H}^{1k} {\cal Z}_{H}^{2k}
\Big(h_{b}x_{j{\rm w}}{\cal Z}_{\tilde{D}^1}^{2\delta}
{\cal Z}_{\tilde{U}^i}^{1\alpha}+h_{d}x_{i{\rm w}}{\cal Z}_{\tilde{D}^1}^{1\delta}
{\cal Z}_{\tilde{U}^i}^{2\alpha}\Big)
{\cal Z}_{\tilde{D}^1}^{1\delta}{\cal Z}_{\tilde{U}^j}^{1\beta}
\nonumber \\
&&\hspace{1.2cm}\Big(F_{D}^{1b}-F_{D}^{1c}\Big)
(x_{i{\rm w}}, x_{j{\rm w}}, x_{H^{-}_{k}{\rm w}}, 1,
x_{\tilde{U}^i_\alpha {\rm w}}, x_{\tilde{D}^1_\delta {\rm w}},
x_{\tilde{g}{\rm w}})\nonumber \\
&&\hspace{1.2cm}-\frac{8}{3\sin\beta}
\sqrt{x_{\tilde{g}{\rm w}}}x_{i{\rm w}}{\cal Z}_{H}^{2j}
{\cal Z}_{\tilde{D}^3}^{2\gamma}{\cal Z}_{\tilde{D}^1}^{1\delta}
{\cal E}^{ib}\Big(F_{A}^{1b}-F_{A}^{1c}\Big)
(x_{i{\rm w}}, x_{H^{-}_{j}{\rm w}}, 1, x_{\tilde{U}^i_\alpha {\rm w}},
x_{\tilde{D}^3_\gamma {\rm w}}, x_{\tilde{D}^1_\delta {\rm w}},
x_{\tilde{g}{\rm w}})\nonumber \\
&&\hspace{1.2cm}-2h_{b}h_{d}({\cal Z}_{H}^{1k})^2\frac{x_{H^{-}_{k}{\rm w}}^2
(x_{H^{-}_{k}{\rm w}}+x_{i{\rm w}})\ln x_{H^{-}_{k}{\rm w}}}{(-1+
x_{H^{-}_{k}{\rm w}})(x_{H^{-}_{k}{\rm w}}-x_{i{\rm w}})^2 (x_{H^{-}_{k}{\rm w}}-
x_{j{\rm w}})}\nonumber \\
&&\hspace{1.2cm}-\frac{4}{3\sin\beta}h_{d}{\cal Z}_{H}^{1k}{\cal Z}_{H}^{2k}
\frac{(x_{H^{-}_{k}{\rm w}}+2x_{\tilde{g}{\rm w}})x_{H^{-}_{k}{\rm w}}
x_{i{\rm w}}\ln x_{H^{-}_{k}{\rm w}}}{(1 - x_{H^{-}_{k}{\rm w}})(-x_{H^{-}_{k}{\rm w}} + x_{i{\rm w}})
(-x_{H^{-}_{k}{\rm w}} + x_{j{\rm w}})}\nonumber \\
&&\hspace{1.2cm}+2h_{b}h_{d}({\cal Z}_{H}^{1k})^2 \Big(-2x_{H^{-}_{k}{\rm w}}x_{i{\rm w}}
+x_{i{\rm w}}^{2}+x_{H^{-}_{k}{\rm w}}x_{i{\rm w}}^{2}+3x_{H^{-}_{k}{\rm w}}x_{j{\rm w}}-2x_{i{\rm w}}x_{j{\rm w}}
\nonumber \\
&&\hspace{1.2cm}
-2x_{H^{-}_{k}{\rm w}}x_{i{\rm w}}x_{j{\rm w}}+x_{i{\rm w}}^{2}x_{j{\rm w}}\Big)\frac{x_{i{\rm w}}^2\ln x_{i{\rm w}}}{
(1 - x_{H^{-}_{k}{\rm w}})^2 (-x_{H^{-}_{k}{\rm w}} + x_{i{\rm w}})^2
(-x_{H^{-}_{k}{\rm w}} + x_{j{\rm w}})^2}
\nonumber \\
&&\hspace{1.2cm}-2h_{b}h_{d}({\cal Z}_{H}^{1k})^{2}\Big(
x_{H^{-}_{k}{\rm w}}x_{i{\rm w}}-x_{i{\rm w}}^{3}-2x_{H^{-}_{k}{\rm w}}x_{j{\rm w}}
+x_{i{\rm w}}x_{j{\rm w}}\nonumber \\
&&\hspace{1.2cm}
+x_{H^{-}_{k}{\rm w}}x_{i{\rm w}}x_{j{\rm w}}\Big)
\frac{x_{i{\rm w}}^{2}\ln x_{i{\rm w}}}{(x_{H^{-}_{k}{\rm w}}-x_{i{\rm w}})^2
(-1+x_{i{\rm w}})^2 (x_{i{\rm w}}-x_{j{\rm w}})^2} \nonumber \\
&&\hspace{1.2cm}-2h_{b}h_{d}({\cal Z}_{H}^{1k})^{2}
\frac{(x_{i{\rm w}}+x_{j{\rm w}})x_{j{\rm w}}^2 \ln x_{j{\rm w}}}{(x_{i{\rm w}}-x_{j{\rm w}})
(-1+x_{j{\rm w}})(-x_{H^{-}_{k}{\rm w}}+x_{j{\rm w}})}\nonumber \\
&&\hspace{1.2cm}-\frac{4}{3\sin\beta}h_{d}{\cal Z}_{H}^{1k}
{\cal Z}_{H}^{2k}\frac{(-2x_{\tilde{g}{\rm w}}+3x_{j{\rm w}})
x_{i{\rm w}}x_{j{\rm w}} \ln x_{j{\rm w}}}{(x_{i{\rm w}}-x_{j{\rm w}})
(-1+x_{j{\rm w}})(-x_{H^{-}_{k}{\rm w}}+x_{j{\rm w}})}
\nonumber \\
&&\hspace{1.2cm}+(i\leftrightarrow j)\; ,
\label{chwh2}
\end{eqnarray}
\begin{eqnarray}
&&\phi^{wh\tilde{g}}_{3}=-2\phi^{wh\tilde{g}}_{2}\; ,
\label{chwh3}
\end{eqnarray}
\begin{eqnarray}
&&\phi^{wh\tilde{g}}_{4}=-\frac{16}{3\sin\beta}h_{d}x_{i{\rm w}}\sqrt{x_{\tilde{g}{\rm w}}x_{i{\rm w}}}
{\cal Z}_{H}^{1k}{\cal Z}_{H}^{2k}
{\cal Z}_{\tilde{D}^1}^{2\delta}{\cal Z}_{\tilde{U}^i}^{1\alpha}
{\cal Z}_{\tilde{D}^1}^{1\delta}{\cal Z}_{\tilde{U}^j}^{1\beta}
\Big(F_{D}^{1b}
\nonumber \\
&&\hspace{1.2cm}
-F_{D}^{1c}\Big)(x_{i{\rm w}}, x_{j{\rm w}}, x_{H^{-}_{k}{\rm w}}, 1,
x_{\tilde{U}^i_\alpha {\rm w}}, x_{\tilde{D}^1_\delta {\rm w}},
x_{\tilde{g}{\rm w}})\nonumber \\
&&\hspace{1.2cm}+\frac{1}{3\sin\beta}
h_{d} x_{i{\rm w}} {\cal Z}_{H}^{1k} {\cal Z}_{H}^{2k}
{\cal Z}_{\tilde{D}^1}^{2\delta}{\cal Z}_{\tilde{U}^i}^{2\alpha}
{\cal Z}_{\tilde{D}^1}^{1\delta}{\cal Z}_{\tilde{U}^j}^{1\beta}
\Big(3F_{D}^{2a}+11F_{D}^{2b}
\nonumber \\
&&\hspace{1.2cm}
+11F_{D}^{2c}+2F_{D}^{2d}
-14F_{D}^{2e}-22F_{D}^{2f}\Big)
(x_{i{\rm w}}, x_{j{\rm w}}, x_{H^{-}_{k}{\rm w}}, 1, x_{\tilde{U}^i_\alpha {\rm w}},
x_{\tilde{D}^1_\delta {\rm w}}, x_{\tilde{g}{\rm w}})
\nonumber \\
&&\hspace{1.2cm}+ \frac{8}{3\sin\beta}\sqrt{x_{\tilde{g}{\rm w}}}
x_{i{\rm w}} {\cal Z}_{H}^{2j}
{\cal Z}_{\tilde{D}^3}^{1\gamma}{\cal Z}_{\tilde{D}^1}^{2\delta}
{\cal E}^{ib}\Big(F_{A}^{1b}-F_{A}^{1c}\Big)
(x_{i{\rm w}}, x_{H^{-}_{j}{\rm w}}, 1, x_{\tilde{U}^i_\alpha {\rm w}},
x_{\tilde{D}^3_\gamma {\rm w}}, x_{\tilde{D}^1_\delta {\rm w}},
x_{\tilde{g}{\rm w}})
\nonumber \\
&&\hspace{1.2cm}+\frac{1}{\sin\beta}h_{d}x_{i{\rm w}}{\cal Z}_{H}^{1k}
{\cal Z}_{H}^{2k}{\cal Z}_{\tilde{D}^1}^{2\delta}
{\cal Z}_{\tilde{U}^i}^{2\alpha}{\cal Z}_{\tilde{D}^1}^{1\delta}
{\cal Z}_{\tilde{U}^j}^{1\beta}\Big(F_{D}^{2a}+F_{D}^{2b}
+F_{D}^{2c}-2F_{D}^{2d}\nonumber \\
&&\hspace{1.2cm}-2F_{D}^{2e}-2F_{D}^{2f}\Big)
(x_{i{\rm w}}, x_{j{\rm w}}, x_{H^{-}_{k}{\rm w}}, 1,
x_{\tilde{U}^i_\alpha {\rm w}}, x_{\tilde{D}^1_\delta {\rm w}}, x_{\tilde{g}{\rm w}})
\nonumber \\
&&\hspace{1.2cm}+(i\leftrightarrow j)\; ,
\label{chwh4}
\end{eqnarray}
\begin{eqnarray}
&&\phi^{wh\tilde{g}}_{5}=\frac{1}{4}\phi^{wh\tilde{g}}_{4}\; ,
\label{chwh5}
\end{eqnarray}
\begin{eqnarray}
&&\phi^{wh\tilde{g}}_{6}=-\frac{8}{3}h_{b}h_{d}\sqrt{x_{\tilde{g}{\rm w}}x_{j{\rm w}}}
({\cal Z}_{H}^{1k})^{2}\Big(
{\cal Z}_{\tilde{D}^1}^{2\delta}{\cal Z}_{\tilde{U}^i}^{1\alpha}+
{\cal Z}_{\tilde{D}^1}^{2\delta}{\cal Z}_{\tilde{U}^i}^{2\alpha}\Big)
{\cal Z}_{\tilde{D}^1}^{1\delta}{\cal Z}_{\tilde{U}^j}^{1\beta}
\Big(F_{D}^{1a}-3F_{D}^{1b}\nonumber \\
&&\hspace{1.2cm}-F_{D}^{1c}\Big)
(x_{i{\rm w}}, x_{j{\rm w}}, x_{H^{-}_{k}{\rm w}}, 1,
x_{\tilde{U}^i_\alpha {\rm w}}, x_{\tilde{D}^1_\delta {\rm w}},
x_{\tilde{g}{\rm w}})
\nonumber \\
&&\hspace{1.2cm}+(i\leftrightarrow j)\; ,
\label{chwh6}
\end{eqnarray}
\begin{eqnarray}
&&\phi^{wh\tilde{g}}_{7}=\frac{1}{\sin\beta}h_{b}x_{j{\rm w}}{\cal Z}_{H}^{1k}
{\cal Z}_{H}^{2k}{\cal Z}_{\tilde{D}^1}^{1\delta}
{\cal Z}_{\tilde{U}^i}^{1\alpha}{\cal Z}_{\tilde{D}^1}^{1\delta}
{\cal Z}_{\tilde{U}^j}^{1\beta}\Big(2F_{D}^{2a}+\frac{14}{3}F_{D}^{2b}
+\frac{14}{3}F_{D}^{2c}-\frac{4}{3}F_{D}^{2d}\nonumber \\
&&\hspace{1.2cm}-\frac{20}{3}F_{D}^{2e}-\frac{28}{3}F_{D}^{2f}\Big)
(x_{i{\rm w}}, x_{j{\rm w}}, x_{H^{-}_{k}{\rm w}}, 1,
x_{\tilde{U}^i_\alpha {\rm w}}, x_{\tilde{D}^1_\delta {\rm w}},
x_{\tilde{g}{\rm w}})\nonumber \\
&&\hspace{1.2cm}
-\frac{16}{3\sin\beta}h_{b}{\cal Z}_{H}^{1k}
{\cal Z}_{H}^{2k}\sqrt{x_{\tilde{g}{\rm w}}x_{i{\rm w}}}
{\cal Z}_{\tilde{D}^1}^{1\delta}{\cal Z}_{\tilde{U}^i}^{2\alpha}
{\cal Z}_{\tilde{D}^1}^{1\delta}{\cal Z}_{\tilde{U}^j}^{1\beta}
\Big(F_{D}^{1b}-F_{D}^{1c}\Big)(x_{i{\rm w}}, x_{j{\rm w}},
x_{H^{-}_{k}{\rm w}}, 1,x_{\tilde{U}^i_\alpha {\rm w}},
x_{\tilde{D}^1_\delta {\rm w}}, x_{\tilde{g}{\rm w}})\nonumber \\
&&\hspace{1.2cm}
-\frac{35h_{b}{\cal Z}_{H}^{1k} {\cal Z}_{H}^{2k}}{3\sin\beta}
\frac{x_{\tilde{g}{\rm w}}^2 x_{H^{-}_{k}{\rm w}} x_{j{\rm w}}
\ln x_{H^{-}_{k}{\rm w}}}{(x_{\tilde{D}^1_\delta {\rm w}} - x_{\tilde{g}{\rm w}})
(1 - x_{H^{-}_{k}{\rm w}})(-x_{H^{-}_{k}{\rm w}} + x_{i{\rm w}})(-x_{H^{-}_{k}{\rm w}} + x_{j{\rm w}})}
\nonumber \\
&&\hspace{1.2cm}
-\frac{h_{b}{\cal Z}_{H}^{1k} {\cal Z}_{H}^{2k}}{3\sin\beta}
\frac{19x_{\tilde{g}{\rm w}}^2 x_{i{\rm w}} x_{j{\rm w}}\ln x_{i{\rm w}}-3
(x_{\tilde{D}^1_\delta {\rm w}}+x_{\tilde{g}{\rm w}})x_{i{\rm w}}^2x_{j{\rm w}}\ln x_{i{\rm w}}}{
(x_{\tilde{D}^1_\delta {\rm w}} - x_{\tilde{g}{\rm w}})(1 - x_{i{\rm w}})
(x_{H^{-}_{k}{\rm w}} - x_{i{\rm w}}) (-x_{i{\rm w}} + x_{j{\rm w}})}
\nonumber \\
&&\hspace{1.2cm}
-\frac{h_{b}{\cal Z}_{H}^{1k} {\cal Z}_{H}^{2k}}{3\sin\beta}
\frac{19x_{\tilde{g}{\rm w}}^2 x_{j{\rm w}}^2
\ln x_{j{\rm w}}}{3\sin\beta (x_{\tilde{D}^1_\delta {\rm w}} - x_{\tilde{g}{\rm w}})
(1 - x_{j{\rm w}})(x_{H^{-}_{k}{\rm w}} - x_{j{\rm w}}) (x_{i{\rm w}} - x_{j{\rm w}})}\Big)
\nonumber \\
&&\hspace{1.2cm}
+\frac{h_{b}{\cal Z}_{H}^{1k} {\cal Z}_{H}^{2k}}{2\sin\beta}
\frac{32x_{H^{-}_{k}{\rm w}}^2 x_{j{\rm w}}\ln x_{H^{-}_{k}{\rm w}}+
21x_{H^{-}_{k}{\rm w}}x_{\tilde{g}{\rm w}}x_{j{\rm w}}\ln x_{H^{-}_{k}{\rm w}}}{
(1 - x_{H^{-}_{k}{\rm w}})(-x_{H^{-}_{k}{\rm w}} + x_{i{\rm w}})(-x_{H^{-}_{k}{\rm w}} + x_{j{\rm w}})}
\nonumber \\
&&\hspace{1.2cm}
+\frac{h_{b}{\cal Z}_{H}^{1k} {\cal Z}_{H}^{2k}}{2\sin\beta}
\frac{(32x_{i{\rm w}}+25x_{\tilde{g}{\rm w}})x_{i{\rm w}}x_{j{\rm w}}\ln x_{i{\rm w}}}{
(1 - x_{i{\rm w}})(x_{H^{-}_{k}{\rm w}} - x_{i{\rm w}})(x_{j{\rm w}} - x_{i{\rm w}})}
\nonumber \\
&&\hspace{1.2cm}
+\frac{h_{b}{\cal Z}_{H}^{1k} {\cal Z}_{H}^{2k}}{6\sin\beta}
\frac{(166x_{j{\rm w}}+39x_{\tilde{g}{\rm w}})x_{j{\rm w}}^2\ln x_{j{\rm w}}}{
(1 - x_{j{\rm w}})(x_{H^{-}_{k}{\rm w}} - x_{j{\rm w}})(x_{i{\rm w}} - x_{j{\rm w}})}
\nonumber \\
&&\hspace{1.2cm}
-\frac{20h_{b}{\cal Z}_{H}^{1k} {\cal Z}_{H}^{2k}}{\sin\beta}
\frac{x_{H^{-}_{k}{\rm w}}x_{j{\rm w}}\ln x_{H^{-}_{k}{\rm w}}}{
\sin\beta (-x_{H^{-}_{k}{\rm w}} + x_{i{\rm w}})(-x_{H^{-}_{k}{\rm w}} + x_{j{\rm w}})}\nonumber \\
&&\hspace{1.2cm}
+\frac{35h_{b}{\cal Z}_{H}^{1k} {\cal Z}_{H}^{2k}}{\sin\beta}
\frac{x_{i{\rm w}}x_{j{\rm w}}\ln x_{i{\rm w}}}{
\sin\beta (-x_{H^{-}_{k}{\rm w}} + x_{i{\rm w}})(-x_{i{\rm w}} + x_{j{\rm w}})}
\nonumber \\
&&\hspace{1.2cm}
+\frac{35h_{b}{\cal Z}_{H}^{1k} {\cal Z}_{H}^{2k}}{\sin\beta}
\frac{x_{j{\rm w}}^2\ln x_{i{\rm w}}}{
\sin\beta (-x_{H^{-}_{k}{\rm w}} + x_{j{\rm w}})(-x_{i{\rm w}} + x_{j{\rm w}})}
\nonumber \\
&&\hspace{1.2cm}
+(i\leftrightarrow j)\; ,
\label{chwh7}
\end{eqnarray}
\begin{eqnarray}
&&\phi^{wh\tilde{g}}_{8}=\frac{1}{4}\phi^{wh\tilde{g}}_{7}\; ,
\label{chwh8}
\end{eqnarray}
\begin{eqnarray}
&&\phi^{hh\tilde{g}}_{1}=\frac{8}{3\sin^3\beta}x_{i{\rm w}}^{5\over 2}x_{j{\rm w}}
{\cal Z}_{H}^{2k}({\cal Z}_{H}^{2l})^{2}
{\cal Z}_{\tilde{D}^1}^{1\gamma}
{\cal Z}_{\tilde{U}^i}^{1\alpha}
{\cal E}^{id}\Big(F_{D}^{1a}+F_{D}^{1b}
-F_{D}^{1c}\Big)
(x_{i{\rm w}}, x_{j{\rm w}}, x_{H_{k}^{-}{\rm w}}, x_{H_{l}^{-}{\rm w}},
x_{\tilde{U}^i_\alpha {\rm w}}, x_{\tilde{D}^1_\gamma {\rm w}},
x_{\tilde{g}{\rm w}})\nonumber \\
&&\hspace{1.2cm}+\frac{4}{3\sin^4\beta}
x_{i{\rm w}}^3 x_{j{\rm w}}^2 ({\cal Z}_{H}^{2k})^{2} ({\cal Z}_{H}^{2l})^{2}
({\cal Z}_{\tilde{U}^i}^{1\alpha})^2\Big(F_{C}^{1a}+F_{C}^{1b}
-F_{C}^{1c}\Big)
(x_{j{\rm w}}, x_{i{\rm w}}, x_{i{\rm w}}, x_{H_{k}^{-}{\rm w}},
x_{H_{l}^{-}{\rm w}}, x_{\tilde{U}^i_\alpha {\rm w}}, x_{\tilde{g}{\rm w}})
\nonumber \\
&&\hspace{1.2cm}-\frac{16}{3\sin^3\beta}
\sqrt{x_{\tilde{g}{\rm w}}} x_{i{\rm w}}^2 x_{j{\rm w}}{\cal Z}_{H}^{2k}
({\cal Z}_{H}^{2l})^{2}{\cal Z}_{\tilde{D}^1}^{1\gamma}{\cal E}^{id}
{\cal Z}_{\tilde{U}^i}^{2\alpha}
F_{D}^{1a}(x_{i{\rm w}}, x_{j{\rm w}}, x_{H_{k}^{-}{\rm w}}, x_{H_{l}^{-}{\rm w}},
x_{\tilde{U}^i_\alpha {\rm w}}, x_{\tilde{D}^1_\gamma {\rm w}},
x_{\tilde{g}{\rm w}})
\nonumber \\
&&\hspace{1.2cm}-\frac{16}{3\sin^4\beta}
x_{i{\rm w}}^2 \sqrt{x_{\tilde{g}{\rm w}} x_{i{\rm w}}} x_{j{\rm w}}^2
({\cal Z}_{H}^{2k})^{2} ({\cal Z}_{H}^{2l})^{2}
{\cal Z}_{\tilde{U}^i}^{1\alpha}
{\cal Z}_{\tilde{U}^i}^{2\alpha}
F_{C}^{1a}(x_{j{\rm w}}, x_{i{\rm w}}, x_{i{\rm w}}, x_{H_{k}^{-}{\rm w}},
x_{H_{l}^{-}{\rm w}},x_{\tilde{U}^i_\alpha {\rm w}}, x_{\tilde{g}{\rm w}})
\nonumber \\
&&\hspace{1.2cm}+\frac{4}{3\sin^4\beta}x_{i{\rm w}}^2 x_{j{\rm w}}^2
({\cal Z}_{H}^{2k})^{2}({\cal Z}_{H}^{2l})^{2}
({\cal Z}_{\tilde{U}^i}^{2\alpha})^2\Big(
F_{C}^{2a}+F_{D}^{2d}-F_{D}^{2e}\Big)
(x_{j{\rm w}}, x_{i{\rm w}}, x_{i{\rm w}}, x_{H_{k}^{-}{\rm w}},
x_{H_{l}^{-}{\rm w}}, x_{\tilde{U}^i_\alpha {\rm w}}, x_{\tilde{g}{\rm w}})
\nonumber \\
&&\hspace{1.2cm}+\frac{4}{3}h_{b}h_{d}{\cal Z}_{H}^{1i}
{\cal Z}_{H}^{1k}{\cal Z}_{\tilde{D}^3}^{1\delta}
{\cal Z}_{\tilde{D}^1}^{1\gamma}
{\cal E}^{ib} {\cal E}^{jd}
\Big(F_{A}^{1a}+F_{A}^{1b}-F_{A}^{1c}\Big)
(x_{j{\rm w}}, x_{H^-_l{\rm w}}, x_{H_{k}^{-}{\rm w}},
x_{\tilde{U}^i_\alpha {\rm w}}, x_{\tilde{D}^3_\delta {\rm w}},
x_{\tilde{D}^1_\gamma {\rm w}}, x_{\tilde{g}{\rm w}})
\nonumber \\
&&\hspace{1.2cm}+\frac{8}{\sin^3\beta}{\cal Z}_{H}^{2k}({\cal Z}_{H}^{2l})^{2}
{\cal Z}_{\tilde{D}^1}^{1\gamma}
{\cal Z}_{\tilde{U}^i}^{1\alpha}{\cal E}^{id}
\frac{x_{H_{k}^{-}{\rm w}}x_{i{\rm w}}^{5\over 2} x_{j{\rm w}}
\ln x_{H_{k}^{-}{\rm w}}}{(-x_{H_{k}^{-}{\rm w}}+x_{H_{l}^{-}{\rm w}})
(-x_{H_{k}^{-}{\rm w}}+x_{i{\rm w}})(-x_{H_{k}^{-}{\rm w}}+x_{j{\rm w}})}
\nonumber \\
&&\hspace{1.2cm}-\frac{2}{\sin^4\beta}({\cal Z}_{H}^{2k})^{2}
({\cal Z}_{H}^{2l})^{2}\frac{x_{H_{k}^{-}{\rm w}}^2 x_{i{\rm w}}^3 x_{j{\rm w}}^2
\ln x_{H_{k}^{-}{\rm w}}}{(-x_{H_{k}^{-}{\rm w}} + x_{H_{l}^{-}{\rm w}})
(-x_{H_{k}^{-}{\rm w}} + x_{i{\rm w}})^2 (-x_{H_{k}^{-}{\rm w}} + x_{j{\rm w}})}
\nonumber \\
&&\hspace{1.2cm}+\frac{2}{\sin^4\beta}\bigg(
\frac{({\cal Z}_{H}^{2k})^{2}x_{H_{k}^{-}{\rm w}}^2 x_{i{\rm w}}^2 x_{j{\rm w}}^2
\ln x_{H_{k}^{-}{\rm w}}}{(-x_{H_{k}^{-}{\rm w}} + x_{i{\rm w}})^2
(-x_{H_{k}^{-}{\rm w}} + x_{j{\rm w}})}-
\frac{({\cal Z}_{H}^{2k})^{2} ({\cal Z}_{H}^{2l})^{2}
x_{H_{k}^{-}{\rm w}}^2 x_{i{\rm w}}^2 x_{j{\rm w}}^2
\ln x_{H_{k}^{-}{\rm w}}}{(-x_{H_{k}^{-}{\rm w}}+x_{H_{l}^{-}{\rm w}})
(-x_{H_{k}^{-}{\rm w}} + x_{i{\rm w}})^2 (-x_{H_{k}^{-}{\rm w}} + x_{j{\rm w}})}\bigg)
\nonumber \\
&&\hspace{1.2cm}+\frac{8}{\sin^3\beta}{\cal Z}_{H}^{2k}
({\cal Z}_{H}^{2l})^{2}{\cal Z}_{\tilde{D}^1}^{1\gamma}
{\cal E}^{id}\frac{x_{H_{l}^{-}{\rm w}}x_{i{\rm w}}^{5\over 2} x_{j{\rm w}}
\ln x_{H_{l}^{-}{\rm w}}}{(x_{H_{k}^{-}{\rm w}}-x_{H_{l}^{-}{\rm w}})
(-x_{H_{l}^{-}{\rm w}} + x_{i{\rm w}})(-x_{H_{l}^{-}{\rm w}} + x_{j{\rm w}})}
\nonumber \\
&&\hspace{1.2cm}-\frac{2}{\sin^4\beta}({\cal Z}_{H}^{2k})^{2}
({\cal Z}_{H}^{2l})^{2}\frac{x_{H_{l}^{-}{\rm w}}^2 x_{i{\rm w}}^3 x_{j{\rm w}}^2
\ln x_{H_{l}^{-}{\rm w}}}{(x_{H_{k}^{-}{\rm w}}-x_{H_{l}^{-}{\rm w}})
(-x_{H_{l}^{-}{\rm w}} + x_{i{\rm w}})^2 (-x_{H_{l}^{-}{\rm w}} + x_{j{\rm w}})}
\nonumber \\
&&\hspace{1.2cm}-\frac{2}{\sin^4\beta}
({\cal Z}_{H}^{2k})^{2}({\cal Z}_{H}^{2l})^{2}
\frac{x_{H_{l}^{-}{\rm w}}^3 x_{i{\rm w}}^2 x_{j{\rm w}}^2
\ln x_{H_{l}^{-}{\rm w}}}{(x_{H_{k}^{-}{\rm w}}-x_{H_{l}^{-}{\rm w}})
(-x_{H_{l}^{-}{\rm w}} + x_{i{\rm w}})^2 (-x_{H_{l}^{-}{\rm w}} + x_{j{\rm w}})}
\nonumber \\
&&\hspace{1.2cm}+\frac{8}{\sin^3\beta}{\cal Z}_{H}^{2k}({\cal Z}_{H}^{2l})^{2}
{\cal Z}_{\tilde{D}^1}^{1\gamma}{\cal Z}_{\tilde{U}^i}^{1\alpha}
{\cal E}^{id}\frac{x_{i{\rm w}}^{7\over 2} x_{j{\rm w}}\ln x_{i{\rm w}}}{
(x_{H_{k}^{-}{\rm w}}-x_{i{\rm w}})(x_{H_{l}^{-}{\rm w}}-x_{i{\rm w}})(-x_{i{\rm w}}+x_{j{\rm w}})}
\nonumber \\
&&\hspace{1.2cm}+\frac{2}{\sin^4\beta}({\cal Z}_{H}^{2k})^{2}
({\cal Z}_{H}^{2l})^{2}\bigg(\frac{x_{i{\rm w}}^5 x_{j{\rm w}}^2 \ln x_{i{\rm w}}}
{(x_{H_{k}^{-}{\rm w}}-x_{i{\rm w}})(x_{H_{l}^{-}{\rm w}}-x_{i{\rm w}})
(-x_{i{\rm w}} + x_{j{\rm w}})^2}\nonumber \\
&&\hspace{1.2cm}+\frac{x_{i{\rm w}}^4 x_{j{\rm w}}^2\ln x_{i{\rm w}}}
{(x_{H_{k}^{-}{\rm w}}-x_{i{\rm w}})(x_{H_{l}^{-}{\rm w}}-x_{i{\rm w}})(-x_{i{\rm w}} + x_{j{\rm w}})}
+\frac{x_{i{\rm w}}^5 x_{j{\rm w}}^2\ln x_{i{\rm w}}}{
(x_{H_{k}^{-}{\rm w}}-x_{i{\rm w}})(x_{H_{l}^{-}{\rm w}}-x_{i{\rm w}})^2
(-x_{i{\rm w}} + x_{j{\rm w}})}\nonumber \\
&&\hspace{1.2cm}+\frac{x_{H_{l}^{-}{\rm w}}x_{i{\rm w}}^4 x_{j{\rm w}}^2\ln x_{i{\rm w}}}
{(x_{H_{k}^{-}{\rm w}}-x_{i{\rm w}})^2 (x_{H_{l}^{-}{\rm w}}-x_{i{\rm w}})(-x_{i{\rm w}} + x_{j{\rm w}})}
\bigg)
\nonumber \\
&&\hspace{1.2cm}+\frac{2}{\sin^4\beta}\bigg(
\frac{({\cal Z}_{H}^{2k})^{2}x_{i{\rm w}}^4 x_{j{\rm w}}^2\ln x_{i{\rm w}}}
{(x_{H_{k}^{-}{\rm w}} - x_{i{\rm w}})(-x_{i{\rm w}} + x_{j{\rm w}})^2}
+\frac{({\cal Z}_{H}^{2k})^{2}({\cal Z}_{H}^{2l})^{2}
x_{i{\rm w}}^4 x_{j{\rm w}}^2 \ln x_{i{\rm w}}}{(x_{H_{k}^{-}{\rm w}}-x_{i{\rm w}})
(x_{H_{l}^{-}{\rm w}}-x_{i{\rm w}})(-x_{i{\rm w}} + x_{j{\rm w}})^2}
\nonumber \\
&&\hspace{1.2cm}+
\frac{({\cal Z}_{H}^{2k})^{2}({\cal Z}_{H}^{2l})^{2}
x_{i{\rm w}}^3 x_{j{\rm w}}^2\ln x_{i{\rm w}}}{(x_{H_{k}^{-}{\rm w}} - x_{i{\rm w}})
(-x_{i{\rm w}} + x_{j{\rm w}})}
+\frac{({\cal Z}_{H}^{2k})^{2}({\cal Z}_{H}^{2l})^{2}
x_{i{\rm w}}^3 x_{j{\rm w}}^2\ln x_{i{\rm w}}}{(x_{H_{k}^{-}{\rm w}} - x_{i{\rm w}})
(x_{H_{l}^{-}{\rm w}} - x_{i{\rm w}})(-x_{i{\rm w}} + x_{j{\rm w}})}
\bigg)
\nonumber \\
&&\hspace{1.2cm}+\frac{8}{\sin^3\beta}{\cal Z}_{H}^{2k}
({\cal Z}_{H}^{2l})^{2}{\cal Z}_{\tilde{D}^1}^{1\gamma}
{\cal Z}_{\tilde{U}^i}^{1\alpha}{\cal E}^{id}
\frac{x_{i{\rm w}}^{5\over 2} x_{j{\rm w}}^2\ln x_{j{\rm w}}}{
(x_{H_{k}^{-}{\rm w}} - x_{j{\rm w}})(x_{H_{l}^{-}{\rm w}} - x_{j{\rm w}})
(x_{i{\rm w}} - x_{j{\rm w}})}
\nonumber \\
&&\hspace{1.2cm}-\frac{2}{\sin^4\beta}({\cal Z}_{H}^{2k})^{2}
({\cal Z}_{H}^{2l})^{2}\frac{(1+x_{i{\rm w}})x_{i{\rm w}}^2 x_{j{\rm w}}^4\ln x_{j{\rm w}}}
{(x_{H_{k}^{-}{\rm w}}-x_{j{\rm w}})(x_{H_{l}^{-}{\rm w}}-x_{j{\rm w}})(x_{i{\rm w}} - x_{j{\rm w}})^2}
\nonumber \\
&&\hspace{1.2cm}+\frac{2}{\sin^4\beta}
({\cal Z}_{H}^{2k})^{2}({\cal Z}_{H}^{2l})^{2}
\frac{x_{i{\rm w}}^2 x_{j{\rm w}}^4\ln x_{j{\rm w}}}{(x_{H_{k}^{-}{\rm w}}-x_{j{\rm w}})
(x_{i{\rm w}} - x_{j{\rm w}})^2}+(i\leftrightarrow j)\; ,
\label{chhh1}
\end{eqnarray}
\begin{eqnarray}
&&\phi^{hh\tilde{g}}_{2}=\frac{4}{3\sin\beta}h_{b}h_{d}x_{i{\rm w}}^{3\over 2}
{\cal Z}_{H}^{1k}{\cal Z}_{H}^{1l}{\cal Z}_{H}^{2l}
{\cal Z}_{\tilde{D}^1}^{1\gamma}
{\cal Z}_{\tilde{U}^i}^{1\alpha}{\cal E}^{id}\Big(
F_{D}^{1a}+F_{D}^{1b}
-F_{D}^{1c}\Big)
(x_{i{\rm w}}, x_{j{\rm w}}, x_{H_{k}^{-}{\rm w}}, x_{H_{l}^{-}{\rm w}},
x_{\tilde{U}^i_\alpha {\rm w}}, x_{\tilde{D}^1_\gamma {\rm w}},
x_{\tilde{g}{\rm w}})\nonumber \\
&&\hspace{1.2cm}-\frac{8}{3\sin^2\beta}h_{b}\sqrt{x_{\tilde{g}{\rm w}}}
x_{i{\rm w}}x_{j{\rm w}}{\cal Z}_{H}^{1l}{\cal Z}_{H}^{2k}{\cal Z}_{H}^{2l}
{\cal Z}_{\tilde{D}^1}^{2\gamma}
{\cal Z}_{\tilde{U}^i}^{1\alpha}
{\cal E}^{id}\Big(
F_{D}^{1a}+F_{D}^{1b}\nonumber \\
&&\hspace{1.2cm}
-F_{D}^{1c}\Big)
(x_{i{\rm w}}, x_{j{\rm w}}, x_{H_{k}^{-}{\rm w}}, x_{H_{l}^{-}{\rm w}},
x_{\tilde{U}^i_\alpha {\rm w}},x_{\tilde{D}^1_\gamma {\rm w}}, x_{\tilde{g}{\rm w}})
\nonumber \\
&&\hspace{1.2cm}+\frac{2}{3\sin^2\beta}h_{b}h_{d}x_{i{\rm w}}^3
{\cal Z}_{H}^{1k}{\cal Z}_{H}^{1l}
{\cal Z}_{H}^{2k}{\cal Z}_{H}^{2l}
({\cal Z}_{\tilde{U}^i}^{1\alpha})^2\Big(F_{C}^{1a}+
F_{C}^{1b}-F_{C}^{1c}\Big)
(x_{j{\rm w}}, x_{i{\rm w}}, x_{i{\rm w}}, x_{H_{k}^{-}{\rm w}},
x_{H_{l}^{-}{\rm w}}, x_{\tilde{U}^i_\alpha {\rm w}},
x_{\tilde{g}{\rm w}})\nonumber \\
&&\hspace{1.2cm}+\frac{2}{3\sin^2\beta}h_{b}h_{d}x_{j{\rm w}}^2
{\cal Z}_{H}^{1k}{\cal Z}_{H}^{1l}{\cal Z}_{H}^{2k}
{\cal Z}_{H}^{2l}
({\cal Z}_{\tilde{U}^i}^{1\alpha})^2\Big(F_{C}^{2a}+
F_{C}^{2d}-F_{C}^{2e}\Big)
(x_{j{\rm w}}, x_{i{\rm w}}, x_{i{\rm w}}, x_{H_{k}^{-}{\rm w}},
x_{H_{l}^{-}{\rm w}}, x_{\tilde{U}^i_\alpha {\rm w}},
x_{\tilde{g}{\rm w}})
\nonumber \\
&&\hspace{1.2cm}-\frac{8}{3\sin\beta}h_{b}h_{d}\sqrt{x_{\tilde{g}{\rm w}}}
x_{i{\rm w}}{\cal Z}_{H}^{1k}{\cal Z}_{H}^{1l}{\cal Z}_{H}^{2l}
{\cal Z}_{\tilde{D}^1}^{1\gamma}
{\cal Z}_{\tilde{U}^i}^{2\alpha}{\cal E}^{id}\Big(F_{D}^{1a}
+F_{D}^{1b}\nonumber \\
&&\hspace{1.2cm}
-F_{D}^{1c}\Big)(
x_{i{\rm w}}, x_{j{\rm w}}, x_{H_{k}^{-}{\rm w}}, x_{H_{l}^{-}{\rm w}},
x_{\tilde{U}^i_\alpha {\rm w}},x_{\tilde{D}^1_\gamma {\rm w}}, x_{\tilde{g}{\rm w}})
\nonumber \\
&&\hspace{1.2cm}+\frac{4}{3\sin^2\beta}h_{b}x_{i{\rm w}}^{3\over 2}x_{j{\rm w}}
{\cal Z}_{H}^{1l}{\cal Z}_{H}^{2k}{\cal Z}_{H}^{2l}
{\cal Z}_{\tilde{D}^1}^{2\gamma}
{\cal Z}_{\tilde{U}^i}^{2\alpha}
{\cal E}^{id}\Big(F_{D}^{1a}+F_{D}^{1b}
-F_{D}^{1c}\Big)
(x_{i{\rm w}}, x_{j{\rm w}}, x_{H_{k}^{-}{\rm w}}, x_{H_{l}^{-}{\rm w}},
x_{\tilde{U}^i_\alpha {\rm w}}, x_{\tilde{D}^1_\gamma {\rm w}},
x_{\tilde{g}{\rm w}})
\nonumber \\
&&\hspace{1.2cm}-\frac{8}{3\sin^2\beta}h_{b}h_{d}
(x_{i{\rm w}}^2+x_{j{\rm w}}^2)\sqrt{x_{\tilde{g}{\rm w}}x_{i{\rm w}}}
{\cal Z}_{H}^{1k}{\cal Z}_{H}^{1l}{\cal Z}_{H}^{2k}
{\cal Z}_{H}^{2l}{\cal Z}_{\tilde{U}^i}^{1\alpha}
{\cal Z}_{\tilde{U}^i}^{2\alpha}
F_{C}^{1a}(x_{j{\rm w}}, x_{i{\rm w}}, x_{i{\rm w}}, x_{H_{k}^{-}{\rm w}},
x_{H_{l}^{-}{\rm w}}, x_{\tilde{U}^i_\alpha {\rm w}}, x_{\tilde{g}{\rm w}})
\nonumber \\
&&\hspace{1.2cm}+\frac{2}{3\sin^2\beta}h_{b}h_{d}x_{i{\rm w}} x_{j{\rm w}}^2
{\cal Z}_{H}^{1k}{\cal Z}_{H}^{1l}{\cal Z}_{H}^{2k}
{\cal Z}_{H}^{2l}
({\cal Z}_{\tilde{U}^i}^{2\alpha})^2
\Big(F_{C}^{1a}+F_{C}^{1b}-F_{C}^{1c}\Big)
(x_{j{\rm w}}, x_{i{\rm w}}, x_{i{\rm w}}, x_{H_{k}^{-}{\rm w}}, x_{H_{l}^{-}{\rm w}},
x_{\tilde{U}^i_\alpha {\rm w}},x_{\tilde{g}{\rm w}})
\nonumber \\
&&\hspace{1.2cm}+\frac{2}{3\sin^2\beta}
h_{b}h_{d}x_{i{\rm w}}^2 {\cal Z}_{H}^{1k}{\cal Z}_{H}^{1l}
{\cal Z}_{H}^{2k}{\cal Z}_{H}^{2l}
({\cal Z}_{\tilde{U}^i}^{2\alpha})^2
\Big(F_{C}^{2a}+F_{C}^{2d}-F_{C}^{2e}\Big)
(x_{j{\rm w}}, x_{i{\rm w}}, x_{i{\rm w}}, x_{H_{k}^{-}{\rm w}}, x_{H_{l}^{-}{\rm w}},
x_{\tilde{U}^i_\alpha {\rm w}},x_{\tilde{g}{\rm w}})
\nonumber \\
&&\hspace{1.2cm}+\frac{2}{3\sin^2\beta}x_{j{\rm w}}^2
{\cal Z}_{H}^{2i}{\cal Z}_{H}^{2k}
{\cal Z}_{\tilde{D}^3}^{1\delta}
{\cal Z}_{\tilde{D}^1}^{1\gamma}
{\cal E}^{ib}{\cal E}^{jd}\Big(F_{A}^{1a}+F_{A}^{1b}
-F_{A}^{1c}\Big)
(x_{j{\rm w}}, x_{H_{i}^{-}{\rm w}}, x_{H_{k}^{-}{\rm w}}, x_{\tilde{U}^i_\alpha {\rm w}},
x_{\tilde{D}^1_\gamma {\rm w}}, x_{\tilde{D}^3_\delta {\rm w}}, x_{\tilde{g}{\rm w}})
\nonumber \\
&&\hspace{1.2cm}
-\frac{8}{3\sin^2\beta}h_{d}x_{i{\rm w}} x_{j{\rm w}}^{3\over 2}
{\cal Z}_{H}^{1l}{\cal Z}_{H}^{2k}{\cal Z}_{H}^{2l}
{\cal Z}_{\tilde{D}^1}^{1\gamma}
{\cal Z}_{\tilde{U}^i}^{1\alpha}
{\cal E}^{id}\Big(F_{D}^{1a}+F_{D}^{1b}-F_{D}^{1c}\Big)
(x_{i{\rm w}}, x_{j{\rm w}}, x_{H_{k}^{-}{\rm w}}, x_{H_{l}^{-}{\rm w}},
x_{\tilde{U}^i_\alpha {\rm w}}, x_{\tilde{D}^1_\gamma {\rm w}},
x_{\tilde{g}{\rm w}})\nonumber \\
&&\hspace{1.2cm}+\frac{16}{3\sin^2\beta}h_{b}x_{i{\rm w}}^2
\sqrt{x_{\tilde{g}{\rm w}}x_{i{\rm w}}x_{j{\rm w}}}
{\cal Z}_{H}^{1k}({\cal Z}_{H}^{2l})^{2}
{\cal Z}_{\tilde{D}^1}^{2\gamma}
{\cal Z}_{\tilde{U}^i}^{1\alpha}{\cal E}^{id}
F_{D}^{0}(x_{i{\rm w}}, x_{j{\rm w}}, x_{H_{k}^{-}{\rm w}}, x_{H_{l}^{-}{\rm w}},
x_{\tilde{U}^i_\alpha {\rm w}},x_{\tilde{D}^1_\gamma {\rm w}},
x_{\tilde{g}{\rm w}})\nonumber \\
&&\hspace{1.2cm}-\frac{4}{3\sin^2\beta}h_{b}h_{d}(x_{i{\rm w}}x_{j{\rm w}})^{3\over 2}
\big(({\cal Z}_{H}^{1l})^2 ({\cal Z}_{H}^{2k})^{2}+
({\cal Z}_{H}^{2l})^2 ({\cal Z}_{H}^{1k})^{2}\big)
({\cal Z}_{\tilde{U}^i}^{1\alpha})^2
\Big(F_{C}^{1a}+F_{C}^{1b}\nonumber \\
&&\hspace{1.2cm}-F_{C}^{1c}\Big)
(x_{j{\rm w}}, x_{i{\rm w}}, x_{i{\rm w}}, x_{H_{k}^{-}{\rm w}},
x_{H_{l}^{-}{\rm w}}, x_{\tilde{U}^i_\alpha {\rm w}},
x_{\tilde{g}{\rm w}})\nonumber \\
&&\hspace{1.2cm}+\frac{16}{3\sin\beta}h_{b}h_{d}
x_{j{\rm w}}\sqrt{x_{\tilde{g}{\rm w}}x_{i{\rm w}}x_{j{\rm w}}}
({\cal Z}_{H}^{1l})^2 {\cal Z}_{H}^{2k}
{\cal Z}_{\tilde{D}^1}^{1\gamma}
{\cal Z}_{\tilde{U}^i}^{2\alpha}{\cal E}^{id}
F_{D}^{0}(x_{i{\rm w}}, x_{j{\rm w}}, x_{H_{k}^{-}{\rm w}}, x_{H_{l}^{-}{\rm w}},
x_{\tilde{U}^i_\alpha {\rm w}},x_{\tilde{D}^1_\gamma {\rm w}},
x_{\tilde{g}{\rm w}})\nonumber \\
&&\hspace{1.2cm}-\frac{8}{3\sin\beta}h_{b}^2 x_{i{\rm w}}^{3\over 2}
{\cal Z}_{H}^{1k}{\cal Z}_{H}^{1l}{\cal Z}_{H}^{2l}
{\cal Z}_{\tilde{D}^1}^{2\gamma}
{\cal Z}_{\tilde{U}^i}^{2\alpha}{\cal E}^{id}
\Big(F_{D}^{1a}+F_{D}^{1b}-F_{D}^{1c}\Big)
(x_{i{\rm w}}, x_{j{\rm w}}, x_{H_{k}^{-}{\rm w}}, x_{H_{l}^{-}{\rm w}},
x_{\tilde{U}^i_\alpha {\rm w}}, x_{\tilde{D}^1_\gamma {\rm w}},
x_{\tilde{g}{\rm w}})\nonumber \\
&&\hspace{1.2cm}+\frac{8}{3\sin^2\beta}h_{b}h_{d}x_{i{\rm w}}^{5\over 2}
x_{j{\rm w}}^{3\over 2} \big(({\cal Z}_{H}^{1l})^2 ({\cal Z}_{H}^{2k})^{2}
+ ({\cal Z}_{H}^{2l})^2 ({\cal Z}_{H}^{1k})^{2}\big)
{\cal Z}_{\tilde{U}^i}^{1\alpha}
{\cal Z}_{\tilde{U}^i}^{2\alpha}\nonumber \\
&&\hspace{1.2cm}
F_{C}^{0}(x_{j{\rm w}}, x_{i{\rm w}}, x_{i{\rm w}}, x_{H_{k}^{-}{\rm w}},
x_{H_{l}^{-}{\rm w}}, x_{\tilde{U}^i_\alpha {\rm w}}, x_{\tilde{g}{\rm w}})
\nonumber \\
&&\hspace{1.2cm}+\frac{8}{3\sin^2\beta}h_{b}h_{d}x_{i{\rm w}}x_{j{\rm w}}
\sqrt{x_{\tilde{g}{\rm w}}x_{j{\rm w}}}
\big(({\cal Z}_{H}^{1l})^2({\cal Z}_{H}^{2k})^{2}+
({\cal Z}_{H}^{2l})^2({\cal Z}_{H}^{1k})^{2}\big)
{\cal Z}_{\tilde{U}^i}^{1\alpha}
{\cal Z}_{\tilde{U}^i}^{2\alpha}\nonumber \\
&&\hspace{1.2cm}F_{C}^{1a}
(x_{j{\rm w}}, x_{i{\rm w}}, x_{i{\rm w}}, x_{H_{k}^{-}{\rm w}},
x_{H_{l}^{-}{\rm w}}, x_{\tilde{U}^i_\alpha {\rm w}}, x_{\tilde{g}{\rm w}})
\nonumber \\
&&\hspace{1.2cm}-\frac{4}{3\sin^2\beta}h_{b}h_{d}(x_{i{\rm w}}x_{j{\rm w}})^{3\over 2}
\Big(({\cal Z}_{H}^{1l})^2 ({\cal Z}_{H}^{2k})^{2}+
({\cal Z}_{H}^{2l})^2 ({\cal Z}_{H}^{1k})^{2}\Big)
({\cal Z}_{\tilde{U}^i}^{2\alpha})^2
\Big(F_{C}^{1a}+F_{C}^{1b}\nonumber \\
&&\hspace{1.2cm}-F_{C}^{1c}\Big)
(x_{j{\rm w}}, x_{i{\rm w}}, x_{i{\rm w}}, x_{H_{k}^{-}{\rm w}},
x_{H_{l}^{-}{\rm w}}, x_{\tilde{U}^i_\alpha {\rm w}}, x_{\tilde{g}{\rm w}})
\nonumber \\
&&\hspace{1.2cm}+\frac{8}{3\sin\beta}h_{b}x_{j{\rm w}}
\sqrt{x_{\tilde{g}{\rm w}} x_{j{\rm w}}}\big({\cal Z}_{H}^{1i}{\cal Z}_{H}^{2k}
+{\cal Z}_{H}^{2i}{\cal Z}_{H}^{1k}\big)
{\cal Z}_{\tilde{D}^3}^{2\delta}
{\cal Z}_{\tilde{D}^1}^{1\gamma}
{\cal E}^{ib} {\cal E}^{jd}\nonumber \\
&&\hspace{1.2cm}
F_{A}^{0}(x_{j{\rm w}}, x_{H^{-}_{i}{\rm w}}, x_{H_{k}^{-}{\rm w}},
x_{\tilde{U}^i_\alpha {\rm w}}, x_{\tilde{\tiny D}^{1}_{m}{\rm w}},
x_{\tilde{D}^1_\gamma {\rm w}},x_{\tilde{g}{\rm w}})
\nonumber \\
&&\hspace{1.2cm}
-\frac{4}{3\sin\beta}h_{b}h_{d}
{\cal Z}_{H}^{1k}{\cal Z}_{H}^{1l}{\cal Z}_{H}^{2l}
{\cal Z}_{\tilde{D}^1}^{1\gamma}
{\cal E}^{id}\frac{x_{H_{k}^{-}{\rm w}}x_{i{\rm w}}^2(
{\cal Z}_{\tilde{U}^i}^{1\alpha}\sqrt{x_{i{\rm w}}}
-2{\cal Z}_{\tilde{U}^i}^{2\alpha}\sqrt{x_{\tilde{g}{\rm w}}})
\ln x_{H_{k}^{-}{\rm w}}}
{(-x_{H_{k}^{-}{\rm w}}+x_{H_{l}^{-}{\rm w}})(-x_{H_{k}^{-}{\rm w}}+x_{i{\rm w}})
(-x_{H_{k}^{-}{\rm w}} + x_{j{\rm w}})}\nonumber \\
&&\hspace{1.2cm}
+\frac{4}{3\sin^2\beta}h_{b}
{\cal Z}_{H}^{1l} {\cal Z}_{H}^{2k}{\cal Z}_{H}^{2l}
{\cal Z}_{\tilde{D}^1}^{2\gamma}
{\cal E}^{id}\frac{(2{\cal Z}_{\tilde{U}^i}^{1\alpha}
\sqrt{x_{\tilde{g}{\rm w}}}
-{\cal Z}_{\tilde{U}^i}^{2\alpha}\sqrt{x_{i{\rm w}}})
x_{H_{k}^{-}{\rm w}} x_{i{\rm w}} x_{j{\rm w}}
 \ln x_{H_{k}^{-}{\rm w}}}{(-x_{H_{k}^{-}{\rm w}}+x_{H_{l}^{-}{\rm w}})
(-x_{H_{k}^{-}{\rm w}}+x_{i{\rm w}})(-x_{H_{k}^{-}{\rm w}}+x_{j{\rm w}})}\nonumber \\
&&\hspace{1.2cm}
-\frac{1}{3\sin^2\beta}h_{b}h_{d}{\cal Z}_{H}^{1k}
{\cal Z}_{H}^{1l}{\cal Z}_{H}^{2k}{\cal Z}_{H}^{2l}
\frac{(x_{i{\rm w}}^2+x_{j{\rm w}}^2)(x_{H_{k}^{-}{\rm w}}^2 x_{i{\rm w}}+x_{H_{k}^{-}{\rm w}}^3)
\ln x_{H_{k}^{-}{\rm w}}}{(-x_{H_{k}^{-}{\rm w}} + x_{H_{l}^{-}{\rm w}})
(-x_{H_{k}^{-}{\rm w}} + x_{i{\rm w}})^2 (-x_{H_{k}^{-}{\rm w}} + x_{j{\rm w}})}
\nonumber \\
&&\hspace{1.2cm}
+\frac{1}{3\sin^2\beta}h_{b}h_{d}
{\cal Z}_{H}^{1k} {\cal Z}_{H}^{1l} {\cal Z}_{H}^{2k} {\cal Z}_{H}^{2l}
\frac{x_{H_{k}^{-}{\rm w}}^2 x_{i{\rm w}}^2\ln(x_{H_{k}^{-}{\rm w}})}
{(-x_{H_{k}^{-}{\rm w}} + x_{i{\rm w}})^2 (-x_{H_{k}^{-}{\rm w}} + x_{j{\rm w}})}
\nonumber \\
&&\hspace{1.2cm}
-\frac{4}{3\sin\beta}h_{b}h_{d}{\cal Z}_{H}^{1k}
{\cal Z}_{H}^{1l}{\cal Z}_{H}^{2l}
{\cal Z}_{\tilde{D}^1}^{1\gamma}
{\cal E}^{id}\frac{({\cal Z}_{\tilde{U}^i}^{1\alpha}\sqrt{x_{i{\rm w}}}
-2{\cal Z}_{\tilde{U}^i}^{2\alpha}\sqrt{x_{\tilde{g}{\rm w}}})
x_{H_{l}^{-}{\rm w}}x_{i{\rm w}}\ln x_{H_{l}^{-}{\rm w}}}{
(x_{H_{k}^{-}{\rm w}}-x_{H_{l}^{-}{\rm w}})(-x_{H_{l}^{-}{\rm w}}+x_{i{\rm w}})
(-x_{H_{l}^{-}{\rm w}} + x_{j{\rm w}})}\nonumber \\
&&\hspace{1.2cm}
+\frac{4}{3\sin^2\beta}h_{b}
{\cal Z}_{H}^{1l}{\cal Z}_{H}^{2k}{\cal Z}_{H}^{2l}
{\cal Z}_{\tilde{D}^1}^{2\gamma}
{\cal E}^{id}\frac{(2{\cal Z}_{\tilde{U}^i}^{1\alpha}
\sqrt{x_{\tilde{g}{\rm w}}}-{\cal Z}_{\tilde{U}^i}^{2\alpha}
\sqrt{x_{i{\rm w}}})x_{H_{l}^{-}{\rm w}}x_{i{\rm w}} x_{j{\rm w}}
\ln x_{H_{l}^{-}{\rm w}}}{(x_{H_{k}^{-}{\rm w}}-x_{H_{l}^{-}{\rm w}})
(-x_{H_{l}^{-}{\rm w}} + x_{i{\rm w}})(-x_{H_{l}^{-}{\rm w}} + x_{j{\rm w}})}
\nonumber \\
&&\hspace{1.2cm}
-\frac{1}{3\sin^2\beta}h_{b}h_{d}
{\cal Z}_{H}^{1k}{\cal Z}_{H}^{1l}{\cal Z}_{H}^{2k}
{\cal Z}_{H}^{2l}\frac{(x_{i{\rm w}}^2+x_{j{\rm w}}^2)(x_{H_{l}^{-}{\rm w}}^2 x_{i{\rm w}}
+x_{H_{l}^{-}{\rm w}}^3)\ln x_{H_{l}^{-}{\rm w}}}{
(x_{H_{k}^{-}{\rm w}}-x_{H_{l}^{-}{\rm w}})(-x_{H_{l}^{-}{\rm w}}+x_{i{\rm w}})^2
(-x_{H_{l}^{-}{\rm w}} + x_{j{\rm w}})}\nonumber \\
&&\hspace{1.2cm}
-\frac{4}{3\sin\beta}h_{b}h_{d}
{\cal Z}_{H}^{1k}{\cal Z}_{H}^{1l}{\cal Z}_{H}^{2l}
{\cal Z}_{\tilde{D}^1}^{1\gamma}
{\cal E}^{id}\frac{({\cal Z}_{\tilde{U}^i}^{1\alpha}
\sqrt{x_{i{\rm w}}}-2{\cal Z}_{\tilde{U}^i}^{2\alpha}
\sqrt{x_{\tilde{g}{\rm w}}})x_{i{\rm w}}^2\ln x_{i{\rm w}}}{(x_{H_{k}^{-}{\rm w}}-x_{i{\rm w}})
(x_{H_{l}^{-}{\rm w}}-x_{i{\rm w}})(-x_{i{\rm w}}+x_{j{\rm w}})}
\nonumber \\
&&\hspace{1.2cm}
+\frac{8}{3\sin^2\beta}h_{b}
{\cal Z}_{H}^{1l}{\cal Z}_{H}^{2k}{\cal Z}_{H}^{2l}
{\cal Z}_{\tilde{D}^1}^{2\gamma}
{\cal Z}_{\tilde{U}^i}^{1\alpha}
{\cal E}^{id}\frac{\sqrt{x_{\tilde{g}{\rm w}}}x_{i{\rm w}}^2 x_{j{\rm w}}
\ln x_{i{\rm w}}}{(x_{H_{k}^{-}{\rm w}}-x_{i{\rm w}})(x_{H_{l}^{-}{\rm w}}-x_{i{\rm w}})
(-x_{i{\rm w}} + x_{j{\rm w}})}
\nonumber \\
&&\hspace{1.2cm}-\frac{1}{3\sin^2\beta}h_{b} h_{d}{\cal Z}_{H}^{1k}
{\cal Z}_{H}^{1l}{\cal Z}_{H}^{2k}{\cal Z}_{H}^{2l}\Big(3x_{H_{k}^{-}{\rm w}}
x_{H_{l}^{-}{\rm w}}x_{i{\rm w}}^5 -x_{j{\rm w}}^5(x_{i{\rm w}}^2 +x_{j{\rm w}}^2)
-5x_{H_{k}^{-}{\rm w}}x_{H_{l}^{-}{\rm w}}x_{i{\rm w}}^4
x_{j{\rm w}}\nonumber \\
&&\hspace{1.2cm}+3(x_{H_{k}^{-}{\rm w}}+x_{H_{k}^{-}{\rm w}})x_{i{\rm w}}^5 x_{j{\rm w}}
+3x_{H_{k}^{-}{\rm w}}x_{H_{l}^{-}{\rm w}}
x_{i{\rm w}}^3 x_{j{\rm w}}^2-(x_{H_{k}^{-}{\rm w}}+x_{H_{l}^{-}{\rm w}})x_{i{\rm w}}^4 (x_{i{\rm w}}^2+x_{j{\rm w}}^2)
\nonumber \\
&&\hspace{1.2cm}-5x_{H_{k}^{-}{\rm w}}x_{H_{l}^{-}{\rm w}}x_{i{\rm w}}^2 x_{j{\rm w}}^3
+3x_{H_{k}^{-}{\rm w}}x_{i{\rm w}}^3 x_{j{\rm w}}^3+3x_{H_{l}^{-}{\rm w}}x_{i{\rm w}}^3 x_{j{\rm w}}^3
-x_{i{\rm w}}^4 x_{j{\rm w}}(x_{i{\rm w}}^2+x_{j{\rm w}}^2)\Big)\nonumber \\
&&\hspace{1.2cm}\frac{\ln x_{i{\rm w}}}{(x_{H_{k}^{-}{\rm w}}-x_{i{\rm w}})^2
(x_{H_{l}^{-}{\rm w}}-x_{i{\rm w}})^2 (-x_{i{\rm w}} + x_{j{\rm w}})^2}
\nonumber \\
&&\hspace{1.2cm}-\frac{4}{3\sin^2\beta}h_{b}
{\cal Z}_{H}^{1l}{\cal Z}_{H}^{2k}{\cal Z}_{H}^{2l}
{\cal Z}_{\tilde{D}^1}^{2\gamma}
{\cal E}^{id}\frac{({\cal Z}_{\tilde{U}^i}^{2\alpha}
\sqrt{x_{i{\rm w}}}-2{\cal Z}_{\tilde{U}^i}^{1\alpha}
\sqrt{x_{\tilde{g}{\rm w}}})x_{i{\rm w}}^2x_{j{\rm w}}\ln x_{i{\rm w}}}{
(x_{H_{k}^{-}{\rm w}}-x_{i{\rm w}})(x_{H_{l}^{-}{\rm w}}-x_{i{\rm w}})(-x_{i{\rm w}}+x_{j{\rm w}})}
\nonumber \\
&&\hspace{1.2cm}
-\frac{4}{3\sin\beta}h_{b}h_{d}
{\cal Z}_{H}^{1k} {\cal Z}_{H}^{1l} {\cal Z}_{H}^{2l}
{\cal Z}_{\tilde{D}^1}^{1\gamma}
{\cal E}^{id}\frac{({\cal Z}_{\tilde{U}^i}^{1\alpha}
\sqrt{x_{i{\rm w}}}-2{\cal Z}_{\tilde{U}^i}^{2\alpha}
\sqrt{x_{\tilde{g}{\rm w}}})x_{i{\rm w}}x_{j{\rm w}}\ln x_{j{\rm w}}}{
(x_{H_{k}^{-}{\rm w}}-x_{j{\rm w}})(x_{H_{l}^{-}{\rm w}}-x_{j{\rm w}})(x_{i{\rm w}}-x_{j{\rm w}})}
\nonumber \\
&&\hspace{1.2cm}
-\frac{4}{3\sin^2\beta}h_{b}
{\cal Z}_{H}^{1l}{\cal Z}_{H}^{2k}{\cal Z}_{H}^{2l}
{\cal Z}_{\tilde{D}^1}^{2\gamma}
{\cal Z}_{\tilde{U}^i}^{2\alpha}{\cal E}^{id}
\frac{x_{i{\rm w}}^{3\over 2}x_{j{\rm w}}^2\ln x_{j{\rm w}}}{
(x_{H_{k}^{-}{\rm w}} - x_{j{\rm w}})(x_{H_{l}^{-}{\rm w}} - x_{j{\rm w}})(x_{i{\rm w}} - x_{j{\rm w}})}
\nonumber \\
&&\hspace{1.2cm}+\frac{1}{3\sin^2\beta}h_{b}h_{d}
{\cal Z}_{H}^{1k} {\cal Z}_{H}^{1l} {\cal Z}_{H}^{2k}
{\cal Z}_{H}^{2l}\frac{(x_{i{\rm w}}^3+x_{i{\rm w}}^2 x_{j{\rm w}}+x_{i{\rm w}}x_{j{\rm w}}^2
+x_{j{\rm w}}^3)x_{j{\rm w}}^2\ln x_{j{\rm w}}}{(x_{H_{k}^{-}{\rm w}} - x_{j{\rm w}})
(x_{H_{l}^{-}{\rm w}} - x_{j{\rm w}})(x_{i{\rm w}} - x_{j{\rm w}})^2}
\nonumber \\
&&\hspace{1.2cm}+(i\leftrightarrow j)\; ,
\label{chhh2}
\end{eqnarray}
\begin{eqnarray}
&&\phi^{hh\tilde{g}}_{3}=-2\phi^{hh\tilde{g}}_{2}\; ,
\label{chhh3}
\end{eqnarray}
\begin{eqnarray}
&&\phi^{hh\tilde{g}}_{4}=-\frac{16}{3\sin^2\beta}h_{d}(x_{i{\rm w}}x_{j{\rm w}})^{3\over 2}
\sqrt{x_{\tilde{g}{\rm w}}}{\cal Z}_{H}^{1l}{\cal Z}_{H}^{2k}
{\cal Z}_{H}^{2l}
{\cal Z}_{\tilde{D}^1}^{2\gamma}
{\cal Z}_{\tilde{U}^i}^{1\alpha}{\cal E}^{id}
F_{D}^{0}(x_{i{\rm w}}, x_{j{\rm w}}, x_{H_{k}^{-}{\rm w}},
x_{H_{l}^{-}{\rm w}},x_{\tilde{U}^i_\alpha {\rm w}},
x_{\tilde{D}^1_\gamma {\rm w}},x_{\tilde{g}{\rm w}})\nonumber \\
&&\hspace{1.2cm}+\frac{4}{3\sin^2\beta}h_{d}^2 (x_{i{\rm w}}x_{j{\rm w}})^{3\over 2}
{\cal Z}_{H}^{1k}{\cal Z}_{H}^{1l}{\cal Z}_{H}^{2k}{\cal Z}_{H}^{2l}
\Big(({\cal Z}_{\tilde{U}^i}^{1\alpha})^2+
({\cal Z}_{\tilde{U}^i}^{2\alpha})^2\Big)
\Big(F_{C}^{1a}+F_{C}^{1b}\nonumber \\
&&\hspace{1.2cm}-F_{C}^{1c}\Big)
(x_{j{\rm w}}, x_{i{\rm w}}, x_{i{\rm w}}, x_{H_{k}^{-}{\rm w}}, x_{H_{l}^{-}{\rm w}},
x_{\tilde{U}^i_\alpha {\rm w}}, x_{\tilde{g}{\rm w}})
\nonumber \\
&&\hspace{1.2cm}-\frac{8}{3\sin^2\beta}h_{d}^2 x_{i{\rm w}} x_{j{\rm w}}
\sqrt{x_{\tilde{g}{\rm w}}x_{j{\rm w}}}{\cal Z}_{H}^{1k}{\cal Z}_{H}^{1l}
{\cal Z}_{H}^{2k}{\cal Z}_{H}^{2l}
{\cal Z}_{\tilde{U}^i}^{1\alpha}
{\cal Z}_{\tilde{U}^i}^{2\alpha}F_{C}^{1a}(x_{j{\rm w}}, x_{i{\rm w}}, x_{i{\rm w}},
x_{H_{k}^{-}{\rm w}}, x_{H_{l}^{-}{\rm w}},
x_{\tilde{U}^i_\alpha {\rm w}}, x_{\tilde{g}{\rm w}})
\nonumber \\
&&\hspace{1.2cm}+\frac{8}{3\sin^2\beta}h_{d}x_{i{\rm w}}x_{j{\rm w}}^{3\over 2}
{\cal Z}_{H}^{1l}{\cal Z}_{H}^{2k}{\cal Z}_{H}^{2l}
{\cal Z}_{\tilde{D}^1}^{2\gamma}
{\cal Z}_{\tilde{U}^i}^{2\alpha}{\cal E}^{id}
\Big(F_{D}^{1a}+F_{D}^{1b}
-F_{D}^{1c}\Big)(x_{i{\rm w}}, x_{j{\rm w}}, x_{H_{k}^{-}{\rm w}}, x_{H_{l}^{-}{\rm w}},
x_{\tilde{U}^i_\alpha {\rm w}}, x_{\tilde{D}^1_\gamma {\rm w}}, x_{\tilde{g}{\rm w}})
\nonumber \\
&&\hspace{1.2cm}-\frac{8}{3\sin^2\beta}h_{d}^2 x_{i{\rm w}}^{5\over 2}
x_{j{\rm w}}^{3\over 2}{\cal Z}_{H}^{1k} {\cal Z}_{H}^{1l}
{\cal Z}_{H}^{2k} {\cal Z}_{H}^{2l}{\cal Z}_{\tilde{U}^i}^{1\alpha}
{\cal Z}_{\tilde{U}^i}^{2\alpha}
F_{C}^{0}(x_{j{\rm w}}, x_{i{\rm w}}, x_{i{\rm w}}, x_{H_{k}^{-}{\rm w}},
x_{H_{l}^{-}{\rm w}}, x_{\tilde{U}^i_\alpha {\rm w}}, x_{\tilde{g}{\rm w}})
\nonumber \\
&&\hspace{1.2cm}-\frac{8}{3\sin\beta}h_{b}x_{j{\rm w}}^{3\over 2}
\sqrt{x_{\tilde{g}{\rm w}}} {\cal Z}_{H}^{1i} {\cal Z}_{H}^{2k}
{\cal Z}_{\tilde{D}^3}^{1\delta}
{\cal Z}_{\tilde{D}^1}^{2\gamma}
{\cal E}^{ib} {\cal E}^{jd}
F_{A}^{0}(x_{j{\rm w}}, x_{H^{-}_{i}{\rm w}}, x_{H_{k}^{-}{\rm w}},
x_{\tilde{U}^i_\alpha {\rm w}}, x_{\tilde{\tiny D}^{1}_{m}{\rm w}},
x_{\tilde{D}^1_\gamma {\rm w}}, x_{\tilde{g}{\rm w}})
\nonumber \\
&&\hspace{1.2cm}+(i\leftrightarrow j)\; ,
\label{chhh4}
\end{eqnarray}
\begin{eqnarray}
&&\phi^{hh\tilde{g}}_{5}=\frac{1}{4}\phi^{hh\tilde{g}}_{4}\; ,
\label{chhh5}
\end{eqnarray}
\begin{eqnarray}
&&\phi^{hh\tilde{g}}_{6}=-\frac{16}{3}h_{b}^2 h_{d}\sqrt{x_{\tilde{g}{\rm w}}}
{\cal Z}_{H}^{1k}({\cal Z}_{H}^{1l})^2
{\cal Z}_{\tilde{D}^1}^{2\gamma}
{\cal Z}_{\tilde{U}^i}^{1\alpha}{\cal E}^{id}
F_{D}^{1a}(x_{i{\rm w}}, x_{j{\rm w}}, x_{H_{k}^{-}{\rm w}}, x_{H_{l}^{-}{\rm w}},
x_{\tilde{U}^i_\alpha {\rm w}}, x_{\tilde{D}^1_\gamma {\rm w}},
x_{\tilde{g}{\rm w}})\nonumber \\
&&\hspace{1.2cm}-\frac{16}{3}h_{b}^2 h_{d}^2 \sqrt{x_{\tilde{g}{\rm w}}x_{i{\rm w}}}
({\cal Z}_{H}^{1k})^2 ({\cal Z}_{H}^{1l})^2
{\cal Z}_{\tilde{U}^i}^{1\alpha}
{\cal Z}_{\tilde{U}^i}^{2\alpha}
F_{C}^{1a}(x_{j{\rm w}}, x_{i{\rm w}}, x_{i{\rm w}}, x_{H_{k}^{-}{\rm w}}, x_{H_{l}^{-}{\rm w}},
    x_{\tilde{U}^i_\alpha {\rm w}}, x_{\tilde{g}{\rm w}})
\nonumber \\
&&\hspace{1.2cm}+\frac{4}{3}h_{b}^2 h_{d}^2
({\cal Z}_{H}^{1k})^2 ({\cal Z}_{H}^{1l})^2
({\cal Z}_{\tilde{U}^i}^{1\alpha})^2
\Big(F_{C}^{2a}+F_{C}^{2d}-F_{C}^{2e}\Big)
(x_{j{\rm w}}, x_{i{\rm w}}, x_{i{\rm w}}, x_{H_{k}^{-}{\rm w}},
x_{H_{l}^{-}{\rm w}}, x_{\tilde{U}^i_\alpha {\rm w}}, x_{\tilde{g}{\rm w}})
\nonumber \\
&&\hspace{1.2cm}+\frac{8}{3}h_{b}^2 h_{d}\sqrt{x_{i{\rm w}}}
{\cal Z}_{H}^{1k}({\cal Z}_{H}^{1l})^2
{\cal Z}_{\tilde{D}^1}^{2\gamma}
{\cal Z}_{\tilde{U}^i}^{2\alpha}{\cal E}^{id}
\Big(F_{D}^{1a}+F_{D}^{1b}-F_{D}^{1c}\Big)
(x_{i{\rm w}}, x_{j{\rm w}}, x_{H_{k}^{-}{\rm w}}, x_{H_{l}^{-}{\rm w}},
x_{\tilde{U}^i_\alpha {\rm w}}, x_{\tilde{D}^1_\gamma {\rm w}}, x_{\tilde{g}{\rm w}})
\nonumber \\
&&\hspace{1.2cm}+\frac{4}{3}h_{b}^2 h_{d}^2 x_{i{\rm w}}
({\cal Z}_{H}^{1k})^2 ({\cal Z}_{H}^{1l})^2
({\cal Z}_{\tilde{U}^i}^{2\alpha})^2
\Big(F_{C}^{1a}+F_{C}^{1b}-F_{C}^{1c}\Big)
(x_{j{\rm w}}, x_{i{\rm w}}, x_{i{\rm w}}, x_{H_{k}^{-}{\rm w}},
x_{H_{l}^{-}{\rm w}}, x_{\tilde{U}^i_\alpha {\rm w}}, x_{\tilde{g}{\rm w}})
\nonumber \\
&&\hspace{1.2cm}+\frac{4}{3\sin^2\beta}x_{j{\rm w}}^2
{\cal Z}_{H}^{2i}{\cal Z}_{H}^{2k}
{\cal Z}_{\tilde{D}^3}^{2\delta}
{\cal Z}_{\tilde{D}^1}^{2\gamma}
{\cal E}^{ib} {\cal E}^{jd}\Big(F_{A}^{1a}+F_{A}^{1b}
-F_{A}^{1c}\Big)
(x_{j{\rm w}}, x_{H^{-}_{i}{\rm w}}, x_{H_{k}^{-}{\rm w}},
x_{\tilde{U}^i_\alpha {\rm w}}, x_{\tilde{\tiny D}^{1}_{m}{\rm w}},
x_{\tilde{D}^1_\gamma {\rm w}}, x_{\tilde{g}{\rm w}})
\nonumber \\
&&\hspace{1.2cm}-2h_{b}^2 h_{d}^2
({\cal Z}_{H}^{1k})^2 ({\cal Z}_{H}^{1l})^2
\frac{x_{H_{k}^{-}{\rm w}}^3 \ln x_{H_{k}^{-}{\rm w}}}
{(-x_{H_{k}^{-}{\rm w}} + x_{H_{l}^{-}{\rm w}})
(-x_{H_{k}^{-}{\rm w}} + x_{i{\rm w}})^2 (-x_{H_{k}^{-}{\rm w}} + x_{j{\rm w}})}
\nonumber \\
&&\hspace{1.2cm}-2h_{b}^2 h_{d}^2
({\cal Z}_{H}^{1k})^2 ({\cal Z}_{H}^{1l})^2
\frac{x_{H_{l}^{-}{\rm w}}^3 \ln x_{H_{l}^{-}{\rm w}}}
{(-x_{H_{l}^{-}{\rm w}} + x_{H_{k}^{-}{\rm w}})
(-x_{H_{l}^{-}{\rm w}} + x_{i{\rm w}})^2 (-x_{H_{l}^{-}{\rm w}} + x_{j{\rm w}})}
\nonumber \\
&&\hspace{1.2cm}-2h_{b}^2 h_{d}^2 ({\cal Z}_{H}^{1k})^2 ({\cal Z}_{H}^{1l})^2
\frac{x_{H_{k}^{-}{\rm w}}^2 x_{i{\rm w}}\ln x_{H_{k}^{-}{\rm w}}}
{(-x_{H_{k}^{-}{\rm w}} + x_{H_{l}^{-}{\rm w}})(-x_{H_{k}^{-}{\rm w}} + x_{i{\rm w}})^2
(-x_{H_{k}^{-}{\rm w}} + x_{j{\rm w}})}\nonumber \\
&&\hspace{1.2cm}-2h_{b}^2 h_{d}^2 ({\cal Z}_{H}^{1k})^2 ({\cal Z}_{H}^{1l})^2
\frac{x_{H_{l}^{-}{\rm w}}^2 x_{i{\rm w}}\ln x_{H_{l}^{-}{\rm w}}}
{(-x_{H_{l}^{-}{\rm w}} + x_{H_{k}^{-}{\rm w}})(-x_{H_{l}^{-}{\rm w}} + x_{i{\rm w}})^2
(-x_{H_{l}^{-}{\rm w}} + x_{j{\rm w}})}\nonumber \\
&&\hspace{1.2cm}+2h_{b}^2 h_{d}^2 ({\cal Z}_{H}^{1k})^2 ({\cal Z}_{H}^{1l})^2
\Big(-3x_{H_{k}^{-}{\rm w}}x_{H_{l}^{-}{\rm w}}x_{i{\rm w}}^3 +x_{H_{k}^{-}{\rm w}}x_{i{\rm w}}^4
+x_{H_{l}^{-}{\rm w}}x_{i{\rm w}}^4\nonumber \\
&&\hspace{1.2cm}+x_{i{\rm w}}^5 + 5x_{H_{k}^{-}{\rm w}}x_{H_{l}^{-}{\rm w}}
x_{i{\rm w}}^2 x_{j{\rm w}}-3x_{H_{k}^{-}{\rm w}}x_{i{\rm w}}^3 x_{j{\rm w}}-3x_{H_{l}^{-}{\rm w}}x_{i{\rm w}}^3 x_{j{\rm w}}
+x_{i{\rm w}}^4 x_{j{\rm w}}\Big)\nonumber \\
&&\hspace{1.2cm}\frac{\ln x_{H_{k}^{-}{\rm w}}}
{(-x_{H_{l}^{-}{\rm w}} + x_{H_{k}^{-}{\rm w}})^2(-x_{H_{l}^{-}{\rm w}} + x_{i{\rm w}})^2
(-x_{H_{l}^{-}{\rm w}} + x_{j{\rm w}})^2}\nonumber \\
&&\hspace{1.2cm}-2h_{b}^2 h_{d}^2({\cal Z}_{H}^{1k})^2({\cal Z}_{H}^{1l})^2
\frac{(x_{i{\rm w}}x_{j{\rm w}}^2+x_{j{\rm w}}^3)\ln x_{j{\rm w}}}
{(x_{H_{k}^{-}{\rm w}} - x_{j{\rm w}})(x_{H_{l}^{-}{\rm w}} - x_{j{\rm w}})
(x_{i{\rm w}} - x_{j{\rm w}})^2}\nonumber \\
&&\hspace{1.2cm}+8h_{b}^2 h_{d} {\cal Z}_{H}^{1k}
({\cal Z}_{H}^{1l})^2 {\cal Z}_{\tilde{D}^1}^{2\gamma}
{\cal Z}_{\tilde{U}^i}^{2\alpha}
{\cal E}^{id}\frac{x_{H_{l}^{-}{\rm w}}\sqrt{x_{i{\rm w}}}\ln x_{H_{l}^{-}{\rm w}}}
{(x_{H_{k}^{-}{\rm w}}-x_{H_{l}^{-}{\rm w}})(-x_{H_{l}^{-}{\rm w}}+ x_{i{\rm w}})
(-x_{H_{l}^{-}{\rm w}} + x_{j{\rm w}})}\nonumber \\
&&\hspace{1.2cm}+8h_{b}^2 h_{d} {\cal Z}_{H}^{1k}
({\cal Z}_{H}^{1l})^2 {\cal Z}_{\tilde{D}^1}^{2\gamma}
{\cal Z}_{\tilde{U}^i}^{2\alpha}
{\cal E}^{id}\frac{x_{H_{k}^{-}{\rm w}}\sqrt{x_{i{\rm w}}}\ln x_{H_{k}^{-}{\rm w}}}
{(x_{H_{l}^{-}{\rm w}}-x_{H_{k}^{-}{\rm w}})(-x_{H_{k}^{-}{\rm w}}+ x_{i{\rm w}})
(-x_{H_{k}^{-}{\rm w}} + x_{j{\rm w}})}\nonumber \\
&&\hspace{1.2cm}+8h_{b}^2 h_{d} {\cal Z}_{H}^{1k}
({\cal Z}_{H}^{1l})^2 {\cal Z}_{\tilde{D}^1}^{2\gamma}
{\cal Z}_{\tilde{U}^i}^{2\alpha}
{\cal E}^{id}\frac{x_{i{\rm w}}^{3\over 2}\ln x_{H_{l}^{-}{\rm w}}}
{(x_{H_{k}^{-}{\rm w}}-x_{i{\rm w}})(-x_{H_{l}^{-}{\rm w}}+ x_{i{\rm w}})
(-x_{i{\rm w}} + x_{j{\rm w}})}\nonumber \\
&&\hspace{1.2cm}+8h_{b}^2 h_{d}
{\cal Z}_{H}^{1k} ({\cal Z}_{H}^{1l})^2
{\cal Z}_{\tilde{D}^1}^{2\gamma}
{\cal Z}_{\tilde{U}^i}^{2\alpha}{\cal E}^{id}
\frac{\sqrt{x_{i{\rm w}}}x_{j{\rm w}}\ln x_{j{\rm w}}}
{(x_{H_{k}^{-}{\rm w}} - x_{j{\rm w}})(x_{H_{l}^{-}{\rm w}}-x_{j{\rm w}})
(x_{i{\rm w}} - x_{j{\rm w}})}
\nonumber \\
&&\hspace{1.2cm}+(i\leftrightarrow j)\; ,
\label{chhh6}
\end{eqnarray}
\begin{eqnarray}
&&\phi^{hh\tilde{g}}_{7}=\frac{8}{3\sin\beta}h_{b}^2
x_{i{\rm w}} \sqrt{x_{j{\rm w}}}
{\cal Z}_{H}^{1k} {\cal Z}_{H}^{1l} {\cal Z}_{H}^{2l}
{\cal Z}_{\tilde{D}^1}^{1\gamma}
{\cal Z}_{\tilde{U}^i}^{1\alpha}{\cal E}^{id}
\Big(F_{D}^{1a}+F_{D}^{1b}-F_{D}^{1c}\Big)
(x_{i{\rm w}}, x_{j{\rm w}}, x_{H_{k}^{-}{\rm w}}, x_{H_{l}^{-}{\rm w}},
x_{\tilde{U}^i_\alpha {\rm w}}, x_{\tilde{D}^1_\gamma {\rm w}},
x_{\tilde{g}{\rm w}})\nonumber \\
&&\hspace{1.2cm}+\frac{4}{3\sin^2\beta}h_{b}^2
(x_{i{\rm w}}x_{j{\rm w}})^{3\over 2}
{\cal Z}_{H}^{1k} {\cal Z}_{H}^{1l} {\cal Z}_{H}^{2k} {\cal Z}_{H}^{2l}
\Big(F_{C}^{1a}+F_{C}^{1b}-F_{C}^{1c}\Big)
(x_{j{\rm w}}, x_{i{\rm w}}, x_{i{\rm w}}, x_{H_{k}^{-}{\rm w}}, x_{H_{l}^{-}{\rm w}},
x_{\tilde{U}^i_\alpha {\rm w}}, x_{\tilde{g}{\rm w}})\nonumber \\
&&\hspace{1.2cm}-\frac{8}{3\sin^2\beta}h_{b}^2
x_{i{\rm w}} x_{j{\rm w}}\sqrt{x_{\tilde{g}{\rm w}}x_{j{\rm w}}}
{\cal Z}_{H}^{1k}{\cal Z}_{H}^{1l}{\cal Z}_{H}^{2k}
{\cal Z}_{H}^{2l}{\cal Z}_{\tilde{U}^i}^{1\alpha}
{\cal Z}_{\tilde{U}^i}^{2\alpha}
F_{C}^{1a}(x_{j{\rm w}}, x_{i{\rm w}}, x_{i{\rm w}}, x_{H_{k}^{-}{\rm w}},
x_{H_{l}^{-}{\rm w}}, x_{\tilde{U}^i_\alpha {\rm w}}, x_{\tilde{g}{\rm w}})
\nonumber \\
&&\hspace{1.2cm}-\frac{16}{3\sin\beta}h_{b}^2
x_{i{\rm w}}\sqrt{x_{\tilde{g}{\rm w}} x_{i{\rm w}} x_{j{\rm w}}}
{\cal Z}_{H}^{1k}{\cal Z}_{H}^{1l}{\cal Z}_{H}^{2l}
{\cal Z}_{\tilde{D}^1}^{1\gamma}
{\cal Z}_{\tilde{U}^i}^{2\alpha}{\cal E}^{id}
F_{D}^{0}(x_{i{\rm w}}, x_{j{\rm w}}, x_{H_{k}^{-}{\rm w}}, x_{H_{l}^{-}{\rm w}},
x_{\tilde{U}^i_\alpha {\rm w}}, x_{\tilde{D}^1_\gamma {\rm w}}, x_{\tilde{g}{\rm w}})
\nonumber \\
&&\hspace{1.2cm}-\frac{8}{3\sin^2\beta}h_{b}^2
x_{i{\rm w}}^{5\over 2} x_{j{\rm w}}^{3\over 2}{\cal Z}_{H}^{1k}
{\cal Z}_{H}^{1l} {\cal Z}_{H}^{2k} {\cal Z}_{H}^{2l}
{\cal Z}_{\tilde{U}^i}^{1\alpha}
{\cal Z}_{\tilde{U}^i}^{2\alpha}
F_{C}^{0}(x_{j{\rm w}}, x_{i{\rm w}}, x_{i{\rm w}}, x_{H_{k}^{-}{\rm w}},
x_{H_{l}^{-}{\rm w}}, x_{\tilde{U}^i_\alpha {\rm w}}, x_{\tilde{g}{\rm w}})
\nonumber \\
&&\hspace{1.2cm}-\frac{8}{3\sin\beta}
h_{d} x_{j{\rm w}}\sqrt{x_{\tilde{g}{\rm w}} x_{j{\rm w}}}
{\cal Z}_{H}^{1k} {\cal Z}_{H}^{2i}
{\cal Z}_{\tilde{D}^3}^{2\delta}
{\cal Z}_{\tilde{D}^1}^{1\gamma}
{\cal E}^{ib} {\cal E}^{jd}
F_{A}^{0}(x_{j{\rm w}}, x_{H^{-}_{i}{\rm w}}, x_{H_{k}^{-}{\rm w}},
x_{\tilde{U}^i_\alpha {\rm w}}, x_{\tilde{\tiny D}^{1}_{m}{\rm w}},
x_{\tilde{D}^1_\gamma {\rm w}}, x_{\tilde{g}{\rm w}})
\nonumber \\
&&\hspace{1.2cm}+(i\leftrightarrow j)\; ,
\label{chhh7}
\end{eqnarray}
\begin{eqnarray}
&&\phi^{hh\tilde{g}}_{8}=\frac{1}{4}\phi^{hh\tilde{g}}_{7}\; ,
\label{chhh8}
\end{eqnarray}
\begin{eqnarray}
&&\phi^{sw\tilde{g}}_{1}=\frac{4}{3}{\cal Z}_{\tilde{D}^1}^{1\delta}
{\cal Z}_{\tilde{D}^{3}}^{1m}\bigg(
{\cal Z}_{\tilde{D}^1}^{1\delta}{\cal Z}_{-}^{1\eta}+
\frac{h_{b}{\cal Z}_{\tilde{D}^1}^{2\delta}{\cal Z}_{-}^{2\eta}}
{\sqrt{2}}\bigg)\bigg(
{\cal Z}_{\tilde{D}^1}^{1\delta} {\cal Z}_{-}^{1\eta}+
\frac{h_{d}{\cal Z}_{\tilde{D}^1}^{2\delta}{\cal Z}_{-}^{2\eta}}
{\sqrt{2}}\bigg)\nonumber \\
&&\hspace{1.2cm}\Big(F_{A}^{2b}+F_{A}^{2c}-F_{A}^{2d}-F_{A}^{2e}-
2F_{A}^{2f}\Big)(x_{i{\rm w}}, x_{j{\rm w}}, 1, x_{\kappa^-_\eta {\rm w}},
x_{\tilde{g}{\rm w}}, x_{\tilde{\tiny D}^{1}_{l}{\rm w}},
x_{\tilde{\tiny D}^{3}_{m}{\rm w}})
\nonumber \\
&&\hspace{1.2cm}+\frac{4}{3}x_{i{\rm w}}\sqrt{x_{\kappa^-_\eta {\rm w}} x_{i{\rm w}}}
{\cal Z}_{\tilde{D}^3}^{1\gamma}{\cal Z}_{\tilde{D}^1}^{1\delta}
\bigg({\cal Z}_{\tilde{D}^1}^{1\delta} {\cal Z}_{-}^{1\eta}+
\frac{h_{d}{\cal Z}_{\tilde{D}^1}^{2\delta}{\cal Z}_{-}^{2\eta}}
{\sqrt{2}} \bigg)
\frac{{\cal Z}_{\tilde{D}^3}^{1\gamma} {\cal Z}_{+}^{2\eta}}
{\sqrt{2}\sin\beta}\nonumber \\
&&\hspace{1.2cm}\Big(F_{A}^{1a}+F_{A}^{1b}-F_{A}^{1c}\Big)
(x_{i{\rm w}}, x_{j{\rm w}}, 1, x_{\kappa^-_\eta {\rm w}}, x_{\tilde{g}{\rm w}},
x_{\tilde{\tiny D}^{1}_{l}{\rm w}}, x_{\tilde{\tiny D}^{3}_{m}{\rm w}})
\nonumber \\
&&\hspace{1.2cm}-\frac{4}{3}x_{i{\rm w}}x_{j{\rm w}}\Big(\sqrt{x_{\kappa^-_\eta {\rm w}}x_{j{\rm w}}}+
\sqrt{x_{i{\rm w}} x_{j{\rm w}}}\Big)
{\cal Z}_{\tilde{D}^3}^{1\gamma}{\cal Z}_{\tilde{D}^1}^{1\delta}
\frac{{\cal Z}_{\tilde{D}^3}^{1\gamma}{\cal Z}_{+}^{2\eta}}
{\sqrt{2}\sin\beta}
\frac{{\cal Z}_{\tilde{D}^1}^{1\delta}{\cal Z}_{+}^{2\eta}}
{\sqrt{2}\sin\beta}\nonumber \\
&&\hspace{1.2cm}\Big(F_{A}^{1a}+F_{A}^{1b}-F_{A}^{1c}\Big)
(x_{i{\rm w}}, x_{j{\rm w}}, 1, x_{\kappa^-_\eta {\rm w}}, x_{\tilde{g}{\rm w}},
x_{\tilde{\tiny D}^{1}_{l}{\rm w}}, x_{\tilde{\tiny D}^{3}_{m}{\rm w}})
\nonumber \\
&&\hspace{1.2cm}+\frac{4}{3}{\cal Z}_{\tilde{U}^i}^{1\alpha}
{\cal Z}_{\tilde{U}^j}^{1\beta}\bigg(
-{\cal Z}_{\tilde{U}^i}^{1\alpha}{\cal Z}_{+}^{1\eta}+
\frac{m_{u^i}{\cal Z}_{\tilde{U}^i}^{2\alpha}{\cal Z}_{+}^{2\eta}}
{\sqrt{2}m_{\rm w} \sin\beta}\bigg)\bigg(
-{\cal Z}_{\tilde{U}^j}^{1\beta} {\cal Z}_{+}^{1\eta}+
\frac{m_{u^j}{\cal Z}_{\tilde{U}^j}^{2\beta}{\cal Z}_{+}^{2\eta}}
{\sqrt{2}m_{\rm w} \sin\beta}\bigg)\nonumber \\
&&\hspace{1.2cm}\Big(F_{A}^{2b}+F_{A}^{2c}-F_{A}^{2d}-F_{A}^{2e}-
2F_{A}^{2f}\Big)(x_{i{\rm w}}, x_{j{\rm w}}, 1, x_{\tilde{g}{\rm w}},
x_{\kappa^-_\eta {\rm w}},x_{\tilde{U}^j_\beta {\rm w}},
x_{\tilde{U}^i_\alpha {\rm w}})\nonumber \\
&&\hspace{1.2cm}+\frac{4}{3}
\bigg(\sqrt{x_{\tilde{g}{\rm w}}x_{i{\rm w}}}{\cal Z}_{\tilde{U}^i}^{2\alpha}
{\cal Z}_{\tilde{U}^j}^{1\beta}
+ \sqrt{x_{\tilde{g}{\rm w}}x_{j{\rm w}}}{\cal Z}_{\tilde{U}^i}^{1\alpha}
{\cal Z}_{\tilde{U}^j}^{2\beta}\bigg)\nonumber \\
&&\hspace{1.2cm}
\bigg(-{\cal Z}_{\tilde{U}^i}^{1\alpha} {\cal Z}_{+}^{1\eta}+
\frac{m_{u^i}{\cal Z}_{\tilde{U}^i}^{2\alpha}{\cal Z}_{+}^{2\eta}}
{\sqrt{2}m_{\rm w} \sin\beta}\bigg)
\bigg(-{\cal Z}_{\tilde{U}^j}^{1\beta} {\cal Z}_{+}^{1\eta}+
\frac{m_{u^j}{\cal Z}_{\tilde{U}^j}^{2\beta}{\cal Z}_{+}^{2\eta}}
{\sqrt{2}m_{\rm w} \sin\beta}\bigg)\nonumber \\
&&\hspace{1.2cm}\Big(F_{A}^{1a}+F_{A}^{1b}-F_{A}^{1c}\Big)
(x_{i{\rm w}}, x_{j{\rm w}}, 1, x_{\tilde{g}{\rm w}}, x_{\kappa^-_\eta {\rm w}},
x_{\tilde{U}^j_\beta {\rm w}}, x_{\tilde{U}^i_\alpha {\rm w}})
\nonumber \\
&&\hspace{1.2cm}-\frac{4}{3}\sqrt{x_{i{\rm w}}x_{j{\rm w}}}
{\cal Z}_{\tilde{U}^i}^{2\alpha}{\cal Z}_{\tilde{U}^j}^{2\beta}
\bigg(-{\cal Z}_{\tilde{U}^i}^{1\alpha}{\cal Z}_{+}^{1\eta}+
\frac{m_{u^i}{\cal Z}_{\tilde{U}^i}^{2\alpha}{\cal Z}_{+}^{2\eta}}
{\sqrt{2}m_{\rm w} \sin\beta}\bigg)
\bigg(-{\cal Z}_{\tilde{U}^j}^{1\beta}{\cal Z}_{+}^{1\eta}+
\frac{m_{u^j}{\cal Z}_{\tilde{U}^j}^{2\beta}{\cal Z}_{+}^{2\eta}}
{\sqrt{2}m_{\rm w} \sin\beta}\bigg)\nonumber \\
&&\hspace{1.2cm}\Big(F_{A}^{1a}-F_{A}^{1b}-F_{A}^{1c}\Big)
(x_{i{\rm w}}, x_{j{\rm w}}, 1, x_{\tilde{g}{\rm w}}, x_{\kappa^-_\eta {\rm w}},
x_{\tilde{U}^j_\beta {\rm w}}, x_{\tilde{U}^i_\alpha {\rm w}})
\nonumber \\
&&\hspace{1.2cm}+(i\leftrightarrow j)\; ,
\label{sw1}
\end{eqnarray}
\begin{eqnarray}
&&\phi^{sw\tilde{g}}_{2}=\frac{2}{3}\Bigg({\cal Z}_{\tilde{D}^3}^{2\gamma}
{\cal Z}_{\tilde{D}^1}^{2\delta}\bigg(
{\cal Z}_{\tilde{D}^1}^{1\delta} {\cal Z}_{-}^{1\eta}+
\frac{h_{b}{\cal Z}_{\tilde{D}^1}^{2\delta}{\cal Z}_{-}^{2\eta}}
{\sqrt{2}}\bigg)\bigg(
{\cal Z}_{\tilde{D}^1}^{1\delta} {\cal Z}_{-}^{1\eta}+
\frac{h_{d}{\cal Z}_{\tilde{D}^1}^{2\delta}{\cal Z}_{-}^{2\eta}}
{\sqrt{2}}\bigg)\nonumber \\
&&\hspace{1.2cm}+{\cal Z}_{\tilde{U}^i}^{1\alpha}
{\cal Z}_{\tilde{U}^j}^{1\beta}
\frac{h_{b}{\cal Z}_{\tilde{U}^i}^{1\alpha}{\cal Z}_{-}^{2l}}
{\sqrt{2}}
\frac{h_{d}{\cal Z}_{\tilde{U}^j}^{1\beta}{\cal Z}_{-}^{2l}}
{\sqrt{2}}\Bigg)\nonumber \\
&&\hspace{1.2cm}\Big(F_{A}^{2b}+F_{A}^{2c}-F_{A}^{2d}-F_{A}^{2e}
-2F_{A}^{2f}\Big)
(x_{i{\rm w}}, x_{j{\rm w}}, 1, x_{\kappa^-_\eta {\rm w}}, x_{\tilde{g}{\rm w}},
x_{\tilde{\tiny D}^{1}_{l}{\rm w}}, x_{\tilde{\tiny D}^{3}_{m}{\rm w}})
\nonumber \\
&&\hspace{1.2cm}+\frac{2}{3}\bigg(\sqrt{x_{\tilde{g}{\rm w}} x_{i{\rm w}}}
{\cal Z}_{\tilde{U}^i}^{2\alpha}{\cal Z}_{\tilde{U}^j}^{1\beta}+
\sqrt{x_{\tilde{g}{\rm w}} x_{j{\rm w}}}
{\cal Z}_{\tilde{U}^i}^{1\alpha}{\cal Z}_{\tilde{U}^j}^{2\beta}
\bigg)\frac{
h_{b} {\cal Z}_{\tilde{U}^i}^{1\alpha}{\cal Z}_{-}^{2l}}
{\sqrt{2}}\frac{
h_{d} {\cal Z}_{\tilde{U}^j}^{1\beta}{\cal Z}_{-}^{2l}}
{\sqrt{2}}\nonumber \\
&&\hspace{1.2cm}\Big(F_{A}^{1a}+F_{A}^{1b}-F_{A}^{1c}\Big)
(x_{i{\rm w}}, x_{j{\rm w}}, 1, x_{\tilde{g}{\rm w}},
x_{\kappa^-_\eta {\rm w}}, x_{\tilde{U}^j_\beta {\rm w}},
x_{\tilde{U}^i_\alpha {\rm w}})\nonumber \\
&&\hspace{1.2cm}-\frac{2}{3}\sqrt{x_{i{\rm w}}x_{j{\rm w}}}
{\cal Z}_{\tilde{U}^i}^{2\alpha}
{\cal Z}_{\tilde{U}^j}^{2\beta}
\frac{h_{b}{\cal Z}_{\tilde{U}^i}^{1\alpha}{\cal Z}_{-}^{2l}}
{\sqrt{2}}
\frac{h_{d}{\cal Z}_{\tilde{U}^j}^{1\beta}{\cal Z}_{-}^{2l}}
{\sqrt{2}}\nonumber \\
&&\hspace{1.2cm}\Big(F_{A}^{1a}-F_{A}^{1b}-F_{A}^{1c}\Big)
(x_{i{\rm w}}, x_{j{\rm w}}, 1, x_{\tilde{g}{\rm w}},
x_{\kappa^-_\eta {\rm w}}, x_{\tilde{U}^j_\beta {\rm w}},
x_{\tilde{U}^i_\alpha {\rm w}})\nonumber \\
&&\hspace{1.2cm}+\frac{2}{3}x_{i{\rm w}}\sqrt{x_{\kappa^-_\eta {\rm w}}x_{i{\rm w}}}
{\cal Z}_{\tilde{D}^3}^{2\gamma}{\cal Z}_{\tilde{D}^1}^{2\delta}
\bigg({\cal Z}_{\tilde{D}^1}^{1\delta} {\cal Z}_{-}^{1\eta}+
\frac{h_{d}{\cal Z}_{\tilde{D}^1}^{2\delta}{\cal Z}_{-}^{2\eta}}
{\sqrt{2}}\bigg)
\frac{{\cal Z}_{\tilde{D}^3}^{1\gamma}{\cal Z}_{+}^{2\eta}}
{\sqrt{2} \sin\beta}\nonumber \\
&&\hspace{1.2cm}\Big(F_{A}^{1a}+F_{A}^{1b}-F_{A}^{1c}\Big)
(x_{i{\rm w}}, x_{j{\rm w}}, 1, x_{\kappa^-_\eta {\rm w}},
x_{\tilde{g}{\rm w}}, x_{\tilde{\tiny D}^{1}_{l}{\rm w}},
x_{\tilde{\tiny D}^{3}_{m}{\rm w}})
\nonumber \\
&&\hspace{1.2cm}-\frac{2}{3}x_{i{\rm w}}x_{j{\rm w}}
\bigg(\sqrt{x_{\kappa^-_\eta {\rm w}}x_{j{\rm w}}}+
\sqrt{x_{i{\rm w}}x_{j{\rm w}}}\bigg){\cal Z}_{\tilde{D}^3}^{2\gamma}
{\cal Z}_{\tilde{D}^1}^{2\delta}
\frac{{\cal Z}_{\tilde{D}^3}^{1\gamma}{\cal Z}_{+}^{2\eta}}
{\sqrt{2}\sin\beta}\frac{
{\cal Z}_{\tilde{D}^1}^{1\delta}{\cal Z}_{+}^{2\eta}}
{\sqrt{2} \sin\beta}\nonumber \\
&&\hspace{1.2cm}\Big(F_{A}^{1a}+F_{A}^{1b}-F_{A}^{1c}\Big)
(x_{i{\rm w}}, x_{j{\rm w}}, 1, x_{\kappa^-_\eta {\rm w}}, x_{\tilde{g}{\rm w}},
x_{\tilde{\tiny D}^{1}_{l}{\rm w}}, x_{\tilde{\tiny D}^{3}_{m}{\rm w}})
\nonumber \\
&&\hspace{1.2cm}+(i\leftrightarrow j)\; ,
\label{sw2}
\end{eqnarray}
\begin{eqnarray}
&&\phi^{sw\tilde{g}}_{3}=-2\phi^{sw\tilde{g}}_{2}\; ,
\label{sw3}
\end{eqnarray}
\begin{eqnarray}
&&\phi^{sh\tilde{g}}_{1}=\frac{2}{3\sin^2\beta}x_{i{\rm w}}x_{j{\rm w}}\sqrt{x_{i{\rm w}} x_{j{\rm w}}}
({\cal Z}_{H}^{2k})^{2}{\cal Z}_{\tilde{D}^3}^{1\gamma}
{\cal Z}_{\tilde{D}^1}^{1\delta}\bigg(
{\cal Z}_{\tilde{D}^1}^{1\delta}{\cal Z}_{-}^{1\eta}+
\frac{h_{b}{\cal Z}_{\tilde{D}^1}^{2\delta}{\cal Z}_{-}^{2\eta}}
{\sqrt{2}}\bigg)\bigg(
{\cal Z}_{\tilde{D}^1}^{1\delta}{\cal Z}_{-}^{1\eta}+
\frac{h_{d}{\cal Z}_{\tilde{D}^1}^{2\delta}{\cal Z}_{-}^{2\eta}}
{\sqrt{2}}\bigg)\nonumber \\
&&\hspace{1.2cm}\Big(F_{A}^{1a}-F_{A}^{1b}-F_{A}^{1c}\Big)
(x_{i{\rm w}}, x_{j{\rm w}}, x_{H_k^-{\rm w}}, x_{\kappa^-_{\eta}{\rm w}},
x_{\tilde{g}{\rm w}}, x_{\tilde{D}^1_\delta {\rm w}},
x_{\tilde{D}^3_\gamma {\rm w}})\nonumber \\
&&\hspace{1.2cm}+\frac{2}{3 \sin^2\beta}x_{i{\rm w}}x_{j{\rm w}}
\sqrt{x_{\kappa^-_{\eta}{\rm w}}x_{j{\rm w}}}
({\cal Z}_{H}^{2k})^{2}{\cal Z}_{\tilde{D}^3}^{1\gamma}
{\cal Z}_{\tilde{D}^1}^{1\delta}
\bigg(-{\cal Z}_{\tilde{D}^1}^{1\delta} {\cal Z}_{-}^{1\eta}-
\frac{h_{d}{\cal Z}_{\tilde{D}^1}^{2\delta}{\cal Z}_{-}^{2\eta}}
{\sqrt{2}}\bigg)
\frac{x_{i{\rm w}} {\cal Z}_{\tilde{D}^3}^{1\gamma} {\cal Z}_{+}^{2\eta}}
{\sqrt{2} \sin\beta}\nonumber \\
&&\hspace{1.2cm}\Big(F_{A}^{1a}+F_{A}^{1b}-F_{A}^{1c}\Big)
(x_{i{\rm w}}, x_{j{\rm w}}, x_{H_k^-{\rm w}}, x_{\kappa^-_{\eta}{\rm w}},
x_{\tilde{g}{\rm w}}, x_{\tilde{D}^1_\delta {\rm w}},
x_{\tilde{D}^3_\gamma {\rm w}})\nonumber \\
&&\hspace{1.2cm}+\frac{2}{3\sin^2\beta}x_{i{\rm w}}\sqrt{x_{\kappa^-_{\eta}{\rm w}}
x_{i{\rm w}}}x_{j{\rm w}}({\cal Z}_{H}^{2k})^{2}
{\cal Z}_{\tilde{D}^3}^{1\gamma}{\cal Z}_{\tilde{D}^1}^{1\delta}
\bigg(-{\cal Z}_{\tilde{D}^1}^{1\delta}{\cal Z}_{-}^{1\eta}-
\frac{h_{b}{\cal Z}_{\tilde{D}^1}^{2\delta}{\cal Z}_{-}^{2\eta}}
{\sqrt{2}}\bigg)\frac{
x_{j{\rm w}}{\cal Z}_{\tilde{D}^1}^{1\delta}{\cal Z}_{+}^{2\eta}}
{\sqrt{2}\sin\beta}\nonumber \\
&&\hspace{1.2cm}\Big(F_{A}^{1a}+F_{A}^{1b}-F_{A}^{1c}\Big)
(x_{i{\rm w}}, x_{j{\rm w}}, x_{H_k^-{\rm w}}, x_{\kappa^-_{\eta}{\rm w}},
x_{\tilde{g}{\rm w}},x_{\tilde{D}^1_\delta {\rm w}},
x_{\tilde{D}^3_\gamma {\rm w}}) \nonumber \\
&&\hspace{1.2cm}-\frac{2}{3 \sin^2\beta}x_{i{\rm w}} x_{j{\rm w}}
({\cal Z}_{H}^{2k})^{2}{\cal Z}_{\tilde{D}^3}^{1\gamma}
{\cal Z}_{\tilde{D}^1}^{1\delta}
\frac{x_{i{\rm w}}{\cal Z}_{\tilde{D}^3}^{1\gamma}{\cal Z}_{+}^{2\eta}}
{\sqrt{2}\sin\beta}
\frac{x_{j{\rm w}}{\cal Z}_{\tilde{D}^1}^{1\delta}{\cal Z}_{+}^{2\eta}}
{\sqrt{2}\sin\beta}\nonumber \\
&&\hspace{1.2cm}\Big(F_{A}^{2b}+F_{A}^{2c}-F_{A}^{2d}-F_{A}^{2e}
-2F_{A}^{2f}\Big)
(x_{i{\rm w}}, x_{j{\rm w}}, x_{H_k^-{\rm w}}, x_{\kappa^-_{\eta}{\rm w}},
x_{\tilde{g}{\rm w}}, x_{\tilde{D}^1_\delta {\rm w}},
x_{\tilde{D}^3_\gamma {\rm w}})\nonumber \\
&&\hspace{1.2cm}-\frac{2}{3 \sin^2\beta}x_{i{\rm w}}x_{j{\rm w}}
\sqrt{x_{i{\rm w}}x_{j{\rm w}}}({\cal Z}_{H}^{2k})^{2}
{\cal Z}_{\tilde{U}^i}^{1\alpha}
{\cal Z}_{\tilde{U}^j}^{1\beta}
\bigg(-{\cal Z}_{\tilde{U}^i}^{1\alpha}{\cal Z}_{+}^{1\eta}+
\frac{m_{u^i}{\cal Z}_{\tilde{U}^i}^{2\alpha}{\cal Z}_{+}^{2\eta}}
{\sqrt{2}m_{\rm w} \sin\beta}\bigg)
\bigg(-{\cal Z}_{\tilde{U}^j}^{1\beta}{\cal Z}_{+}^{1\eta}+
\frac{m_{u^j}{\cal Z}_{\tilde{U}^j}^{2\beta}{\cal Z}_{+}^{2\eta}}
{\sqrt{2}m_{\rm w} \sin\beta}\bigg)\nonumber \\
&&\hspace{1.2cm}\Big(F_{A}^{1a}-F_{A}^{1b}-F_{A}^{1c}\Big)
(x_{i{\rm w}}, x_{j{\rm w}}, x_{H_k^-{\rm w}}, x_{\tilde{g}{\rm w}},
x_{\kappa^-_{\eta}{\rm w}},x_{\tilde{U}^j_\beta {\rm w}}, x_{\tilde{U}^i_\alpha {\rm w}})\nonumber \\
&&\hspace{1.2cm}+ \frac{2}{3\sin^2\beta}x_{i{\rm w}}x_{j{\rm w}}\sqrt{x_{\tilde{g}{\rm w}}x_{j{\rm w}}}
({\cal Z}_{H}^{2k})^{2}{\cal Z}_{\tilde{U}^i}^{2\alpha}
{\cal Z}_{\tilde{U}^j}^{1\beta}
\bigg(-{\cal Z}_{\tilde{U}^i}^{1\alpha}{\cal Z}_{+}^{1\eta}+
\frac{m_{u^i}{\cal Z}_{\tilde{U}^i}^{2\alpha}
{\cal Z}_{+}^{2\eta}}{\sqrt{2}m_{\rm w} \sin\beta}\bigg)
\bigg(-{\cal Z}_{\tilde{U}^j}^{1\beta}{\cal Z}_{+}^{1\eta}+
\frac{m_{u^j}{\cal Z}_{\tilde{U}^j}^{2\beta}{\cal Z}_{+}^{2\eta}}
{\sqrt{2}m_{\rm w} \sin\beta}\bigg)\nonumber \\
&&\hspace{1.2cm}\Big(F_{A}^{1a}+F_{A}^{1b}-F_{A}^{1c}\Big)
(x_{i{\rm w}}, x_{j{\rm w}}, x_{H_k^-{\rm w}}, x_{\tilde{g}{\rm w}}, x_{\kappa^-_{\eta}{\rm w}},
    x_{\tilde{U}^j_\beta {\rm w}}, x_{\tilde{U}^i_\alpha {\rm w}})
\nonumber \\
&&\hspace{1.2cm}+\frac{2}{3\sin^2\beta}x_{i{\rm w}}x_{j{\rm w}}\sqrt{x_{\tilde{g}{\rm w}}x_{i{\rm w}}}
({\cal Z}_{H}^{2k})^{2}{\cal Z}_{\tilde{U}^i}^{1\alpha}
{\cal Z}_{\tilde{U}^j}^{2\beta}
\bigg(-{\cal Z}_{\tilde{U}^i}^{1\alpha}{\cal Z}_{+}^{1\eta}+
\frac{m_{u^i}{\cal Z}_{\tilde{U}^i}^{2\alpha}{\cal Z}_{+}^{2\eta}}
{\sqrt{2}m_{\rm w} \sin\beta}\bigg)
\bigg(-{\cal Z}_{\tilde{U}^j}^{1\beta}{\cal Z}_{+}^{1\eta}+
\frac{m_{u^j}{\cal Z}_{\tilde{U}^j}^{2\beta}
{\cal Z}_{+}^{2\eta}}{\sqrt{2}m_{\rm w} \sin\beta}\bigg)\nonumber \\
&&\hspace{1.2cm}\Big(F_{A}^{1a}+F_{A}^{1b}-F_{A}^{1c}\Big)
(x_{i{\rm w}}, x_{j{\rm w}}, x_{H_k^-{\rm w}}, x_{\tilde{g}{\rm w}},
x_{\kappa^-_{\eta}{\rm w}},x_{\tilde{U}^j_\beta {\rm w}},
x_{\tilde{U}^i_\alpha {\rm w}})\nonumber \\
&&\hspace{1.2cm}+\frac{2}{3\sin^2\beta}x_{i{\rm w}}x_{j{\rm w}}
({\cal Z}_{H}^{2k})^{2}{\cal Z}_{\tilde{U}^i}^{2\alpha}
{\cal Z}_{\tilde{U}^j}^{2\beta}\bigg(
-{\cal Z}_{\tilde{U}^i}^{1\alpha}{\cal Z}_{+}^{1\eta}+
\frac{m_{u^i}{\cal Z}_{\tilde{U}^i}^{2\alpha}{\cal Z}_{+}^{2\eta}}
{\sqrt{2}m_{\rm w} \sin\beta}\bigg)\bigg(
-{\cal Z}_{\tilde{U}^j}^{1\beta}{\cal Z}_{+}^{1\eta}+
\frac{m_{u^j}{\cal Z}_{\tilde{U}^j}^{2\beta}{\cal Z}_{+}^{2\eta}}
{\sqrt{2}m_{\rm w} \sin\beta}\bigg)\nonumber \\
&&\hspace{1.2cm}\Big(F_{A}^{2b}+F_{A}^{2c}-F_{A}^{2d}-F_{A}^{2e}
-2F_{A}^{2f}\Big)
(x_{i{\rm w}}, x_{j{\rm w}}, x_{H_k^-{\rm w}}, x_{\tilde{g}{\rm w}}, x_{\kappa^-_{\eta}{\rm w}},
x_{\tilde{U}^j_\beta {\rm w}}, x_{\tilde{U}^i_\alpha {\rm w}})
\nonumber \\
&&\hspace{1.2cm}+(i\leftrightarrow j)\; ,
\label{sh1}
\end{eqnarray}

\begin{eqnarray}
&&\phi^{sh\tilde{g}}_{3}=-2\phi^{sh\tilde{g}}_{2}\; ,
\label{sh3}
\end{eqnarray}
\begin{eqnarray}
&&\phi^{sh\tilde{g}}_{4}=\frac{2}{3\sin\beta}h_{d}x_{i{\rm w}}\sqrt{x_{\tilde{g}{\rm w}}x_{i{\rm w}}}
{\cal Z}_{H}^{1k}{\cal Z}_{H}^{2k}
{\cal Z}_{\tilde{D}^3}^{1\gamma}{\cal Z}_{\tilde{D}^1}^{2\delta}
\bigg({\cal Z}_{\tilde{D}^1}^{1\delta}{\cal Z}_{-}^{1\eta}+
\frac{h_{b}{\cal Z}_{\tilde{D}^1}^{2\delta}{\cal Z}_{-}^{2\eta}}
{\sqrt{2}}\bigg)
\bigg({\cal Z}_{\tilde{D}^1}^{1\delta}{\cal Z}_{-}^{1\eta}+
\frac{h_{d}{\cal Z}_{\tilde{D}^1}^{2\delta}{\cal Z}_{-}^{2\eta}}
{\sqrt{2}}\bigg)\nonumber \\
&&\hspace{1.2cm}\Big(F_{A}^{1a}-F_{A}^{1b}+F_{A}^{1c}\Big)
(x_{i{\rm w}}, x_{j{\rm w}}, x_{H_k^-{\rm w}}, x_{\kappa^-_{\eta}{\rm w}},
x_{\tilde{g}{\rm w}}, x_{\tilde{D}^1_\delta {\rm w}},
x_{\tilde{D}^3_\gamma {\rm w}})\nonumber \\
&&\hspace{1.2cm}-\frac{2}{3\sin\beta}h_{d}x_{i{\rm w}}^2
\sqrt{x_{\kappa^-_{\eta}{\rm w}} x_{\tilde{g}{\rm w}}}{\cal Z}_{H}^{1k}
{\cal Z}_{H}^{2k}{\cal Z}_{\tilde{D}^3}^{1\gamma}
{\cal Z}_{\tilde{D}^1}^{2\delta}
\bigg({\cal Z}_{\tilde{D}^1}^{1\delta}{\cal Z}_{-}^{1\eta}+
\frac{h_{d}{\cal Z}_{\tilde{D}^1}^{2\delta}{\cal Z}_{-}^{2\eta}}
{\sqrt{2}}\bigg)\frac{
{\cal Z}_{\tilde{D}^3}^{1\gamma}{\cal Z}_{+}^{2\eta}}
{\sqrt{2}\sin\beta}\nonumber \\
&&\hspace{1.2cm}F_{A}^{1a}(x_{i{\rm w}}, x_{j{\rm w}}, x_{H_k^-{\rm w}},
x_{\kappa^-_{\eta}{\rm w}},x_{\tilde{g}{\rm w}}, x_{\tilde{D}^1_\delta {\rm w}},
x_{\tilde{D}^3_\gamma {\rm w}})\nonumber \\
&&\hspace{1.2cm}-\frac{4}{3\sin\beta}h_{d}(x_{i{\rm w}}x_{j{\rm w}})^{\frac{3}{2}}
\sqrt{x_{\kappa^-_{\eta}{\rm w}}x_{\tilde{g}{\rm w}}}{\cal Z}_{H}^{1k}{\cal Z}_{H}^{2k}
{\cal Z}_{\tilde{D}^3}^{1\gamma}{\cal Z}_{\tilde{D}^1}^{2\delta}
\bigg({\cal Z}_{\tilde{D}^1}^{1\delta}{\cal Z}_{-}^{1\eta}+
\frac{h_{b}{\cal Z}_{\tilde{D}^1}^{2\delta}{\cal Z}_{-}^{2\eta}}
{\sqrt{2}}\bigg)\frac{
{\cal Z}_{\tilde{D}^1}^{1\delta}{\cal Z}_{+}^{2\eta}}
{\sqrt{2}\sin\beta}\nonumber \\
&&\hspace{1.2cm}F_{A}^{0}(x_{i{\rm w}}, x_{j{\rm w}}, x_{H_k^-{\rm w}},
x_{\kappa^-_{\eta}{\rm w}},x_{\tilde{g}{\rm w}}, x_{\tilde{D}^1_\delta {\rm w}},
x_{\tilde{D}^3_\gamma {\rm w}})\nonumber \\
&&\hspace{1.2cm}+\frac{4}{3\sin\beta}h_{d}x_{i{\rm w}}^2 x_{j{\rm w}}
\sqrt{x_{\tilde{g}{\rm w}} x_{j{\rm w}}}{\cal Z}_{H}^{1k}{\cal Z}_{H}^{2k}
{\cal Z}_{\tilde{D}^3}^{1\gamma}{\cal Z}_{\tilde{D}^1}^{2\delta}
\frac{{\cal Z}_{\tilde{D}^3}^{1\gamma}{\cal Z}_{+}^{2\eta}}
{\sqrt{2}\sin\beta}\frac{
{\cal Z}_{\tilde{D}^1}^{1\delta}{\cal Z}_{+}^{2\eta}}
{\sqrt{2}\sin\beta}\nonumber \\
&&\hspace{1.2cm}\Big(F_{A}^{1a}-F_{A}^{1b}+F_{A}^{1c}\Big)
(x_{i{\rm w}}, x_{j{\rm w}}, x_{H_k^-{\rm w}}, x_{\kappa^-_{\eta}{\rm w}}, x_{\tilde{g}{\rm w}},
x_{\tilde{D}^1_\delta {\rm w}}, x_{\tilde{D}^3_\gamma {\rm w}})
\nonumber \\
&&\hspace{1.2cm}+\frac{2}{3\sin\beta}h_{d}x_{j{\rm w}}
\sqrt{x_{\kappa^-_{\eta}{\rm w}} x_{j{\rm w}}}{\cal Z}_{H}^{1k}
{\cal Z}_{H}^{2k}{\cal Z}_{\tilde{U}^i}^{1\alpha}
{\cal Z}_{\tilde{U}^j}^{1\beta}
\frac{h_{d}{\cal Z}_{\tilde{U}^j}^{1\beta}{\cal Z}_{-}^{2\eta}}
{\sqrt{2}}\bigg(
-{\cal Z}_{\tilde{U}^i}^{1\alpha} {\cal Z}_{+}^{1\eta}+
\frac{m_{u^i}{\cal Z}_{\tilde{U}^i}^{2\alpha}{\cal Z}_{+}^{2\eta}}
{\sqrt{2}m_{\rm w} \sin\beta}\bigg)\nonumber \\
&&\hspace{1.2cm}\Big(F_{A}^{1a}-F_{A}^{1b}+F_{A}^{1c}\Big)
(x_{i{\rm w}}, x_{j{\rm w}}, x_{H_k^-{\rm w}}, x_{\tilde{g}{\rm w}},
x_{\kappa^-_{\eta}{\rm w}}, x_{\tilde{U}^j_\beta {\rm w}},
x_{\tilde{U}^i_\alpha {\rm w}})\nonumber \\
&&\hspace{1.2cm}-\frac{4}{3\sin\beta}
h_{d}x_{j{\rm w}}\sqrt{x_{\kappa^{-}_{i}{\rm w}} x_{\kappa^-_{\eta}{\rm w}}x_{\tilde{g}{\rm w}}x_{l{\rm w}}}
{\cal Z}_{H}^{1k}{\cal Z}_{H}^{2k}{\cal Z}_{\tilde{U}^i}^{2\alpha}
{\cal Z}_{\tilde{U}^j}^{1\beta}\frac{
h_{d}{\cal Z}_{\tilde{U}^j}^{1\beta}{\cal Z}_{-}^{2\eta}}
{\sqrt{2}}\bigg(
-{\cal Z}_{\tilde{U}^i}^{1\alpha} {\cal Z}_{+}^{1\eta}+
\frac{m_{u^i}{\cal Z}_{\tilde{U}^i}^{2\alpha}{\cal Z}_{+}^{2\eta}}
{\sqrt{2}m_{\rm w} \sin\beta}\bigg)\nonumber \\
&&\hspace{1.2cm}F_{A}^{0}(x_{i{\rm w}}, x_{j{\rm w}}, x_{H_k^-{\rm w}}, x_{\tilde{g}{\rm w}},
x_{\kappa^-_{\eta}{\rm w}}, x_{\tilde{U}^{j}_{\beta}{\rm w}},
x_{\tilde{U}^i_\alpha {\rm w}})\nonumber \\
&&\hspace{1.2cm}-\frac{4}{3\sin\beta}
h_{d}x_{j{\rm w}}\sqrt{x_{\tilde{g}{\rm w}}x_{l{\rm w}}}{\cal Z}_{H}^{1k}
{\cal Z}_{H}^{2k}{\cal Z}_{\tilde{U}^i}^{1\alpha}
{\cal Z}_{\tilde{U}^j}^{2\beta}
\frac{h_{d}{\cal Z}_{\tilde{U}^j}^{1\beta}{\cal Z}_{-}^{2\eta}}
{\sqrt{2}}\bigg(
-{\cal Z}_{\tilde{U}^i}^{1\alpha} {\cal Z}_{+}^{1\eta}+
\frac{m_{u^i}{\cal Z}_{\tilde{U}^i}^{2\alpha}{\cal Z}_{+}^{2\eta}}
{\sqrt{2}m_{\rm w}\sin\beta}\bigg)\nonumber \\
&&\hspace{1.2cm}F_{A}^{1a}(x_{i{\rm w}}, x_{j{\rm w}}, x_{H_k^-{\rm w}},
x_{\tilde{g}{\rm w}}, x_{\kappa^-_{\eta}{\rm w}},
x_{\tilde{U}^j_\beta {\rm w}}, x_{\tilde{U}^i_\alpha {\rm w}})
\nonumber\\
&&\hspace{1.2cm}+\frac{2}{3\sin\beta}
h_{d}\sqrt{x_{\kappa^-_{\eta}{\rm w}}x_{i{\rm w}}}x_{j{\rm w}}{\cal Z}_{H}^{1k}
{\cal Z}_{H}^{2k}{\cal Z}_{\tilde{U}^i}^{2\alpha}
{\cal Z}_{\tilde{U}^j}^{2\beta}
\frac{h_{d}{\cal Z}_{\tilde{U}^j}^{1\beta}{\cal Z}_{-}^{2\eta}}
{\sqrt{2}}\bigg(
-{\cal Z}_{\tilde{U}^i}^{1\alpha}{\cal Z}_{+}^{1\eta}+
\frac{m_{u^i}{\cal Z}_{\tilde{U}^i}^{2\alpha}{\cal Z}_{+}^{2\eta}}
{\sqrt{2}m_{\rm w} \sin\beta}\bigg)\nonumber \\
&&\hspace{1.2cm}\Big(F_{A}^{1a}-F_{A}^{1b}+F_{A}^{1c}\Big)
(x_{i{\rm w}}, x_{j{\rm w}}, x_{H_k^-{\rm w}}, x_{\tilde{g}{\rm w}},
x_{\kappa^-_{\eta}{\rm w}}, x_{\tilde{U}^j_\beta {\rm w}},
x_{\tilde{U}^i_\alpha {\rm w}})
\nonumber \\
&&\hspace{1.2cm}+(i\leftrightarrow j)\; ,
\label{sh4}
\end{eqnarray}
\begin{eqnarray}
&&\phi^{sh\tilde{g}}_{5}=\frac{1}{4}\phi^{sh\tilde{g}}_{4}\; ,
\label{sh5}
\end{eqnarray}
\begin{eqnarray}
&&\phi^{sh\tilde{g}}_{6}=-\frac{2}{3}h_{b}h_{d}({\cal Z}_{H}^{1k})^{2}
{\cal Z}_{\tilde{D}^3}^{2\gamma}{\cal Z}_{\tilde{D}^1}^{2\delta}
\bigg({\cal Z}_{\tilde{D}^1}^{1\delta} {\cal Z}_{-}^{1\eta}+
\frac{h_{b}{\cal Z}_{\tilde{D}^1}^{2\delta}{\cal Z}_{-}^{2\eta}}
{\sqrt{2}}\bigg)\bigg(
{\cal Z}_{\tilde{D}^1}^{1\delta} {\cal Z}_{-}^{1\eta}+
\frac{h_{d}{\cal Z}_{\tilde{D}^1}^{2\delta}{\cal Z}_{-}^{2\eta}}
{\sqrt{2}}\bigg)\nonumber \\
&&\hspace{1.2cm}\Big(F_{A}^{2b}+F_{A}^{2c}-F_{A}^{2d}-F_{A}^{2e}-
2F_{A}^{2f}\Big)
(x_{i{\rm w}}, x_{j{\rm w}}, x_{H_k^-{\rm w}}, x_{\kappa^-_{\eta}{\rm w}},
x_{\tilde{g}{\rm w}}, x_{\tilde{D}^1_\delta {\rm w}},
x_{\tilde{D}^3_\gamma {\rm w}})\nonumber \\
&&\hspace{1.2cm}+\frac{2}{3}h_{b}h_{d}({\cal Z}_{H}^{1k})^{2}
{\cal Z}_{\tilde{U}^i}^{1\alpha}{\cal Z}_{\tilde{U}^j}^{1\beta}
\frac{h_{b}{\cal Z}_{\tilde{U}^i}^{1\alpha}{\cal Z}_{-}^{2\eta}}
{\sqrt{2}}
\frac{h_{d}{\cal Z}_{\tilde{U}^j}^{1\beta}{\cal Z}_{-}^{2\eta}}
{\sqrt{2}}\nonumber \\
&&\hspace{1.2cm}\Big(F_{A}^{2b}+F_{A}^{2c}-F_{A}^{2d}-F_{A}^{2e}-
2F_{A}^{2f}\Big)
(x_{i{\rm w}}, x_{j{\rm w}}, x_{H_k^-{\rm w}}, x_{\tilde{g}{\rm w}},
x_{\kappa^-_{\eta}{\rm w}}, x_{\tilde{U}^j_\beta {\rm w}},
x_{\tilde{U}^i_\alpha {\rm w}})\nonumber \\
&&\hspace{1.2cm}+\frac{2}{3}h_{b}h_{d}\sqrt{x_{\tilde{g}{\rm w}}x_{i{\rm w}}}
({\cal Z}_{H}^{1k})^{2}
\bigg({\cal Z}_{\tilde{U}^i}^{2\alpha}
{\cal Z}_{\tilde{U}^j}^{1\beta}
+{\cal Z}_{\tilde{U}^i}^{1\alpha}
{\cal Z}_{\tilde{U}^j}^{2\beta}-
{\cal Z}_{\tilde{U}^i}^{2\alpha}
{\cal Z}_{\tilde{U}^j}^{2\beta}\bigg)
\frac{h_{b}{\cal Z}_{\tilde{U}^i}^{1\alpha}{\cal Z}_{-}^{2\eta}}
{\sqrt{2}}
\frac{h_{d}{\cal Z}_{\tilde{U}^j}^{1\beta}{\cal Z}_{-}^{2\eta}}
{\sqrt{2}}\nonumber \\
&&\hspace{1.2cm}\Big(F_{A}^{1a}+F_{A}^{1b}-F_{A}^{1c}\Big)
(x_{i{\rm w}}, x_{j{\rm w}}, x_{H_k^-{\rm w}}, x_{\tilde{g}{\rm w}},
x_{\kappa^-_{\eta}{\rm w}}, x_{\tilde{U}^j_\beta {\rm w}},
x_{\tilde{U}^i_\alpha {\rm w}})\nonumber \\
&&\hspace{1.2cm}-\frac{2}{3}h_{b} h_{d}x_{i{\rm w}}\sqrt{x_{\kappa^-_{\eta}{\rm w}}x_{i{\rm w}}}
({\cal Z}_{H}^{1k})^{2}{\cal Z}_{\tilde{D}^3}^{2\gamma}
{\cal Z}_{\tilde{D}^1}^{2\delta}\bigg(
{\cal Z}_{\tilde{D}^1}^{1\delta}{\cal Z}_{-}^{1\eta}+
\frac{h_{d}{\cal Z}_{\tilde{D}^1}^{2\delta}{\cal Z}_{-}^{2\eta}}
{\sqrt{2}}\bigg)
\frac{{\cal Z}_{\tilde{D}^3}^{1\gamma}{\cal Z}_{+}^{2\eta}}
{\sqrt{2}\sin\beta}\nonumber \\
&&\hspace{1.2cm}\Big(F_{A}^{1a}+F_{A}^{1b}-F_{A}^{1c}\Big)
(x_{i{\rm w}}, x_{j{\rm w}}, x_{H_k^-{\rm w}}, x_{\kappa^-_{\eta}{\rm w}},
x_{\tilde{g}{\rm w}}, x_{\tilde{D}^1_\delta {\rm w}},
x_{\tilde{D}^3_\gamma {\rm w}})\nonumber \\
&&\hspace{1.2cm}-\frac{2}{3}h_{b}h_{d}x_{j{\rm w}}\sqrt{x_{\kappa^-_{\eta}{\rm w}}x_{j{\rm w}}}
({\cal Z}_{H}^{1k})^{2}{\cal Z}_{\tilde{D}^3}^{2\gamma}
{\cal Z}_{\tilde{D}^1}^{2\delta}
\bigg({\cal Z}_{\tilde{D}^1}^{1\delta} {\cal Z}_{-}^{1\eta}+
\frac{h_{b}{\cal Z}_{\tilde{D}^1}^{2\delta}{\cal Z}_{-}^{2\eta}}
{\sqrt{2}}\bigg)
\frac{{\cal Z}_{\tilde{D}^1}^{1\delta}{\cal Z}_{+}^{2\eta}}
{\sqrt{2}\sin\beta}\nonumber \\
&&\hspace{1.2cm}\Big(F_{A}^{1a}+F_{A}^{1b}-F_{A}^{1c}\Big)
(x_{i{\rm w}}, x_{j{\rm w}}, x_{H_k^-{\rm w}}, x_{\kappa^-_{\eta}{\rm w}},
x_{\tilde{g}{\rm w}}, x_{\tilde{D}^1_\delta {\rm w}},
x_{\tilde{D}^3_\gamma {\rm w}})\nonumber \\
&&\hspace{1.2cm}+\frac{2}{3}h_{b}h_{d}(x_{i{\rm w}}x_{j{\rm w}})^{\frac{3}{2}}
({\cal Z}_{H}^{1k})^{2}{\cal Z}_{\tilde{D}^3}^{2\gamma}
{\cal Z}_{\tilde{D}^1}^{2\delta}
\frac{{\cal Z}_{\tilde{D}^3}^{1\gamma} {\cal Z}_{+}^{2\eta}}
{\sqrt{2} \sin\beta}
\frac{{\cal Z}_{\tilde{D}^1}^{1\delta}{\cal Z}_{+}^{2\eta}}
{\sqrt{2} \sin\beta}\nonumber \\
&&\hspace{1.2cm}\Big(F_{A}^{1a}-F_{A}^{1b}-F_{A}^{1c}\Big)
(x_{i{\rm w}}, x_{j{\rm w}}, x_{H_k^-{\rm w}}, x_{\kappa^-_{\eta}{\rm w}},
x_{\tilde{g}{\rm w}}, x_{\tilde{D}^1_\delta {\rm w}},
x_{\tilde{D}^3_\gamma {\rm w}})
\nonumber \\
&&\hspace{1.2cm}+(i\leftrightarrow j)\; ,
\label{sh6}
\end{eqnarray}
\begin{eqnarray}
&&\phi^{sh\tilde{g}}_{7}=\frac{4}{3\sin\beta}h_{b}x_{j{\rm w}}\sqrt{x_{\tilde{g}{\rm w}}x_{j{\rm w}}}
{\cal Z}_{H}^{1k} {\cal Z}_{H}^{2k}
{\cal Z}_{\tilde{D}^3}^{2\gamma}{\cal Z}_{\tilde{D}^1}^{1\delta}
\bigg({\cal Z}_{\tilde{D}^1}^{1\delta} {\cal Z}_{-}^{1\eta}+
\frac{h_{b}{\cal Z}_{\tilde{D}^1}^{2\delta}{\cal Z}_{-}^{2\eta}}
{\sqrt{2}}\bigg)\bigg(
{\cal Z}_{\tilde{D}^1}^{1\delta}{\cal Z}_{-}^{1\eta}+
\frac{h_{d}{\cal Z}_{\tilde{D}^1}^{2\delta}{\cal Z}_{-}^{2\eta}}
{\sqrt{2}}\bigg)\nonumber \\
&&\hspace{1.2cm}\Big(F_{A}^{1a}-F_{A}^{1b}+F_{A}^{1c}\Big)
(x_{i{\rm w}}, x_{j{\rm w}}, x_{H_k^-{\rm w}}, x_{\kappa^-_{\eta}{\rm w}},
x_{\tilde{g}{\rm w}}, x_{\tilde{D}^1_\delta {\rm w}},
x_{\tilde{D}^3_\gamma {\rm w}})\nonumber \\
&&\hspace{1.2cm}-\frac{4}{3\sin\beta}h_{b}(x_{i{\rm w}}x_{j{\rm w}})^2
\sqrt{x_{\kappa^-_{\eta}{\rm w}} x_{\tilde{g}{\rm w}}}
{\cal Z}_{H}^{1k} {\cal Z}_{H}^{2k}
{\cal Z}_{\tilde{D}^3}^{2\gamma}
{\cal Z}_{\tilde{D}^1}^{1\delta}
\bigg({\cal Z}_{\tilde{D}^1}^{1\delta} {\cal Z}_{-}^{1\eta}+
\frac{h_{d}{\cal Z}_{\tilde{D}^1}^{2\delta}{\cal Z}_{-}^{2\eta}}
{\sqrt{2}}\bigg)
\frac{{\cal Z}_{\tilde{D}^3}^{1\gamma}{\cal Z}_{+}^{2\eta}}
{\sqrt{2}\sin\beta}\nonumber \\
&&\hspace{1.2cm}F_{A}^{0}(x_{i{\rm w}}, x_{j{\rm w}}, x_{H_k^-{\rm w}},
x_{\kappa^-_{\eta}{\rm w}}, x_{\tilde{g}{\rm w}},x_{\tilde{D}^1_\delta {\rm w}},
x_{\tilde{D}^3_\gamma {\rm w}})\nonumber \\
&&\hspace{1.2cm}-\frac{2}{3\sin\beta}h_{b}
x_{j{\rm w}}^2 \sqrt{x_{\kappa^-_{\eta}{\rm w}}x_{\tilde{g}{\rm w}}}
{\cal Z}_{H}^{1k} {\cal Z}_{H}^{2k}
{\cal Z}_{\tilde{D}^3}^{2\gamma}
{\cal Z}_{\tilde{D}^1}^{1\delta}
\bigg({\cal Z}_{\tilde{D}^1}^{1\delta} {\cal Z}_{-}^{1\eta}+
\frac{h_{b}{\cal Z}_{\tilde{D}^1}^{2\delta}{\cal Z}_{-}^{2\eta}}
{\sqrt{2}}\bigg)
\frac{{\cal Z}_{\tilde{D}^1}^{1\delta}{\cal Z}_{+}^{2\eta}}
{\sqrt{2}\sin\beta}\nonumber \\
&&\hspace{1.2cm}F_{A}^{1a}(x_{i{\rm w}}, x_{j{\rm w}}, x_{H_k^-{\rm w}},
x_{\kappa^-_{\eta}{\rm w}},x_{\tilde{g}{\rm w}}, x_{\tilde{D}^1_\delta {\rm w}},
x_{\tilde{D}^3_\gamma {\rm w}})\nonumber \\
&&\hspace{1.2cm}+\frac{2}{3\sin\beta}h_{b}x_{i{\rm w}}x_{j{\rm w}}^2
\sqrt{x_{\tilde{g}{\rm w}} x_{i{\rm w}}}{\cal Z}_{H}^{1k}{\cal Z}_{H}^{2k}
{\cal Z}_{\tilde{D}^3}^{2\gamma}
{\cal Z}_{\tilde{D}^1}^{1\delta}
\frac{{\cal Z}_{\tilde{D}^3}^{1\gamma}{\cal Z}_{+}^{2\eta}}
{\sqrt{2}\sin\beta}
\frac{{\cal Z}_{\tilde{D}^1}^{1\delta}{\cal Z}_{+}^{2\eta}}
{\sqrt{2} \sin\beta}\nonumber \\
&&\hspace{1.2cm}\Big(F_{A}^{1a}-F_{A}^{1b}+F_{A}^{1c}\Big)
(x_{i{\rm w}}, x_{j{\rm w}}, x_{H_k^-{\rm w}}, x_{\kappa^-_{\eta}{\rm w}}, x_{\tilde{g}{\rm w}},
x_{\tilde{D}^1_\delta {\rm w}}, x_{\tilde{D}^3_\gamma {\rm w}})
\nonumber \\
&&\hspace{1.2cm}+\frac{2}{3\sin\beta}h_{b}x_{i{\rm w}}
\sqrt{x_{\kappa^-_{\eta}{\rm w}}x_{i{\rm w}}}{\cal Z}_{H}^{1k}
{\cal Z}_{H}^{2k}\bigg(
{\cal Z}_{\tilde{U}^i}^{1\alpha}
{\cal Z}_{\tilde{U}^j}^{1\beta}-
2{\cal Z}_{\tilde{U}^i}^{2\alpha}
{\cal Z}_{\tilde{U}^j}^{1\beta}+
{\cal Z}_{\tilde{U}^i}^{2\alpha}
{\cal Z}_{\tilde{U}^j}^{2\beta}\bigg)\nonumber \\
&&\hspace{1.2cm}
\frac{h_{b}{\cal Z}_{\tilde{U}^i}^{1\alpha}{\cal Z}_{-}^{2\eta}}
{\sqrt{2}}\bigg(
-{\cal Z}_{\tilde{U}^j}^{1\beta}{\cal Z}_{+}^{1\eta}+
\frac{m_{u^j}{\cal Z}_{\tilde{U}^j}^{2\beta}{\cal Z}_{+}^{2\eta}}
{\sqrt{2}m_{\rm w} \sin\beta}\bigg)\nonumber \\
&&\hspace{1.2cm}\Big(F_{A}^{1a}-F_{A}^{1b}+F_{A}^{1c}\Big)
(x_{i{\rm w}}, x_{j{\rm w}}, x_{H_k^-{\rm w}}, x_{\tilde{g}{\rm w}},
x_{\kappa^-_{\eta}{\rm w}}, x_{\tilde{U}^j_\beta {\rm w}},
x_{\tilde{U}^i_\alpha {\rm w}})
\nonumber \\
&&\hspace{1.2cm}+(i\leftrightarrow j)\; ,
\label{sh7}
\end{eqnarray}
\begin{eqnarray}
&&\phi^{sh\tilde{g}}_{8}=\frac{1}{4}\phi^{sh\tilde{g}}_{7}\; ,
\label{sh8}
\end{eqnarray}
\begin{eqnarray}
&&\phi^{p\tilde{g}}_{1}=-\frac{16}{3}\Bigg({\cal Z}_{\tilde{D}^3}^{1\gamma}{\cal Z}_{\tilde{D}^1}^{1\delta} \bigg(-{\cal Z}_{\tilde{D}^3}^{1\gamma}
{\cal Z}_{-}^{1\lambda} + \frac{h_{b}{\cal Z}_{\tilde{D}^3}^{2\gamma}
{\cal Z}_{-}^{2\lambda}}{\sqrt{2}}\bigg)
\bigg(-({\cal Z}_{\tilde{D}^1}^{1\delta}{\cal Z}_{-}^{1\eta}) +
\frac{(h_{d}{\cal Z}_{\tilde{D}^1}^{2\delta}{\cal Z}_{-}^{2\eta})}
{\sqrt{2}}\bigg) \nonumber \\
&&\hspace{1.2cm}
\bigg(-({\cal Z}_{\tilde{U}^{i}}^{1\alpha} {\cal Z}_{+}^{1\lambda}) +
\frac{m_{u^i}{\cal Z}_{\tilde{U}^{i}}^{2\alpha} {\cal Z}_{+}^{2\lambda}}
{\sqrt{2}m_{\rm w} \sin\beta}\bigg)
\bigg(-{\cal Z}_{\tilde{U}^{i}}^{1\alpha}{\cal Z}_{+}^{1\eta} + \frac{
m_{u^j}{\cal Z}_{\tilde{U}^{i}}^{2\alpha} {\cal Z}_{+}^{2\eta}}{\sqrt{2}m_{\rm w} \sin\beta}\bigg)\Bigg)\nonumber \\
&&\hspace{1.2cm}\Big(F_{A}^{2b}+F_{A}^{2c}-F_{A}^{2d}-F_{A}^{2e}
-2F_{A}^{2f}\Big)(x_{\kappa^-_{\lambda}{\rm w}},
 x_{\tilde{U}^i_{\alpha}{\rm w}}, x_{\kappa^-_{\eta}{\rm w}}, x_{j{\rm w}},
 x_{\tilde{g}{\rm w}}, x_{\tilde{D}^3_{\gamma}{\rm w}},
x_{\tilde{D}^1_{\delta}{\rm w}}) \nonumber \\
&&\hspace{1.2cm} + \frac{16}{3}
\sqrt{x_{\kappa^-_{\lambda}{\rm w}} x_{j{\rm w}}}
{\cal Z}_{\tilde{D}^3}^{1\gamma}{\cal Z}_{\tilde{D}^1}^{1\delta}
\bigg(-{\cal Z}_{\tilde{D}^1}^{1\delta}{\cal Z}_{-}^{1\eta} +
\frac{h_{d}{\cal Z}_{\tilde{D}^1}^{2\delta}{\cal Z}_{-}^{2\eta}}{\sqrt{2}}\bigg)
\frac{m_{u^j}{\cal Z}_{\tilde{D}^3}^{1\gamma}{\cal Z}_{+}^{2\lambda}}{\sqrt{2}m_{\rm w}\sin\beta}\nonumber \\
&&\hspace{1.2cm}
\bigg(-{\cal Z}_{\tilde{U}^{i}}^{1\alpha} {\cal Z}_{+}^{1\lambda} +
\frac{m_{u^i}{\cal Z}_{\tilde{U}^{i}}^{2\alpha} {\cal Z}_{+}^{2\lambda}}{\sqrt{2}m_{\rm w}\sin\beta}\bigg)
\bigg(-{\cal Z}_{\tilde{U}^{i}}^{1\alpha}{\cal Z}_{+}^{1\eta} +
\frac{m_{u^j}{\cal Z}_{\tilde{U}^{i}}^{2\alpha}{\cal Z}_{+}^{2\eta}}{\sqrt{2}m_{\rm w}\sin\beta}\bigg)
\nonumber \\
&&\hspace{1.2cm}\Big(F_{A}^{1a}+F_{A}^{1b}-F_{A}^{1c}\Big)
(x_{\kappa^-_{\lambda}{\rm w}}, x_{\tilde{U}^i_{\alpha}{\rm w}},
x_{\kappa^-_{\eta}{\rm w}}, x_{j{\rm w}}, x_{\tilde{g}{\rm w}}, x_{\tilde{D}^3_{\gamma}{\rm w}},
x_{\tilde{D}^1_{\delta}{\rm w}}) \nonumber \\
&&\hspace{1.2cm}+\frac{16}{3}
\sqrt{x_{\kappa^-_{\eta}{\rm w}} x_{j{\rm w}}}{\cal Z}_{\tilde{D}^3}^{1\gamma}
{\cal Z}_{\tilde{D}^1}^{1\delta}\bigg(-{\cal Z}_{\tilde{D}^3}^{1\gamma}
{\cal Z}_{-}^{1\lambda} + \frac{h_{b}{\cal Z}_{\tilde{D}^3}^{2\gamma}
{\cal Z}_{-}^{2\lambda}}{\sqrt{2}}\bigg)\bigg(-{\cal Z}_{\tilde{U}^{i}}^{1\alpha}
{\cal Z}_{+}^{1\lambda}+\frac{m_{u^i}{\cal Z}_{\tilde{U}^{i}}^{2\alpha}
{\cal Z}_{+}^{2\lambda})}{\sqrt{2}m_{\rm w}\sin\beta}\bigg) \nonumber \\
&&\hspace{1.2cm}
\frac{m_{u^j}{\cal Z}_{\tilde{D}^1}^{1\delta}
{\cal Z}_{+}^{2\eta}}{\sqrt{2}m_{\rm w} \sin\beta}
\bigg(-{\cal Z}_{\tilde{U}^{i}}^{1\alpha} {\cal Z}_{+}^{1\eta} +
\frac{m_{u^j}{\cal Z}_{\tilde{U}^{i}}^{2\alpha}{\cal Z}_{+}^{2\eta}}{
\sqrt{2} m_{\rm w} \sin\beta}\bigg)\nonumber \\
&&\hspace{1.2cm}\Big(F_{A}^{1a}+F_{A}^{1b}-F_{A}^{1c}
\Big)(x_{\kappa^-_{\lambda}{\rm w}},x_{\tilde{U}^i_{\alpha}{\rm w}},
x_{\kappa^-_{\eta}{\rm w}}, x_{j{\rm w}}, x_{\tilde{g}{\rm w}},
x_{\tilde{D}^3_{\gamma}{\rm w}}, x_{\tilde{D}^1_{\delta}{\rm w}}) \nonumber \\
&&\hspace{1.2cm}+\frac{16}{3}
\sqrt{x_{\kappa^-_{\lambda}{\rm w}} x_{\kappa^-_{\eta}{\rm w}}}
{\cal Z}_{\tilde{D}^3}^{1\gamma}{\cal Z}_{\tilde{D}^1}^{1\delta}
\frac{m_{u^j}{\cal Z}_{\tilde{D}^3}^{1\gamma}{\cal Z}_{+}^{2\lambda}}{\sqrt{2}m_{\rm w} \sin\beta}
\bigg(-({\cal Z}_{\tilde{U}^{i}}^{1\alpha}{\cal Z}_{+}^{1\lambda})+
\frac{m_{u^i}{\cal Z}_{\tilde{U}^{i}}^{2\alpha}
{\cal Z}_{+}^{2\lambda}}{\sqrt{2}m_{\rm w} \sin\beta}\bigg) \nonumber \\
&&\hspace{1.2cm}\frac{m_{u^j}{\cal Z}_{\tilde{D}^1}^{1\delta}
{\cal Z}_{+}^{2\eta}}{\sqrt{2}m_{\rm w} \sin\beta}
\bigg(-({\cal Z}_{\tilde{U}^{i}}^{1\alpha} {\cal Z}_{+}^{1\eta}) +
\frac{m_{u^j}{\cal Z}_{\tilde{U}^{i}}^{2\alpha}
{\cal Z}_{+}^{2\eta}}{\sqrt{2}m_{\rm w} \sin\beta}\bigg)\nonumber \\
&&\hspace{1.2cm}\Big(F_{A}^{1a}-F_{A}^{1b}
-F_{A}^{1c}\Big)(x_{\kappa^-_{\lambda}{\rm w}},
x_{\tilde{U}^i_{\alpha}{\rm w}}, x_{\kappa^-_{\eta}{\rm w}}, x_{j{\rm w}}
, x_{\tilde{g}{\rm w}}, x_{\tilde{D}^3_{\gamma}{\rm w}},
x_{\tilde{D}^1_{\delta}{\rm w}}) \nonumber \\
&&\hspace{1.2cm}-\frac{64}{3}\sqrt{x_{\tilde{g}{\rm w}} x_{i{\rm w}}}
\bigg({\cal Z}_{\tilde{U}^i}^{2\alpha_1}{\cal Z}_{\tilde{U}^i}^{1\alpha_2}
+{\cal Z}_{\tilde{U}^i}^{1\alpha_1}{\cal Z}_{\tilde{U}^i}^{2\alpha_2}\bigg)
\bigg(-({\cal Z}_{\tilde{U}^i}^{1\alpha_1} {\cal Z}_{+}^{1\lambda})+
\frac{m_{u^i}{\cal Z}_{\tilde{U}^i}^{2\alpha_1}{\cal Z}_{+}^{2\lambda}}
{\sqrt{2}m_{\rm w} \sin\beta}\bigg)
\bigg(-{\cal Z}_{\tilde{U}^i}^{1\alpha_2} {\cal Z}_{+}^{1\lambda}
\nonumber \\
&&\hspace{1.2cm} +
\frac{m_{u^i}{\cal Z}_{\tilde{U}^i}^{2\alpha_2}{\cal Z}_{+}^{2\lambda}}
{\sqrt{2}m_{\rm w} \sin\beta}\bigg)\bigg(-{\cal Z}_{\tilde{U}^j}^{1\beta}
{\cal Z}_{+}^{1\eta} +
\frac{m_{u^j}{\cal Z}_{\tilde{U}^j}^{2\beta}{\cal Z}_{+}^{2\eta}}
{\sqrt{2}m_{\rm w} \sin\beta}\bigg)
\bigg(-{\cal Z}_{\tilde{U}^j}^{1\beta} {\cal Z}_{+}^{1\eta}+
\frac{m_{u^j}{\cal Z}_{\tilde{U}^j}^{2\beta}
{\cal Z}_{+}^{2\eta}}{\sqrt{2}m_{\rm w}\sin\beta}\bigg)\nonumber \\
&&\hspace{1.2cm}F_{C}^{1a}
(x_{\tilde{U}^i_{\alpha_1}{\rm w}}, x_{\kappa^-_\lambda {\rm w}},
x_{\kappa^-_{\eta}{\rm w}}, x_{\tilde{U}^j_{\beta}{\rm w}},
x_{i{\rm w}}, x_{\tilde{U}^i_{\alpha_2}{\rm w}}, x_{\tilde{g}{\rm w}}) \nonumber \\
&&\hspace{1.2cm}-32x_{\kappa^-_\lambda {\rm w}} \sqrt{x_{\tilde{g}{\rm w}} x_{i{\rm w}}}
\bigg({\cal Z}_{\tilde{U}^i}^{2\alpha_1}{\cal Z}_{
\tilde{\tiny U}^{m}}^{1n}+{\cal Z}_{\tilde{U}^i}^{1\alpha_1}{\cal Z}_{
\tilde{\tiny U}^{m}}^{2n}\bigg)
\bigg(-{\cal Z}_{\tilde{U}^i}^{1\alpha_1}{\cal Z}_{+}^{1\lambda}+
\frac{m_{u^i}{\cal Z}_{\tilde{U}^i}^{1\alpha_1}{\cal Z}_{+}^{2\lambda}}
{\sqrt{2}m_{\rm w} \sin\beta} \bigg)
\nonumber \\
&&\hspace{1.2cm}\bigg(-{\cal Z}_{\tilde{U}^i}^{1\alpha_2}
{\cal Z}_{+}^{1\lambda} +
\frac{m_{u^i}{\cal Z}_{\tilde{U}^i}^{2\alpha_2}{\cal Z}_{+}^{2\lambda}}
{\sqrt{2}m_{\rm w} \sin\beta}\bigg)
\bigg(-{\cal Z}_{\tilde{U}^j}^{1\beta}{\cal Z}_{+}^{1\eta} +
\frac{m_{u^j}{\cal Z}_{\tilde{U}^j}^{2\beta}{\cal Z}_{+}^{2\eta}}
{\sqrt{2}m_{\rm w} \sin\beta}\bigg)
\bigg(-{\cal Z}_{\tilde{U}^j}^{1\beta}{\cal Z}_{+}^{1\eta} +
\frac{m_{u^j}{\cal Z}_{\tilde{U}^j}^{2\beta}{\cal Z}_{+}^{2\eta}}
{\sqrt{2}m_{\rm w} \sin\beta}\bigg)\nonumber \\
&&\hspace{1.2cm}\Big(\frac{\ln x_{\kappa^-_\lambda {\rm w}}}{
(-x_{\kappa^-_\lambda {\rm w}} + x_{\kappa^-_{\eta}{\rm w}})
(-x_{\kappa^-_\lambda {\rm w}} + x_{\tilde{U}^i_{\alpha}{\rm w}})
(-x_{\kappa^-_\lambda {\rm w}} + x_{\tilde{U}^j_{\beta}{\rm w}})}\nonumber \\
&&\hspace{1.2cm}-\frac{x_{i{\rm w}}\ln x_{\kappa^-_\lambda {\rm w}}}{
(-x_{\kappa^-_\lambda {\rm w}} + x_{\kappa^-_{\eta}{\rm w}})
(-x_{\kappa^-_\lambda {\rm w}} + x_{i{\rm w}})(-x_{\kappa^-_\lambda {\rm w}}
+x_{\tilde{U}^i_{\alpha}{\rm w}})(-x_{\kappa^-_\lambda {\rm w}}
+ x_{\tilde{U}^j_{\beta}{\rm w}})}\Big) \nonumber \\
&&\hspace{1.2cm}-32x_{\kappa^-_{\eta}{\rm w}}\sqrt{x_{\tilde{g}{\rm w}}x_{i{\rm w}}}
\bigg({\cal Z}_{\tilde{U}^i}^{2\alpha_1}{\cal Z}_{\tilde{U}^i}^{1\alpha_2}+
{\cal Z}_{\tilde{U}^i}^{1\alpha_1}{\cal Z}_{\tilde{U}^i}^{2\alpha_2}\bigg)
\bigg(-{\cal Z}_{\tilde{U}^i}^{1\alpha_1} {\cal Z}_{+}^{1\lambda}+
\frac{m_{u^i}{\cal Z}_{\tilde{U}^i}^{1\alpha_1}{\cal Z}_{+}^{2\lambda}}
{\sqrt{2}m_{\rm w} \sin\beta}\bigg) \nonumber \\
&&\hspace{1.2cm}\bigg(-({\cal Z}_{\tilde{U}^i}^{1\alpha_2}{\cal Z}_{+}^{1\lambda}) +
\frac{m_{u^i}{\cal Z}_{\tilde{U}^i}^{2\alpha_2}{\cal Z}_{+}^{2\lambda}}
{\sqrt{2}m_{\rm w} \sin\beta}\bigg)
\bigg(-{\cal Z}_{\tilde{U}^j}^{1\beta} {\cal Z}_{+}^{1\eta} +
\frac{m_{u^j}{\cal Z}_{\tilde{U}^j}^{2\beta}{\cal Z}_{+}^{2\eta}}
{\sqrt{2}m_{\rm w} \sin\beta}\bigg)
\bigg(-{\cal Z}_{\tilde{U}^j}^{1\beta} {\cal Z}_{+}^{1\eta} +
\frac{m_{u^j}{\cal Z}_{\tilde{U}^j}^{2\beta}{\cal Z}_{+}^{2\eta}}
{\sqrt{2}m_{\rm w} \sin\beta}\bigg)\nonumber \\
&&\hspace{1.2cm}
\Big(\frac{\ln x_{\kappa^-_{\eta}{\rm w}}}{(x_{\kappa^-_\lambda {\rm w}} - x_{\kappa^-_{\eta}{\rm w}})
(-x_{\kappa^-_{\eta}{\rm w}} + x_{\tilde{U}^i_{\alpha}{\rm w}})
(-x_{\kappa^-_{\eta}{\rm w}} + x_{\tilde{U}^j_{\beta}{\rm w}})}\nonumber \\
&&\hspace{1.2cm}-\frac{x_{i{\rm w}}\ln x_{\kappa^-_{\eta}{\rm w}}}{
(x_{\kappa^-_\lambda {\rm w}} - x_{\kappa^-_{\eta}{\rm w}})
(-x_{\kappa^-_{\eta}{\rm w}} + x_{i{\rm w}})(-x_{\kappa^-_{\eta}{\rm w}}
+ x_{\tilde{U}^i_{\alpha}{\rm w}})(-x_{\kappa^-_{\eta}{\rm w}}
+x_{\tilde{U}^j_{\beta}{\rm w}})}\Big)\nonumber \\
&&\hspace{1.2cm}+32x_{i{\rm w}}^2 \sqrt{x_{\tilde{g}{\rm w}}x_{i{\rm w}}}
\bigg({\cal Z}_{\tilde{U}^i}^{2\alpha_1}{\cal Z}_{\tilde{
\tiny U}^{m}}^{1n}+{\cal Z}_{\tilde{U}^i}^{1\alpha_1}
{\cal Z}_{\tilde{U}^i}^{2\alpha_2}\bigg)
\bigg(-{\cal Z}_{\tilde{U}^i}^{1\alpha_1}{\cal Z}_{+}^{1\lambda}+
\frac{m_{u^i}{\cal Z}_{\tilde{U}^i}^{1\alpha_1}{\cal Z}_{+}^{2\lambda}}
{\sqrt{2}m_{\rm w} \sin\beta}\bigg) \nonumber \\
&&\hspace{1.2cm}
\bigg(-{\cal Z}_{\tilde{U}^i}^{1\alpha_2} {\cal Z}_{+}^{1\lambda}+
\frac{m_{u^i}{\cal Z}_{\tilde{U}^i}^{2\alpha_2}{\cal Z}_{+}^{2\lambda}}
{\sqrt{2}m_{\rm w} \sin\beta}\bigg)
\bigg(-{\cal Z}_{\tilde{U}^j}^{1\beta}{\cal Z}_{+}^{1\eta} +
\frac{m_{u^j}{\cal Z}_{\tilde{U}^j}^{2\beta} {\cal Z}_{+}^{2\eta}}
{\sqrt{2}m_{\rm w} \sin\beta}\bigg)
\bigg(-{\cal Z}_{\tilde{U}^j}^{1\beta}{\cal Z}_{+}^{1\eta}+
\frac{m_{u^j}{\cal Z}_{\tilde{U}^j}^{2\beta}{\cal Z}_{+}^{2\eta}}
{\sqrt{2}m_{\rm w} \sin\beta}\bigg) \nonumber \\
&&\hspace{1.2cm}\frac{\ln x_{i{\rm w}}}{(x_{\kappa^-_\lambda {\rm w}} - x_{i{\rm w}})
(x_{\kappa^-_{\eta}{\rm w}} - x_{i{\rm w}})(-x_{i{\rm w}}+x_{\tilde{U}^i_{\alpha}{\rm w}})
(-x_{i{\rm w}}+x_{\tilde{U}^j_{\beta}{\rm w}})}\nonumber \\
&&\hspace{1.2cm}-32(\sqrt{x_{\tilde{g}{\rm w}} x_{i{\rm w}}}
x_{\tilde{U}^i_{\alpha}{\rm w}}\bigg({\cal Z}_{\tilde{U}^i}^{2\alpha_1}
{\cal Z}_{\tilde{U}^i}^{1\alpha_2}+{\cal Z}_{\tilde{U}^i}^{1\alpha_1}
{\cal Z}_{\tilde{U}^i}^{2\alpha_2}\bigg)
\bigg(-{\cal Z}_{\tilde{U}^i}^{1\alpha_1}{\cal Z}_{+}^{1\lambda}+
\frac{m_{u^i}{\cal Z}_{\tilde{U}^i}^{2\alpha_1}
{\cal Z}_{+}^{2\lambda}}{\sqrt{2}m_{\rm w} \sin\beta}\bigg) \nonumber \\
&&\hspace{1.2cm}
\bigg(-{\cal Z}_{\tilde{U}^i}^{1\alpha_2} {\cal Z}_{+}^{1\lambda}+
\frac{m_{u^i}{\cal Z}_{\tilde{U}^i}^{2\alpha_2}
{\cal Z}_{+}^{2\lambda}}{\sqrt{2}m_{\rm w} \sin\beta}\bigg)
\bigg(-{\cal Z}_{\tilde{U}^j}^{1\beta}{\cal Z}_{+}^{1\eta}+
\frac{m_{u^j}{\cal Z}_{\tilde{U}^j}^{2\beta}{\cal Z}_{+}^{2\eta}}
{\sqrt{2}m_{\rm w} \sin\beta}\bigg)
\bigg(-{\cal Z}_{\tilde{U}^j}^{1\beta}{\cal Z}_{+}^{1\eta}+
\frac{m_{u^j}{\cal Z}_{\tilde{U}^j}^{2\beta}
{\cal Z}_{+}^{2\eta}}{\sqrt{2}m_{\rm w} \sin\beta}\bigg)\nonumber \\
&&\hspace{1.2cm}\Big(\frac{
\ln x_{\tilde{U}^i_{\alpha}{\rm w}}}{(x_{\kappa^-_\lambda {\rm w}}
- x_{\tilde{U}^i_{\alpha}{\rm w}})(x_{\kappa^-_{\eta}{\rm w}} -
x_{\tilde{U}^i_{\alpha}{\rm w}})(-x_{\tilde{U}^i_{\alpha}{\rm w}}
 + x_{\tilde{U}^j_{\beta}{\rm w}})}\nonumber \\
&&\hspace{1.2cm}-\frac{x_{i{\rm w}}\ln x_{\tilde{U}^i_{\alpha}{\rm w}}}
{(x_{\kappa^-_\lambda {\rm w}} - x_{\tilde{U}^i_{\alpha}{\rm w}})
(x_{\kappa^-_{\eta}{\rm w}} - x_{\tilde{U}^i_{\alpha}{\rm w}})
(x_{i{\rm w}} - x_{\tilde{U}^i_{\alpha}{\rm w}})
(-x_{\tilde{U}^i_{\alpha}{\rm w}}+x_{\tilde{U}^j_{\beta}{\rm w}})}\Big)
\nonumber \\
&&\hspace{1.2cm}-32\sqrt{x_{\tilde{g}{\rm w}}x_{i{\rm w}}}x_{\tilde{
\tiny U}^{j}_{\beta}{\rm w}}\bigg({\cal Z}_{\tilde{U}^i}^{2\alpha_1}
{\cal Z}_{\tilde{U}^i}^{1\alpha_2}+{\cal Z}_{\tilde{
\tiny U}^{m}}^{1i}{\cal Z}_{\tilde{U}^i}^{2\alpha_2}\bigg)
\bigg(-{\cal Z}_{\tilde{U}^i}^{1\alpha_1} {\cal Z}_{+}^{1\lambda}+
\frac{m_{u^i}{\cal Z}_{\tilde{U}^i}^{1\alpha_1} {\cal Z}_{+}^{2\lambda}}
{\sqrt{2}m_{\rm w} \sin\beta}\bigg)
\nonumber \\
&&\hspace{1.2cm}
\bigg(-{\cal Z}_{\tilde{U}^i}^{1\alpha_2} {\cal Z}_{+}^{1\lambda}+
\frac{m_{u^i}{\cal Z}_{\tilde{U}^i}^{2\alpha_2}{\cal Z}_{+}^{2\lambda}}
{\sqrt{2}m_{\rm w} \sin\beta}\bigg)
\bigg(-{\cal Z}_{\tilde{U}^j}^{1\beta}{\cal Z}_{+}^{1\eta}+
\frac{m_{u^j}{\cal Z}_{\tilde{U}^j}^{2\beta}
{\cal Z}_{+}^{2\eta}}{\sqrt{2}m_{\rm w} \sin\beta}\bigg)
\nonumber \\
&&\hspace{1.2cm}
\bigg(-{\cal Z}_{\tilde{U}^j}^{1\beta} {\cal Z}_{+}^{1\eta} +
\frac{m_{u^j}{\cal Z}_{\tilde{U}^j}^{2\beta}{\cal Z}_{+}^{2\eta}}
{\sqrt{2}m_{\rm w} \sin\beta}\bigg)\Big(\frac{\ln x_{\tilde{U}^j_{\beta}{\rm w}}}
{(x_{\kappa^-_\lambda {\rm w}} - x_{\tilde{U}^j_{\beta}{\rm w}})
(x_{\kappa^-_{\eta}{\rm w}} - x_{\tilde{U}^j_{\beta}{\rm w}})
(x_{\tilde{U}^i_{\alpha}{\rm w}} - x_{\tilde{U}^j_{\beta}{\rm w}})}
\nonumber \\
&&\hspace{1.2cm}-\frac{x_{i{\rm w}}\ln x_{\tilde{U}^j_{\beta}{\rm w}}}
{(x_{\kappa^-_\lambda {\rm w}} - x_{\tilde{U}^j_{\beta}{\rm w}})
(x_{\kappa^-_{\eta}{\rm w}} - x_{\tilde{U}^j_{\beta}{\rm w}})
(x_{i{\rm w}} - x_{\tilde{U}^j_{\beta}{\rm w}})(x_{\tilde{U}^i_{\alpha}{\rm w}}
- x_{\tilde{U}^j_{\beta}{\rm w}})}\Big)
+(i\leftrightarrow j)\; ,
\label{pcg1}
\end{eqnarray}

\begin{eqnarray}
&&\phi^{p\tilde{g}}_{3}=-2\phi^{p\tilde{g}}_{2}\; ,
\label{pcg3}
\end{eqnarray}
\begin{eqnarray}
&&\phi^{p\tilde{g}}_{4}=-\frac{16}{3}\sqrt{x_{\kappa^-_{\lambda}{\rm w}}x_{g{\rm w}}}
{\cal Z}_{\tilde{\tiny D}^{3}}^{1n}{\cal Z}_{\tilde{\tiny D}^{1}}^{2m}
\frac{h_{d}{\cal Z}_{\tilde{U}^{i}}^{1\alpha}{\cal Z}_{-}^{2\lambda}}
{\sqrt{2}}
\bigg(-{\cal Z}_{\tilde{\tiny D}^{3}}^{1m}{\cal Z}_{-}^{1\lambda}+
\frac{h_{b}{\cal Z}_{\tilde{\tiny D}^{3}}^{2m}{\cal Z}_{-}^{2\lambda}}
{\sqrt{2}}\bigg) \nonumber \\
&&\hspace{1.2cm}
\bigg(-{\cal Z}_{\tilde{\tiny D}^{1}}^{1n}{\cal Z}_{-}^{1\eta}+
\frac{h_{d}{\cal Z}_{\tilde{\tiny D}^{1}}^{2n}{\cal Z}_{-}^{2\eta}}
{\sqrt{2}}\bigg)
\bigg(-{\cal Z}_{\tilde{U}^{i}}^{1\alpha}{\cal Z}_{+}^{1\eta}+
\frac{m_{u^j}{\cal Z}_{\tilde{U}^{i}}^{2\alpha}{\cal Z}_{+}^{2\eta}}
{\sqrt{2}m_{\rm w} \sin\beta}\bigg)\nonumber \\
&&\hspace{1.2cm}\Big(F_{A}^{1a}-F_{A}^{1b}+F_{A}^{1c}\Big)
(x_{\kappa^-_{\lambda}{\rm w}}, x_{\tilde{U}^i_{\alpha}{\rm w}},
x_{\kappa^-_{\eta}{\rm w}}, x_{j{\rm w}}, x_{g{\rm w}},
x_{\tilde{D}^3_{\gamma}{\rm w}}, x_{\tilde{D}^1_{\delta}{\rm w}})\nonumber \\
&&\hspace{1.2cm} -\frac{32}{3}\sqrt{x_{g{\rm w}}x_{j{\rm w}}}
{\cal Z}_{\tilde{\tiny D}^{3}}^{1n}
{\cal Z}_{\tilde{\tiny D}^{1}}^{2m}
\frac{h_{d}{\cal Z}_{\tilde{U}^{i}}^{1\alpha}{\cal Z}_{-}^{2\lambda}}
{\sqrt{2}}\bigg(-{\cal Z}_{\tilde{\tiny D}^{1}}^{1n}{\cal Z}_{-}^{1\eta}+
\frac{h_{d}{\cal Z}_{\tilde{\tiny D}^{1}}^{2n}{\cal Z}_{-}^{2\eta}}
{\sqrt{2}}\bigg)\nonumber \\
&&\hspace{1.2cm}
\frac{m_{u^j}{\cal Z}_{\tilde{\tiny D}^{3}}^{1m}{\cal Z}_{+}^{2\lambda}}{
\sqrt{2}m_{\rm w} \sin\beta}\bigg(
-{\cal Z}_{\tilde{U}^{i}}^{1\alpha}{\cal Z}_{+}^{1\eta}+
\frac{m_{u^j}{\cal Z}_{\tilde{U}^{i}}^{2\alpha}{\cal Z}_{+}^{2\eta}}
{\sqrt{2}m_{\rm w} \sin\beta}\bigg)\nonumber \\
&&\hspace{1.2cm}F_{A}^{1a}(x_{\kappa^-_{\lambda}{\rm w}},
x_{\tilde{U}^i_{\alpha}{\rm w}}, x_{\kappa^-_{\eta}{\rm w}},
x_{j{\rm w}}, x_{g{\rm w}}, x_{\tilde{D}^3_{\gamma}{\rm w}},
x_{\tilde{D}^1_{\delta}{\rm w}})\nonumber \\
&&\hspace{1.2cm}-\frac{32}{3}\sqrt{x_{\kappa^-_{\lambda}{\rm w}}
x_{\kappa^-_{\eta}{\rm w}} x_{g{\rm w}} x_{j{\rm w}}}
{\cal Z}_{\tilde{\tiny D}^{3}}^{1n}{\cal Z}_{\tilde{\tiny D}^{1}}^{2m}
\frac{h_{d}{\cal Z}_{\tilde{U}^{i}}^{1\alpha}{\cal Z}_{-}^{2\lambda}}
{\sqrt{2}}\bigg(
-{\cal Z}_{\tilde{\tiny D}^{3}}^{1m}{\cal Z}_{-}^{1\lambda}+
\frac{h_{b}{\cal Z}_{\tilde{\tiny D}^{3}}^{2m}{\cal Z}_{-}^{2\lambda}}
{\sqrt{2}}\bigg)\nonumber \\
&&\hspace{1.2cm}\frac{m_{u^j}{\cal Z}_{\tilde{\tiny D}^{1}}^{1n}
{\cal Z}_{+}^{2\eta}}{\sqrt{2}m_{\rm w} \sin\beta}\bigg(
-{\cal Z}_{\tilde{U}^{i}}^{1\alpha}{\cal Z}_{+}^{1\eta}+
\frac{m_{u^j}{\cal Z}_{\tilde{U}^{i}}^{2\alpha}{\cal Z}_{+}^{2\eta}}
{\sqrt{2}m_{\rm w} \sin\beta}\bigg)\nonumber \\
&&\hspace{1.2cm}F_{A}^{0}(x_{\kappa^-_{\lambda}{\rm w}},
x_{\tilde{U}^i_{\alpha}{\rm w}}, x_{\kappa^-_{\eta}{\rm w}},
x_{j{\rm w}}, x_{g{\rm w}}, x_{\tilde{D}^3_{\gamma}{\rm w}},
x_{\tilde{D}^1_{\delta}{\rm w}})\nonumber \\
&&\hspace{1.2cm}-\frac{16}{3}\sqrt{x_{\kappa^-_{\eta}{\rm w}}x_{g{\rm w}}}
{\cal Z}_{\tilde{\tiny D}^{3}}^{1n}{\cal Z}_{\tilde{\tiny D}^{1}}^{2m}
\frac{h_{d}{\cal Z}_{\tilde{U}^{i}}^{1\alpha}{\cal Z}_{-}^{2\lambda}}
{\sqrt{2}}
\frac{m_{u^j}{\cal Z}_{\tilde{\tiny D}^{3}}^{1m}{\cal Z}_{+}^{2\lambda}}{
\sqrt{2}m_{\rm w} \sin\beta}\nonumber \\
&&\hspace{1.2cm}\frac{m_{u^j}{\cal Z}_{\tilde{\tiny D}^{1}}^{1n}
{\cal Z}_{+}^{2\eta}}{\sqrt{2}m_{\rm w} \sin\beta}\bigg(
-{\cal Z}_{\tilde{U}^{i}}^{1\alpha}{\cal Z}_{+}^{1\eta}+
\frac{m_{u^j}{\cal Z}_{\tilde{U}^{i}}^{2\alpha}{\cal Z}_{+}^{2\eta}}
{\sqrt{2}m_{\rm w} \sin\beta}\nonumber \\
&&\hspace{1.2cm}\Big(F_{A}^{1a}- F_{A}^{1b}+F_{A}^{1c}\Big)
(x_{\kappa^-_{\lambda}{\rm w}}, x_{\tilde{U}^i_{\alpha}{\rm w}},
x_{\kappa^-_{\eta}{\rm w}}, x_{j{\rm w}}, x_{g{\rm w}},
x_{\tilde{D}^3_{\gamma}{\rm w}}, x_{\tilde{D}^1_{\delta}{\rm w}})
+(i\leftrightarrow j)\; ,
\label{pcg4}
\end{eqnarray}
\begin{eqnarray}
&&\phi^{p\tilde{g}}_{5}=\frac{1}{4}\phi^{p\tilde{g}}_{4}\; ,
\label{pcg5}
\end{eqnarray}
\begin{eqnarray}
&&\phi^{p\tilde{g}}_{6}=\frac{16}{3}\sqrt{x_{\kappa^-_{\lambda}{\rm w}}x_{\kappa^-_{\eta}{\rm w}}}
{\cal Z}_{\tilde{\tiny D}^{3}}^{2n}{\cal Z}_{\tilde{\tiny D}^{1}}^{2m}
\frac{h_{d}{\cal Z}_{\tilde{U}^{i}}^{1\alpha}{\cal Z}_{-}^{2\lambda}}
{\sqrt{2}}\bigg(
-{\cal Z}_{\tilde{\tiny D}^{3}}^{1m} {\cal Z}_{-}^{1\lambda}+
\frac{h_{b}{\cal Z}_{\tilde{\tiny D}^{3}}^{2m}{\cal Z}_{-}^{2\lambda}}
{\sqrt{2}}\bigg)\nonumber \\
&&\hspace{1.2cm}\frac{h_{b}{\cal Z}_{\tilde{U}^{i}}^{1\alpha}
{\cal Z}_{-}^{2\eta}}{\sqrt{2}}\bigg(
-{\cal Z}_{\tilde{\tiny D}^{1}}^{1n} {\cal Z}_{-}^{1\eta}+
\frac{h_{d}{\cal Z}_{\tilde{\tiny D}^{1}}^{2n}{\cal Z}_{-}^{2\eta}}
{\sqrt{2}}\bigg)\nonumber \\
&&\hspace{1.2cm}\Big(F_{A}^{1a}-F_{A}^{1b}-F_{A}^{1c}\Big)
(x_{\kappa^-_{\lambda}{\rm w}}, x_{\tilde{U}^i_{\alpha}{\rm w}},
x_{\kappa^-_{\eta}{\rm w}}, x_{j{\rm w}}, x_{g{\rm w}},
x_{\tilde{D}^3_{\gamma}{\rm w}}, x_{\tilde{D}^1_{\delta}{\rm w}})
\nonumber \\
&&\hspace{1.2cm}-\frac{64}{3}\sqrt{x_{g{\rm w}}x_{i{\rm w}}}\bigg(
{\cal Z}_{\tilde{U}^i}^{2\alpha_1}{\cal Z}_{\tilde{U}^i}^{1\alpha_2}+
{\cal Z}_{\tilde{U}^i}^{1\alpha_1}{\cal Z}_{\tilde{U}^i}^{2\alpha_2}\bigg)
\frac{h_{d}{\cal Z}_{\tilde{U}^i}^{1\alpha_1}{\cal Z}_{-}^{2\lambda}}
{\sqrt{2}}
\frac{h_{b}{\cal Z}_{\tilde{U}^i}^{1\alpha_2}{\cal Z}_{-}^{2\lambda}}
{\sqrt{2}}
\frac{h_{b}{\cal Z}_{\tilde{U}^j}^{1\beta}{\cal Z}_{-}^{2\eta}}
{\sqrt{2}}
\frac{h_{d}{\cal Z}_{\tilde{U}^j}^{1\beta}{\cal Z}_{-}^{2\eta}}
{\sqrt{2}}\nonumber \\
&&\hspace{1.2cm}F_{C}^{1a}
(x_{\tilde{U}^i_{\alpha_2}{\rm w}}, x_{\kappa^-_\lambda {\rm w}},
 x_{\kappa^-_{\eta}{\rm w}}, x_{\tilde{U}^i_{\alpha_1}{\rm w}},
x_{i{\rm w}}, x_{\tilde{U}^i_\alpha {\rm w}}, x_{g{\rm w}}) \nonumber \\
&&\hspace{1.2cm}+\frac{16}{3}\sqrt{x_{\kappa^-_{\eta}{\rm w}}x_{j{\rm w}}}
{\cal Z}_{\tilde{\tiny D}^{3}}^{2n}{\cal Z}_{\tilde{\tiny D}^{1}}^{2m}
\frac{h_{d}{\cal Z}_{\tilde{U}^{i}}^{1\alpha}{\cal Z}_{-}^{2\lambda}}
{\sqrt{2}}
\frac{h_{b}{\cal Z}_{\tilde{U}^{i}}^{1\alpha}{\cal Z}_{-}^{2\eta}}
{\sqrt{2}}\nonumber \\
&&\hspace{1.2cm}\bigg(-{\cal Z}_{\tilde{\tiny D}^{1}}^{1n}
{\cal Z}_{-}^{1\eta}+\frac{h_{d}{\cal Z}_{\tilde{\tiny D}^{1}}^{2n}
{\cal Z}_{-}^{2\eta}}{\sqrt{2}}\bigg)
\frac{m_{u^j}{\cal Z}_{\tilde{\tiny D}^{3}}^{1m}{\cal Z}_{+}^{2\lambda}}
{\sqrt{2}m_{\rm w} \sin\beta}\nonumber \\
&&\hspace{1.2cm}\Big(F_{A}^{1a}+F_{A}^{1b}-F_{A}^{1c}\Big)
(x_{\kappa^-_{\lambda}{\rm w}}, x_{\tilde{U}^i_{\alpha}{\rm w}},
x_{\kappa^-_{\eta}{\rm w}}, x_{j{\rm w}}, x_{g{\rm w}}, x_{\tilde{D}^3_{\gamma}{\rm w}},
x_{\tilde{D}^1_{\delta}{\rm w}})
\nonumber \\
&&\hspace{1.2cm}+\frac{16}{3}\sqrt{x_{\kappa^-_{\lambda}{\rm w}}x_{j{\rm w}}}
{\cal Z}_{\tilde{\tiny D}^{3}}^{2n}{\cal Z}_{\tilde{\tiny D}^{1}}^{2m}
\frac{h_{d}{\cal Z}_{\tilde{U}^{i}}^{1\alpha}{\cal Z}_{-}^{2\lambda}}
{\sqrt{2}}\bigg(-{\cal Z}_{\tilde{\tiny D}^{3}}^{1m}
{\cal Z}_{-}^{1\lambda}+\frac{h_{b}{\cal Z}_{\tilde{\tiny D}^{3}}^{2m}
{\cal Z}_{-}^{2\lambda}}{\sqrt{2}}\bigg)\nonumber \\
&&\hspace{1.2cm}\frac{h_{b}{\cal Z}_{\tilde{U}^{i}}^{1\alpha}
{\cal Z}_{-}^{2\eta}}{\sqrt{2}}
\frac{m_{u^j}{\cal Z}_{\tilde{\tiny D}^{1}}^{1n}{\cal Z}_{+}^{2\eta}}
{\sqrt{2}m_{\rm w} \sin\beta}\Big(F_{A}^{1a}+F_{A}^{1b}\nonumber \\
&&\hspace{1.2cm}-F_{A}^{1c}\Big)
(x_{\kappa^-_{\lambda}{\rm w}}, x_{\tilde{U}^i_{\alpha}{\rm w}},
x_{\kappa^-_{\eta}{\rm w}}, x_{j{\rm w}}, x_{g{\rm w}},
x_{\tilde{D}^3_{\gamma}{\rm w}}, x_{\tilde{D}^1_{\delta}{\rm w}})\nonumber \\
&&\hspace{1.2cm}-\frac{16}{3}{\cal Z}_{\tilde{\tiny D}^{3}}^{2n}
{\cal Z}_{\tilde{\tiny D}^{1}}^{2m}\frac{h_{d}{\cal Z}_{\tilde{
\tiny U}^{I}}^{1j}{\cal Z}_{-}^{2\lambda}}{\sqrt{2}}
\frac{h_{b}{\cal Z}_{\tilde{U}^{i}}^{1\alpha}{\cal Z}_{-}^{2\eta}}
{\sqrt{2}}
\frac{m_{u^j}{\cal Z}_{\tilde{\tiny D}^{3}}^{1m} {\cal Z}_{+}^{2\lambda}}
{\sqrt{2}m_{\rm w} \sin\beta}
\frac{m_{u^j}{\cal Z}_{\tilde{\tiny D}^{1}}^{1n}{\cal Z}_{+}^{2\eta}}
{\sqrt{2}m_{\rm w} \sin\beta}\nonumber \\
&&\hspace{1.2cm}\Big(F_{A}^{2b}+ F_{A}^{2c}-F_{A}^{2d}-F_{A}^{2e}
-2F_{A}^{2f}\Big)
x_{\kappa^-_{\lambda}{\rm w}}, x_{\tilde{U}^i_{\alpha}{\rm w}},
x_{\kappa^-_{\eta}{\rm w}}, x_{j{\rm w}}, x_{g{\rm w}}, x_{\tilde{D}^{3}_{\gamma}{\rm w}},
 x_{\tilde{D}^1_{\delta}{\rm w}})
\nonumber \\
&&\hspace{1.2cm}-32x_{\kappa^-_\lambda {\rm w}}\sqrt{x_{g{\rm w}}x_{i{\rm w}}}
\bigg({\cal Z}_{\tilde{U}^i}^{2\alpha_1}
{\cal Z}_{\tilde{U}^i}^{1\alpha_2}+
{\cal Z}_{\tilde{U}^i}^{1\alpha_1}
{\cal Z}_{\tilde{U}^i}^{2\alpha_2}\bigg)
\frac{h_{d}{\cal Z}_{\tilde{U}^i}^{1\alpha_1}{\cal Z}_{-}^{2\lambda}}
{\sqrt{2}}
\frac{h_{b}{\cal Z}_{\tilde{U}^i}^{1\alpha_2} {\cal Z}_{-}^{2\lambda}}
{\sqrt{2}}
\frac{h_{b}{\cal Z}_{\tilde{U}^j}^{1\beta}{\cal Z}_{-}^{2\eta}}
{\sqrt{2}}
\frac{h_{d}{\cal Z}_{\tilde{U}^j}^{1\beta}{\cal Z}_{-}^{2\eta}}
{\sqrt{2}}\nonumber \\
&&\hspace{1.2cm}\Big(\frac{\ln x_{\kappa^-_\lambda {\rm w}}}{
(-x_{\kappa^-_\lambda {\rm w}} + x_{\kappa^-_{\eta}{\rm w}})(-x_{\kappa^-_\lambda {\rm w}}
+ x_{\tilde{U}^i_{\alpha_2}{\rm w}}) (-x_{\kappa^-_\lambda {\rm w}} +
x_{\tilde{U}^i_{\alpha_1}{\rm w}})}
\nonumber \\
&&\hspace{1.2cm}-\frac{x_{i{\rm w}}\ln x_{\kappa^-_\lambda {\rm w}}}
{(-x_{\kappa^-_\lambda {\rm w}} + x_{\kappa^-_{\eta}{\rm w}})
(-x_{\kappa^-_\lambda {\rm w}} + x_{i{\rm w}})(-x_{\kappa^-_\lambda {\rm w}}
+ x_{\tilde{U}^i_{\alpha_2}{\rm w}})(-x_{\kappa^-_\lambda {\rm w}}
+ x_{\tilde{U}^i_{\alpha_1}{\rm w}})}\Big)\nonumber \\
&&\hspace{1.2cm}-32x_{\kappa^-_{\eta}{\rm w}}\sqrt{x_{g{\rm w}}x_{i{\rm w}}}\bigg(
{\cal Z}_{\tilde{U}^i}^{2\alpha_1}{\cal Z}_{\tilde{U}^i}^{1\alpha_2}+
{\cal Z}_{\tilde{U}^i}^{1\alpha_1}{\cal Z}_{\tilde{U}^i}^{2\alpha_2}
\bigg)
\frac{h_{d}{\cal Z}_{\tilde{U}^i}^{1\alpha_1}{\cal Z}_{-}^{2\lambda}}
{\sqrt{2}}
\frac{h_{b}{\cal Z}_{\tilde{U}^i}^{1\alpha_2}{\cal Z}_{-}^{2\lambda}}
{\sqrt{2}}
\frac{h_{b}{\cal Z}_{\tilde{U}^j}^{1\beta}{\cal Z}_{-}^{2\eta}}
{\sqrt{2}}
\frac{h_{d}{\cal Z}_{\tilde{U}^j}^{1\beta}{\cal Z}_{-}^{2\eta}}
{\sqrt{2}}\nonumber \\
&&\hspace{1.2cm}\Big(\frac{\ln x_{\kappa^-_{\eta}{\rm w}}}{
(x_{\kappa^-_\lambda {\rm w}} - x_{\kappa^-_{\eta}{\rm w}})(-x_{\kappa^-_{\eta}{\rm w}}
+ x_{\tilde{U}^i_{\alpha_2}{\rm w}}) (-x_{\kappa^-_{\eta}{\rm w}}
+ x_{\tilde{U}^i_{\alpha_1}{\rm w}})}\nonumber \\
&&\hspace{1.2cm}-\frac{x_{i{\rm w}}\ln x_{\kappa^-_{\eta}{\rm w}}}
{(x_{\kappa^-_\lambda {\rm w}}-x_{\kappa^-_{\eta}{\rm w}})(-x_{\kappa^-_{\eta}{\rm w}}+
x_{i{\rm w}})(-x_{\kappa^-_{\eta}{\rm w}}+x_{\tilde{U}^i_{\alpha_2}{\rm w}})
(-x_{\kappa^-_{\eta}{\rm w}} + x_{\tilde{U}^i_{\alpha_1}{\rm w}})}\Big)
\nonumber \\
&&\hspace{1.2cm}+32x_{i{\rm w}}^2 \sqrt{x_{g{\rm w}} x_{i{\rm w}}}
\bigg({\cal Z}_{\tilde{U}^i}^{2\alpha_1}
{\cal Z}_{\tilde{U}^i}^{1\alpha_2}+{\cal Z}_{\tilde{U}^i}^{1\alpha_1}
{\cal Z}_{\tilde{U}^i}^{2\alpha_2}\bigg)
\frac{h_{d}{\cal Z}_{\tilde{U}^i}^{1\alpha_1}{\cal Z}_{-}^{2\lambda}}
{\sqrt{2}}
\frac{h_{b}{\cal Z}_{\tilde{U}^i}^{1\alpha_2}{\cal Z}_{-}^{2\lambda}}
{\sqrt{2}}
\frac{h_{b}{\cal Z}_{\tilde{U}^j}^{1\beta}{\cal Z}_{-}^{2\eta}}
{\sqrt{2}}
\frac{h_{d}{\cal Z}_{\tilde{U}^j}^{1\beta}{\cal Z}_{-}^{2\eta}}
{\sqrt{2}}\nonumber \\
&&\hspace{1.2cm}\frac{\ln x_{i{\rm w}}}{(x_{\kappa^-_\lambda {\rm w}} - x_{i{\rm w}})
(x_{\kappa^-_{\eta}{\rm w}} - x_{i{\rm w}})(-x_{i{\rm w}} + x_{\tilde{U}^i_{\alpha_2}{\rm w}})
(-x_{i{\rm w}} + x_{\tilde{U}^i_{\alpha_1}{\rm w}})}\nonumber \\
&&\hspace{1.2cm}-32\sqrt{x_{g{\rm w}}x_{i{\rm w}}}x_{\tilde{U}^i_{\alpha_2}{\rm w}}
\bigg({\cal Z}_{\tilde{U}^i}^{2\alpha_1}
{\cal Z}_{\tilde{U}^i}^{1\alpha_2}+{\cal Z}_{\tilde{U}^i}^{1\alpha_1}
{\cal Z}_{\tilde{U}^i}^{2\alpha_2}\bigg)
\frac{h_{d}{\cal Z}_{\tilde{U}^i}^{1\alpha_1}{\cal Z}_{-}^{2\lambda}}
{\sqrt{2}}
\frac{h_{b}{\cal Z}_{\tilde{U}^i}^{1\alpha_2}{\cal Z}_{-}^{2\lambda}}
{\sqrt{2}}
\frac{h_{b}{\cal Z}_{\tilde{U}^j}^{1\beta}{\cal Z}_{-}^{2\eta}}
{\sqrt{2}}
\frac{h_{d}{\cal Z}_{\tilde{U}^j}^{1\beta}{\cal Z}_{-}^{2\eta}}
{\sqrt{2}}\nonumber \\
&&\hspace{1.2cm}\Big(\frac{\ln x_{\tilde{U}^i_{\alpha_2}{\rm w}}}{
(x_{\kappa^-_\lambda {\rm w}} - x_{\tilde{U}^i_{\alpha_2}{\rm w}})
(x_{\kappa^-_{\eta}{\rm w}} - x_{\tilde{U}^i_{\alpha_2}{\rm w}})
(-x_{\tilde{U}^i_{\alpha_2}{\rm w}} +
x_{\tilde{U}^i_{\alpha_1}{\rm w}})}\nonumber \\
&&\hspace{1.2cm}-\frac{x_{i{\rm w}}\ln x_{\tilde{U}^i_{\alpha_2}{\rm w}}}
{(x_{\kappa^-_\lambda {\rm w}} - x_{\tilde{U}^i_{\alpha_2}{\rm w}})
(x_{\kappa^-_{\eta}{\rm w}} - x_{\tilde{U}^i_{\alpha_2}{\rm w}})
(x_{i{\rm w}} - x_{\tilde{U}^i_{\alpha_2}{\rm w}})
(-x_{\tilde{U}^i_{\alpha_2}{\rm w}}+x_{\tilde{U}^i_{\alpha_1}{\rm w}})}\Big)
\nonumber \\
&&\hspace{1.2cm}-32\sqrt{x_{g{\rm w}}x_{i{\rm w}}}x_{\tilde{U}^i_{\alpha_1}{\rm w}}
\bigg({\cal Z}_{\tilde{U}^i}^{2\alpha_1}
{\cal Z}_{\tilde{U}^i}^{1\alpha_2}+
{\cal Z}_{\tilde{U}^i}^{1\alpha_1}
{\cal Z}_{\tilde{U}^i}^{2\alpha_2}\bigg)
\frac{h_{d}{\cal Z}_{\tilde{U}^i}^{1\alpha_1}{\cal Z}_{-}^{2\lambda}}
{\sqrt{2}}
\frac{h_{b}{\cal Z}_{\tilde{U}^i}^{1\alpha_2}{\cal Z}_{-}^{2\lambda}}
{\sqrt{2}}
\frac{h_{b}{\cal Z}_{\tilde{U}^j}^{1\beta}{\cal Z}_{-}^{2\eta}}
{\sqrt{2}}
\frac{h_{d}{\cal Z}_{\tilde{U}^j}^{1\beta}{\cal Z}_{-}^{2\eta}}
{\sqrt{2}}\nonumber \\
&&\hspace{1.2cm}\Big(\frac{\ln x_{\tilde{U}^i_{\alpha_1}{\rm w}}}
{(x_{\kappa^-_\lambda {\rm w}} - x_{\tilde{U}^i_{\alpha_1}{\rm w}})
(x_{\kappa^-_{\eta}{\rm w}} - x_{\tilde{U}^i_{\alpha_1}{\rm w}})
(x_{\tilde{U}^i_{\alpha_2}{\rm w}} - x_{\tilde{U}^i_{\alpha_1}{\rm w}})}
\nonumber \\
&&\hspace{1.2cm}-\frac{x_{i{\rm w}}\ln x_{\tilde{U}^i_{\alpha_1}{\rm w}}}
{(x_{\kappa^-_\lambda {\rm w}} - x_{\tilde{U}^i_{\alpha_1}{\rm w}})
(x_{\kappa^-_{\eta}{\rm w}} - x_{\tilde{U}^i_{\alpha_1}{\rm w}})
(x_{i{\rm w}} - x_{\tilde{U}^i_{\alpha_1}{\rm w}})
(x_{\tilde{U}^i_{\alpha_2}{\rm w}} - x_{\tilde{U}^i_{\alpha_1}{\rm w}})}\Big)
+(i\leftrightarrow j)\; ,
\label{pcg6}
\end{eqnarray}
\begin{eqnarray}
&&\phi^{p\tilde{g}}_{7}=-\frac{16}{3}\sqrt{x_{\kappa^-_{\eta}{\rm w}}x_{g{\rm w}}}
{\cal Z}_{\tilde{\tiny D}^{3}}^{2n}{\cal Z}_{\tilde{\tiny D}^{1}}^{1m}
\bigg(-{\cal Z}_{\tilde{\tiny D}^{3}}^{1m} {\cal Z}_{-}^{1\lambda}+
\frac{h_{b}{\cal Z}_{\tilde{\tiny D}^{3}}^{2m}{\cal Z}_{-}^{2\lambda}}
{\sqrt{2}}\bigg)
\frac{h_{b}{\cal Z}_{\tilde{U}^{i}}^{1\alpha}{\cal Z}_{-}^{2\eta}}
{\sqrt{2}} \nonumber \\
&&\hspace{1.2cm}
\bigg(-{\cal Z}_{\tilde{\tiny D}^{1}}^{1n}{\cal Z}_{-}^{1\eta}+
\frac{h_{d}{\cal Z}_{\tilde{\tiny D}^{1}}^{2n}{\cal Z}_{-}^{2\eta}}
{\sqrt{2}}\bigg)\bigg(
-{\cal Z}_{\tilde{U}^{i}}^{1\alpha} {\cal Z}_{+}^{1\lambda}+
\frac{m_{u^i}{\cal Z}_{\tilde{U}^{i}}^{2\alpha}{\cal Z}_{+}^{2\lambda}}
{\sqrt{2}m_{\rm w} \sin\beta}\bigg)\nonumber \\
&&\hspace{1.2cm}\Big(F_{A}^{1a}-F_{A}^{1b}+F_{A}^{1c}\Big)
(x_{\kappa^-_{\lambda}{\rm w}}, x_{\tilde{U}^i_{\alpha}{\rm w}},
x_{\kappa^-_{\eta}{\rm w}}, x_{j{\rm w}}, x_{g{\rm w}}, x_{\tilde{
\tiny D}^{1}_{m}{\rm w}}, x_{\tilde{D}^1_{\delta}{\rm w}})\nonumber \\
&&\hspace{1.2cm}-\frac{32}{3}\sqrt{x_{\kappa^-_{\lambda}{\rm w}}
x_{\kappa^-_{\eta}{\rm w}} x_{g{\rm w}} x_{j{\rm w}}}
{\cal Z}_{\tilde{\tiny D}^{3}}^{2n}{\cal Z}_{\tilde{\tiny D}^{1}}^{1m}
\frac{h_{b}{\cal Z}_{\tilde{U}^{i}}^{1\alpha}{\cal Z}_{-}^{2\eta}}
{\sqrt{2}}
\bigg(-{\cal Z}_{\tilde{\tiny D}^{1}}^{1n}{\cal Z}_{-}^{1\eta}+
\frac{h_{d}{\cal Z}_{\tilde{\tiny D}^{1}}^{2n}{\cal Z}_{-}^{2\eta}}
{\sqrt{2}}\bigg) \nonumber \\
&&\hspace{1.2cm}
\frac{m_{u^j}{\cal Z}_{\tilde{\tiny D}^{3}}^{1m}{\cal Z}_{+}^{2\lambda}}
{\sqrt{2}m_{\rm w} \sin\beta}
\bigg(-{\cal Z}_{\tilde{U}^{i}}^{1\alpha}{\cal Z}_{+}^{1\lambda}+
\frac{m_{u^i}{\cal Z}_{\tilde{U}^{i}}^{2\alpha}{\cal Z}_{+}^{2\lambda}}
{\sqrt{2}m_{\rm w} \sin\beta}\bigg)\nonumber \\
&&\hspace{1.2cm}F_{A}^{0}(x_{\kappa^-_{\lambda}{\rm w}},
x_{\tilde{U}^i_{\alpha}{\rm w}}, x_{\kappa^-_{\eta}{\rm w}},
x_{j{\rm w}}, x_{g{\rm w}}, x_{\tilde{D}^3_{\gamma}{\rm w}},
x_{\tilde{D}^1_{\delta}{\rm w}})\nonumber \\
&&\hspace{1.2cm}-\frac{32}{3}\sqrt{x_{g{\rm w}}x_{j{\rm w}}}
{\cal Z}_{\tilde{\tiny D}^{3}}^{2n}{\cal Z}_{\tilde{\tiny D}^{1}}^{1m}
\bigg(-{\cal Z}_{\tilde{\tiny D}^{3}}^{1m}{\cal Z}_{-}^{1\lambda}+
\frac{h_{b}{\cal Z}_{\tilde{\tiny D}^{3}}^{2m}{\cal Z}_{-}^{2\lambda}}
{\sqrt{2}}\bigg)
\frac{h_{b}{\cal Z}_{\tilde{U}^{i}}^{1\alpha} {\cal Z}_{-}^{2\eta}}
{\sqrt{2}}\nonumber \\
&&\hspace{1.2cm}\bigg(-{\cal Z}_{\tilde{U}^{i}}^{1\alpha}
{\cal Z}_{+}^{1\lambda} +\frac{
m_{u^i}{\cal Z}_{\tilde{U}^{i}}^{2\alpha}{\cal Z}_{+}^{2\lambda}}
{\sqrt{2}m_{\rm w} \sin\beta}\bigg)
\frac{m_{u^j}{\cal Z}_{\tilde{\tiny D}^{1}}^{1n}
{\cal Z}_{+}^{2\eta}}{\sqrt{2}m_{\rm w} \sin\beta}\nonumber \\
&&\hspace{1.2cm}F_{A}^{1a}(x_{\kappa^-_{\lambda}{\rm w}},
x_{\tilde{U}^i_{\alpha}{\rm w}}, x_{\kappa^-_{\eta}{\rm w}},
x_{j{\rm w}}, x_{g{\rm w}}, x_{\tilde{D}^3_{\gamma}{\rm w}},
x_{\tilde{D}^1_{\delta}{\rm w}})\nonumber \\
&&\hspace{1.2cm}-\frac{16}{3}\sqrt{x_{\kappa^-_{\lambda}{\rm w}}x_{g{\rm w}}}
{\cal Z}_{\tilde{\tiny D}^{3}}^{2n}{\cal Z}_{\tilde{\tiny D}^{1}}^{1m}
\frac{h_{b}{\cal Z}_{\tilde{U}^{i}}^{1\alpha}{\cal Z}_{-}^{2\eta}}
{\sqrt{2}}
\frac{m_{u^j}{\cal Z}_{\tilde{\tiny D}^{3}}^{1m}{\cal Z}_{+}^{2\lambda}}
{\sqrt{2}m_{\rm w} \sin\beta}\nonumber \\
&&\hspace{1.2cm}\bigg(-{\cal Z}_{\tilde{U}^{i}}^{1\alpha}
{\cal Z}_{+}^{1\lambda})+\frac{m_{u^i}{\cal Z}_{\tilde{U}^{i}}^{2\alpha}
{\cal Z}_{+}^{2\lambda}}{\sqrt{2}m_{\rm w} \sin\beta}\bigg)
\frac{m_{u^j}{\cal Z}_{\tilde{\tiny D}^{1}}^{1n}{\cal Z}_{+}^{2\eta}}
{\sqrt{2}m_{\rm w} \sin\beta}\nonumber \\
&&\hspace{1.2cm}\Big(F_{A}^{1a}-F_{A}^{1b}+F_{A}^{1c}\Big)
(x_{\kappa^-_{\lambda}{\rm w}}, x_{\tilde{U}^i_{\alpha}{\rm w}},
x_{\kappa^-_{\eta}{\rm w}}, x_{j{\rm w}}, x_{g{\rm w}}, x_{\tilde{D}^3_{\gamma}{\rm w}},
x_{\tilde{D}^1_{\delta}{\rm w}})
+(i\leftrightarrow j)\; ,
\label{pcg7}
\end{eqnarray}
\begin{eqnarray}
&&\phi^{p\tilde{g}}_{8}=\frac{1}{4}\phi^{p\tilde{g}}_{7}\; .
\label{pcg8}
\end{eqnarray}

\end{document}